\documentclass[sn-aps]{sn-jnl}

\usepackage{placeins}
\usepackage{braket}
\usepackage{subcaption}
\DeclareMathOperator*{\argmax}{arg\,max}

\newcommand{\plotwidth}{0.32}

\jyear{2022}%

\theoremstyle{thmstyleone}%
\theoremstyle{thmstyletwo}%

\theoremstyle{thmstylethree}%

\begin{document}

\title[Discrepancies between POD and Fourier modes on Aperiodic Domains]{On the Discrepancies between POD and Fourier Modes on Aperiodic Domains}


\author*[1]{\fnm{Azur} \sur{Hod\v zi\'c}}\email{azur.hod@gmail.com}

\author[1]{\fnm{Peder} \sur{J. Olesen}}\email{pjool@mek.dtu.dk}
\equalcont{These authors contributed equally to this work.}

\author[1]{\fnm{Clara} \sur{M. Velte}}\email{cmve@dtu.dk}
\equalcont{These authors contributed equally to this work.}

\affil*[1]{\orgdiv{Department of Civil and Mechanical Engineering}, \orgname{Technical University of Denmark}, \orgaddress{\street{Anker Engelunds Vej 1}, \city{Kgs. Lyngby}, \postcode{2800}, \country{Denmark}}}

\abstract{The application of Fourier analysis in combination with the Proper Orthogonal Decomposition (POD) is investigated. In this approach to turbulence decomposition, which has recently been termed Spectral POD (SPOD), Fourier modes are considered as solutions to the corresponding Fredholm integral equation of the second kind along homogeneous-periodic or homogeneous coordinates. In the present work, the notion that the POD modes formally converge to Fourier modes for increasing domain length is challenged. Numerical results indicate that the discrepancy between POD and Fourier modes along \textit{locally} translationally invariant coordinates is coupled to the Taylor macro/micro scale ratio (MMSR) of the kernel in question. Increasing discrepancies are observed for smaller MMSRs, which are characteristic of low Reynolds number flows. It is observed that the asymptotic convergence rate of the eigenspectrum matches the corresponding convergence rate of the exact analytical Fourier spectrum of the kernel in question - even for extremely small domains and small MMSRs where the corresponding DFT spectra suffer heavily from windowing effects. These results indicate that the accumulated discrepancies between POD and Fourier modes play a role in producing the spectral convergence rates expected from Fourier transforms of translationally invariant kernels on infinite domains.
%
%
}

\keywords{Proper Orthogonal Decomposition, Fourier analysis, SPOD, DFT, Translationally invariant kernels, Windowing, Spectral leakage}



\maketitle

\section{Introduction}\label{sec1}
The Proper Orthogonal Decomposition (POD) has been applied extensively by the turbulence community since its introduction in \cite{lumley1967structure}. The method was introduced with the aim of decomposing turbulent signals into a set of orthogonal basis functions in an energy-optimized way. In the same work a particular set of eigenfunctions was identified in the case of translationally invariant kernels, namely the trigonometric polynomials. The work concluded that any kernel that exhibits translational invariance (such as in the cases of stationarity and homogeneity) could be decomposed using a Fourier transform along those directions and combined with the POD along any other coordinate direction along which the flow is inhomogeneous. 

The role of homogeneous and/or stationary flows has been central in the construction of fundamental arguments behind the application of the Fourier-based POD on finite \textit{aperiodic} domains. The assumption and utilization of translational invariance has been advocated in numerous works since its introduction by \cite{lumley1967structure}. The Fourier-based implementation of the POD was spearheaded by the work of \cite{Glauser1987}, and popularly advocated by \cite{George1988}. The seminal work of \cite{Citriniti2000} implemented the Fourier-based decomposition on experimental jet data acquired using a rake consisting of 138 hot-wires. This work was extended by \cite{Jung2004, Gamard2004} to other regions of the jet using the same hot-wire rake. A multi-component implementation of the Fourier-based POD was performed by \cite{Iqbal2007} and later by \cite{Tinney2008a} using stereoscopic PIV measurements. The Fourier-based POD became popularly known as spectral POD (SPOD) - to be distinguished from the method of \cite{Sieber2016} (see also \cite{Noack2016} in this relation) that bears the same name - as a result of the works of \cite{Towne2018} and \cite{Schmidt2017}, in which a four-dimensional space-frequency implementation of the method was applied to the analysis of a Large Eddy Simulation (LES) jet. Its application was followed by \cite{MuralidharJFM2019} in their analysis of a turbulent channel flow, where a Fourier-based decomposition was applied in the lateral coordinate as well as in time. The Fourier decomposition along \textit{locally} translation invariant directions (that is, for kernels that are locally stationary/homogeneous, defined on a finite domain without the periodic boundary condition) combined with a numerical POD has proved to serve a multitude of purposes. Firstly, it has provided enhanced insight into the dynamics of turbulent flows, due to their semi-analytical form. Secondly, the fact that the decomposition is frequency-based provides additional insight into the modal structure of the turbulent flow at hand. Thirdly, the use of an analytical set of orthogonal basis functions along a given coordinate direction allows for a reduction of the memory load of the problem when a numerical implementation of the POD is performed, since a separate POD analysis is performed on a set of cross-correlation matrices for each of the corresponding Fourier coefficients. 

The use of Fourier modes in combination with the POD along aperiodic coordinates is traditionally justified in literature by refering to the works \cite{lumley1967structure} and \cite{Lumley1970}. A much overlooked warning, however, appears in the latter work directed towards the use of Fourier modes on non-stationary\footnote{\textit{Stationarity} here refers strictly to problems characterized by translation invariant kernels defined on the entire real line representing infinite energy, \cite{Lumley1970}} flows, herein those characterized by \textit{locally} translation invariant kernels\footnote{Last paragraph of Section 3.12 in \cite{Lumley1970}}:\begin{quote}
Although Fourier transforms may, of course, be used wherever they exist, this theorem serves as a justification for their use in connection with stationary functions, and as a warning against their use in connection with nonstationary ensembles (in the sense that they have no special appropriateness for such ensembles).
\end{quote} In addition, the work of \cite{Lumley1970} does not explicitly elaborate on the direct use of Fourier modes in combination with the POD for stationary flows, beyond providing analogies between general characteristics of the POD and the Fourier transform of kernels defined on the real line. A somewhat rare re-addressal of this subject is found in \cite{George1999}, where the consequences of equating homogeneous fields with periodic ones in relation to the POD were emphasized by noting that periodicity fixes the phase relations of all scales and affecting mostly the representation of the largest scales. 
%
%
%
In \cite{Holmes2012}, a related problem was only briefly discussed where it was stated that difficulties may arise when the POD was applied to time dependent problems on infinite temporal domains\footnote{Third paragraph on p. 74 in \cite{Holmes2012}}. This comment was, however, not put into context with the use of say SPOD. Important and impactful works have been published on the relationships between the established methods of the POD, Dynamic Mode Decomposition (DMD), SPOD, and Resolvent Analysis, \cite{Chen2012,Towne2018} where clear connections between these methods can be established in the case of periodic domains. Literature on the impact of kernel characteristics on the relations between the POD and Fourier modes along locally translationally invariant \textit{aperiodic} coordinates is, however, significantly more scarce. This is presumably due to the more complex nature of relating the Fourier transform to the POD and the historical use of Fourier analysis in relation to homogeneous/stationary turbulence, going back to the works of \cite{Kolmogorov1941} and \cite{Batchelor1953} (see Appendix \ref{app:infinite_domains}). A limited study on this topic was performed by \cite{Chambers1988} investigating Reynolds number similarity across POD solutions. A comparison of the POD and Fourier spectra obtained from a quasihomogeneous region of a solution to the Burgers' equation resulted in similarities being identified not only between Fourier and POD spectra for a subregion of the eigenvalue/power spectra but also between the modes themselves. 

%

Although complex exponentials may appear to be eigenfunctions to the POD integral equation in the case of translationally invariant kernels over the entire real line, such functions do not satisfy the fundamental requirement of square integrability over the real line, upon which the POD integral equation is conditioned. Complex exponentials are only square integrable over finite domains, but homogeneity precludes the area of integration being finite, \cite{lumley1967structure}. Put in different terms, the inability of the POD to deal with homogeneous fields is due to the fact that the respective kernels are not nuclear, \cite{Towne2018}. Although these paradoxes formally exclude the POD modes from being Fourier modes in the case of translationally invariant kernels defined on the entire real line, they are traditionally ignored and the use of a Fourier basis along finite aperiodic locally translationally invariant coordinate directions is common. Nevertheless, incorporating domain finiteness without assuming periodicity would yield a more parsimonious model for such kernels; this is one approach adopted in the present work. 

Formally equating POD modes to Fourier modes in these cases blurs the relation between the two sets of functions. The ambiguity of their connection is further exemplified by statements in literature often claiming that 1) homogeneity leads to a Fourier basis in relation to the POD, and 2) as a consequence of this the Fourier modes form an optimal basis for infinite homogeneous/stationary turbulent flows. Fourier modes do, in many cases, appear highly optimized in terms of representing energetic structures along locally translationally invariant coordinates, and can be justified based on this to serve as reasonable approximations to POD modes. Nevertheless, given that they differentiate themselves from POD modes by not qualifying as solutions to the POD integral equation (emerging from the underlying optimization problem) for any case other than the periodic one, the focus of the current work is on quantifying their discrepancies, also in terms of the effect on other flow properties than the energetics - as these properties may be significant for transport processes expressed in terms of Galerkin projections of governing equations of fluid flow.

A specific aim of the current work is to characterize the spectral discrepancies between POD and Fourier modes on locally translationally invariant kernels on finite aperiodic domains as a function of kernel characteristics and domain size. We analyze the relation beteen POD modes and Fourier modes for increasing domain sizes - the latter being a commonly used strategy to reduce the effect of "windowing" and "spectral leakage",  \cite{Glauser1992}. In this capacity, we examine some consequences that the use of a Fourier-based decomposition on aperiodic domains may have on the spectral analysis of POD kernels and relate these discrepancies to the macro/micro scale ratio (MMSR).

The results presented in this work may be relevant not only to spectral convergence considerations, but also to the search for analytical solutions to the POD integral where kernels exhibit symmetries other than translational invariance. \footnote{A recent demonstration of an analytical solution to the POD integral in the case of a Gaussian kernel is found on p.\ 22 in \cite{george2017}. The eigenfunction, in this case, was shown to be Gaussian as well. Although purely theoretical, this result is highly significant as it not only demonstrates the existance of an aditional analytical solution to the POD integral, but also exemplifies that POD eigenfunctions do not necessarily need to have compact support.} Subtleties in the choice of domain which may disqualify an analytical candidate solution are therefore discussed. The origins of the extension of SPOD to flows with a symmetry weaker than homogeneity are exemplified by the work of \cite{Ewing2007} where similarity analysis is used to argue for Fourier-based POD solutions to the jet far-field. This was later implemented by \cite{Wanstrom2009,Hodzic2018b,Kuhn2021}, in which SPOD was applied along the streamwise direction of the flow, despite the fact that the flow was not homogeneous along that coordinate. The current investigation is therefore a step in evaluating how far one can extend the SPOD, by considering the most fundamental example first, namely one with one-dimensional translationally invariant kernels.

The paper is structured as follows: in Section \ref{sec:POD} the fundamentals of the POD are defined on the space $L_w^2(\Omega,\mathbb{C}^n)$. In Section \ref{sec:relations_between_Fourier_and_eigenspectra}, a relation between the Fourier and eigenspectra is provided by a Fourier expansion of the POD modes which will be used for a numerical analysis of the coupling between the Fourier and eigenspectrum. The numerical analysis of the spectral properties of two sets of correlation functions is presented in Section \ref{sec:numerical_analysis} where the discrepancies between the POD and Fourier spectra are analyzed. 

\section{Proper Orthogonal Decomposition}\label{sec:POD} 
It is imperative to properly define the vector space in which the candidate basis functions obtained by the POD integral reside. This step is crucial since this vector space defines both the domain and range of the integral operator in the POD integral eigenvalue problem. For the sake of generality we consider here the \textit{weighted} vector space of complex-valued integrable functions defined as
\begin{equation}
L_w^2\left(\Omega,\mathbb{C}^n\right):=\left\{\varphi:\Omega\to\mathbb{C}^n\left\lvert \int_\Omega \vert\varphi(x)\vert^2 w(x)dx<\infty\right.\hspace{0.2cm},\hspace{0.2cm} w(x)> 0\right\}\,,\label{eq:L2_w}
\end{equation}
and the weighted inner product
\begin{equation}
\left(\cdot,\cdot\right)_w:L_w^2\left(\Omega,\mathbb{C}^n\right)\times L_w^2\left(\Omega,\mathbb{C}^n\right)\to\mathbb{C}\,,
\end{equation}
which is antilinear in the second argument
\begin{equation}
\left(\varphi,\psi\right)_w = \int_\Omega \varphi(x) \psi^*(x) w(x)dx\,.\label{eq:inner_product}
\end{equation}
Equipped with the inner product induced norm
\begin{equation}
\left\|\varphi\right\|_w = \sqrt{\left(\varphi,\varphi\right)_w}\,,\label{eq:norm}
\end{equation}
$L_w^2(\Omega,\mathbb{C}^n)$ is a Hilbert space. The following maximization problem is then considered
\begin{equation}
\argmax_{\varphi\in L_w^2(\Omega,\mathbb{C}^n)}\frac{\left\langle \left\{\left\lvert \left(u_n,\varphi\right)_w\right\rvert^2\right\}_{n=1}^N \right\rangle}{\left\|\varphi\right\|_w^2}\,,\label{eq:maximization}
\end{equation}
where the angled brackets designate ensemble averaging (see definition in \eqref{eq:ensemble_averaging} below). This reduces to the following integral eigenvalue problem by means of the calculus of variations
\begin{equation}
\int_\Omega H(x,y)\varphi(y)w(y)dy=\lambda\varphi(x)\,,\quad x\in\Omega\,,\label{eq:Lumley_Decomposition}
\end{equation}
where the following estimator for the autocorrelation function used
\begin{equation}
H(x,y) = \left\langle \left\{u_n(x)u_n(y)\right\}_{n=1}^N\right\rangle=\frac{1}{N}\sum_{n=1}^Nu_n(x)u_n(y)\,,\label{eq:ensemble_averaging}
\end{equation}
and $N$ designates the total number of samples. The formulation \eqref{eq:Lumley_Decomposition} can be considered as the eigenvalue problem of the operator $R:L^2_w\left(\Omega,\mathbb{C}^n\right)\to L^2_w\left(\Omega,\mathbb{C}^n\right)$
\begin{equation}
R\varphi = \lambda\varphi\,.
\end{equation}
In the following, the relation between Fourier and eigenspectra is formulated. The coupling between the aforemention spectra is formulated directly as a function of the POD operator, without the requirement of explicit information about the instantaneous realizations underlying the generation of the POD kernel.

\section{Coupling of Fourier and eigenspectra}\label{sec:relations_between_Fourier_and_eigenspectra}
The general arguments behind the use of SPOD on aperiodic domains, e.g. in its application to truncated aperiodic stationary turbulent signals, are structured around a reduction of windowing effects. One proxy typically used for identifying whether a sufficiently long measurement domain has been achieved is the convergence of the Fourier energy spectrum - the underlying idea being that the Fourier spectrum would converge (by some measure) as the window is continually increased. Given that the Fourier modes defined on the real line do not constitute a basis for a $L ^2(\mathbb{R},\mathbb{C})$, we ask the obvious question: to what extent, if any, do we see a convergence between the POD and Fourier modes on a finite domain as the domain is increased? The effects of integration intervals on the spectral properties of operators are central to consider (see Appendix \ref{app:domain_dependence_on_solutions}). Trigonometric polynomials satisfy the POD eigenvalue problem only in the case of translationally invariant kernels on periodic domains. Since the eigenfunctions are required to be elements in a Hilbert space they cannot be solutions to the POD eigenvalue problem if their domain is chosen to be the entire real line. However, filtering the kernel by introducing a weight/window function into the inner product definition breaks the translational invariance of the kernel as well as the orthogonality of the Fourier modes with respect to that inner product weight, disqualifying the latter from being a complete basis for the pre-filtered field. Aspects of this problem have been discussed in the past by \cite{Buchhave2014,george2017,Schmidt2020b} and others who analyzed the windowing effects on eigenspectra in the case of homogeneous turbulence. The windowing effect is related to the so-called \textit{spectral leakage} where spectral energy is redistributed from lower wavenumbers to higher ones as a result of a reduction of the domain. 

In the following, the deviations between the POD eigenfunctions and Fourier modes are investigated by expanding the eigenfunctions with a Fourier basis, and then expanding the eigenspectrum using the latter. The analysis will be performed for several POD eigenvalue problems across various combinations of kernels and domain lengths where the aim is to quantify the windowing effects, and to analyze the nature of the convergence between the two sets of basis functions. In the numerical study that follows, analytic kernels will be used for the generation of the correlation matrix, allowing us to inspect the effects of kernel characteristics and domain length in the comparison of the two sets of modes.
\subsection{Fourier expansion of POD eigenfunctions and eigenvalues}
The numerical analysis is performed in the Hilbert space $\mathbb{C}^{N}$, with the inner product, $(\cdot,\cdot): \mathbb{C}^{N}\times \mathbb{C}^{N}\rightarrow \mathbb{C}$, defined as the complex canonical inner product
\begin{equation}
(\varphi,\psi) = \sum_{i=1}^N \varphi_i\psi_i^*\,,
\end{equation}
and norm
\begin{equation}
\left\|\varphi\right\| = \sqrt{(\varphi,\varphi)}\,.
\end{equation}
Here the short notation, $\varphi_i = \varphi(x_i)$ and $\psi_i = \psi(x_i)$ is implied. The choice of vector space, $\mathbb{C}^N$, is not arbitrary. For the numerical analysis one might consider imposing a vector space that would result in an integral-based POD, where a quadrature rule would need to be imposed in the definition of the inner product. Any such choice of vector space, however, would imply that the span of the POD modes would be larger than the Fourier modes when the Discrete Fourier Transform (DFT) is implemented, due to the implied condition of periodicity of the domain when using the DFT. To enable a one-to-one comparison of Fourier and POD modes, we restrict the numerical analysis to the vector space $\mathbb{C}^N$ where the number of discrete wavenumbers/frequencies is the same as the number of spatial/temporal grid points and POD modes. This allows us to consider an expansion of the POD modes using Fourier modes, which is useful in determining deviation between the two sets.

Let $\left\{\psi^n\right\}_{n=1}^N$ be a Fourier orthonormal basis for $\mathbb{C}^N$. If $\left\{\varphi^\alpha\right\}_{\alpha=1}^N$ is a POD basis related to the operator $\textbf{R}\in\mathbb{C}^{N\times N}$, then $\mathrm{span}\left\{\varphi^\alpha\right\}_{\alpha=1}^N\subseteq \mathbb{C}^N$. It is therefore possible to expand each member of the POD eigenvectors with the Fourier series basis so each $\varphi^\alpha$ can be written as
\begin{equation}
\varphi^\alpha = \sum_{n=1}^Nc^{\alpha, n}\psi^n\,,\label{eq:psi_expansion_of_phi}
\end{equation}
where 
\begin{equation}
c^{\alpha, n} = (\varphi^\alpha,\psi^n)\,,\quad \alpha,n\in[1:N]\,,
\end{equation}
and $\vert c^{\alpha,n}\vert^2$ represents the Fourier spectrum of the POD mode $\varphi^\alpha$. Note the difference between the formulation in \eqref{eq:psi_expansion_of_phi} and the implied stament in SPOD along the locally translationally invariant coordinate is that $\varphi^n= \psi^m$ for some $m$. In \eqref{eq:psi_expansion_of_phi}, we are allowing each POD mode to consists of multiple Fourier mode components, unlike the case in SPOD where it is implied from the outset that each POD mode corresponds exactly to a single Fourier mode. 

Substituting \eqref{eq:psi_expansion_of_phi} into the corresponding POD eigenvalue problem, the following expansion of the eigenvalues using the Fourier basis is obtained
\begin{equation}
\lambda^\alpha =\sum_{n=1}^N\sum_{m=1}^N c^{\alpha,m}c^{\alpha,n*}\left(\textbf{R}\psi^m,\psi^n\right)\,,\quad\alpha\in[1:N]\,,\label{eq:psi_expansion_of_lambda}
\end{equation}
where we designate
\begin{equation}
H^{\alpha m n } = c^{\alpha,m}c^{\alpha,n*}\left(\textbf{R}\psi^m,\psi^n\right)\,. \label{eq:H}
\end{equation}
Since $\textbf{R}$ is Hermitian, $H^{\alpha m n } = H^{\alpha n m*}$. From \eqref{eq:psi_expansion_of_lambda} we see that if $\psi^m=\varphi^\alpha$ for $\alpha = m$, the right-hand side of \eqref{eq:psi_expansion_of_lambda} would produce a single non-zero term corresponding to the eigenvalue $\lambda^\alpha$. If, on the other hand, $\psi^m\neq\varphi^\alpha$ for $\alpha=m$, multiple terms on the right-hand side of \eqref{eq:psi_expansion_of_lambda} would in general be needed to reconstruct each $\lambda^\alpha$. In this way, the convergence rate of each $\lambda^{\alpha}$ with respect to an increasing $N$ is a measure of the efficiency of the Fourier modes in reconstructing the eigenspectrum and is therefore a proxy for the energy optimality of the Fourier basis compared to the POD basis. 

From \eqref{eq:psi_expansion_of_lambda} it is seen that the contributions to the reconstruction of $\lambda^\alpha$ consists of products of the factors $c^{\alpha,m}c^{\alpha,n*}$ and $\left(\textbf{R}\psi^m,\psi^n\right)$. For $n=m$ the first factor reduces to $\vert c^{\alpha,m}\vert^2$, corresponding to the Fourier energy spectrum for the mode $\varphi^\alpha$. For $n\neq m$ the first factor contributes only if a given $\varphi^\alpha$ is non-orthogonal to both the $m$-th and $n$-th harmonic. The factor $\left(\textbf{R}\psi^m,\psi^n\right)$ may be understood by first considering the expression ${\left(\textbf{R}\varphi^\alpha,\varphi^\beta\right) = \lambda^\beta}$. That is, when $\textbf{R}$ is applied to its eigenfunction, the operation corresponds to a scaling of that function. However, when $\textbf{R}$ is applied to the $n$-th Fourier harmonic, $\psi^n$, it imposes a rotation on that function in addition to a scaling. For this reason the projection of $\textbf{R}\psi^m$ on $\psi^n$ is generally not zero for $n\neq m$ and thereby produces a non-zero contribution in \eqref{eq:psi_expansion_of_lambda}. 

Given that each eigenvalue can be formulated as
\begin{equation}
\lambda^\alpha = \left(\textbf{R}\varphi^\alpha,\varphi^\alpha\right)\,,\quad\alpha\in[1:N]\,,\label{eq:lambda_2}
\end{equation}
the relation between the eigenspectrum and the Fourier spectrum obtained from the locally translationally invariant kernel can be obtained from \eqref{eq:lambda_2} by replacing the eigenfunctions by the Fourier basis subject to the assumption of periodicity of the domain. This naturally means that the right hand-side in \eqref{eq:lambda_2} no longer represents the eigenspectrum, but yields the Fourier energy spectrum, $\sigma^m$, and takes the form
\begin{equation}
\sigma^m=\left(\textbf{R}\psi^m,\psi^m\right)\,,\quad m\in[1:N]\,.
\label{eq:fourier_spectrum}
\end{equation}
This formulation allows a comparison of the Fourier spectrum of the translationally invariant kernel on a periodic domain and the eigenspectrum of the corresponding locally translationally invariant kernel in \eqref{eq:psi_expansion_of_lambda}. It is seen that \eqref{eq:fourier_spectrum} appears in \eqref{eq:psi_expansion_of_lambda} as a factor in the terms where $m=n$. Using \eqref{eq:fourier_spectrum} we can rewrite \eqref{eq:psi_expansion_of_lambda} as
\begin{equation}
\lambda^\alpha =\sum_{m=1}^N \vert c^{\alpha,m}\vert^2\sigma^m + \sum_{n\neq m} c^{\alpha,m}c^{\alpha,n*}\left(\textbf{R}\psi^m,\psi^n\right)\,,\quad\alpha\in[1:N]\,.\label{eq:psi_expansion_of_lambda_fourier}
\end{equation}
The components of the second sum in \eqref{eq:psi_expansion_of_lambda_fourier} correspond to $H^{\alpha m n}$ for $m\neq n$, which are the contributions related to the Fourier cross terms. Since 
\begin{equation}
\sum_{\alpha=1}^N \lambda^\alpha = \sum_{m=1}^N \sigma^m\,,\label{eq:sum_of_sigma_lambda}
\end{equation}
the relation between the Fourier spectrum and eigenvalues in \eqref{eq:psi_expansion_of_lambda_fourier} represents a redistribution of the same energy related to the kernel in question. 

From \eqref{eq:psi_expansion_of_lambda_fourier}, we see that the mapping of the Fourier spectrum to the eigenspectrum is non-linear. Secondly, in order for this mapping to be invertible, the determinant of the matrix corresponding to $\vert c^{\alpha,m}\vert^2$ must be non-zero. Due to the symmetry of the Fourier spectrum of the eigenfunctions represented by $\vert c^{\alpha,m}\vert ^2$, the corresponding matrix can in fact be shown to be singular - which means that the mapping is not invertible. This implies that the Fourier spectrum cannot be obtained from the eigenspectrum due to the assumption of periodicity implicit in the former. 

Element $(i,j)$ of $\textbf{R}$ may be reconstructed using the POD basis by
\begin{equation}
R_{i,j} = \sum_{\alpha=1}^N\lambda^\alpha\varphi^\alpha_i\varphi^{\alpha*}_j\,,\quad i,j\in[1:N]\,,\label{eq:R_recon_POD}
\end{equation}
and combining the above results the reconstruction may be performed using the Fourier basis by
\begin{equation}
R_{i,j} = \sum_{\alpha,m,n,p,q}H^{\alpha mn}c^{\alpha,p}c^{\alpha,q*}\psi^p_i\psi^{q*}_j\,,\quad i,j\in[1:N]\,.\label{eq:R_recon_fourier}
\end{equation}
For translationally invariant correlation functions on periodic domains we naturally have that $\varphi^\alpha=\psi^\alpha$, for all $\alpha$. The resulting correlation function is given by the expression analogous to \eqref{eq:R_recon_POD}
\begin{equation}
R_{\sigma,i,j} = \sum_{\alpha=1}^N\sigma^\alpha\psi^\alpha_i\psi^{\alpha*}_j\,,\quad i,j\in[1:N]\,.\label{eq:R_recon_filtered}
\end{equation}
where $R_{\sigma,i,j}=R_{i,j}$. For locally translationally invariant correlation functions, however, we have that $R_{\sigma,i,j}\neq R_{i,j}$. The assertion, $R_{\sigma,i,j}= R_{i,j}$, in these cases enforces a \textit{periodificiation} of the original correlation function generating a kernel for which the discrete Fourier series are the exact eigenvectors. In the current work, the subscripted symbol $\sigma$ following a second order statistic indicates that the latter was generated using Fourier modes by the suppresion of cross terms for locally translationally invariant correlation functions - analogous to the step from \eqref{eq:R_recon_fourier} to \eqref{eq:R_recon_filtered}. The comparison between $R_{i,j}$ and $R_{\sigma,i,j}$ as a function of MMSR for locally translationally invariant correlation functions is treated in Section \ref{sec:Taylor_micro_scale_reconstruction}. In the following, the deviation between the Fourier and eigenspectra spectra will be analysed for two sets of correlation functions.
\section{Numerical analysis}\label{sec:numerical_analysis}
In the comparison between POD and SPOD results, the following numerical analyses are limited to spectral analyses of discretized versions of two sets of analytical correlation functions. The correlation functions are chosen in order to investigate the spectral responses to modifications of specific correlation function characteristics often used to characterize turbulent flows, namely the Taylor macro and micro scales. More specifically, we investigate the relation between the MMSR and the observed differences in spectral convergence rates between the Fourier and eigenspectra, where a large MMSR is generally expected for high Reynolds number flows. 

In Section \ref{sec:bessel} the spectral responses to variations of the MMSR are investigated for a family of correlation functions constructed from an inverse Fourier transform of a set of analytical Fourier spectra, characterized by asymptotic power law decay rates. The numerical analysis is then extended to a new set of arbitrarily chosen correlation functions in Section \ref{sec:other_correlation_functions}, in order to investigate whether the correlation between the MMSR and the spectral discrepancies between Fourier and eigenspectra can be expected to hold more generally. The contributions of the Fourier modes in the reconstruction of the eigenspectrum are then analyzed in Section \ref{sec:fourier_reconstruction_of_Eigenspectra}, in order to map the effects of window size on the relation between the discrete Fourier and eigenspectra. Finally, the impact of assuming the POD basis to be a Fourier basis on the estimation of the Taylor micro scale is analyzed in Section \ref{sec:Taylor_micro_scale_reconstruction}. It demonstrates how this assumption impacts the representation of the smallest turbulent scales, compared to the corresponding characteristics of the POD modes, and also serves as a quantification of the spectral discrepancies observed in sections \ref{sec:bessel} and \ref{sec:other_correlation_functions}.
\subsection{Spectral responses to Taylor micro scale\label{sec:bessel}}
\begin{figure}[b]
\centering
\begin{subfigure}[h]{0.49\textwidth}
\includegraphics[width=\textwidth]{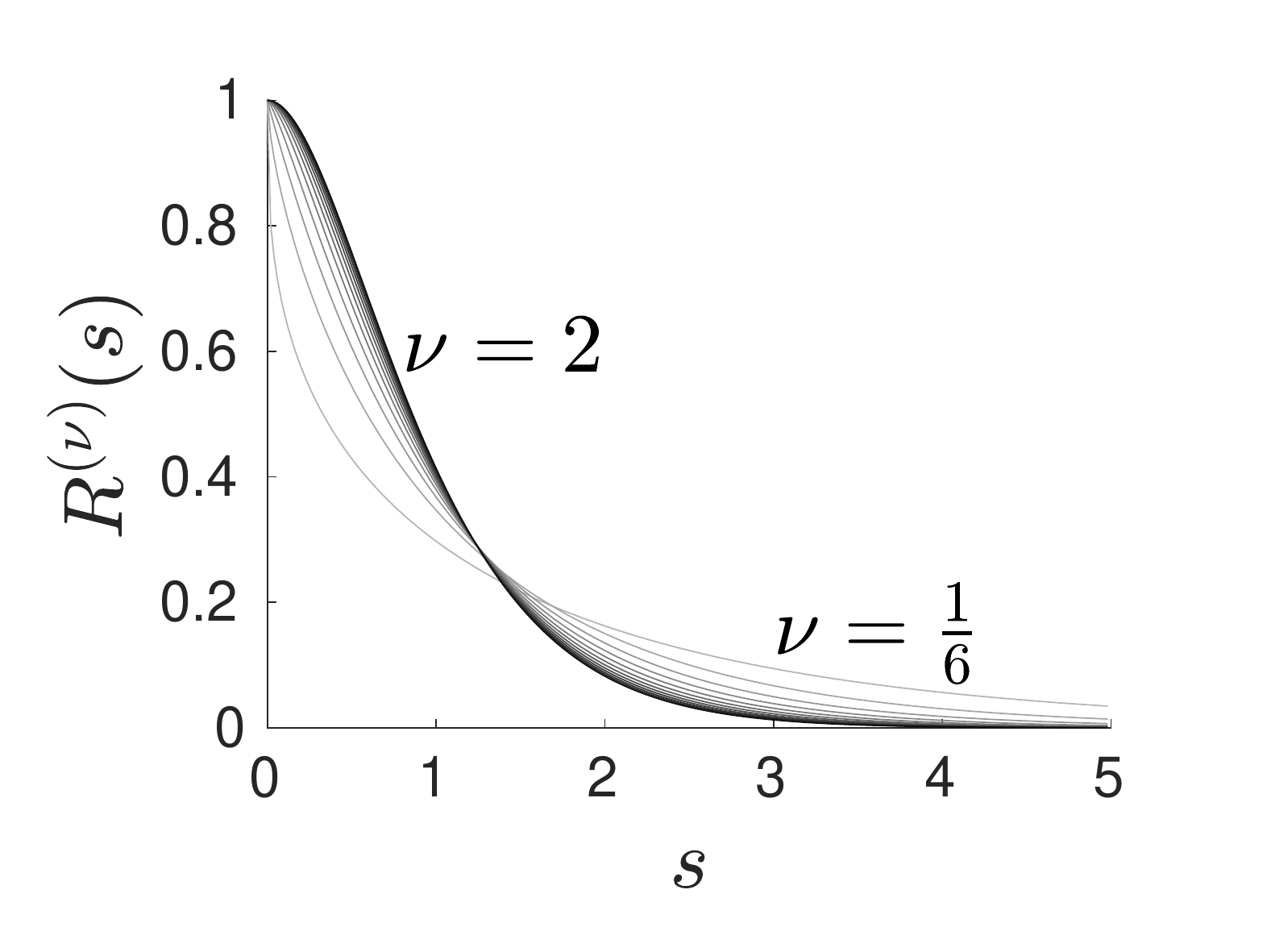}
\caption{\label{fig:correlation_function_bessel_1}}
\end{subfigure}
\begin{subfigure}[h]{0.49\textwidth}
\includegraphics[width=\textwidth]{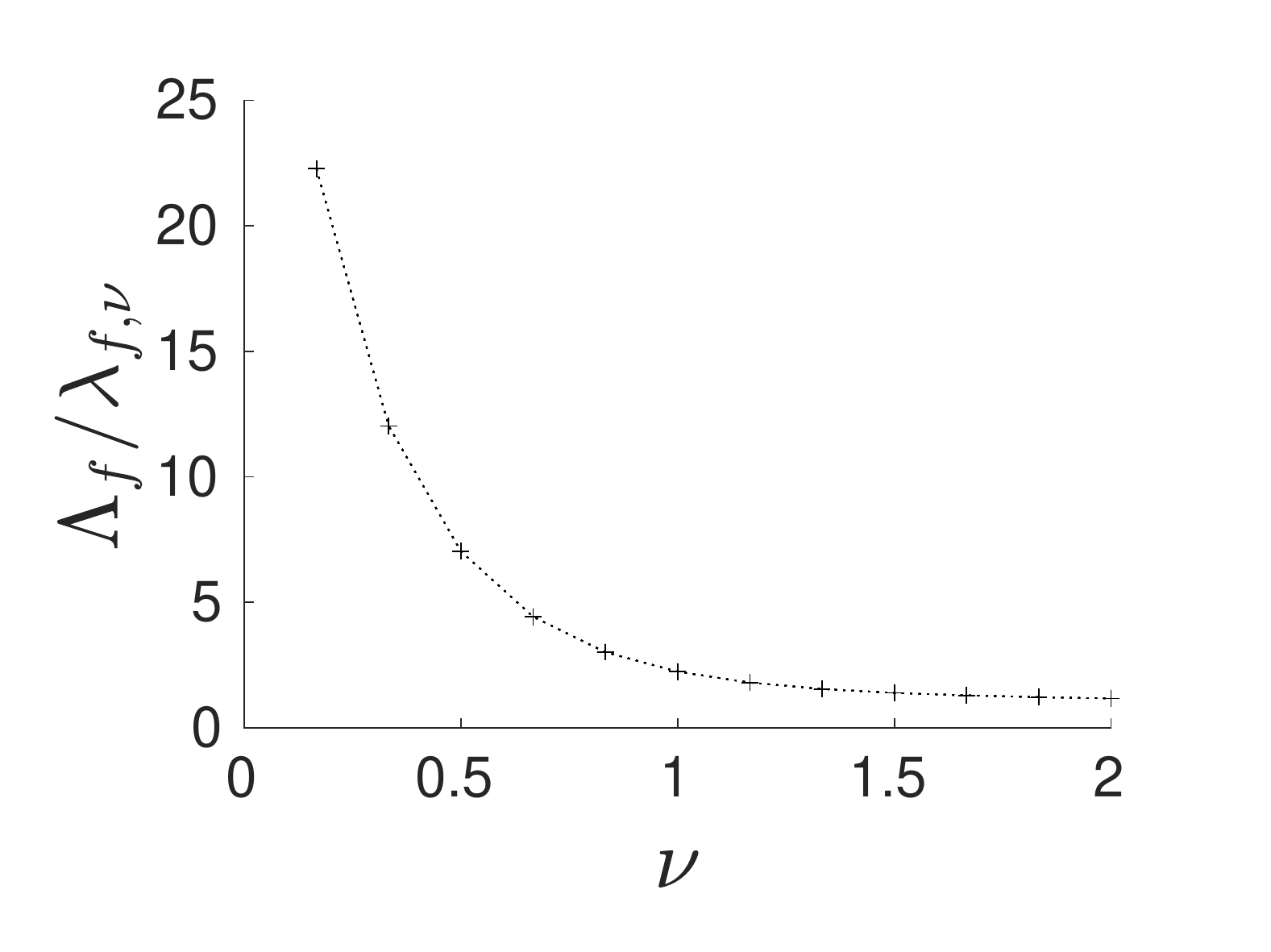}
\caption{\label{fig:micro_macro_ratio}}
\end{subfigure}
\caption{(a): Illustration of $R^{\left(\nu\right)}(s)$, \eqref{eq:R_bessel}, for $\nu=p/6$, $p\in[1:12]$. An increase in $\nu$ corresponds to an increase in the Taylor micro scale while the integral scale remains constant (Table \ref{tab:micro_scale_bessel}), (b): Taylor macro/micro scale ratio as a function of $\nu$.\label{fig:correlation_function_bessel}}
\end{figure}

In the present analysis, we examine the Fourier and eigenspectral responses to the change in the Taylor macro/micro scale obtained from a specific parametrizable family of correlation functions. Here the micro scale is varied independently from the macro scale in order to systematize the spectral response analysis. The analytical kernels chosen for this purpose are given by \cite{Pope2000}
\begin{subequations}
\begin{eqnarray}
R^{\left(\nu\right)}(s) &=& \frac{2}{\Gamma(\nu)}\left(\frac{1}{2}s\right)^\nu K_\nu\left(s\beta\right)\,,\label{eq:R_bessel}\\
\beta &=& \frac{\sqrt{\pi}}{\Gamma\left(\nu\right)}\Gamma\left(\nu+\frac{1}{2}\right)\,,
\end{eqnarray}
\end{subequations}
where $s=\vert x-y\vert$ . $\Gamma$ and $K_\nu$ are the gamma function and the modified Bessel function of the second kind, respectively, defined as \cite{Abramowitz1970}
\begin{subequations}
\begin{eqnarray}
\Gamma\left(\nu\right) &=& \int_0^\infty x^{\nu-1}e^{-x}dx\,,\\
K_\nu(s\beta) &=&  \frac{\pi}{2}\frac{I_{-\nu}(s\beta)-I_\nu(s\beta)}{\sin\left(\nu\pi\right)}\,,
\end{eqnarray}
\end{subequations}
where
\begin{align}
	I_\nu\left(s\right) &= \left\{\begin{array}{l}
		e^{-\frac{1}{2}\nu\pi i}J_\nu\left(se^{\frac{1}{2}\pi i}\right)\,,\quad -\pi<\arg s\leq \frac{\pi}{2}\,, \\ e^{\frac{3}{2}\nu\pi i}J_\nu\left(se^{-\frac{3}{2}\pi i}\right)\,, \quad \frac{\pi}{2}<\arg s\leq \pi\,,
	\end{array}\right.
\end{align}
and $J_\nu$ is the Bessel function of the first kind.  In the current case, we are considering the functions characterized by $\nu=p/6$, $p\in[1:12]$. The analytical Fourier transform of \eqref{eq:R_bessel} possesses the property, \cite{Pope2000}
\begin{equation}
E_\nu \sim \omega^{-\left(1+2\nu\right)}\,,\label{eq:E_nu_asymp}
\end{equation}
for large values of $\omega$. A characteristic trait of \eqref{eq:R_bessel} is that the Taylor macro (integral) scale evaluates to unity for all $\nu$, i.e.
\begin{equation}
\Lambda_f = \int_0^\infty R^{\left(\nu\right)}(s)ds = 1\,.\label{eq:taylor_macro_scale}
\end{equation}
The discretization of \eqref{eq:R_bessel} is performed using an equidistant grid spacing given by $\Delta s = \Delta x = \Delta y = 1/50$ and the grid points are defined by $x_{n}-x_0 = y_{n}-y_0 = n\Delta x\,,\,n\in[0:N-1]$. The results of the analysis using two domains are included in the analysis that follows: one for which $\,x,y\in[-20\Lambda_f:20\Lambda_f]$, and a second where $\,x,y\in[-5\Lambda_f:5\Lambda_f]$. These results were chosen from a more comprehensive set of analyses, which were performed for domains ranging from $5\Lambda_f$ to $80\Lambda_f$. These showed similar tendencies, and were therefore not included in what is to follow. The Taylor micro scale can generally be defined for some correlation function, $R(x,y)$, as
\begin{equation}
\Pi_f = \sqrt{\frac{-2}{\frac{\partial^2R(x,y)}{\partial x^2}\vert_{y=x}}}\,\,.\label{eq:taylor_micro_scale}
\end{equation}	
Figure \ref{fig:correlation_function_bessel_1} illustrates the family of correlation functions in \eqref{eq:R_bessel} for $\nu=p/6$, $p\in[1:12]$, the Taylor micro scales of which are denoted by
\begin{equation}
\Pi_{f,\nu} = \sqrt{\frac{-2}{\frac{d^2R^{\left(\nu\right)}(s)}{ds^2}\vert_{s=0}}}\,,\label{eq:taylor_micro_scale_nu}
\end{equation}	
and are estimated numerically from a parabolic fit to three points around the discretized version of $R^{\left(\nu\right)}(0)$. The evaluations of $\Pi_{f,\nu}$ are shown in Table \ref{tab:micro_scale_bessel}, where a monotonic increase of $\Pi_{f,\nu}$ is seen to follow from an increase in $\nu$. Since the integral scale evaluates to unity for all $\nu$, it means that the ratio of the Taylor macro scale to the micro scale is simply the reciprocal value of the micro scale, which is seen to range from $22.3$ to $1.2$ (see Table \ref{tab:micro_scale_bessel} and Figure \ref{fig:micro_macro_ratio}) over the range of $\nu$-values considered here. 

The corresponding matrix eigenvalue problem related to \eqref{eq:R_bessel} is formulated discretely by the eigenvalue problem related to the correlation matrix operator $\textbf{R}^{\left(\nu\right)}:\mathbb{C}^N\rightarrow\mathbb{C}^N$ given by
\begin{equation}
\textbf{R}^{\left(\nu\right)} \varphi^{\alpha}_{\nu} = \lambda^{\alpha}_{\nu}\varphi^{\alpha}_{\nu}\,,\quad\alpha \in [1:N]\,.
\end{equation}
This is solved numerically using the MATLAB function \textit{eig}, where the kernels are the discretized correlation functions expresed in matrix form as Toeplitz matrices in order to represent \textit{locally translationally invariant} kernels. The normalized discrete $m$-th Fourier modes, $\psi^{m}\in\mathbb{C}^{N}$, are defined as 
\begin{equation}
\psi^{m} = N^{-\frac{1}{2}}\sum_{n=1}^Ne^{2\pi i (m-1)(n-1)/N}\hat{e}_n\,,\quad m\in[1:N]\,,\label{eq:psi_m}
\end{equation}
where $\hat{e}_n$ represents the $n$-th Cartesian basis vector. The corresponding Fourier spectra can be computed from \eqref{eq:fourier_spectrum} by 
\begin{equation}
\sigma_\nu^m = \left(\textbf{R}^{\left(\nu\right)}\psi^m,\psi^m\right)\,,\quad m\in[1:N]\,,
\end{equation}
enabling a comparison of the Fourier and POD decomposition of a given kernel. The Fourier spectrum of a POD kernel is defined by \eqref{eq:fourier_spectrum} and given that \eqref{eq:sum_of_sigma_lambda} holds, the total energy represented in each spectrum is the same. For practical purposes, however, only normalized versions of the spectra are considered which are denoted by a tilde above the respective variables
\begin{subequations}
\begin{eqnarray}
\tilde{\lambda}^\alpha_\nu &=& \frac{\lambda^\alpha_\nu}{\sum_{\alpha=1}^N\lambda^\alpha_\nu}\,,\label{eq:lambda_tilde}\\
\tilde{\sigma}^m_\nu &=& \frac{\sigma^m_\nu}{\sum_{m=1}^N\sigma^m_\nu}\,.\label{eq:sigma_tilde}
\end{eqnarray}
\end{subequations}
%
%
Unlike the POD eigenspectrum, the Fourier spectrum of a POD kernel, defined by \eqref{eq:sigma_tilde}, is symmetric. For a meaningful comparison of the two sets of spectra, the usual Fourier spectrum representation of only the half-spectrum cannot be used directly, as the number of spectral points related to the two sets of bases would not be the same. 
\begin{table}
\begin{center}
\caption{Taylor micro and macro scales related to $R^{\left(\nu\right)}$ as a function of $\nu$.}\label{tab:micro_scale_bessel}
\centering
\begin{tabular}{lcccccccccccc} 
\toprule%
$\nu$ & $\frac{1}{6}$ & $\frac{1}{3}$ & $\frac{1}{2}$ & $\frac{2}{3}$ & $\frac{5}{6}$ & $1$ & $\frac{7}{6}$ & $\frac{4}{3}$ & $\frac{3}{2}$ & $\frac{5}{3}$ & $\frac{11}{6}$ & $2$\\
\midrule
$\Lambda_f$ & $1$ & $1$ & $1$ & $1$ & $1$ & $1$ & $1$ & $1$ & $1$ & $1$ & $1$ & $1$\\
$\Pi_{f,\nu}$ & $0.04$ & $0.08$ & $0.14$ & $  0.23 $ & $ 0.33 $ & $0.45  $ & $  0.55  $ & $  0.64 $ & $0.72 $ & $ 0.77 $ & $0.82 $ & $0.85$ \\ 
$\Lambda_f/\Pi_{f,\nu}$ & $22.27$ & $12.03$ & $7.04$ & $4.43$ & $3.01$ & $2.24$ &     $1.81$   &   $1.55$   &   $1.40$   &   $1.29$   &   $1.23$   &   $1.18$\\
\botrule
\end{tabular}
\end{center}
\end{table}
In the numerical evaluation of the assumption of Fourier modes being POD modes along locally translationally invariant coordinates, a consistent method of comparison between the two types of spectra is based on their respective convergence rates thereby requiring us to make use of both sides of the symmetric Fourier spectrum. 
\noindent
%
In order to achieve this, the Fourier spectral values are sorted in descending order, such that $\sigma^{\alpha+1}_\nu\leq \sigma^\alpha_\nu$, for $\alpha\in[1:N-1]$ and all $\nu$ - as is generally done for the POD eigenspectrum. Due to the symmetry of the Fourier spectrum, this then implies that $\sigma^{\alpha+1}_\nu = \sigma^\alpha_\nu$, for $\alpha=2,4,\dots,N-1$, since all Fourier spectral values have an equal spectral value generated by the corresponding complex conjugate mode (with the exception to the spectral value related to the zeroth harmonic). 

A more detailed analysis of the collapse between Fourier and eigenspectra follows in Section \ref{sec:other_correlation_functions}, where the coupling between the Fourier and eigenspectrum is investigated for the second set of correlation functions. Currently, however, the focus is on the general tendencies of Fourier and eigenspectra related to $\textbf{R}^{\left(\nu\right)}$ in order to investigate more general spectral responses to the variations of the MMSR of the kernels \eqref{eq:R_bessel} and how the spectra are related to the asymptotic behaviour defined by \eqref{eq:E_nu_asymp}.

\begin{figure}[t]
\centering
\begin{subfigure}[h]{0.32\textwidth}
\includegraphics[width=\textwidth]{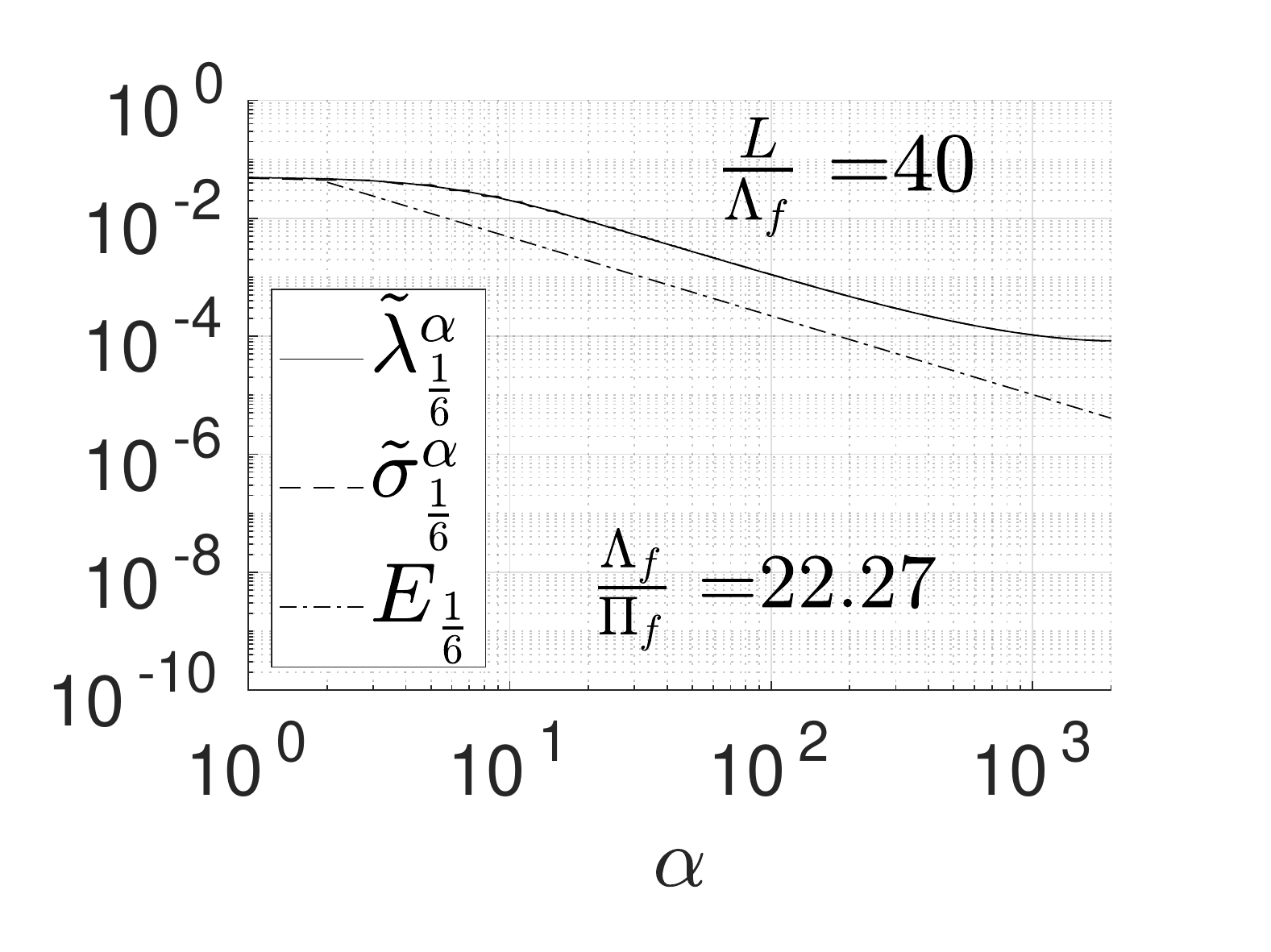}
\caption{\label{fig:e_values_fourier_spectrum_bessel_kernel_1_domain_5}}
\end{subfigure}
\begin{subfigure}[h]{0.32\textwidth}
\includegraphics[width=\textwidth]{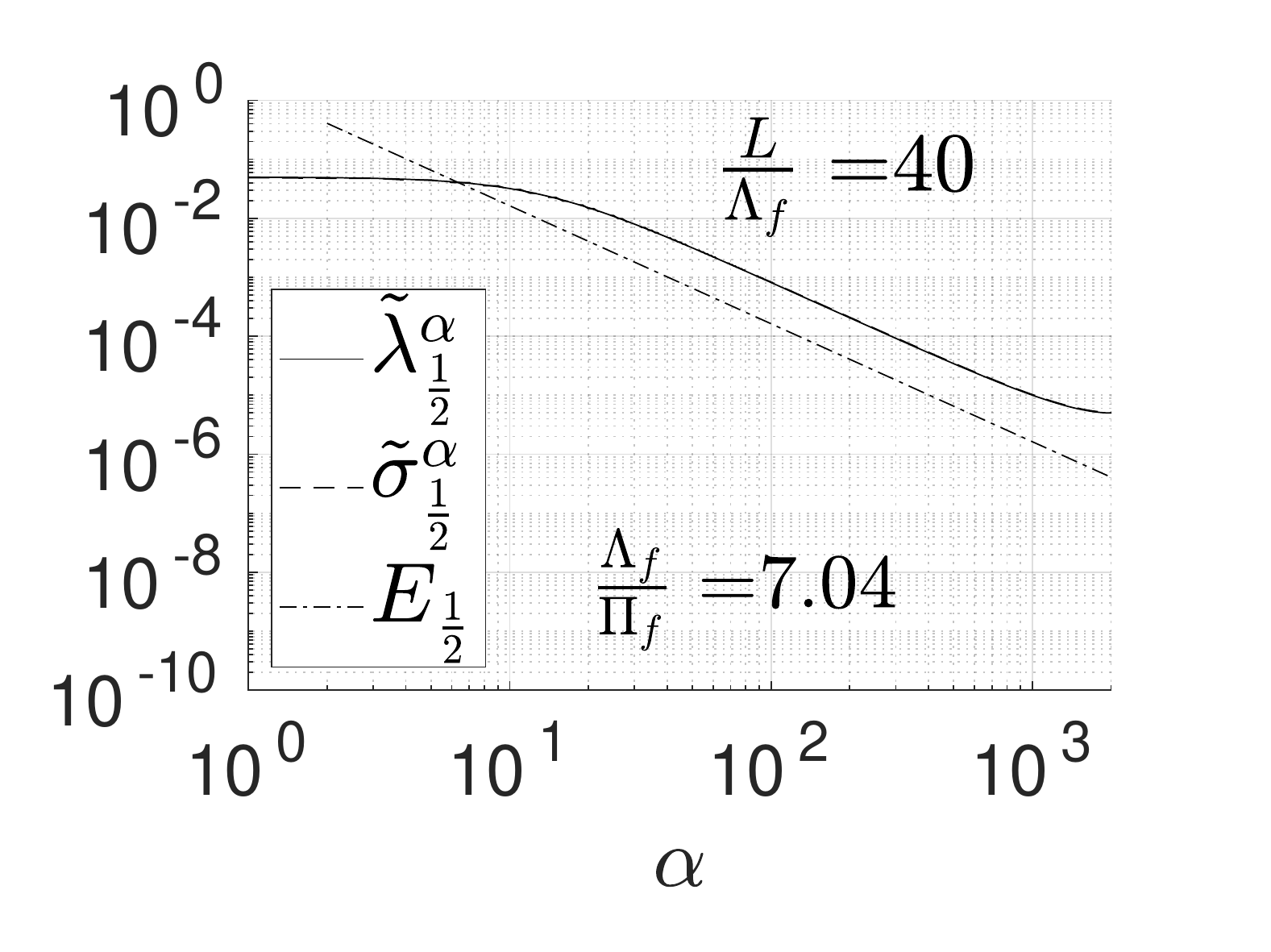}
\caption{\label{fig:e_values_fourier_spectrum_bessel_kernel_3_domain_5}}
\end{subfigure}
\begin{subfigure}[h]{0.32\textwidth}
\includegraphics[width=\textwidth]{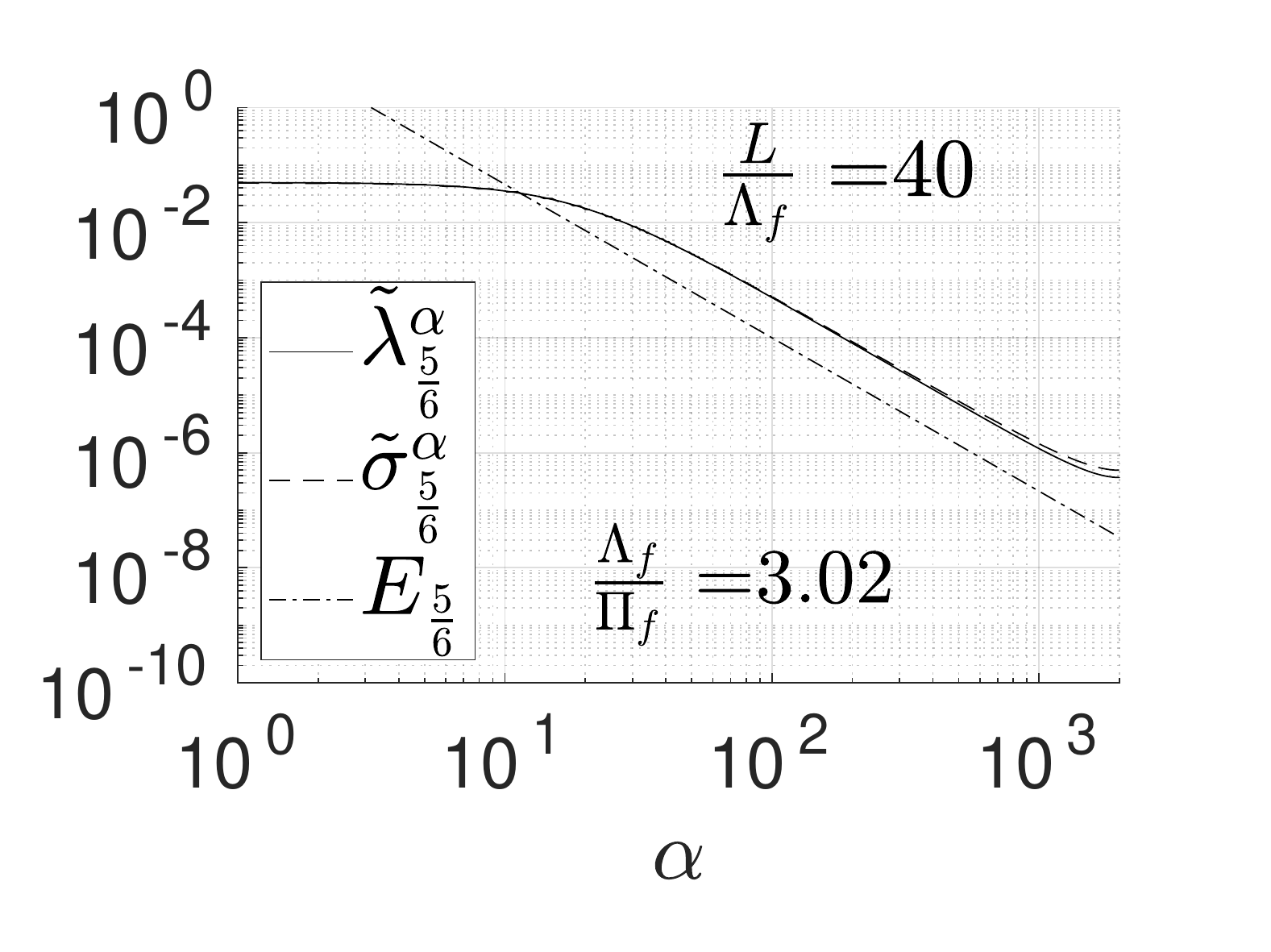}
\caption{\label{fig:e_values_fourier_spectrum_bessel_kernel_5_domain_5}}
\end{subfigure}
\begin{subfigure}[h]{0.32\textwidth}
\includegraphics[width=\textwidth]{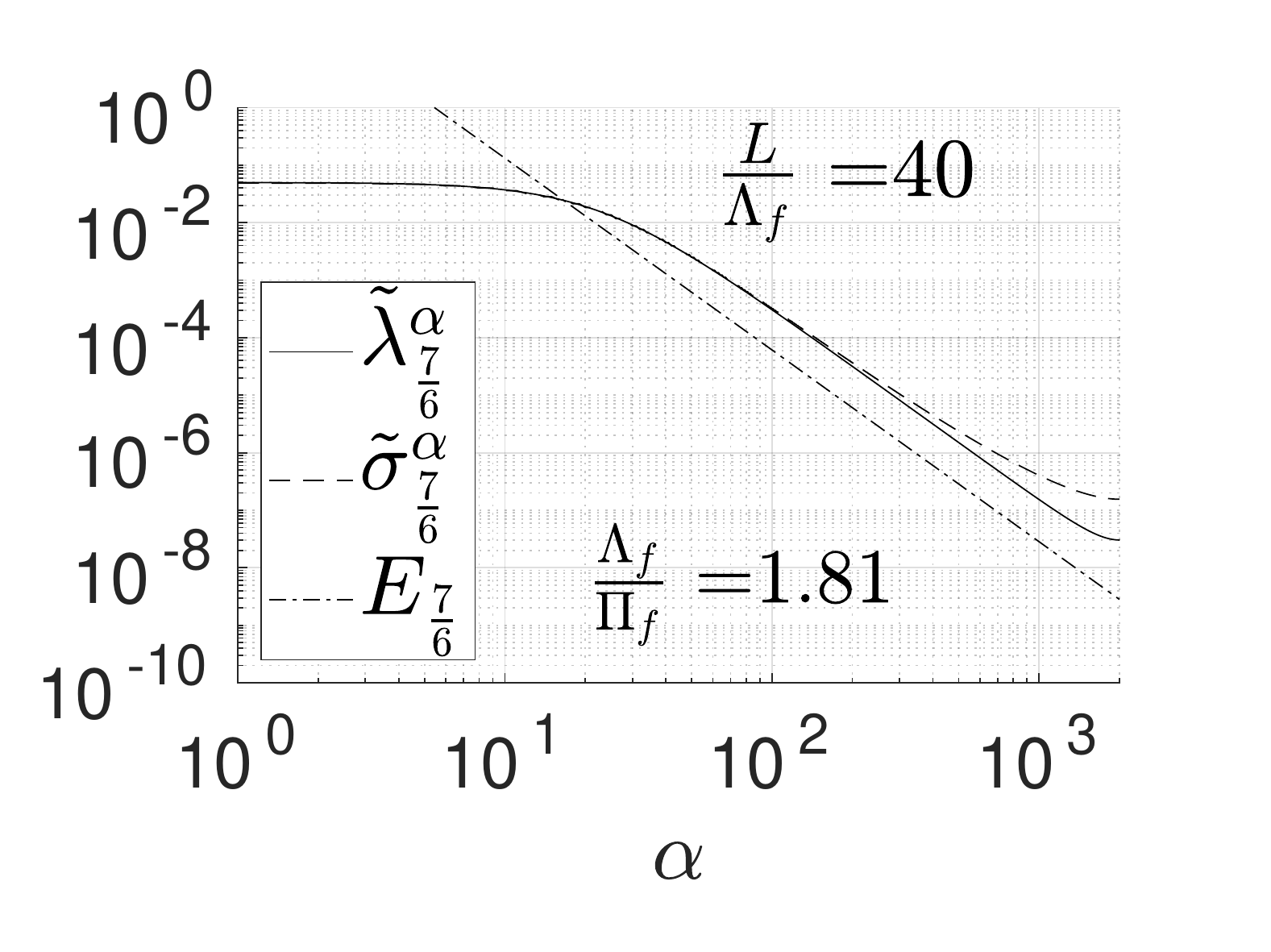}
\caption{\label{fig:e_values_fourier_spectrum_bessel_kernel_7_domain_5}}
\end{subfigure}
\begin{subfigure}[h]{0.32\textwidth}
\includegraphics[width=\textwidth]{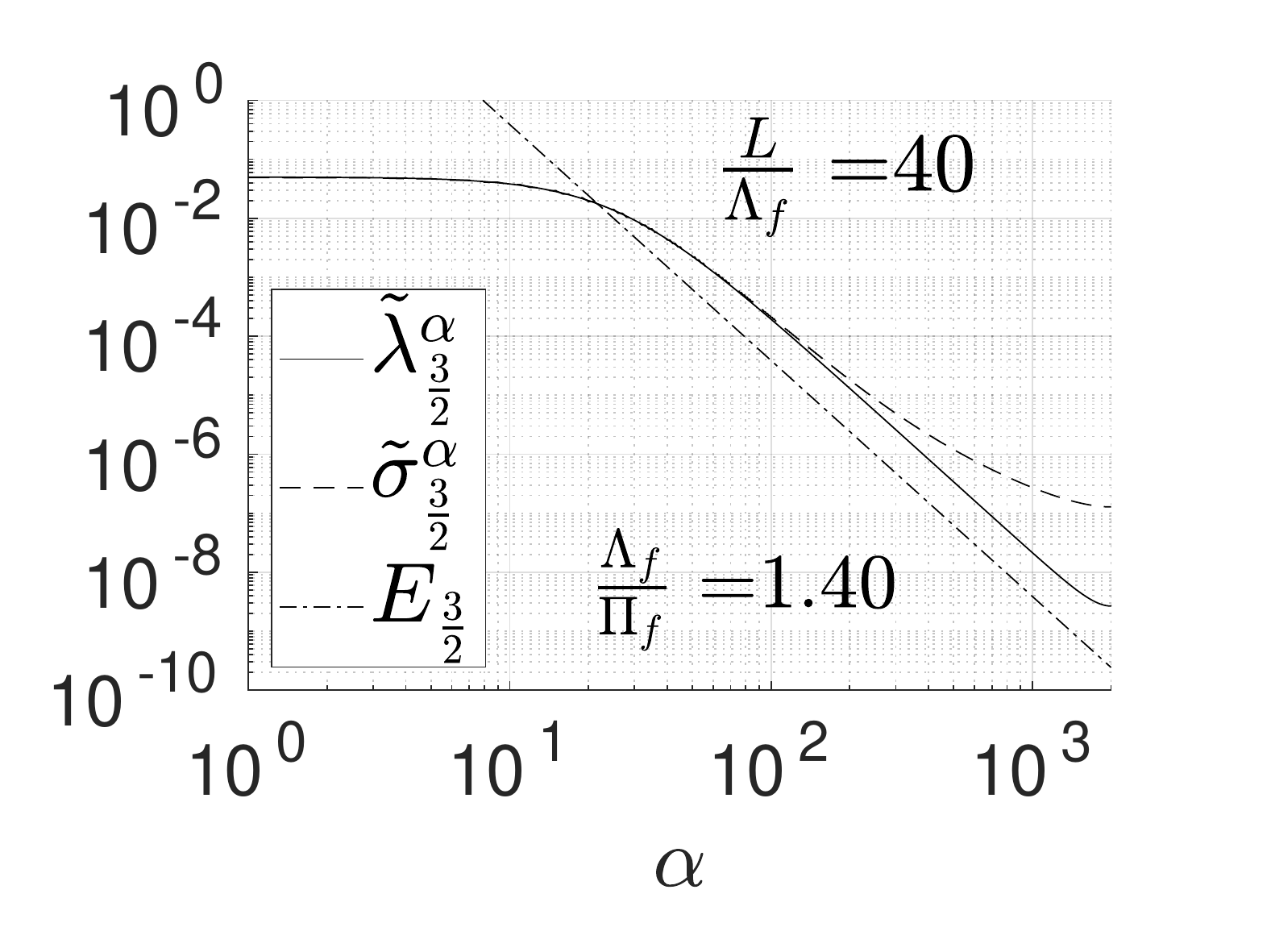}
\caption{\label{fig:e_values_fourier_spectrum_bessel_kernel_9_domain_5}}
\end{subfigure}
\begin{subfigure}[h]{0.32\textwidth}
\includegraphics[width=\textwidth]{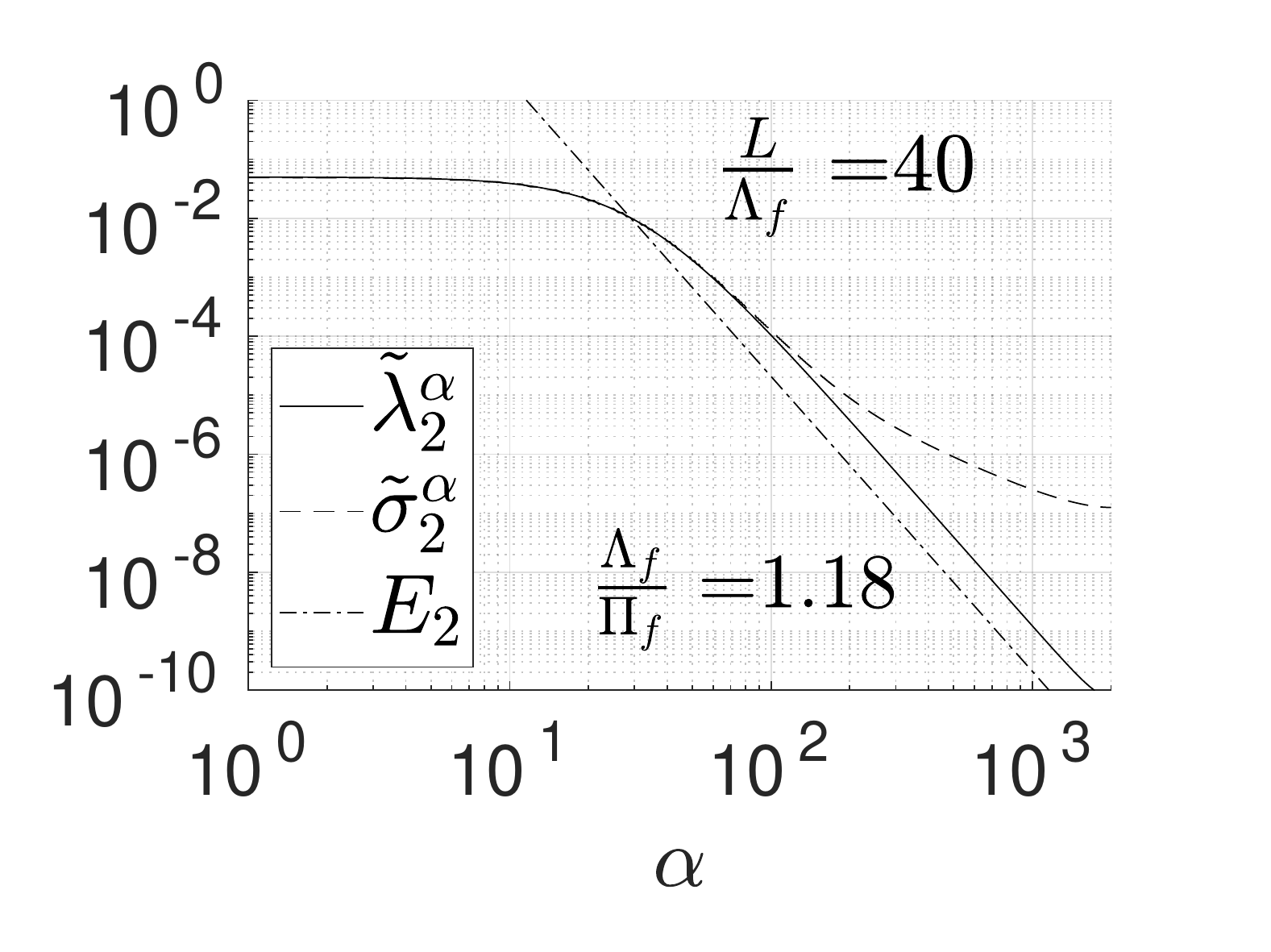}
\caption{\label{fig:e_values_fourier_spectrum_bessel_kernel_12_domain_5}}
\end{subfigure}
\caption{(a)-(f): Normalized eigenvalues, $\tilde{\lambda}_\nu^\alpha$, and sorted Fourier spectra, $\tilde{\sigma}_\nu^\alpha$, related to $\textbf{R}^{\left(\nu\right)}$ corresponding to $\Lambda_f/\Pi_f = [22.27, 7.04, 3.01, 1.81, 1.40, 1.18]$ for a domain length of $40$ integral length scales. $E_\nu$ depicts the asymptotic behaviour of the exact Fourier transform of $R^{\left(\nu\right)}$. \label{fig:e_values_fourier_spectrum_bessel}}
\end{figure}
\noindent
The Fourier and eigenspectra shown in Figure \ref{fig:e_values_fourier_spectrum_bessel} are related to $\textbf{R}^{\left(\nu\right)}$ for $\nu$-values corresponding to $\Lambda_f/\Pi_f = [22.27, 7.04, 3.01, 1.81, 1.40, 1.18]$, respectively. The spectra are numerically evaluated over a domain length of~${L/\Lambda_f = 40}$. As the MMSR-ratio is decreased (by increasing $\nu$), a noticeable difference in the low-energetic regions of the Fourier and eigenspectra is observed. As the convergence rate of the eigenspectrum must be at least as fast as the Fourier spectrum due to the optimality of the POD eigenfunctions, a gradually increasing tail is observed for the Fourier spectra with decreasing MMSR (increasing $\nu$). However, and more interestingly, the eigenspectrum exhibits the same asymptotic power-law behaviour as is expected by the analytical Fourier spectrum, \eqref{eq:E_nu_asymp}.
\begin{figure}[h]
\centering
\begin{subfigure}[h]{0.49\textwidth}
\includegraphics[width=\textwidth]{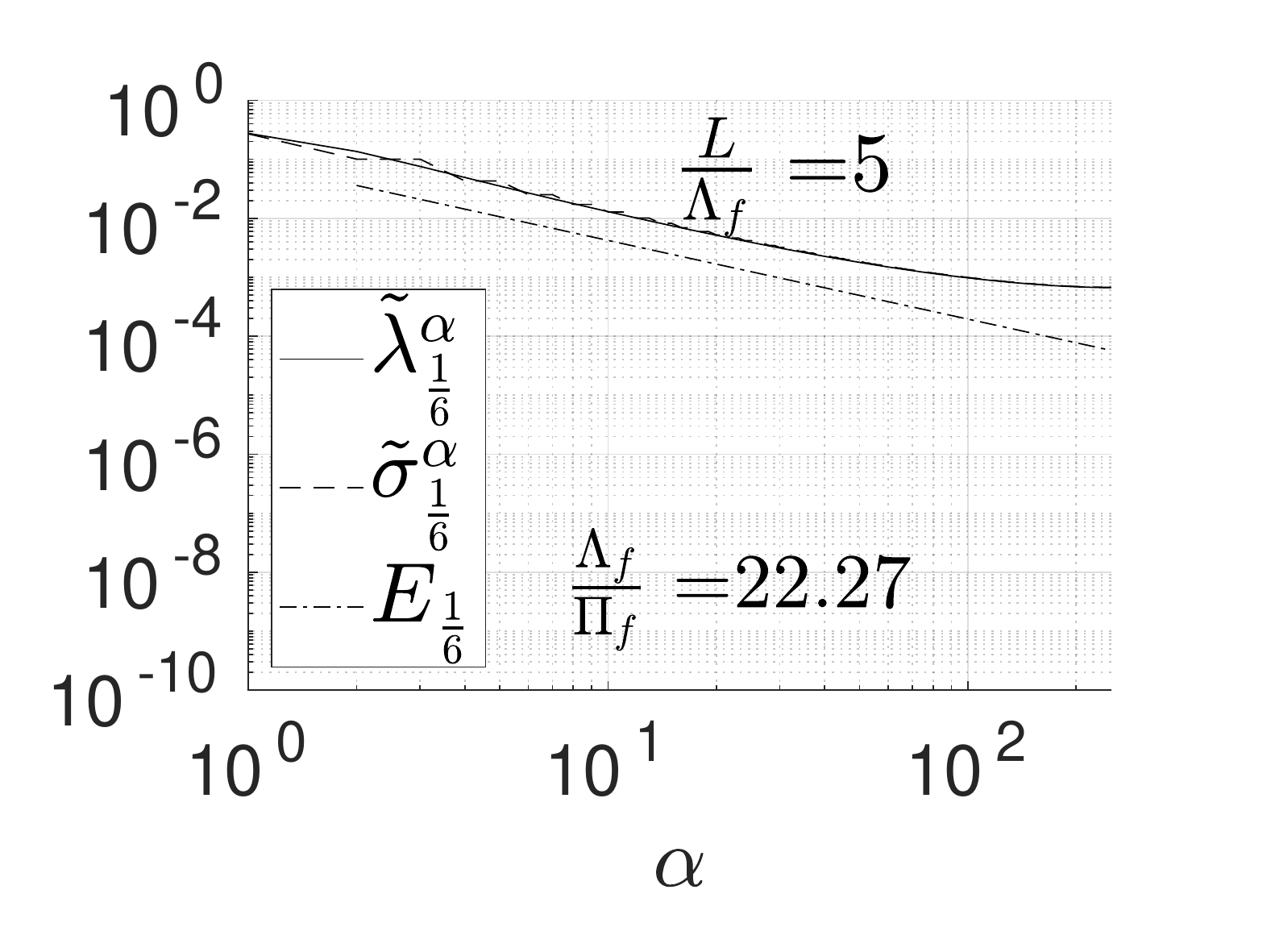}
\caption{\label{fig:e_values_fourier_spectrum_bessel_kernel_1_domain_1}}
\end{subfigure}
\begin{subfigure}[h]{0.49\textwidth}
\includegraphics[width=\textwidth]{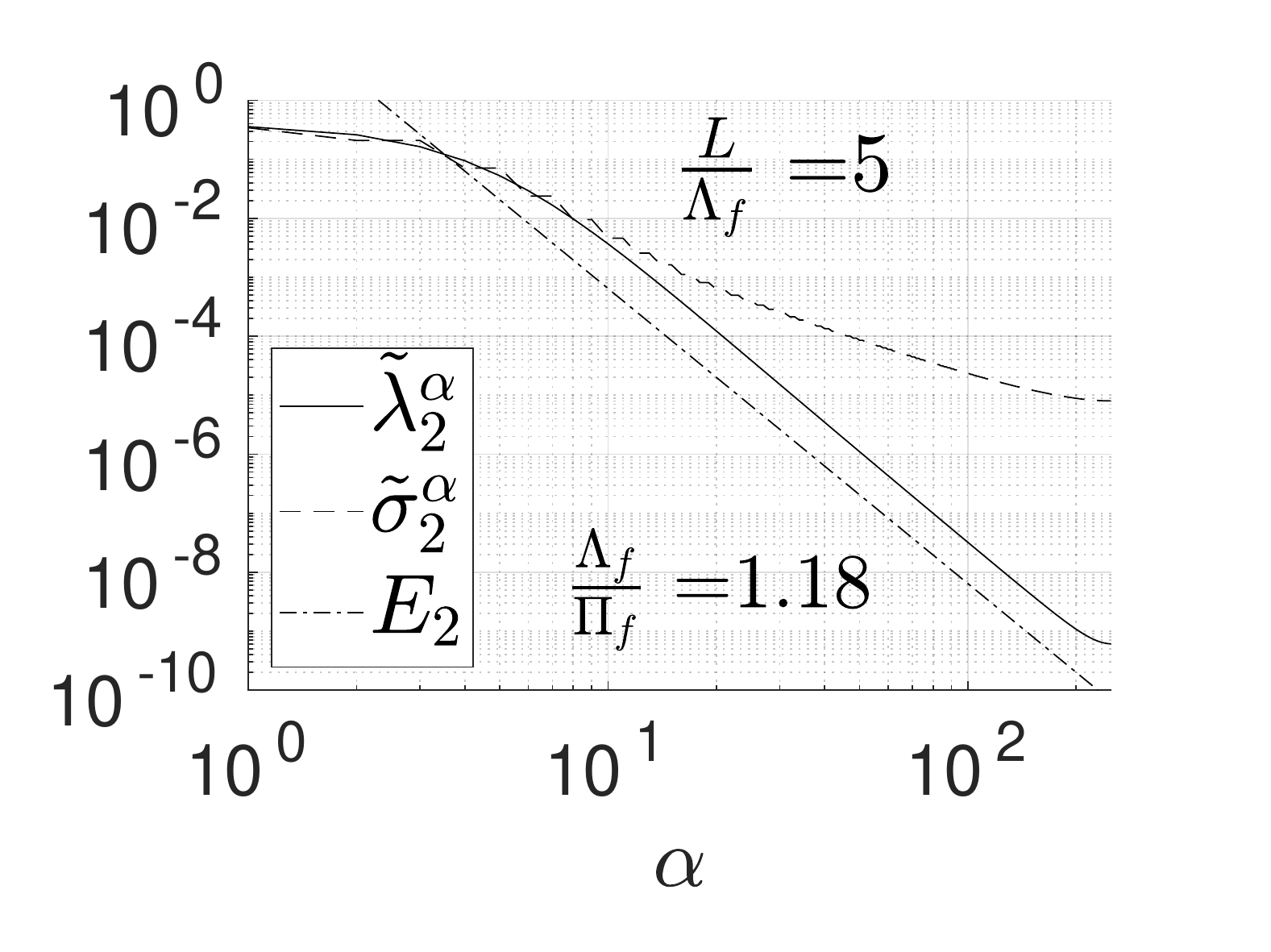}
\caption{\label{fig:e_values_fourier_spectrum_bessel_kernel_12_domain_1}}
\end{subfigure}
\caption{(a)-(b):Normalized eigenvalues and sorted Fourier spectra related to $R^{\left(\nu\right)}$ corresponding to $\Lambda_f/\Pi_f = [22.27, 1.18]$, respectively, for a domain length of $5$ integral length scales. $E_\nu$ depicts the asymptotic behaviour of the exact Fourier transform of $R^{\left(\nu\right)}$. \label{fig:e_values_fourier_spectrum_bessel_12}}
\end{figure}
\noindent

This behaviour is tested for the extreme case in Figure \ref{fig:e_values_fourier_spectrum_bessel_12}, which shows the spectral comparison for the very small window size $L/\Lambda_f = 5$ for MMSR corresponding to the extremes $22.27$ and $1.18$. Even for this very narrow window, which still captures the main correlation signature, the eigenspectrum exhibits the asymptotic spectral behaviour characterizing the exact Fourier spectrum, and down to the very small value of $\Lambda_f/\Pi_f=1.18$. These results suggest that the optimality criterion underlying the POD correlates with a reduced windowing effect on the corresponding spectra (see Section \ref{sec:Taylor_micro_scale_reconstruction} for further consequences of using the DFT as opposed to the POD modes). If so, the fact that the POD modes deviate from Fourier modes is a central ingredient to achieving asymptotic spectral behaviours of the exact Fourier spectra, \eqref{eq:E_nu_asymp}. The underlying reason for this is not entirely clear, but seems to reflect that the asymptotic spectral behaviour of the POD eigenvalues is closer to the asymptotic behaviour of the exact Fourier spectrum than the spectrum resulting from a DFT. As the tail of DFT spectra is a central focus point in the diagnosis of windowing effects and underlying the choice of window functions used to correct for the finiteness of the signal, the current results demonstrate the advantages of using a POD decomposition in place of a the DFT, even for locally translationally invariant kernels as it may reduce the requirements for the window size in order to achieve the same spectral behaviour as expected using a Fourier transform on a fully translationally invariant kernel on an infinite domain. Secondly, if window functions are considered as inner product weights, it is easily demonstrated that the set of discrete Fourier modes, \eqref{eq:psi_m} is not orthogonal with respect to these weighted inner products (see also Appendix \ref{app:weighted_invariant_kernels}). This point is related to the analysis of the spectra, as it implies that although the inner product weight reduces the windowing effect by reducing the tail end of the corresponding Fourier spectrum (see \cite{Welch1967}, \cite{Harris1978a})), it complicates the interpretation of the same Fourier spectrum, since the Fourier modes are not orthogonal with respect to the underlying weighted inner product. A consequence of this non-orthogonality is that the energy related to each wavenumber fails to decouple from the energy related to other wavenumbers. This adds a degree of abstractness to the interpretation of the resulting Fourier spectrum. 

It is worth recalling that the POD does not suffer from this issue, as the POD eigenfuntions are guaranteed to be mutually orthogonal as long as the corresponding operator is Hermitian. Although one could argue that the efficiency of the FFT algorithm is enough to motivate the use of the Fourier decomposition, the corresponding windowing effects may require a much longer measurement/simulation window. This may introduce significant costs in terms of experimental equipment and/or computational costs, both related to the generation of data (in case of a simulation) and to data analysis, compared to the alternative of chosing a smaller domain and applying a POD decomposition to this data set. 

A comparison of the results in Figure \ref{fig:e_values_fourier_spectrum_bessel} furthermore indicates that the differences between the Fourier and POD eigenspectra may be directly affected by the Reynolds number of the turbulent flow, given that the MMSR-ratio increases for increasing Reynolds numbers. The spectral analysis of $\textbf{R}^{\left(\nu\right)}$ indicates that for low Reynolds number flows the difference between the POD and Fourier modes is more profound, given that their spectra are different. In order to determine whether this tendency is exclusively related to the discretized versions of the specific family of functions defined by \eqref{eq:R_bessel} or if there is support for this hypothesis for arbitrary correlation functions, the spectral dependence on the MMSR-ratio will be extended to other types of POD kernels in the following Section.
\FloatBarrier
\subsection{Fourier and eigenspectrum discrepancy dependence on macro/micro scale ratio \label{sec:other_correlation_functions}}
The analysis of the relation between the MMSR and the discrepancies observed between Fourier and eigenspectra is now extended to a new set of arbitrarily chosen correlation functions, in order to investigate whether the MMSR in more general terms can be expected to play a role in the deviations between the aformentioned spectra. These analyses are performed for six domain sizes (using the same grid resolution), in order to evaluate the effects of spectral leakage.

Discretized versions of the following five analytical kernel forms are now considered
\begin{subequations}
\begin{eqnarray}
K_{1j}(x,y) &=& e^{-\vert x-y\vert }\,,\label{eq:kernel_exp}\\
K_{2j}(x,y) &=& \begin{cases} K_{4j}(x,y)-0.1\sin\left(2\pi \vert x-y\vert /L\right)\,,\label{eq:kernel_modified} & \text{if } \vert x-y\vert \leq L \\
                      0\hspace*{5cm},        & \text{if } \vert x-y\vert > L      %
        \end{cases}\,,\\
K_{3j}(x,y) &=& (4(x-y)^2+1)^{-1}\,,\label{eq:kernel_hyp}\\
K_{4j}(x,y) &=& \begin{cases} \left(1-\vert x-y\vert/L\right)^4\left(1+4\vert x-y\vert /L\right)\,,\label{eq:kernel_wendland} & \text{if } \vert x-y\vert \leq L \\
                      0\,,        & \text{if } \vert x-y\vert > L      %
\end{cases}\,,\\
K_{5j}(x,y) &=& e^{-a(x-y)^2}\,,\label{eq:kernel_gauss}
\end{eqnarray}
\end{subequations}
\noindent
where $\,x,y\in\Omega_j$, $j\in[1:6]$, where $j$ indicates the domain. Note that \eqref{eq:kernel_exp} corresponds to $R_\frac{1}{2}$ in \eqref{eq:R_bessel}. $a=1/8$ in \eqref{eq:kernel_gauss} was included as this correlation function models self-preserving decaying homogeneous isotropic low Reynolds number turbulent fields appearing as the solution to the fundamental equation for the propagation of the correlation function, \cite{karman1938}.
\begin{figure}[t]
\centering
\begin{subfigure}[h]{0.32\textwidth}
\includegraphics[width=\textwidth]{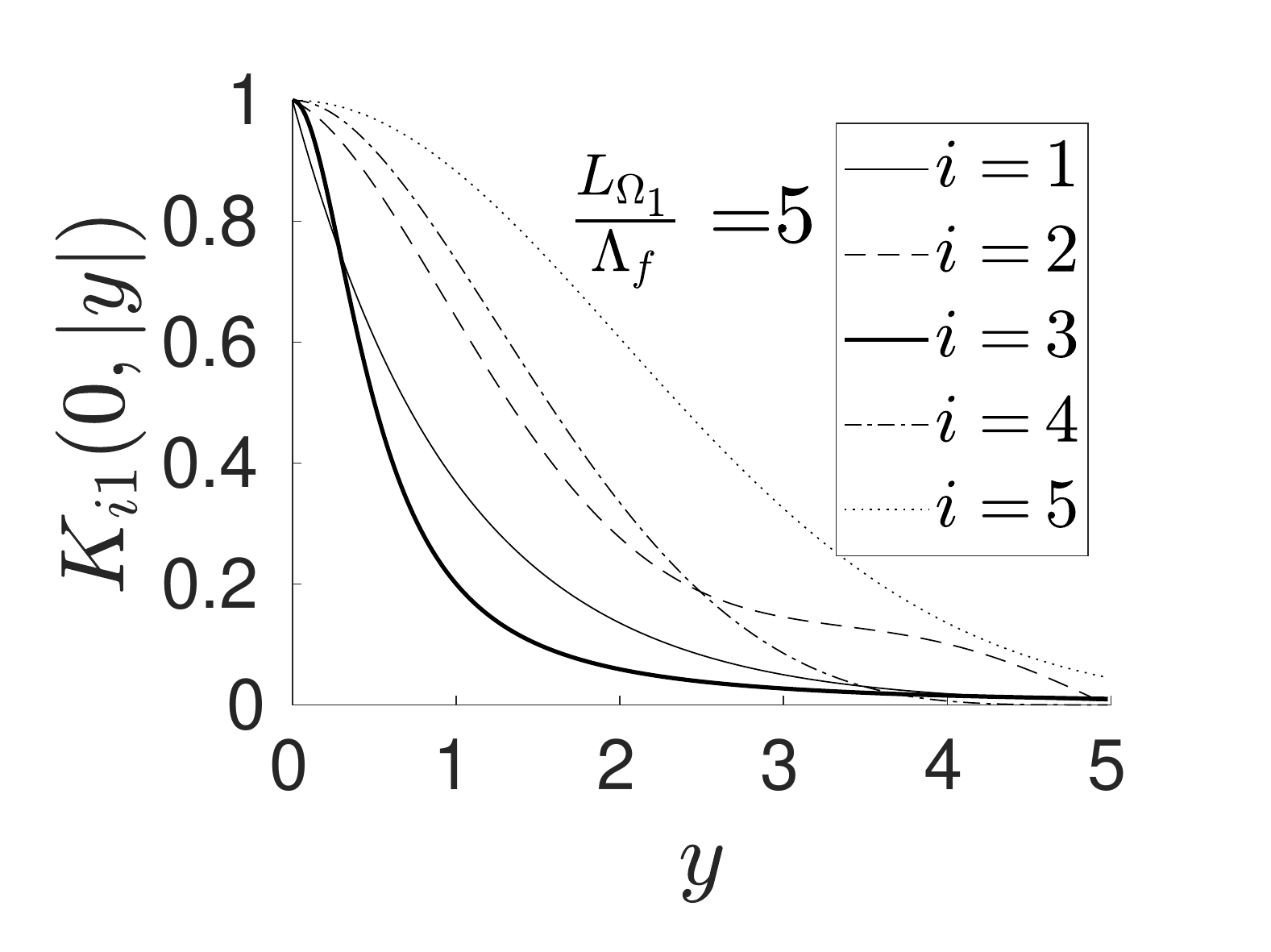}
\caption{\label{fig:correlation_functions_1}}
\end{subfigure}
\begin{subfigure}[h]{0.32\textwidth}
\includegraphics[width=\textwidth]{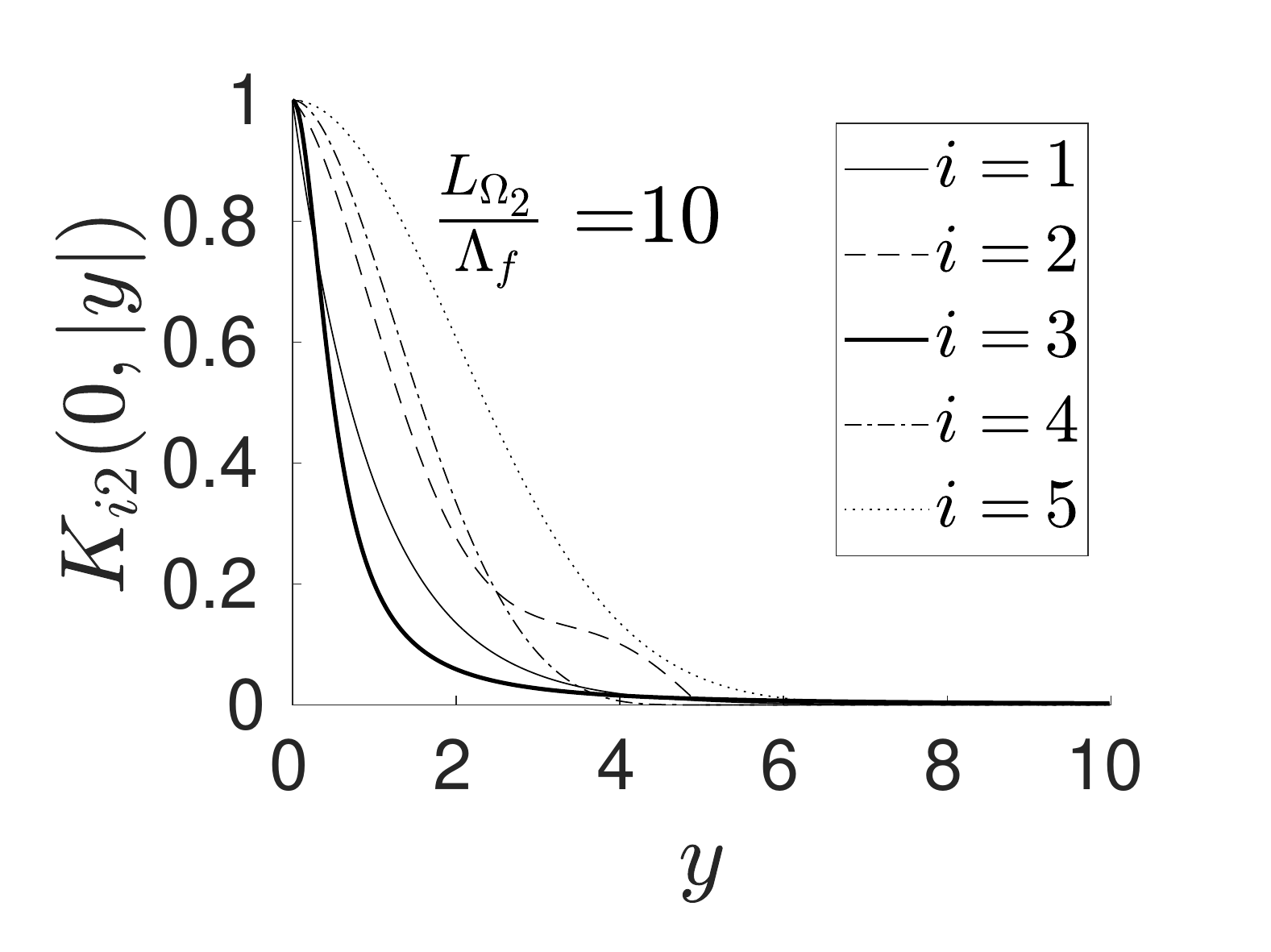}
\caption{\label{fig:correlation_functions_2}}
\end{subfigure}
\begin{subfigure}[h]{0.32\textwidth}
\includegraphics[width=\textwidth]{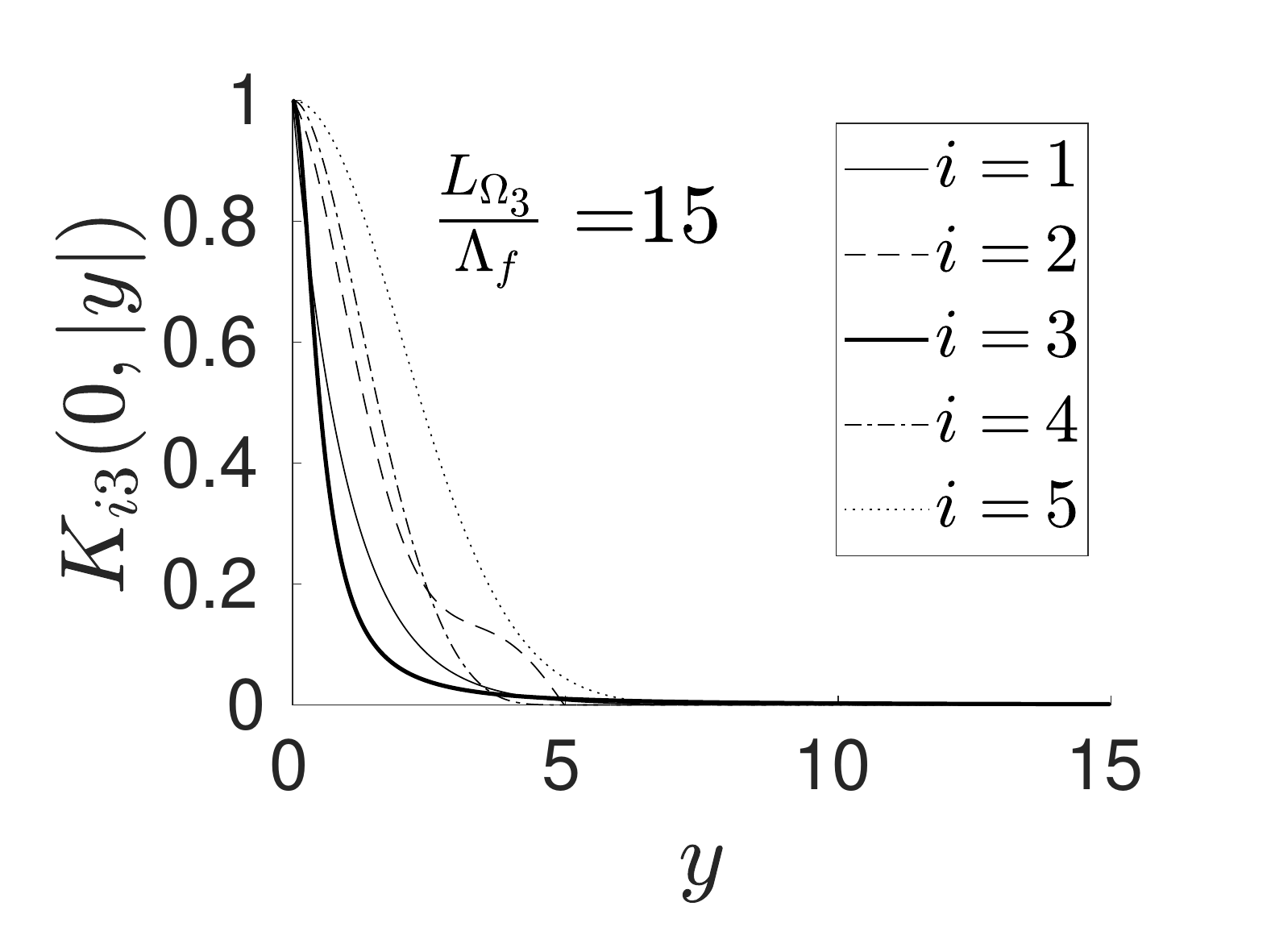}
\caption{\label{fig:correlation_functions_3}}
\end{subfigure}
\begin{subfigure}[h]{0.32\textwidth}
\includegraphics[width=\textwidth]{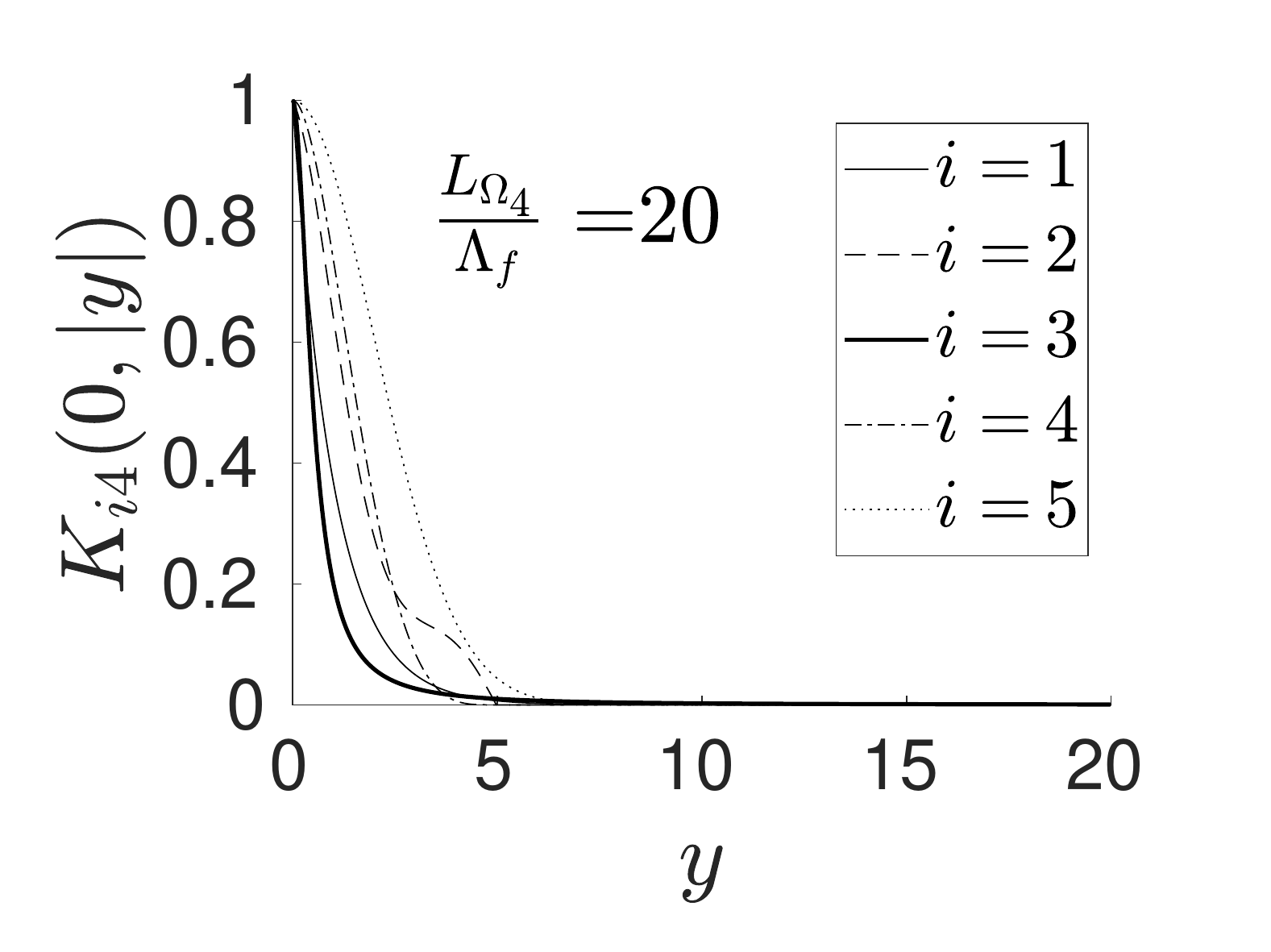}
\caption{\label{fig:correlation_functions_4}}
\end{subfigure}
\begin{subfigure}[h]{0.32\textwidth}
\includegraphics[width=\textwidth]{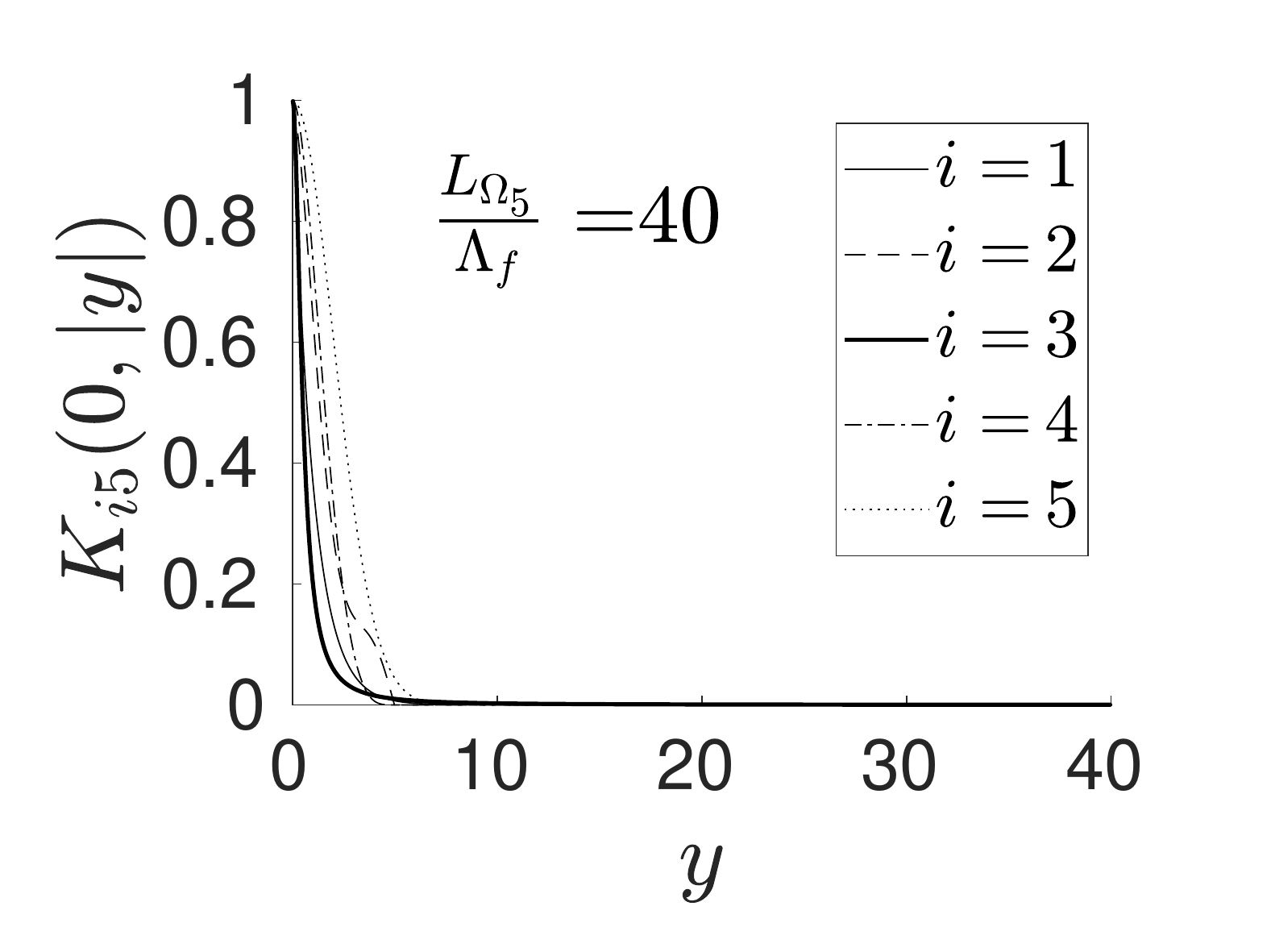}
\caption{\label{fig:correlation_functions_5}}
\end{subfigure}
\begin{subfigure}[h]{0.32\textwidth}
\includegraphics[width=\textwidth]{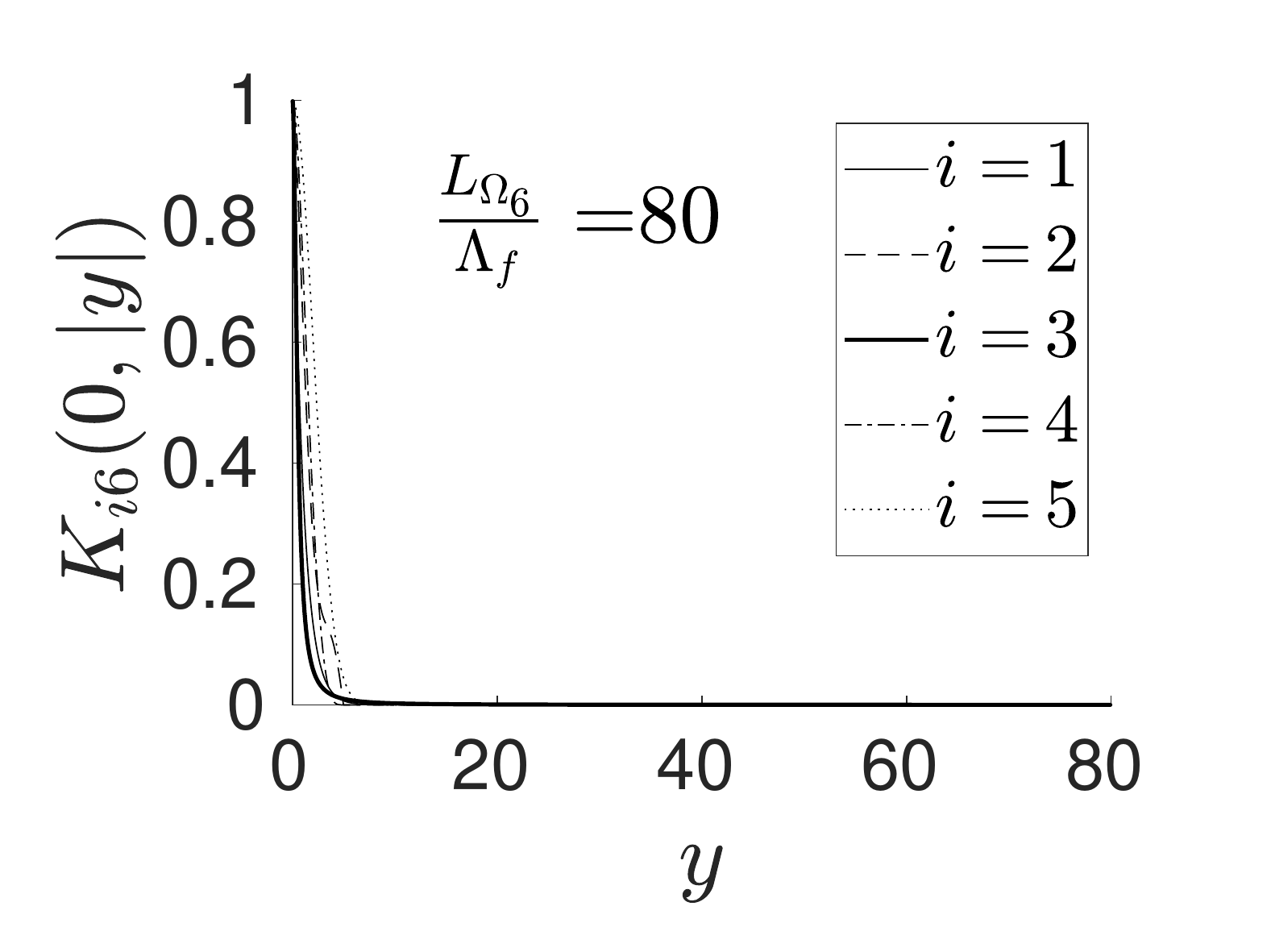}
\caption{\label{fig:correlation_functions_6}}
\end{subfigure}
\caption{(a)-(f): Illustration of the correlation functions, $K_{ij}$, evaluated for $x=0$ on domains with lengths $L_{\Omega_j}$, $j\in[1:6]$, respectively.\label{fig:correlation_functions}}
\end{figure}
%
\noindent
\begin{table}[b]
\begin{center}
\begin{minipage}{1\textwidth}
\caption{Domain specifications where $L_{\Omega_j}$ is the domain length, $\Lambda_f^{ref}$ is the integral length scale of $K_{1,j}$, $N_j$ is the number of grid points, and $L$ designates the support of the functions $K_{4j}$ and $K_{5j}$, $j\in [1:6]$.}\label{tab:grid}%
\centering
\begin{tabular}{@{}lcccccc@{}}
\toprule
$j$ & $1$ & $2$ & $3$ & $4$ & $5$ & $6$ \\ 
\midrule 
$L_{\Omega_j}/\Lambda_f^{ref}$ & 5 & 10 & 15 & 20 & 40 & 80 \\ 
$N_j$ & 250 & 500 & 750 & 1000 & 2000 & 4000 \\ 
$L/\Lambda_f^{ref}$ & 5 & 5 & 5 & 5 & 5 & 5\\
\botrule
\end{tabular}
\end{minipage}
\end{center}
\end{table}
\noindent
The lengths of the domains $\Omega_j$ are denoted by $L_{\Omega_j}$ such that $L\leq L_{\Omega_j}$ and given in table \ref{tab:grid} along 
\begin{figure}[h]
\centering
\begin{subfigure}[h]{0.32\textwidth}
\includegraphics[width=\textwidth]{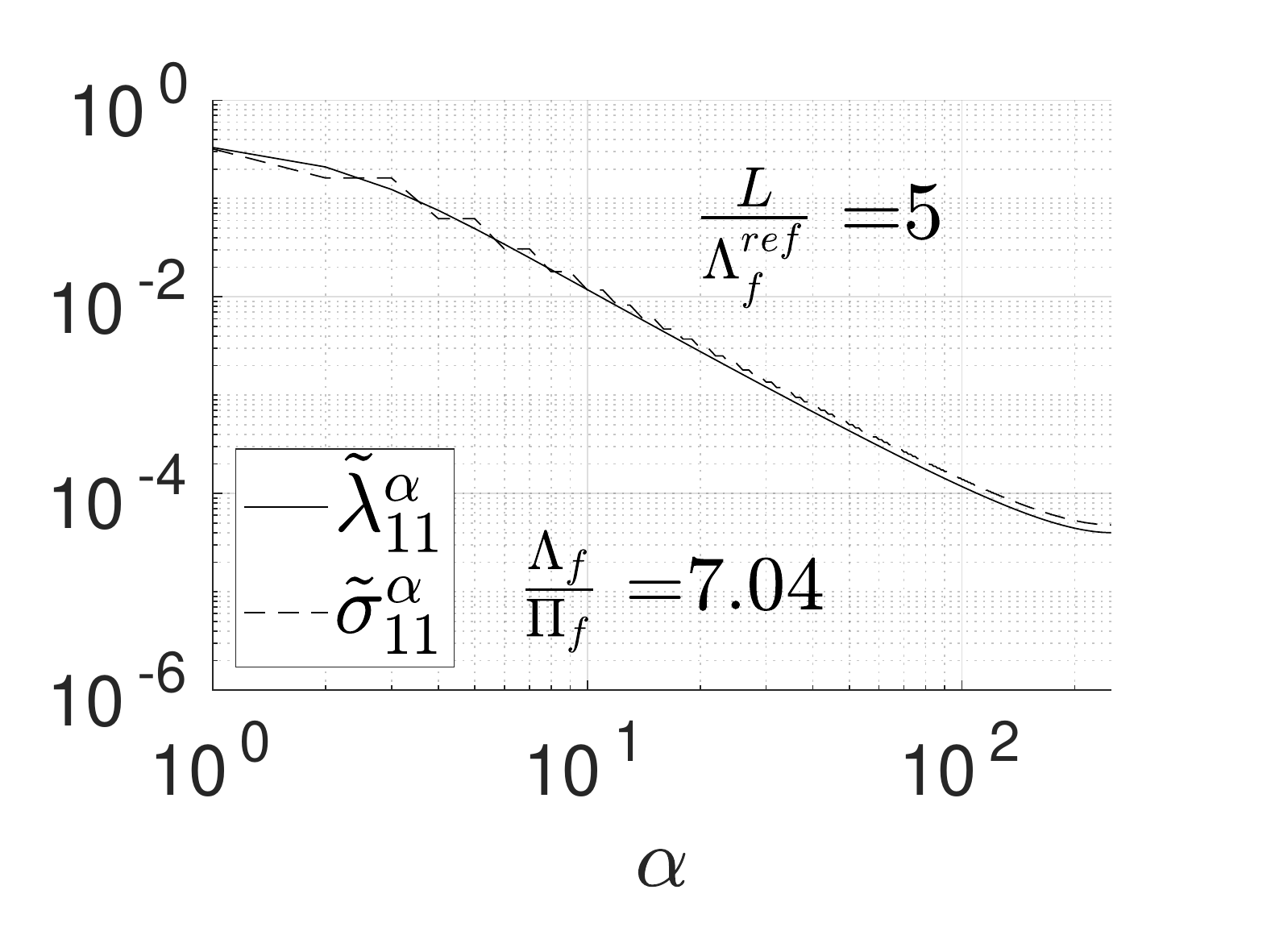}
\caption{\label{fig:e_values_fourier_spectrum_kernel_1_domain_1}}
\end{subfigure}
\begin{subfigure}[h]{0.32\textwidth}
\includegraphics[width=\textwidth]{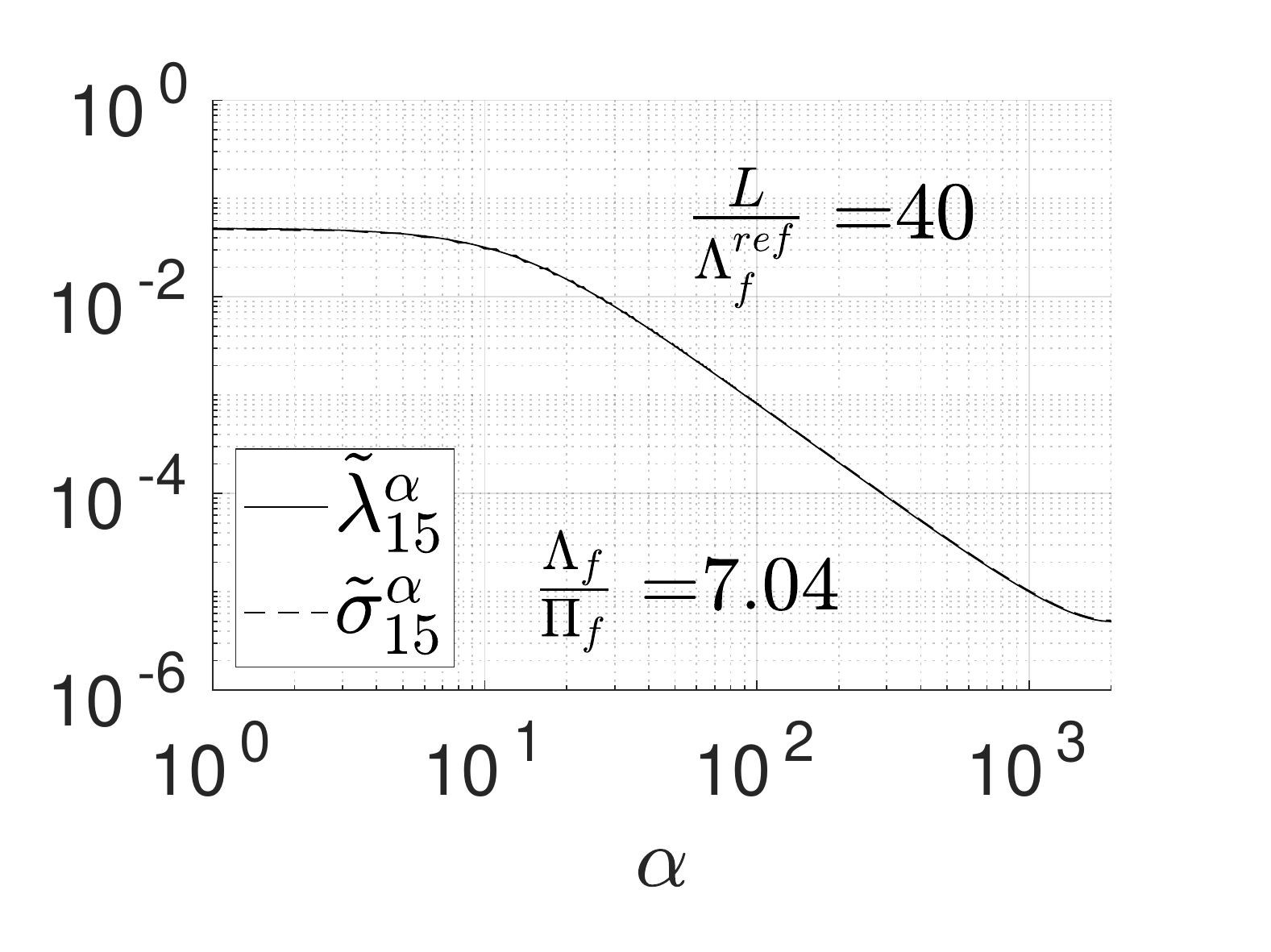}
\caption{\label{fig:e_values_fourier_spectrum_kernel_1_domain_4}}
\end{subfigure}
\begin{subfigure}[h]{0.32\textwidth}
\includegraphics[width=\textwidth]{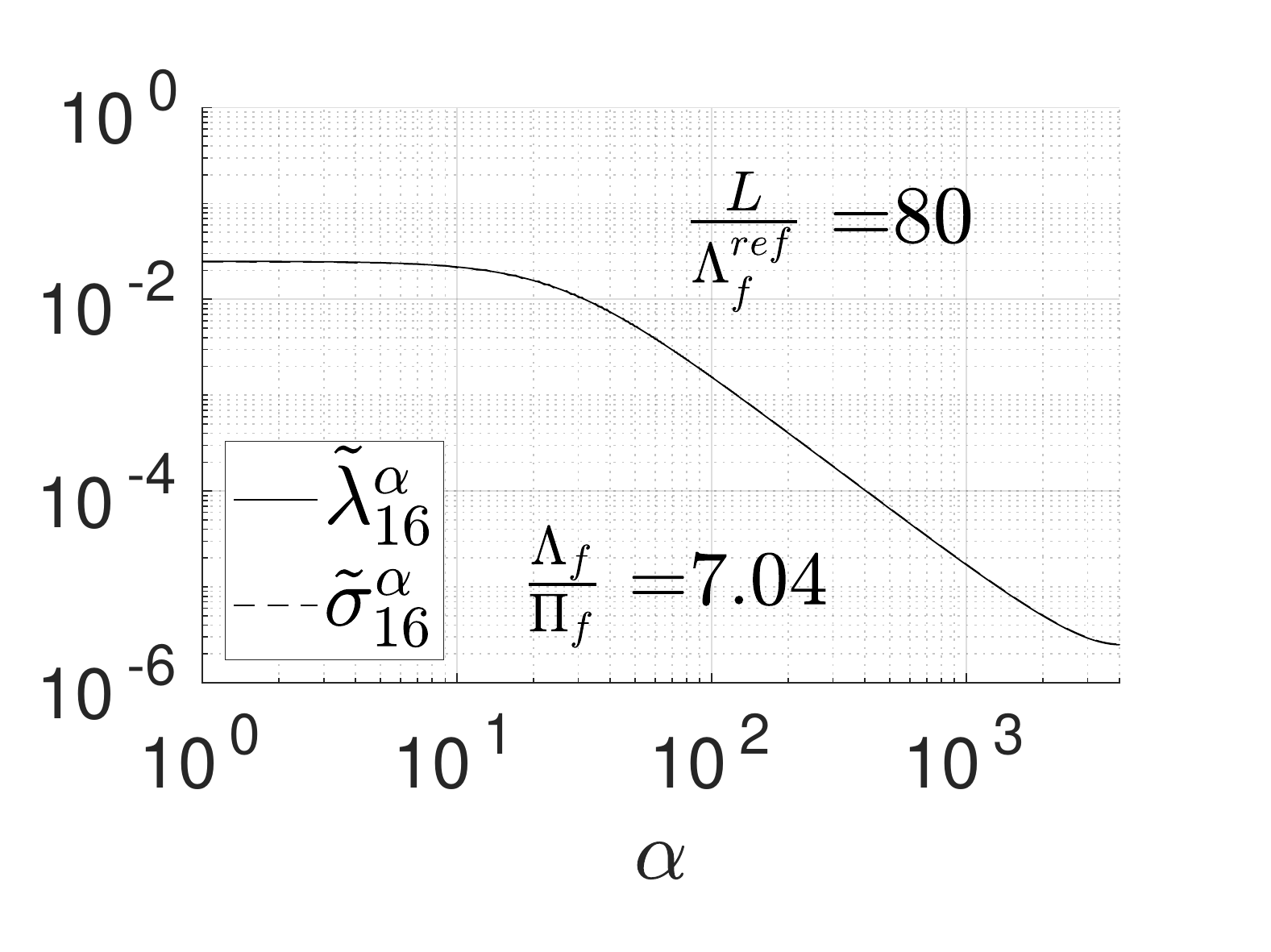}
\caption{\label{fig:e_values_fourier_spectrum_kernel_1_domain_6}}
\end{subfigure}
\begin{subfigure}[h]{0.32\textwidth}
\includegraphics[width=\textwidth]{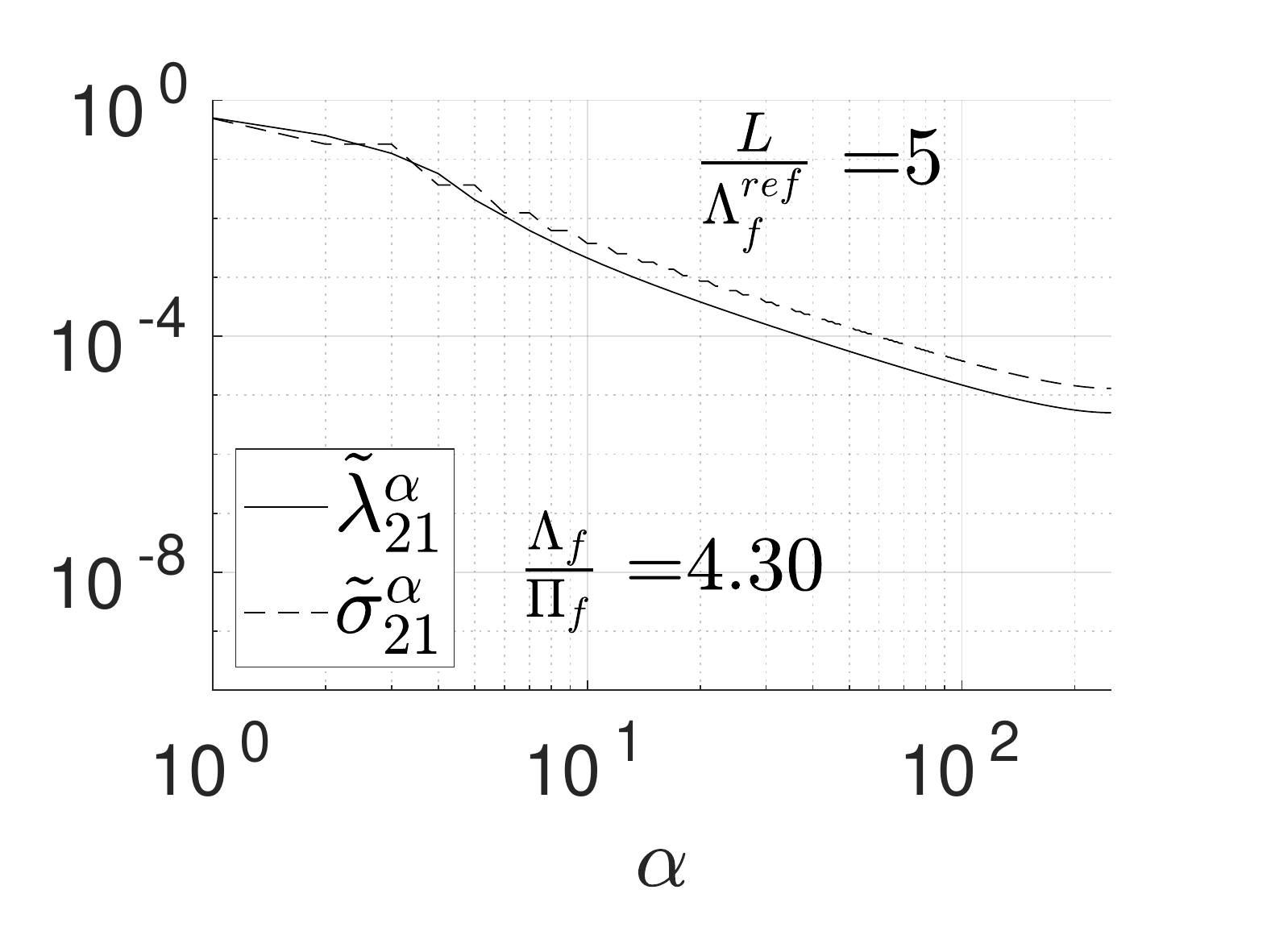}
\caption{\label{fig:e_values_fourier_spectrum_kernel_2_domain_1}}
\end{subfigure}
\begin{subfigure}[h]{0.32\textwidth}
\includegraphics[width=\textwidth]{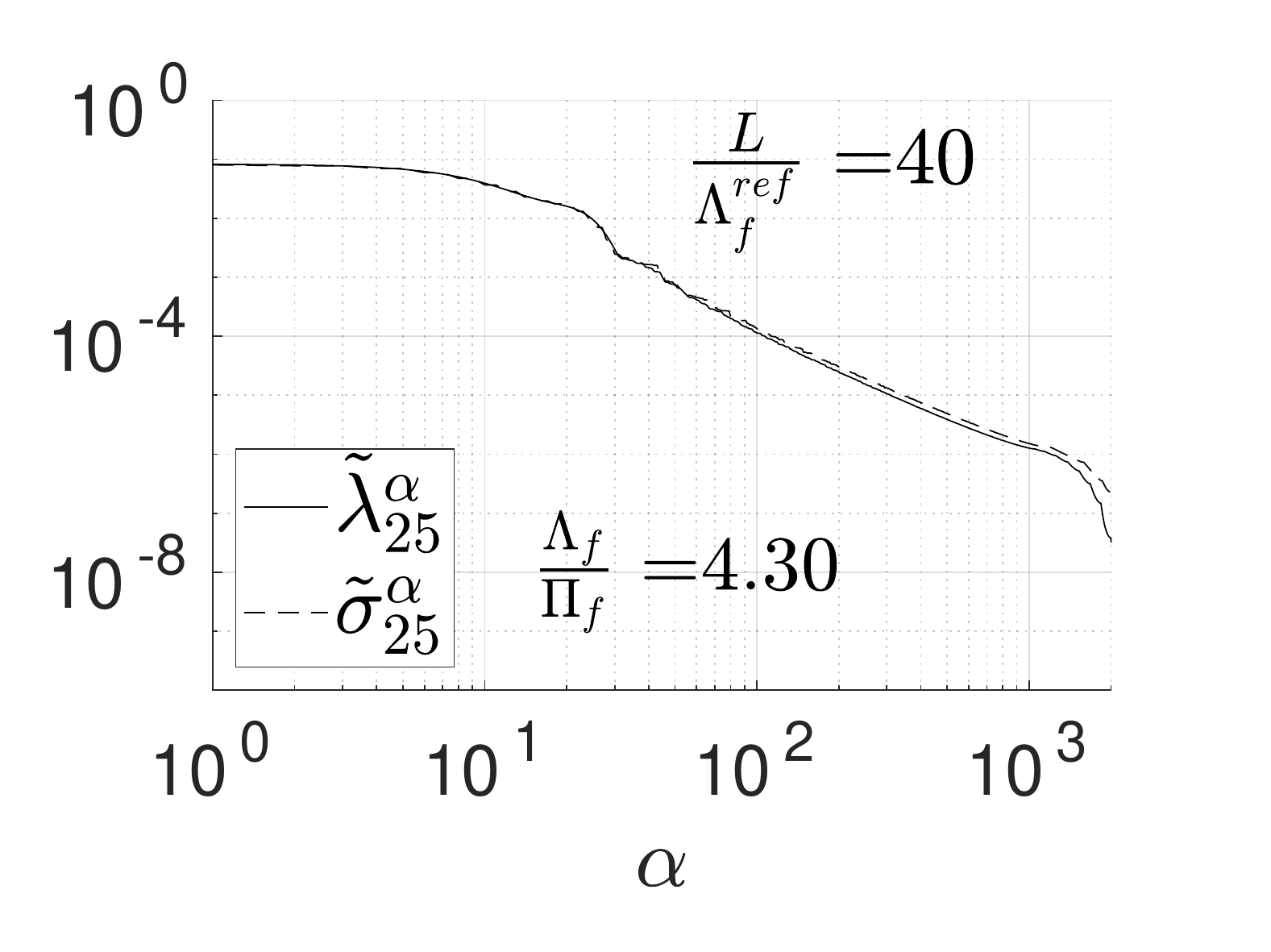}
\caption{\label{fig:e_values_fourier_spectrum_kernel_2_domain_4}}
\end{subfigure}
\begin{subfigure}[h]{0.32\textwidth}
\includegraphics[width=\textwidth]{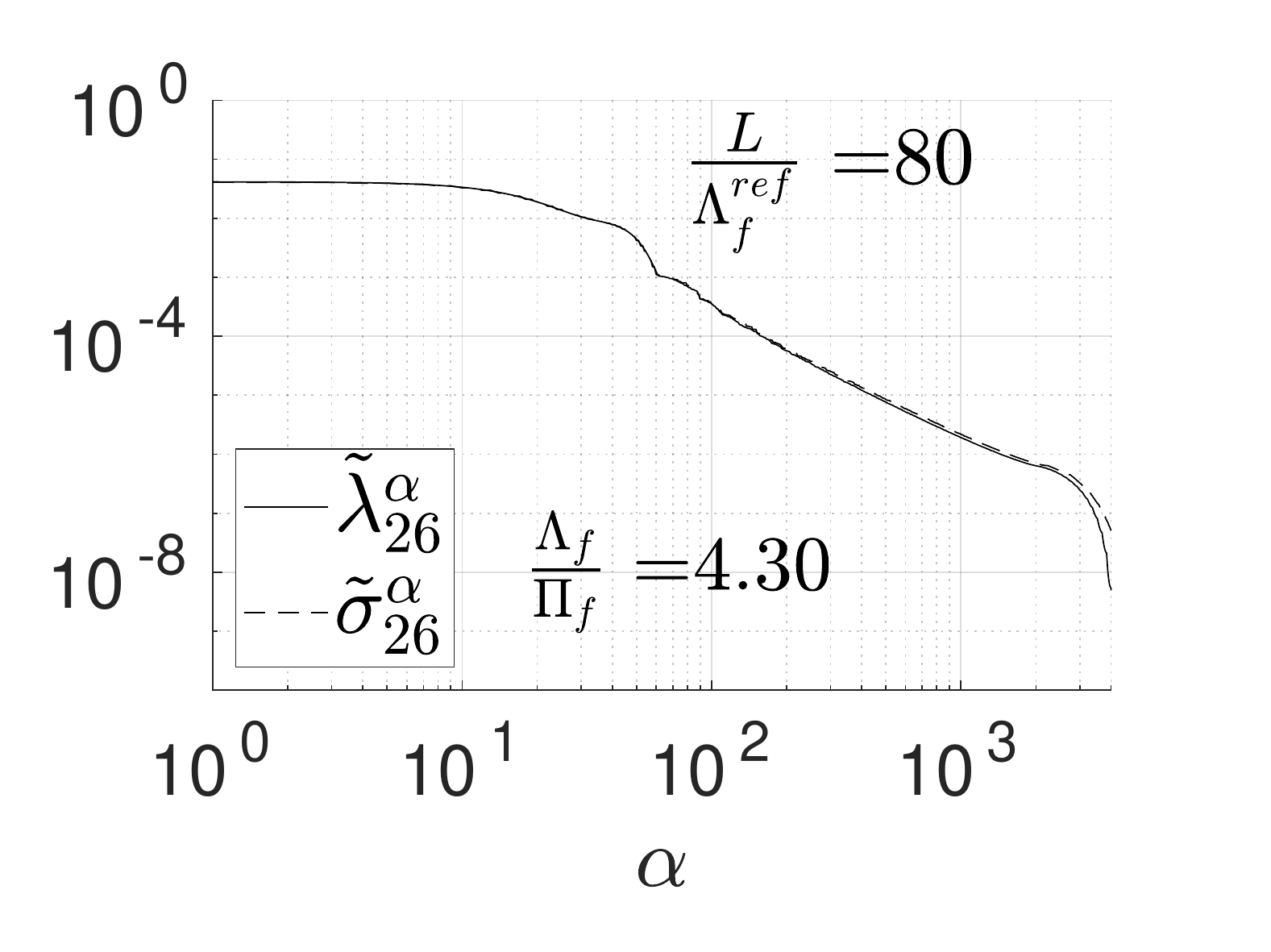}
\caption{\label{fig:e_values_fourier_spectrum_kernel_2_domain_6}}
\end{subfigure}
\begin{subfigure}[h]{0.32\textwidth}
\includegraphics[width=\textwidth]{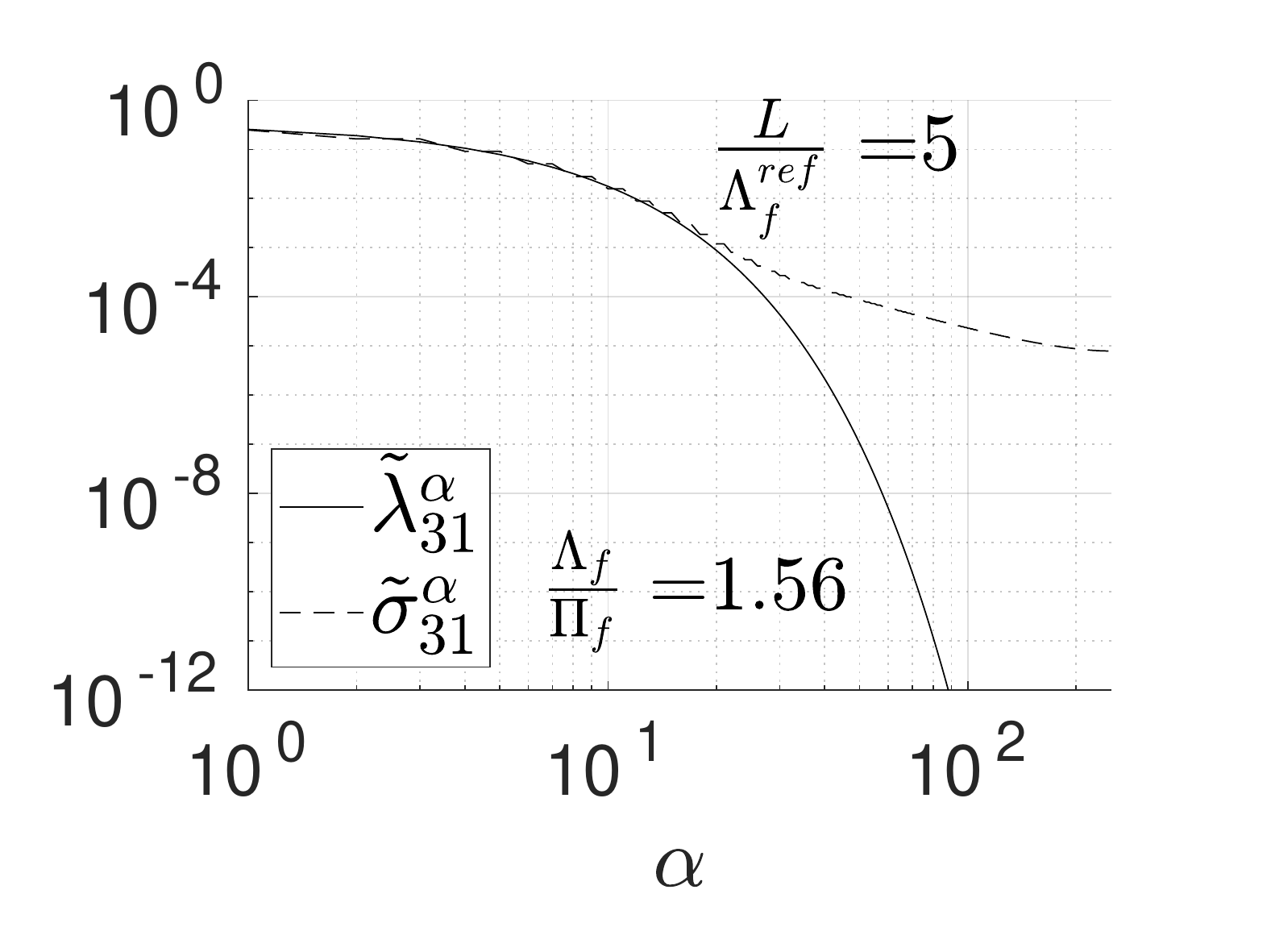}
\caption{\label{fig:e_values_fourier_spectrum_kernel_3_domain_1}}
\end{subfigure}
\begin{subfigure}[h]{0.32\textwidth}
\includegraphics[width=\textwidth]{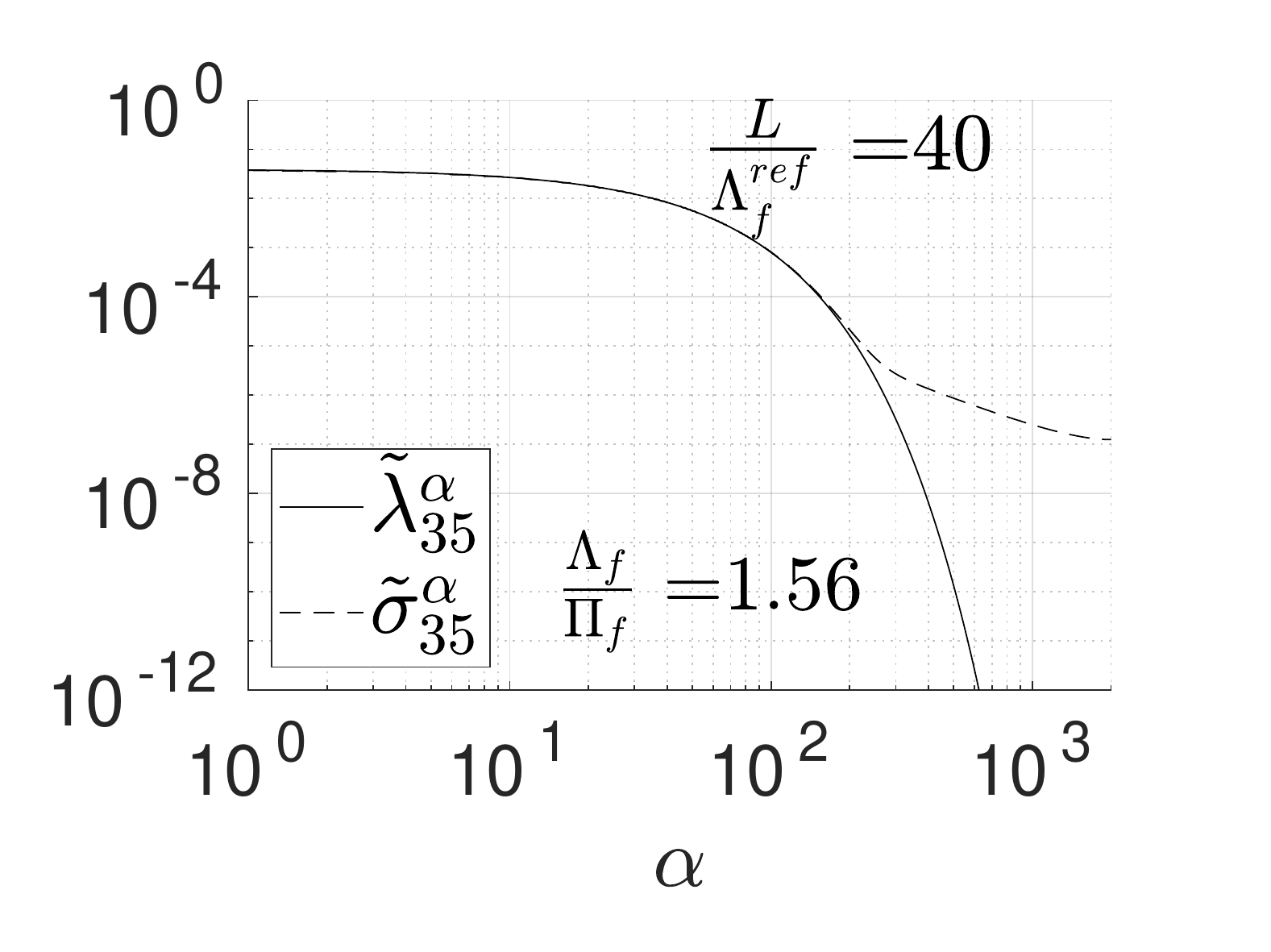}
\caption{\label{fig:e_values_fourier_spectrum_kernel_3_domain_4}}
\end{subfigure}
\begin{subfigure}[h]{0.32\textwidth}
\includegraphics[width=\textwidth]{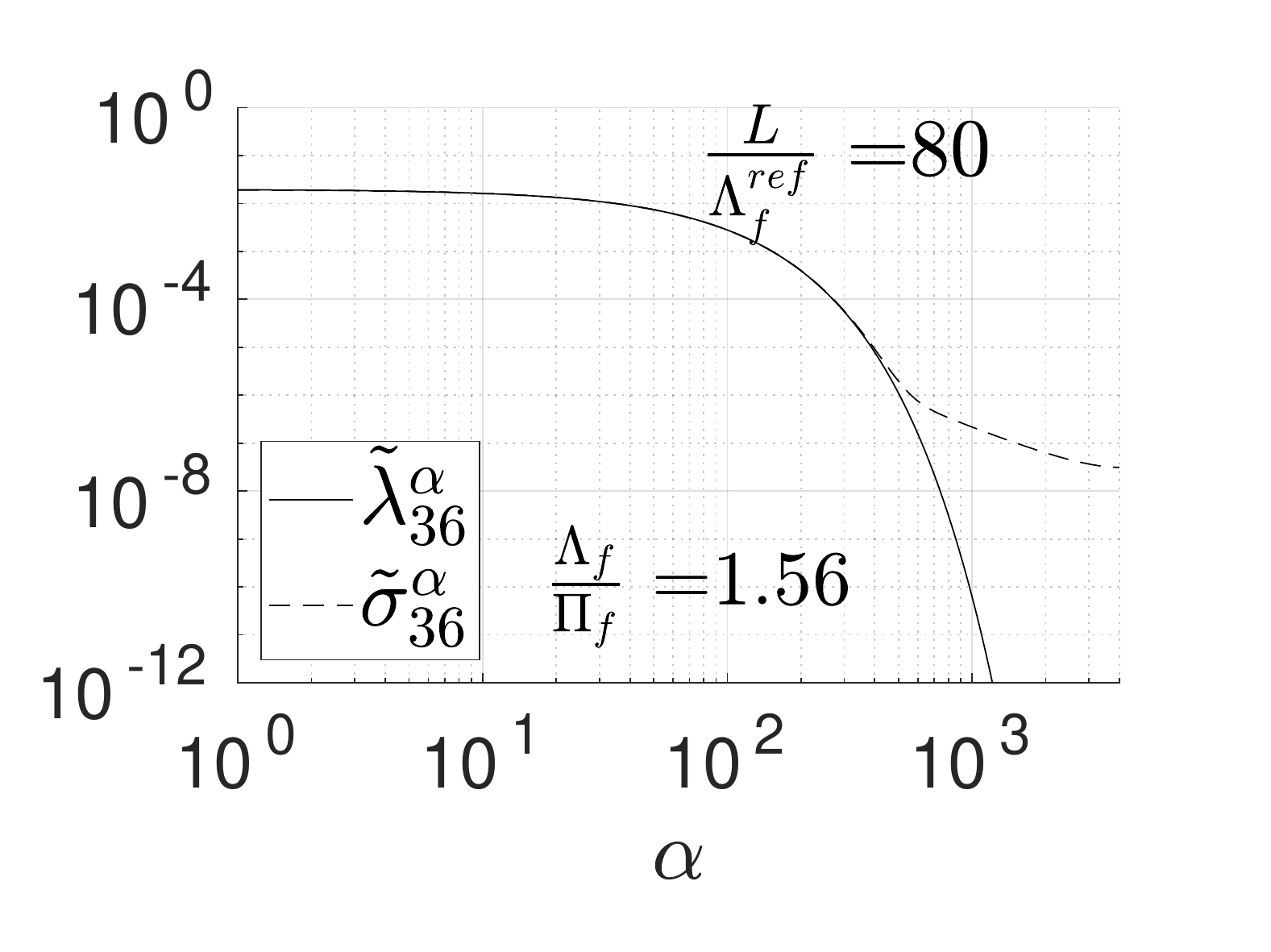}
\caption{\label{fig:e_values_fourier_spectrum_kernel_3_domain_6}}
\end{subfigure}
\begin{subfigure}[h]{0.32\textwidth}
\includegraphics[width=\textwidth]{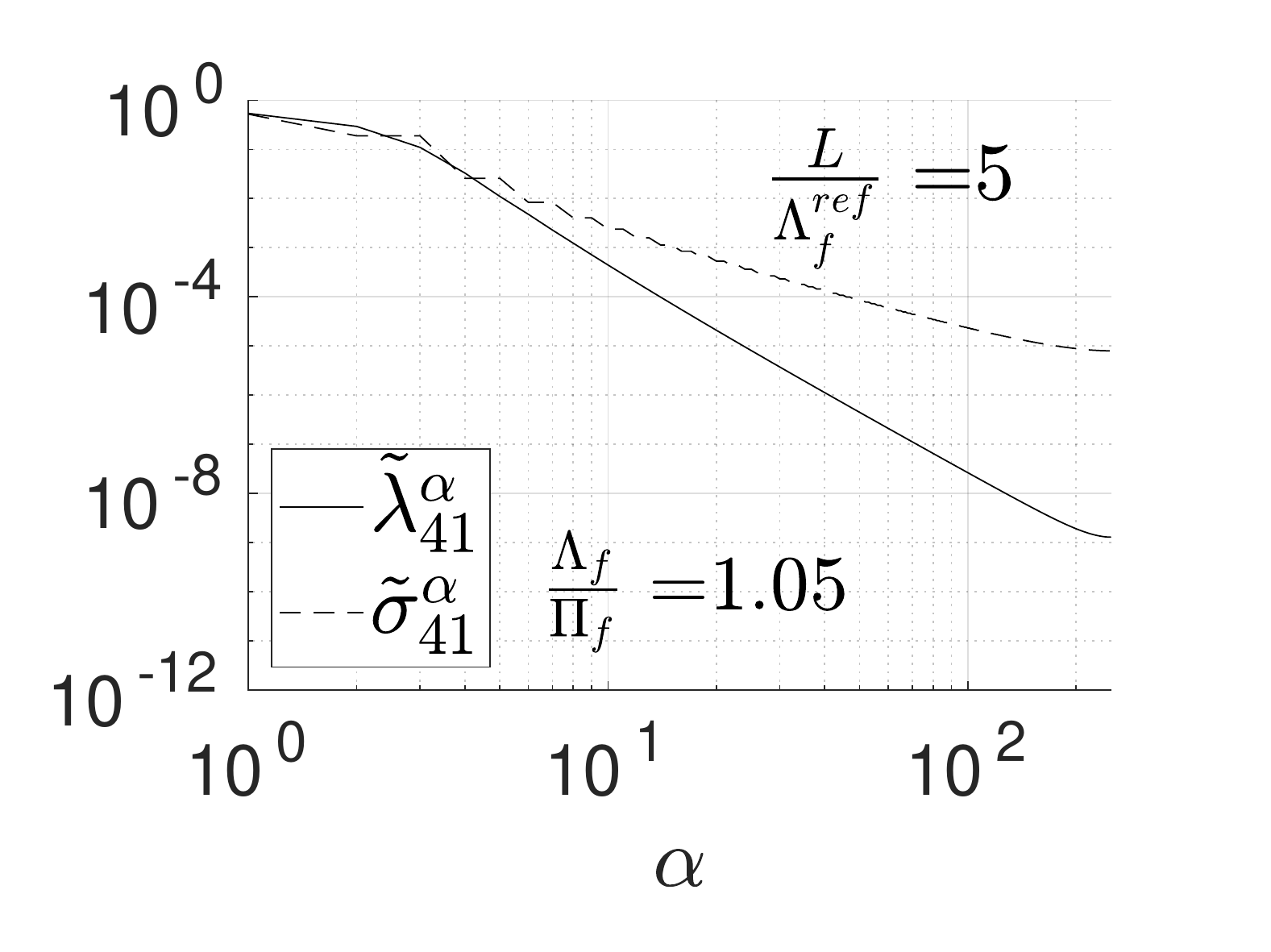}
\caption{\label{fig:e_values_fourier_spectrum_kernel_4_domain_1}}
\end{subfigure}
\begin{subfigure}[h]{0.32\textwidth}
\includegraphics[width=\textwidth]{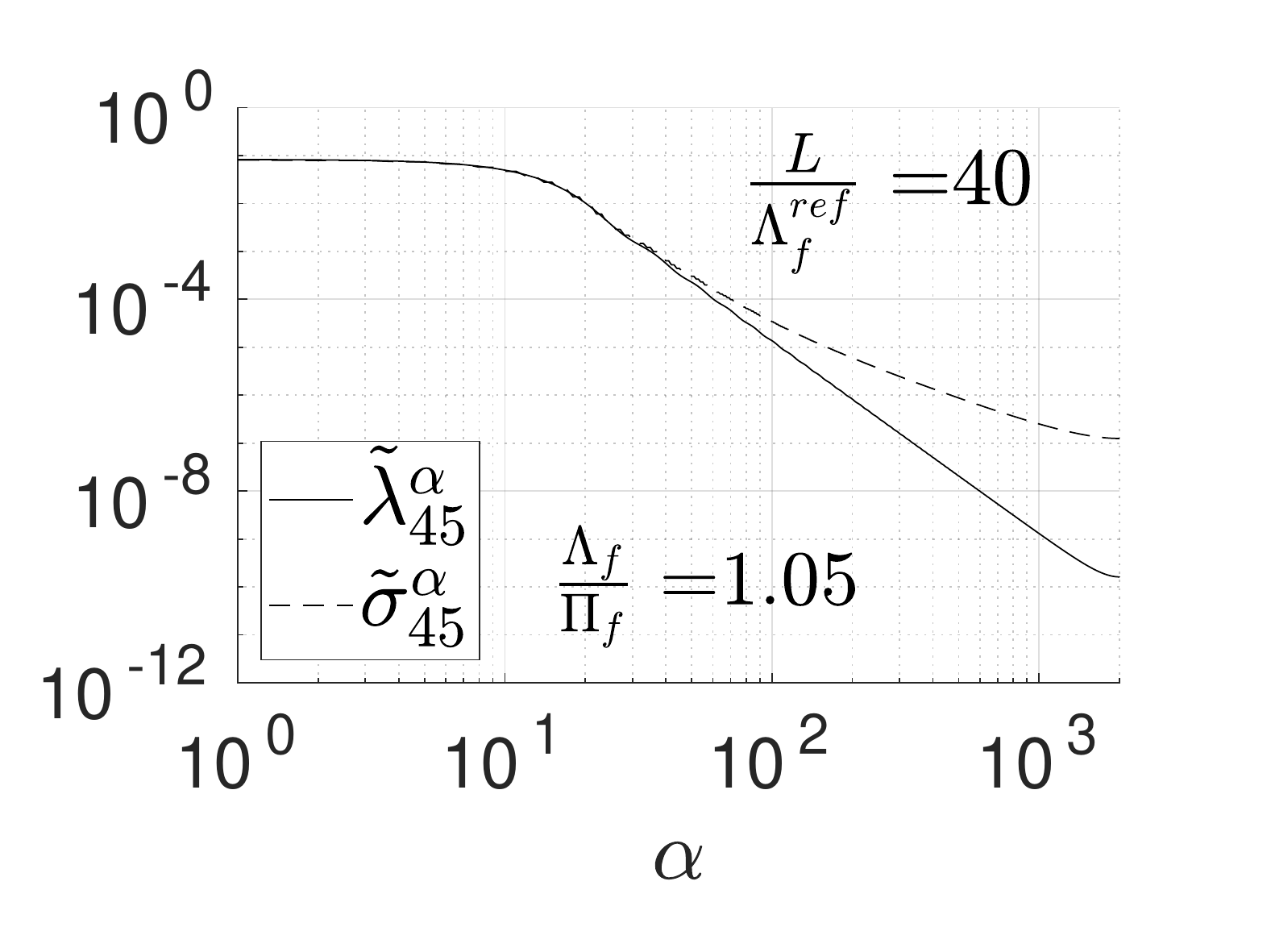}
\caption{\label{fig:e_values_fourier_spectrum_kernel_4_domain_4}}
\end{subfigure}
\begin{subfigure}[h]{0.32\textwidth}
\includegraphics[width=\textwidth]{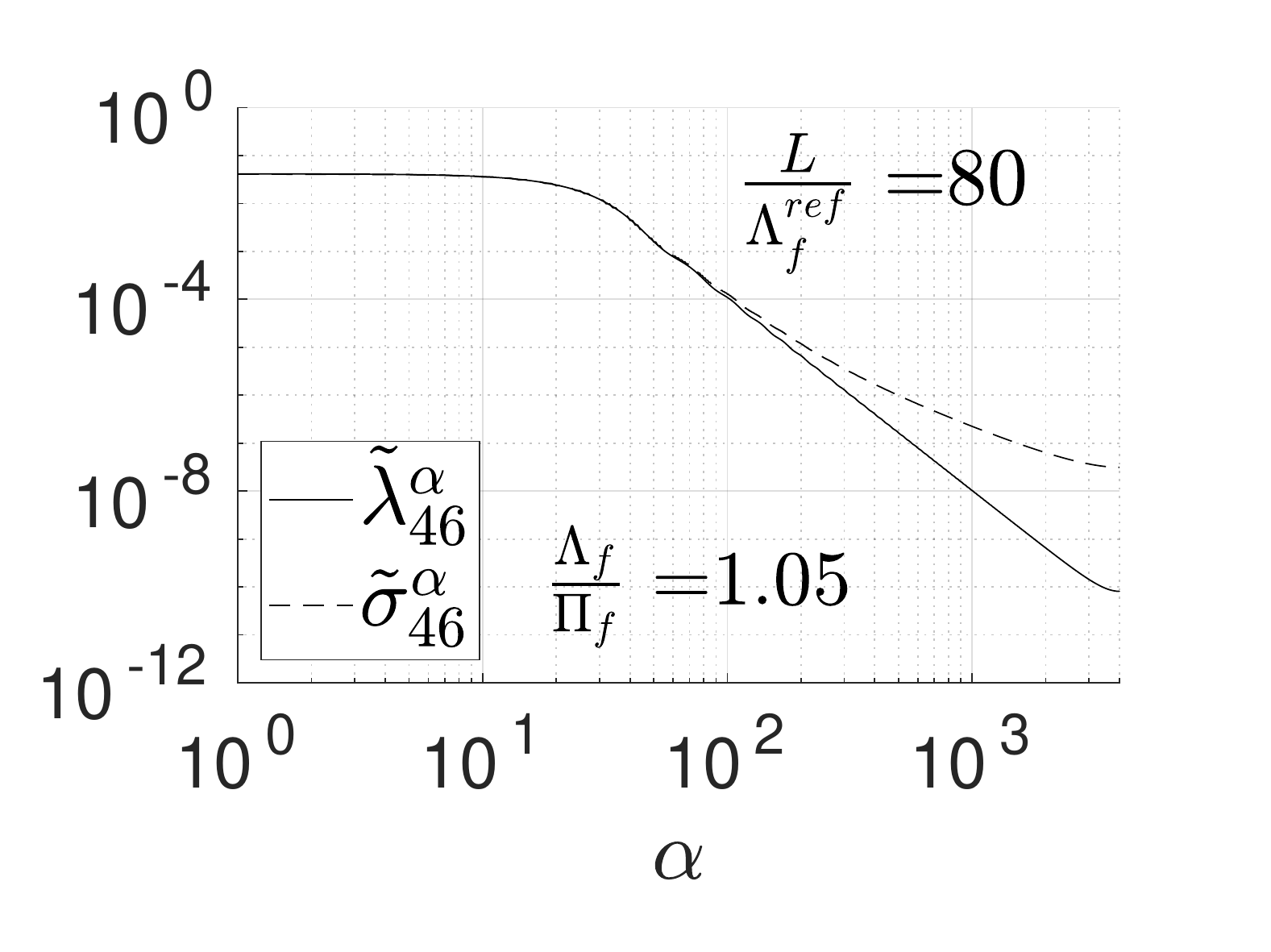}
\caption{\label{fig:e_values_fourier_spectrum_kernel_4_domain_6}}
\end{subfigure}

\begin{subfigure}[h]{0.32\textwidth}
\includegraphics[width=\textwidth]{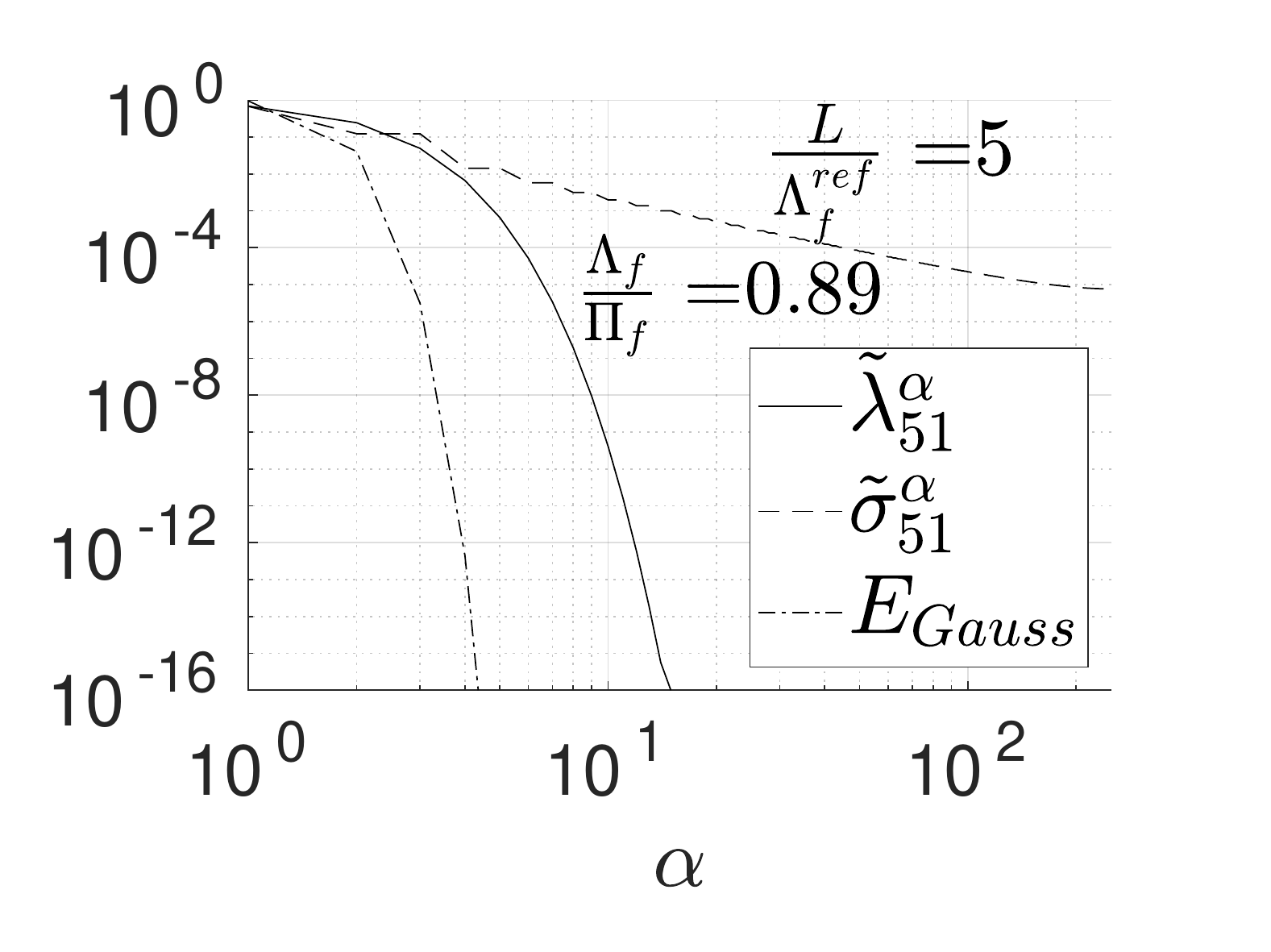}
\caption{\label{fig:e_values_fourier_spectrum_kernel_4_domain_1}}
\end{subfigure}
\begin{subfigure}[h]{0.32\textwidth}
\includegraphics[width=\textwidth]{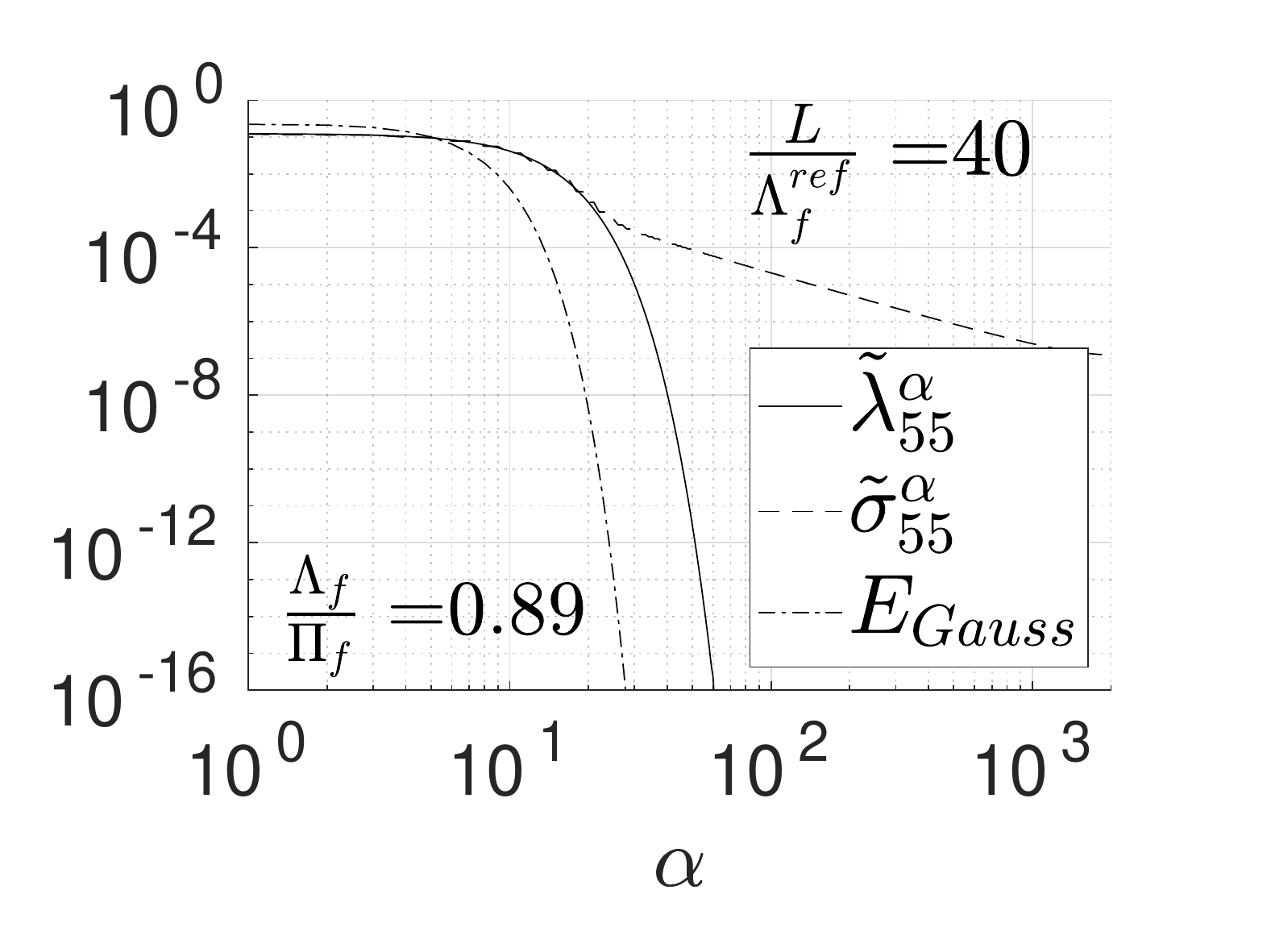}
\caption{\label{fig:e_values_fourier_spectrum_kernel_4_domain_4}}
\end{subfigure}
\begin{subfigure}[h]{0.32\textwidth}
\includegraphics[width=\textwidth]{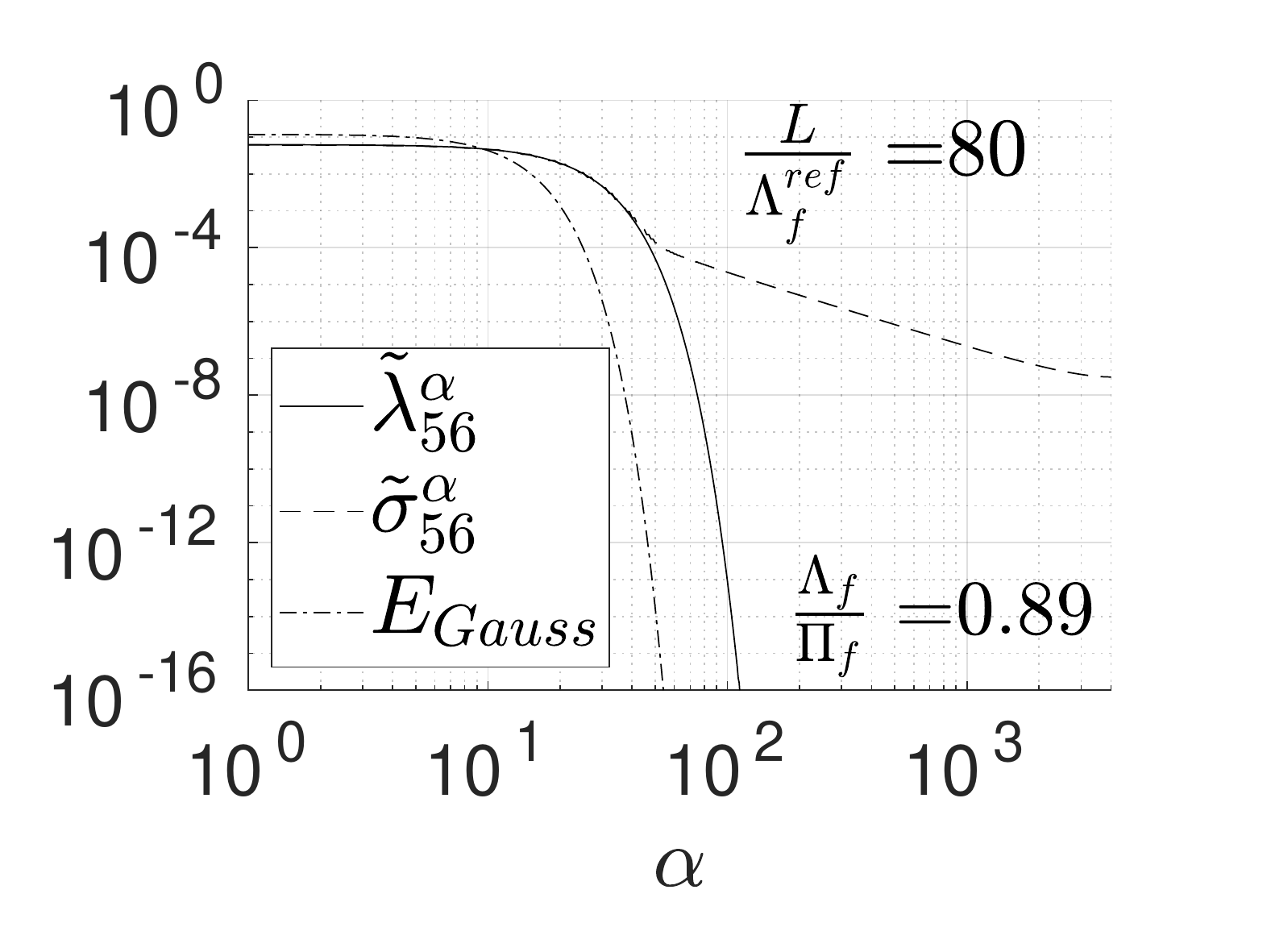}
\caption{\label{fig:e_values_fourier_spectrum_kernel_4_domain_6}}
\end{subfigure}
\caption{(a)-(f): Normalized eigenvalues, $\tilde{\lambda}_{ij}^\alpha$, and sorted Fourier spectra, $\tilde{\sigma}_{ij}^\alpha$, related to the kernels $K_{ij}$, $i\in[1:5]$, for domain lengths corresponding to $j\in[1,5,6]$.\label{fig:e_values_fourier_spectrum_all_kernels}}
\end{figure}
\FloatBarrier
\noindent
with the number of discretization intervals, $N_j$, used for the discretization 
\begin{equation}
\Delta x = \Delta y = \frac{L_{\Omega_j}}{N_j}=\frac{1}{50}\,,\, j\in[1:6]\,,
\end{equation}
with the grid points for each domain defined as
\begin{equation}
x_{n,j}-x_0 = y_{n,j}-y_0 = n\Delta x\,,\,n\in[0:N_j-1]\,,\,j\in[1:6]\,.
\end{equation}
The kernels can be categorized into two distinct groups: those characterized by their support (for a fixed $y$) equalling the domain, and those whose support is strictly smaller than the domain. The former group consists of the kernels $K_{1j}$, $K_{3j}$, and $K_{5j}$ and the latter consists of $K_{2j}$ and $K_{4j}$ for $j>1$. Table \ref{tab:micro_scale_2} shows 

\begin{table}[t]
\begin{center}
\begin{minipage}{1\textwidth}
\caption{Taylor macro and micro scales related to $K_{i6}$, $i\in[1:5]$ as a function of $\nu$ obtained from numerical integrations.}\label{tab:micro_scale_2}%
\centering
\begin{tabular}{@{}lccccc@{}}
\toprule
$i$ & $1$ & $2$ & $3$ & $4$ & $5$\\
\midrule
$\Lambda_f$ & $1.00$ & $1.66$ & $0.78$ & $1.66$  & $2.51$  \\
$\Pi_f$ &  $0.14$ & $ 0.39 $ & $0.50$  & $1.58$ & $2.83$ \\ 
$\Lambda_f/\Pi_f$ & $7.04$& $4.30$ & $1.56$ & $1.05$ & $0.89$\\
\botrule
\end{tabular}
\end{minipage}
\end{center}
\end{table}
\noindent
the Taylor macro/micro scale values related to each correlation function, $K_{ij}$. The five kernels \eqref{eq:kernel_exp}-\eqref{eq:kernel_gauss} are expressed in matrix form as Toeplitz matrices, denoted as $\textbf{K}_{ij}\in\mathbb{R}^{N_j\times N_j}$, where the kernel type is denoted by the subscript $i$ (corresponding to \eqref{eq:kernel_exp}-\eqref{eq:kernel_gauss}), in order to represent discretized versions of \textit{locally translationally invariant} kernels for all domains $\Omega_j$, $j\in[1:6]$. The discretized kernels are seen in Figure \ref{fig:correlation_functions}.  As an operator, the kernel matrices are defined as $\textbf{K}_{ij}:\mathbb{C}^{N_j}\rightarrow\mathbb{C}^{N_j}$, $i\in[1:5]$, $j\in[1:6]$ where the corresponding eigenvalues, $\lambda^{\alpha}_{ij}\in\mathbb{R}_+$, and eigenvectors, $\varphi^{\alpha}_{ij}\in\mathbb{R}^{N_j}$, were obtained numerically for every kernel-domain combination from the following set of equations
\begin{equation}
\textbf{K}_{ij} \varphi^{\alpha}_{ij} = \lambda^{\alpha}_{ij}\varphi^{\alpha}_{ij}\,,\,\alpha \in [1:N_j]\,,\quad i\in[1:5]\,,\,j\in[1:6]\,,
\end{equation}
using the MATLAB function \textit{eig}. The normalized discrete $m$-th Fourier mode related to the $j$-th domain, $\psi^{m}_j\in\mathbb{C}^{N_j}$, is defined as 
\begin{equation}
\psi^{m}_j = N_j^{-\frac{1}{2}}\sum_{n=1}^Ne^{2\pi i (m-1)(n-1)/N_j}\hat{e}_n\,,\quad m\in[1:N_j]\,,\,j\in[1:6]\,.
\end{equation}
The Fourier spectra are then obtained from
\begin{equation}
\sigma_{ij}^m = \left(\textbf{K}_{ij}\psi^m_j,\psi^m_j\right)\,,\quad m\in[1:N_j]\,,\,j\in[1:6]\,,\label{eq:sigma_ij}
\end{equation}
and the normalized spectra are denoted by a tilde over the variable, i.e. $\tilde{\lambda}^{\alpha}_{ij}$ and $\tilde{\sigma}^{\alpha}_{ij}$, analogously to \eqref{eq:lambda_tilde}-\eqref{eq:sigma_tilde}.

Table \ref{tab:micro_scale_2} shows that the MMSR decreases with increasing index, $i$. Based on the previously demonstrated strong correlation between MMSR and spectral discrepancies, the closest match between the Fourier and eigenspectra are expected to occur for \eqref{eq:kernel_exp} and \eqref{eq:kernel_modified}, with \eqref{eq:kernel_wendland} and \eqref{eq:kernel_gauss} expected to exhibit the most significant discrepancies. Figure \ref{fig:e_values_fourier_spectrum_all_kernels} shows the normalized Fourier and eigenvalue spectra of all kernels, \eqref{eq:kernel_exp}-\eqref{eq:kernel_gauss}, and for the domains $\Omega_j\,,j=[1,4,6]$. For the kernel, \eqref{eq:kernel_exp}, with an MMSR of $7.04$ the two sets of spectra in figures (a)-(c), appear to be in very good agreeement for all domain sizes - even for the smallest domain corresponding to $L = 5\Lambda_f$. It is worth noting that the spectra are in fact not equal, despite the fact that they appear to collapse for the case of the largest domain. This deviation will be analyzed in more detail in Section \ref{sec:fourier_reconstruction_of_Eigenspectra}. As the MMSR is decreased, the tail ends of the Fourier and eigenspectra are seen to deviate from each other, as was observed in Figure \ref{fig:e_values_fourier_spectrum_bessel} for the Bessel function-generated kernels. In this context, it is interesting to compare the results for kernels  $K_{2j}$ with $\Lambda_f/\Pi_f=4.30$ and $K_{4j}$ where $\Lambda_f/\Pi_f=1.05$. Despite the substantial differences in the appearance of these two correlation functions (see Figure \ref{fig:correlation_functions}) the more complex looking ones characterized by a sinusoid-generated lobed shape, $K_{2j}$, exhibit the smallest spectral discrepancies - which may appear somewhat surprising. However, the smaller spectral discrepancies of $K_{2j}$ compared to $K_{4j}$ are nevertheless well predicted by the larger MMSR of the former.

The results for the similarity solution for the case of isotropic homogeneous decaying turbulence, $K_{5j}$, from \cite{karman1938} are seen in Figures \ref{fig:e_values_fourier_spectrum_all_kernels} (m)-(o), showing the most significant deviations at the low energy end of the spectrum. The fact that this case represents a low Reynolds number turbulent flow solution for the correlation function is naturally indicated by the small MMSR, $\Lambda_f/\Pi_f=0.89$. This case supports the notion that the correlation between spectral descrepancies and the MMSR holds for correlation functions satisfying the Navier-Stokes equations, which the former correlation functions are not guaranteed to do. Although the asymptotic trends of the eigenspectrum in this case are seen to deviate significantly from the DFT spectra - even for the largest domain - the asymptotic trends between the eigenspectrum and the analytical Fourier spectrum of the corresponding Gaussian are still in very good agreement.

Given that the ratio between the compact support of the kernels and domain lengths, $L_{\Omega_j}$, $j\in[1:6]$, varies for the kernels \eqref{eq:kernel_exp}-\eqref{eq:kernel_gauss}, it is observed that the discrepancy variations between the spectra are consistently well predicted by the MMSR for all kernels. This shows that the support of the various kernels is not, by itself, the decisive factor for the deviation seen between the spectra and reveals that if the SPOD is viewed as an approximation of the POD, the quality of the approximation may be reduced significantly for low Reynolds number turbulent flows for which MMSR approaches unity. For high Reynolds number flows, however, the spectral discrepancies can be expected to be less profound. 

In the following, a more detailed inspection of the relation between Fourier and eigenspectra will be performed. This includes a convergence study of the spectral discrepancies with respect to increasing domain and thereby an inspection of whether we can expect POD modes to converge to Fourier modes for increasing domain sizes.

\FloatBarrier
\subsection{Divergence between POD and Fourier modes \label{sec:fourier_reconstruction_of_Eigenspectra}}
\begin{figure}[t]
\centering
\begin{subfigure}[h]{0.45\textwidth}
\includegraphics[width=\textwidth]{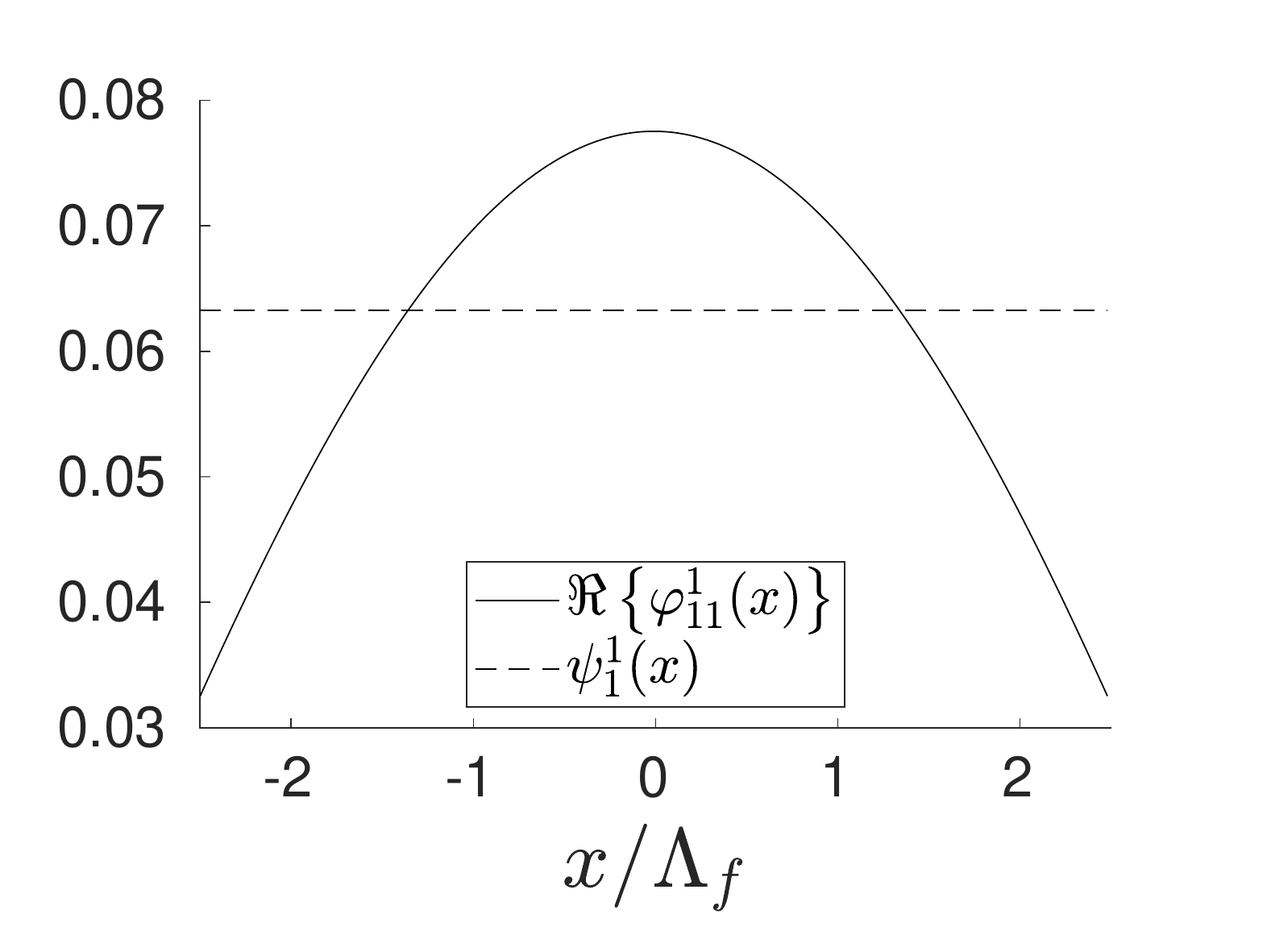}
\caption{\label{fig:modes_domain_1}}
\end{subfigure}
\begin{subfigure}[h]{0.45\textwidth}
\includegraphics[width=\textwidth]{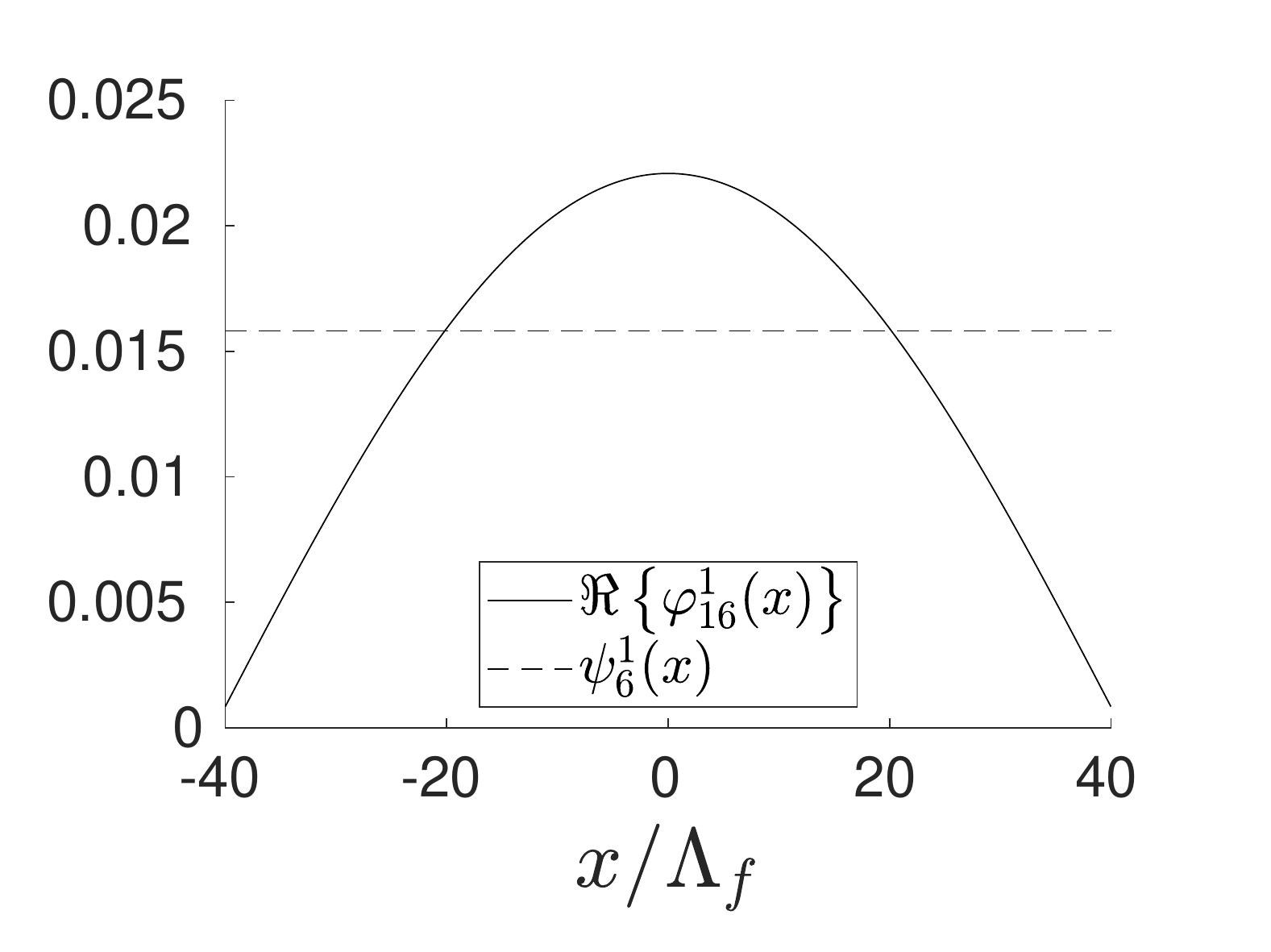}
\caption{\label{fig:modes_domain_6}}
\end{subfigure}
\begin{subfigure}[h]{0.45\textwidth}
\includegraphics[width=\textwidth]{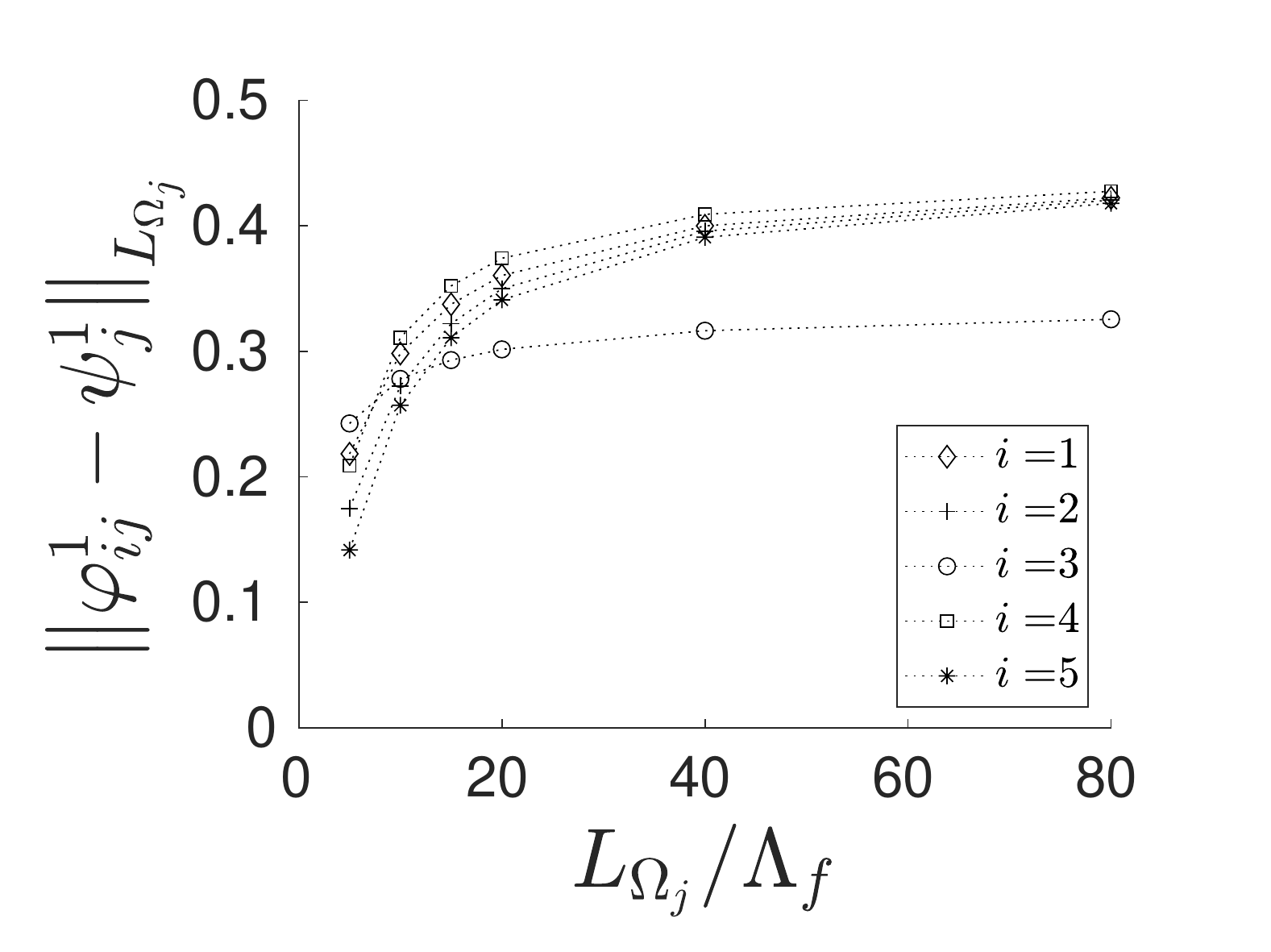}
\caption{\label{fig:resnorm_0}}
\end{subfigure}
\caption{Comparisons between the first POD mode and the zeroth harmonic for $K_{1j}(a):$ as a function of domain size. (a): real part of first POD mode and zeroth harmonic for $L_{\Omega_1}=5\Lambda_f$, (b): real part of first POD mode and zeroth harmonic for $L_{\Omega_1}=80\Lambda_f$, (c): residual norm between the first POD modes of all kernels, $\varphi^1_{ij}\,,i\in[1:5]\,,\,j\in[1:6]$, and the corresponding zeroth harmonics as a function of domains lengths $L_{\Omega_j}$, $j\in[1:6]$.\label{fig:first_mode_comparison}}
\end{figure}

Preceeding the analysis of discrepancies between POD and Fourier spectra, a comparison of POD and Fourier modes may be performed. The first POD mode of \eqref{eq:kernel_exp} is shown along with the zeroth Fourier harmonic in Figures \ref{fig:first_mode_comparison} as a function of domain size. The modes are illustrated for visual comparison in Figures (a) and (b) for $L_{\Omega_1} = 5\Lambda_f$, and $L_{\Omega_6} = 80\Lambda_f$, respectively, where only the real part of the POD modes is shown given that the imaginary part is negligible. These results illustrate that the first POD mode fails to approach the zeroth harmonic as the domain length is increased. This is quantified by the residual norm shown in Figure \ref{fig:resnorm_0} between the first POD mode and the zeroth harmonic. This demonstrates diverging behaviour between the modes as a function of domain size, and is part of a more general trend of divergence observed for all of the kernels, \eqref{eq:kernel_exp}-\eqref{eq:kernel_gauss}. Based on these discrepancies, the coupling between the Fourier and eigenspectra as a function of domain size cannot be expected converge pointwise, as will be demonstrated in the following.
%
\noindent

We now consider the Fourier building blocks of the eigenspectrum according to \eqref{eq:psi_expansion_of_lambda}. The main focus of the analysis is the reconstruction of eigenspectra related to the kernels $K_{1j}$ and $K_{4j}$ as an extension of the analysis in the previous Section. These kernels are chosen as they exhibited significant variations of convergence rates in terms of their spectra (see Figure \ref{fig:e_values_fourier_spectrum_all_kernels}), and in terms of the collapse of their Fourier and eigenspectra. The contributions of various Fourier modes to the reconstruction of eigenvalues can be quantified by the measure
\begin{equation}
\Gamma^{\alpha mn}_{ij}=\Re\left\{\frac{c^{\alpha,n}_{ij}c^{\alpha,m*}_{ij}\left(\textbf{K}_{ij}\psi^n_j,\psi^m_j\right)}{\lambda^\alpha_{ij}}\right\}\,,\label{eq:Gamma}
\end{equation}
since the corresponding imaginary part to the above is negligible in the current cases. In the special case that  $\psi^n_j=\varphi^n_{ij}$ for some $i$ in \eqref{eq:Gamma}, we must have that ${\Gamma^{\alpha mn}_{ij} = \delta_{mn}}$. If the POD and Fourier modes are not the same, however, $\Gamma^{\alpha mn}_{ij}$ can potentially have non-zero values for all index combinations. This would mean that all Fourier modes spanning the Hilbert space in question contribute to the reconstruction of all the eigenvalues. In the following, $\Gamma^{\alpha mn}_{ij}$ will be investigated for certain $\alpha$-values as a function of $m$ and $n$ in order to illustrate some general tendencies that generally arise for all kernels, given the relatively high dimensionality of $\Gamma^{\alpha mn}_{ij}$.

\begin{figure}[t]
\centering
\begin{subfigure}[h]{0.49\textwidth}
\includegraphics[width=\textwidth]{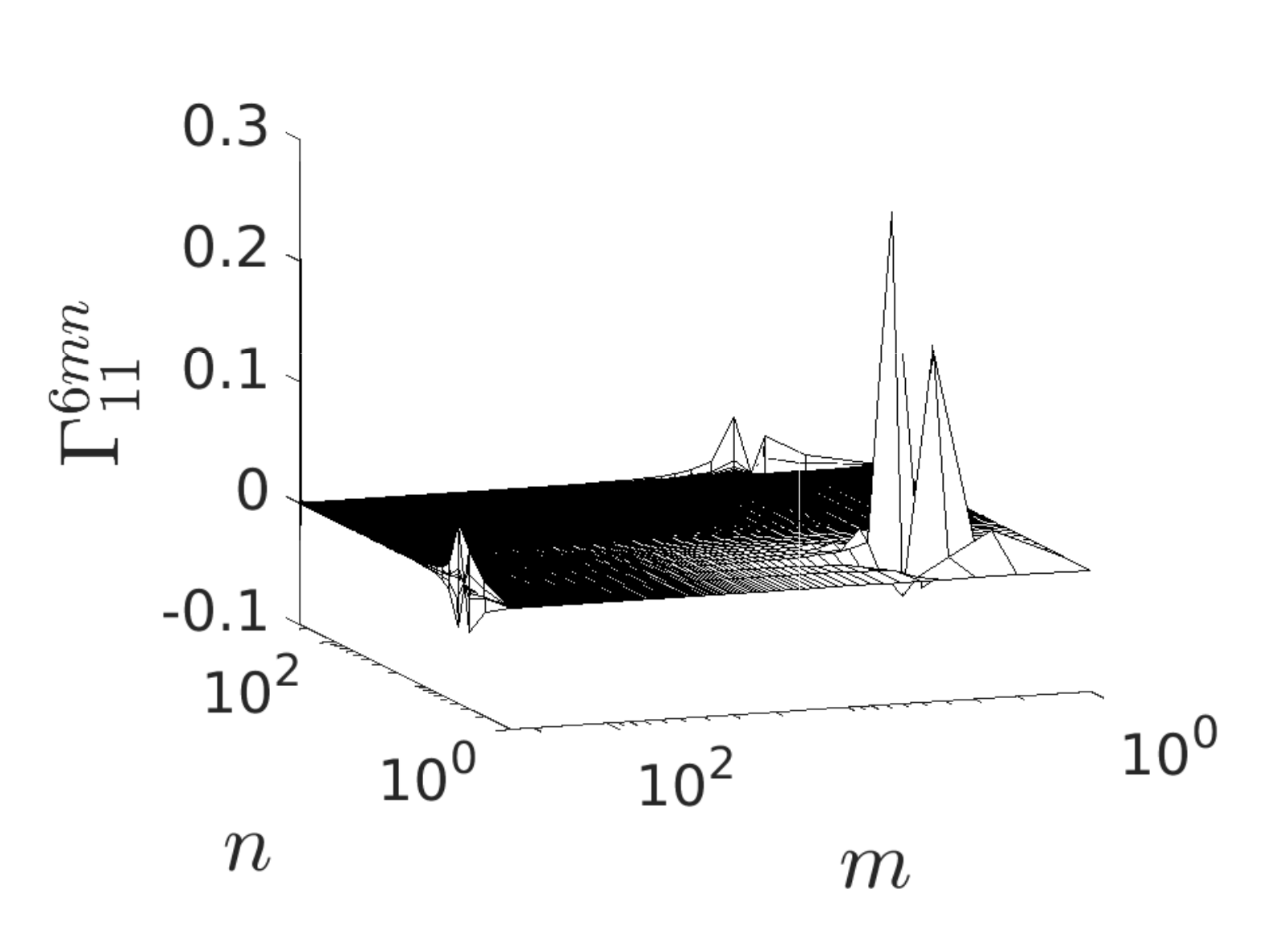}
\caption{\label{fig:evalue_reconstruction_domain_1_kernel_1_alpha_6}}
\end{subfigure}
\begin{subfigure}[h]{0.49\textwidth}
\includegraphics[width=\textwidth]{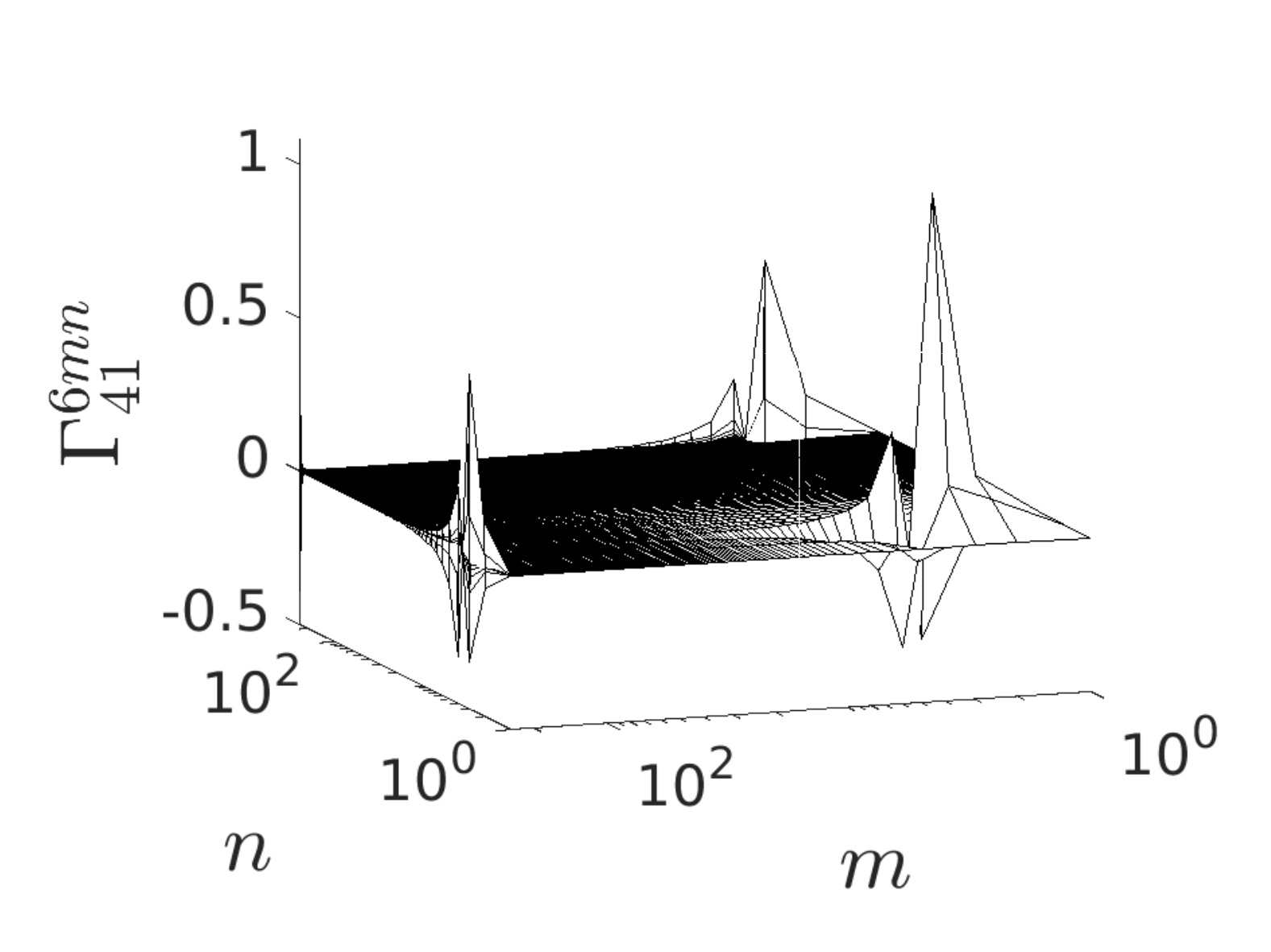}
\caption{\label{fig:evalue_reconstruction_domain_1_kernel_4_alpha_6}}
\end{subfigure}
\caption{Relative contributions, $\Gamma^{6 mn}_{i1}$, to the reconstruction of $\tilde{\lambda}^6_{i1}$ for the domain with length $L_{\Omega_1}=5\Lambda_f$. (a):  Results for $i=1$, (b): results for $i=4$. Multiple Fourier modes are seen to contribute to the reconstruction of the eigenvalue in question. \label{fig:reconstruction_of_eigenvalue_6}}
\end{figure}

Figure \ref{fig:reconstruction_of_eigenvalue_6} shows as an example $\Gamma^{6 mn}_{11}$ and $\Gamma^{6 mn}_{41}$ and thereby the relative Fourier contributions to the reconstructions of $\lambda^6_{11}$ and $\lambda^6_{41}$. We note that the evaluation of \eqref{eq:Gamma} is shown in a double logarithmic representation along $m$ and $n$. Given the symmetries of the Fourier spectrum, there is also a corresponding symmetry \textit{along} the diagonal of $\Gamma^{\alpha mn}_{ij}$ defined by $m=n$, which is not evident in the figures, exhibited for all $m,n\neq 1$. This means that each value of $\Gamma^{\alpha mn}_{ij}$ for $m,n\neq 1$, which is shown in Figure \ref{fig:reconstruction_of_eigenvalue_6}, represents approximately half the reconstructed relative energy of the eigenvalue $\alpha$ of the corresponding complex conjugate Fourier pair. Deviations from a single peak in Figure \ref{fig:reconstruction_of_eigenvalue_6} are obvious and may be expected for a domain length corresponding to merely five integral length scales. Nevertheless, this deviation reveals spectrally the deviation between the Fourier and POD basis given that for both kernels, multiple Fourier modes are needed to reconstruct the given eigenvalue.  Appendix \ref{app:reconstruction_of_eigenspectra_using_Fourier_modes} includes the evaluation of \eqref{eq:Gamma} for all the kernels \eqref{eq:kernel_exp}-\eqref{eq:kernel_gauss} across domain sizes, illustrating that these tendencies do not constitute a special case, but a more general feature for small domain sizes.
\noindent
\begin{figure}[b]
\centering
\begin{subfigure}[h]{0.49\textwidth}
\includegraphics[width=\textwidth]{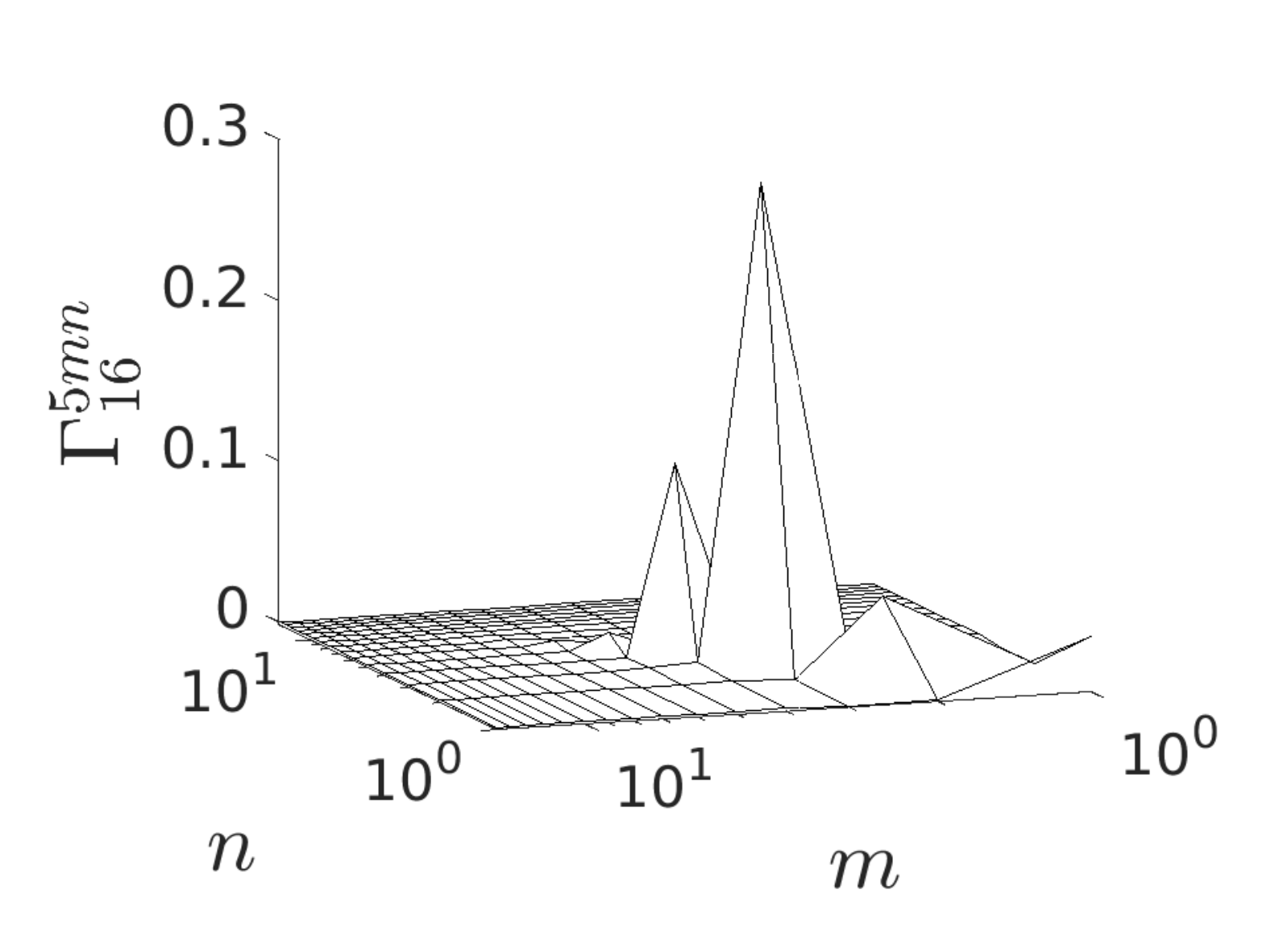}
\caption{\label{fig:evalue_reconstruction_domain_6_kernel_1_alpha_5}}
\end{subfigure}
\begin{subfigure}[h]{0.49\textwidth}
\includegraphics[width=\textwidth]{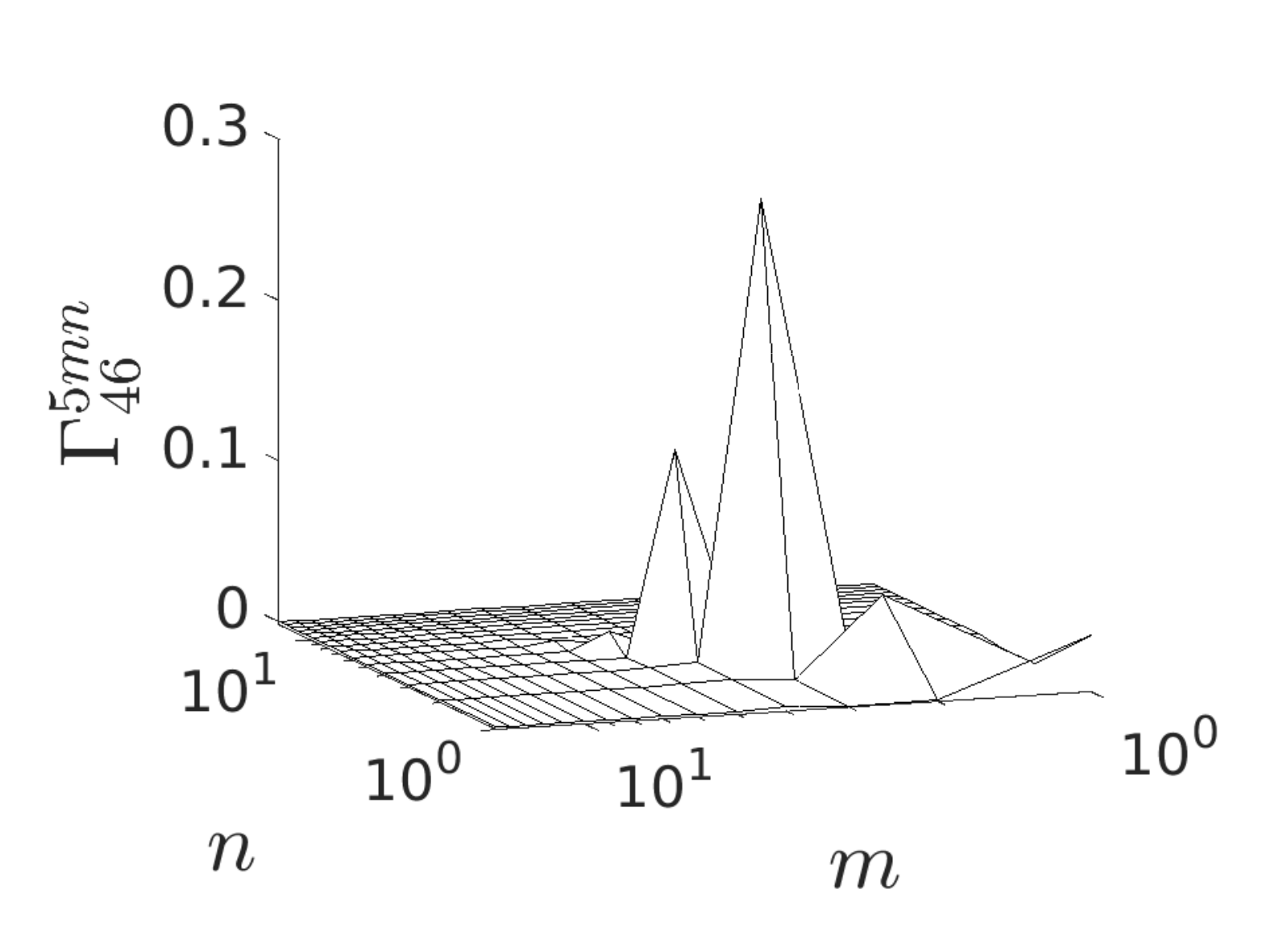}
\caption{\label{fig:evalue_reconstruction_domain_6_kernel_4_alpha_5}}
\end{subfigure}
\caption{Relative contributions, $\Gamma^{5 mn}_{i6}$, to the reconstruction of $\tilde{\lambda}^5_{i6}$ for the domain with length $L_{\Omega_6}=80\Lambda_f$. (a):  Results for $i=1$, (b): results for $i=4$. Multiple modes are seen to contribute to the reconstruction of the eigenvalue in question. \label{fig:reconstruction_of_eigenvalue_5}}
\end{figure}
\noindent
In figures \ref{fig:evalue_reconstruction_domain_6_kernel_1_alpha_1}-\ref{fig:evalue_reconstruction_domain_6_kernel_1_alpha_6}, an alternating pattern between even and odd $\alpha$-values is noted where for even $\alpha$, the reconstruction is dominated by a single Fourier complex conjugate pair. For odd $\alpha$ the reconstruction of these modes is less efficient using the Fourier basis, where approximately $60\%$ of the energy of the given eigenvalue is reconstructed by a single conjugate pair. It is worth noting that since $\sum_{m,n}\Gamma^{\alpha mn}_{ij} = 1$ for all $\alpha$, the existence of $\Gamma^{\alpha mn}_{ij}>0.5$, $m,n\neq 1$, is compensated for by negative contributions to the reconstruction of eigenvalues from the set of remaining Fourier modes. 
\begin{figure}[h]
\centering
\begin{subfigure}[h]{0.32\textwidth}
\includegraphics[width=\textwidth]{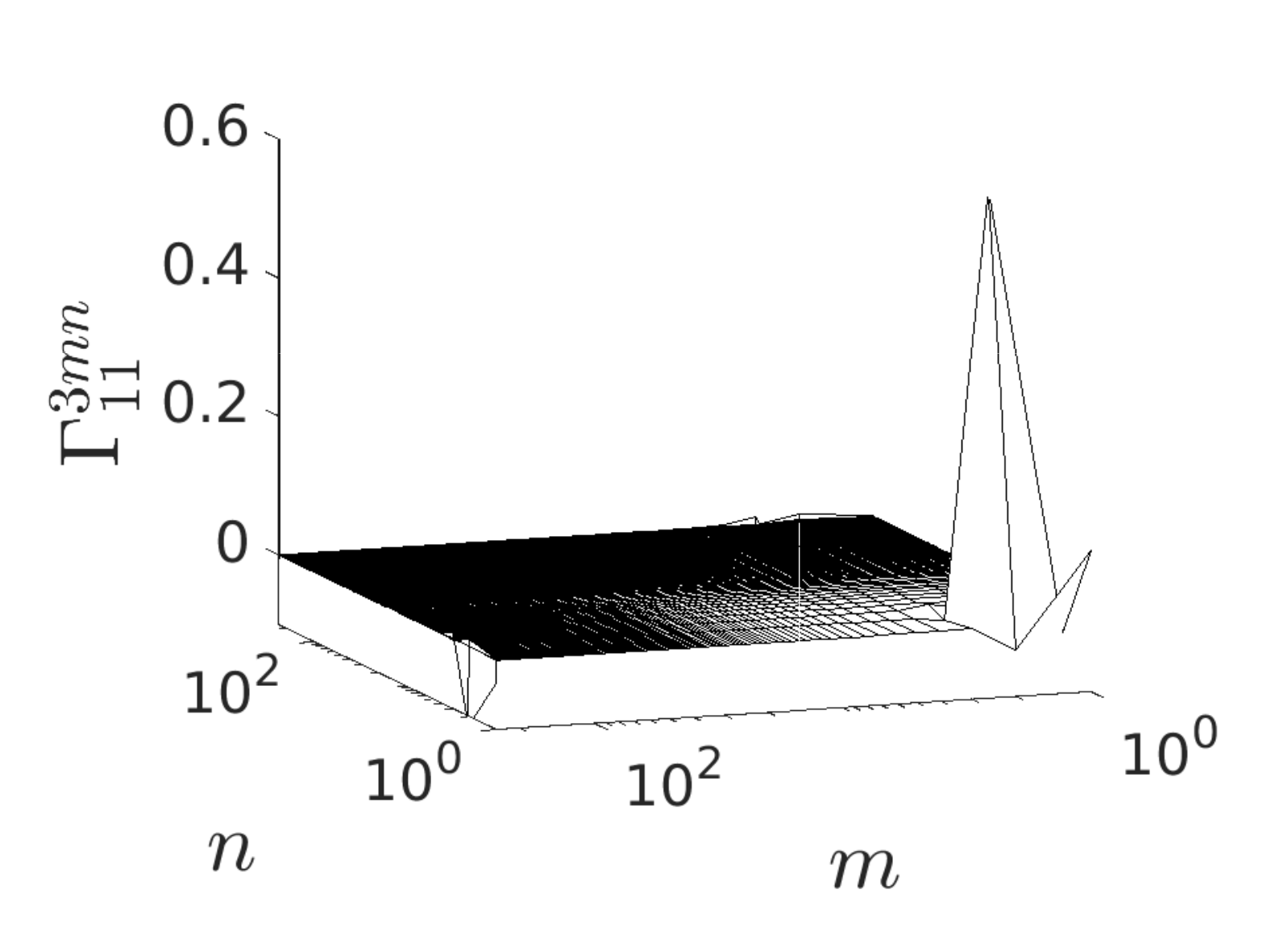}
\caption{\label{fig:evalue_reconstruction_domain_1_kernel_1_alpha_3}}
\end{subfigure}
\begin{subfigure}[h]{0.32\textwidth}
\includegraphics[width=\textwidth]{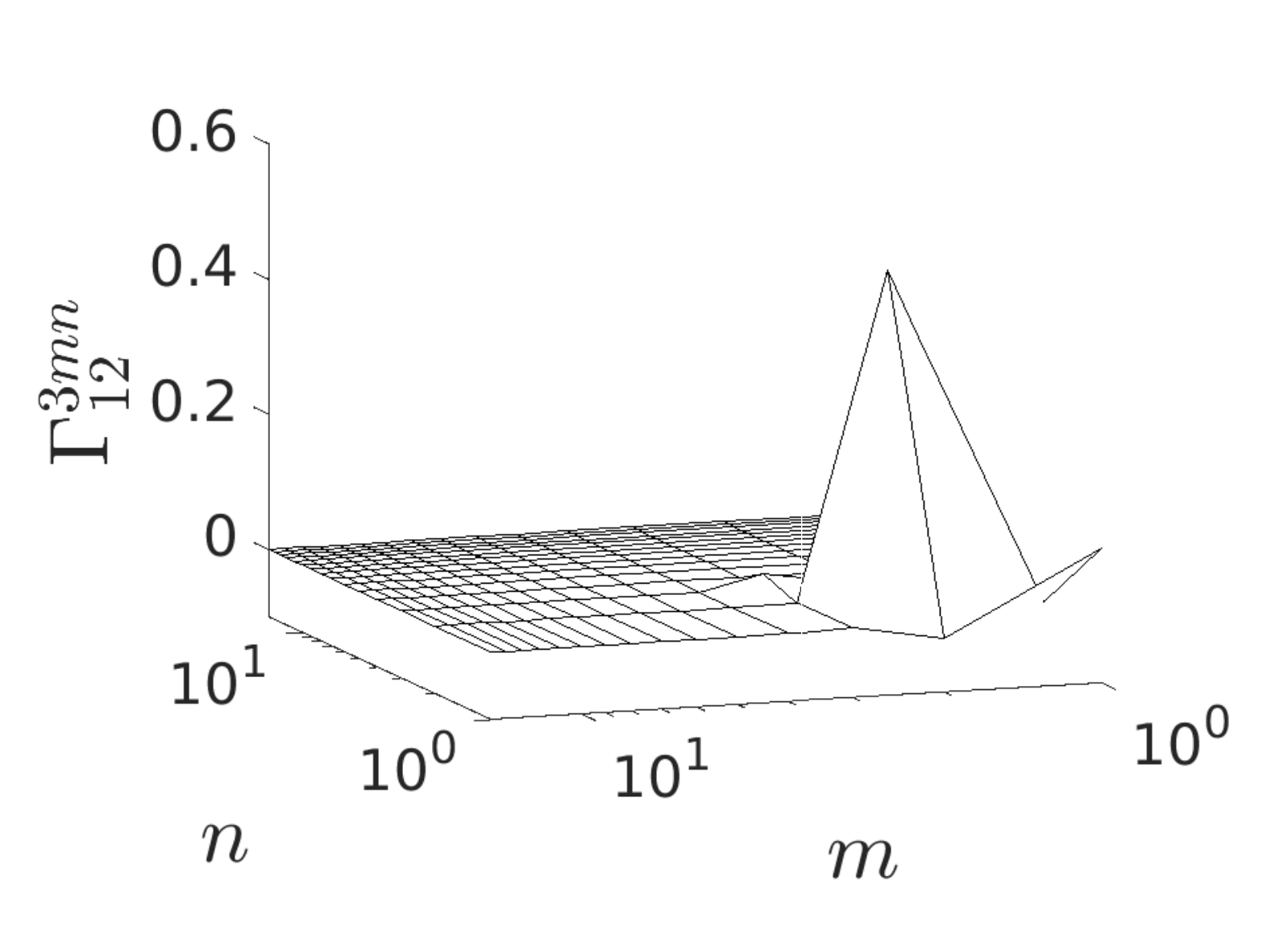}
\caption{\label{fig:evalue_reconstruction_domain_2_kernel_1_alpha_3}}
\end{subfigure}
\begin{subfigure}[h]{0.32\textwidth}
\includegraphics[width=\textwidth]{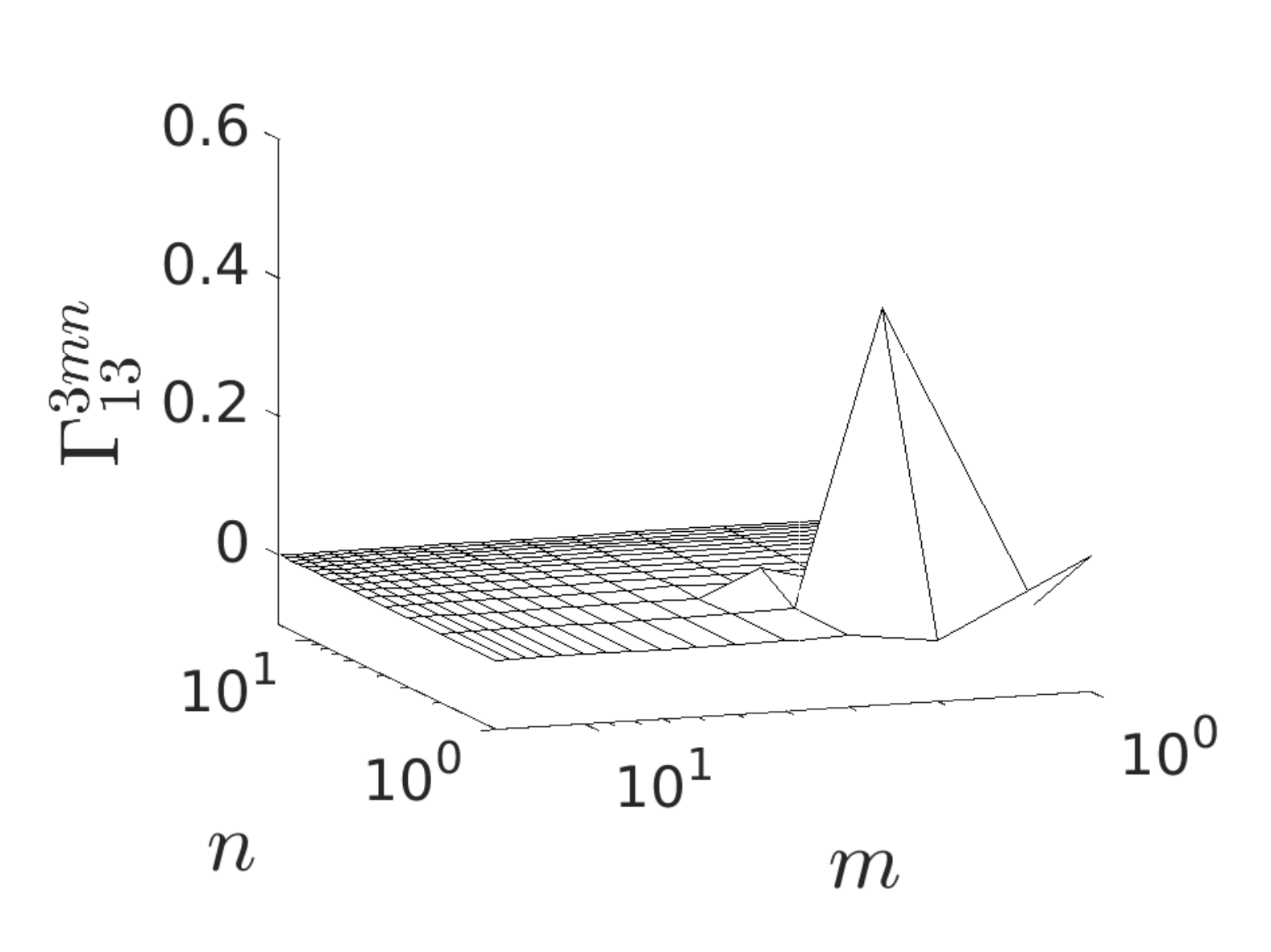}
\caption{\label{fig:evalue_reconstruction_domain_3_kernel_1_alpha_3}}
\end{subfigure}
\begin{subfigure}[h]{0.32\textwidth}
\includegraphics[width=\textwidth]{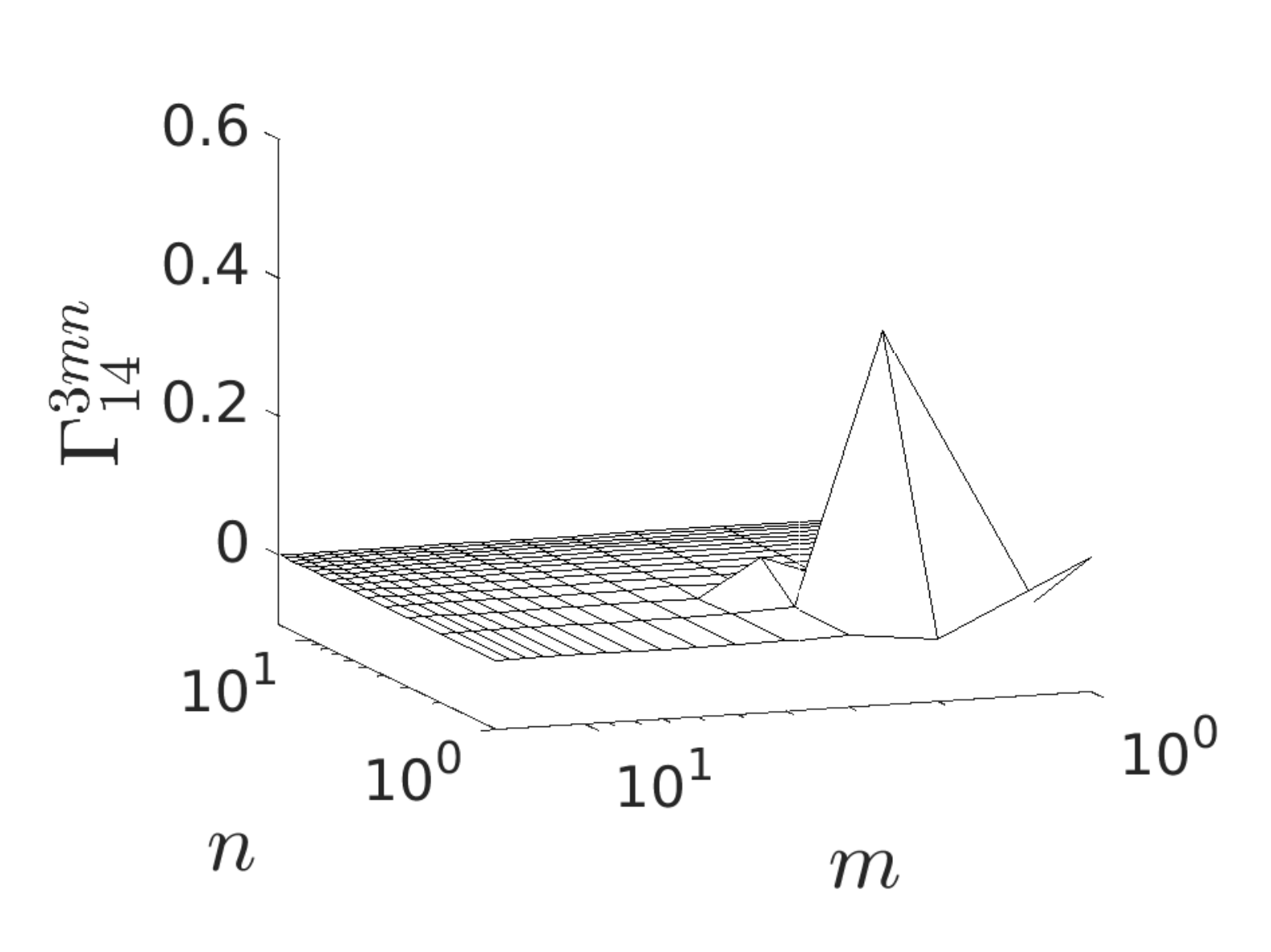}
\caption{\label{fig:evalue_reconstruction_domain_4_kernel_1_alpha_3}}
\end{subfigure}
\begin{subfigure}[h]{0.32\textwidth}
\includegraphics[width=\textwidth]{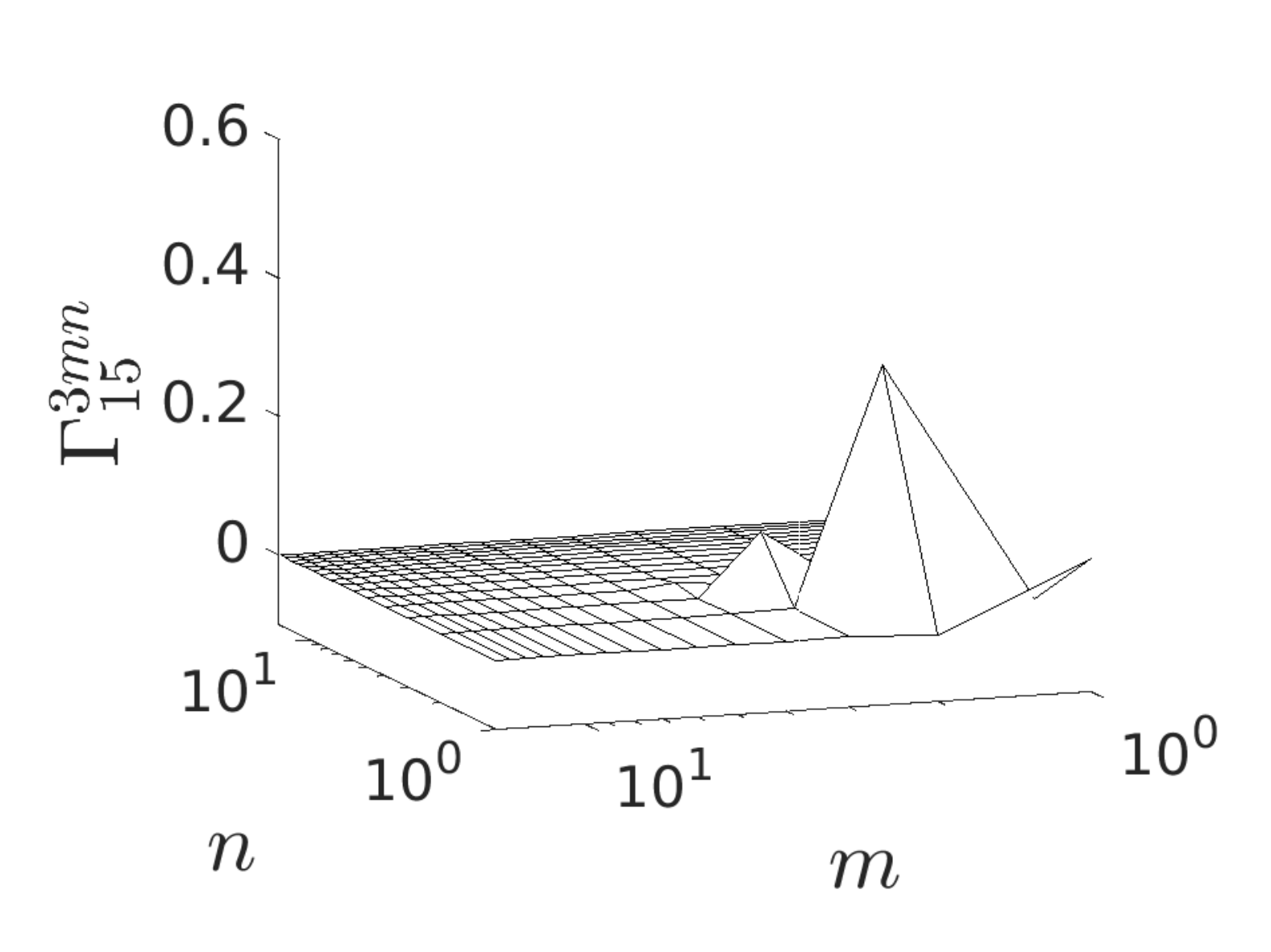}
\caption{\label{fig:evalue_reconstruction_domain_5_kernel_1_alpha_3}}
\end{subfigure}
\begin{subfigure}[h]{0.32\textwidth}
\includegraphics[width=\textwidth]{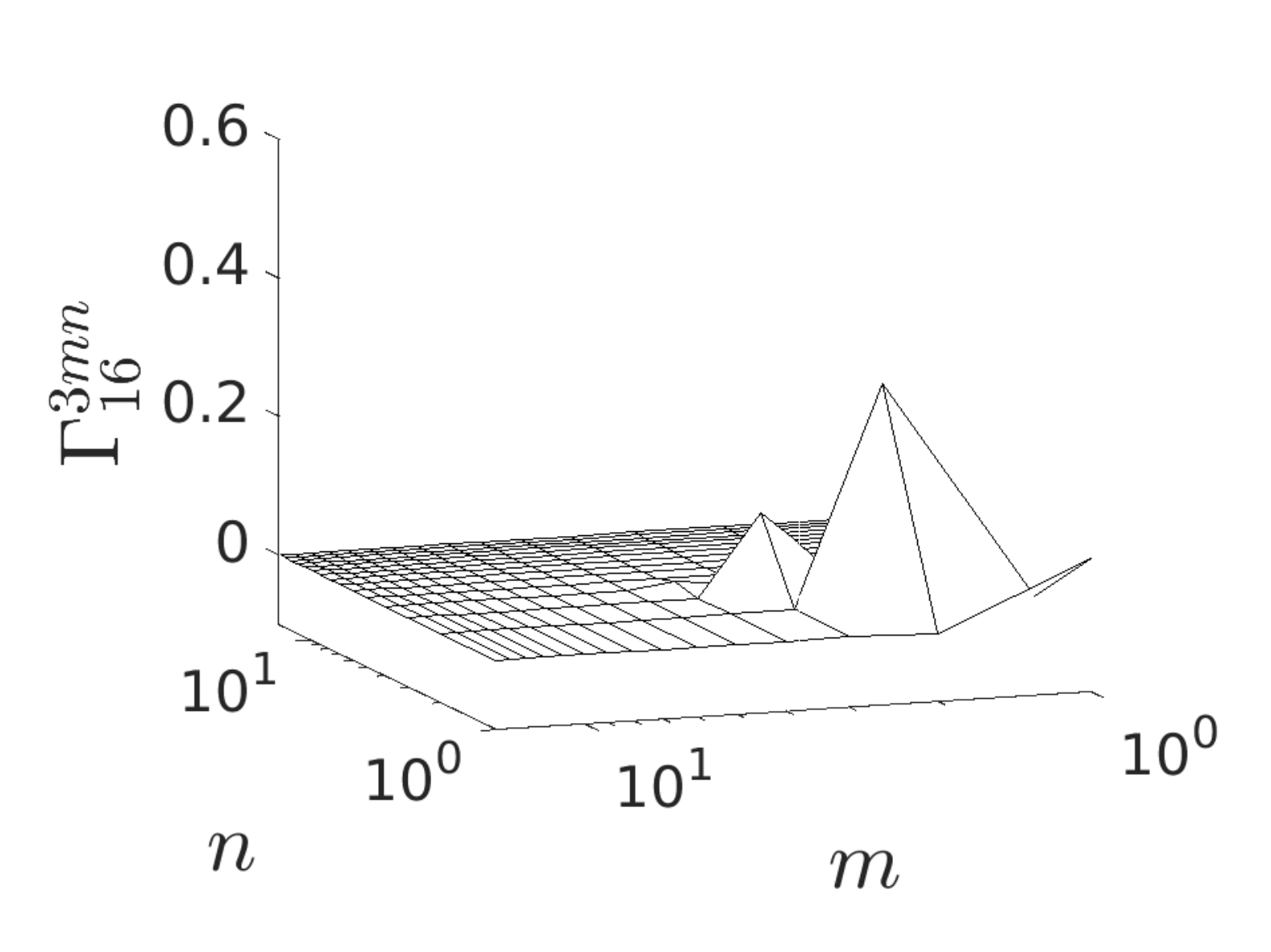}
\caption{\label{fig:evalue_reconstruction_domain_6_kernel_1_alpha_3}}
\end{subfigure}
\caption{(a)-(f): Relative contributions, $\Gamma^{3 mn}_{1j}$, to the reconstruction of $\tilde{\lambda}^3_{1j}$ for the domains with length $L_{\Omega_j}$, $j\in[1:6]$, respectively. The figures illustrate a divergence from a single dominant mode contributing to the reconstruction of $\tilde{\lambda}^3_{1j}$ as the domain is increased, as the contribution of $m=n=2$ is decreasing as the domain is increased.\label{fig:divergence_eigenvalue}}
\end{figure}
\noindent
This leads to negative values of $\Gamma^{\alpha mn}_{ij}$ which are evident in Figure \ref{fig:reconstruction_of_eigenvalue_6}. It is also seen that the deviation from a single dominant peak increases for increasing $\alpha$ in the range $\alpha\in[1:8]$ for both kernels.

The reconstruction components of $\lambda^5_{16}$ and $\lambda^5_{46}$ are shown in Figure \ref{fig:reconstruction_of_eigenvalue_5} for the largest domain of length $80\Lambda_f$. Here, the significant contributions to the reconstruction of the most energetic eigenvalues is dominated by fewer Fourier modes than for the case of $L_{\Omega_1}$, but nevertheless, still only $60\%$ of the energy is reconstructed by a single Fourier mode pair in the case of both the $K_{16}$ and $K_{46}$ kernels for $\alpha\in[1:8]$. This behaviour characterized by multiple Fourier modes being needed to reconstruct a given eigenvalue is consistent despite the appearance of high degree of collapse of the spectra related to the $K_{1j}$ kernel seen in Figure \ref{fig:e_values_fourier_spectrum_all_kernels}. This, similarly to the results in Figure \ref{fig:resnorm_0}, shows that the convergence between the two sets of basis functions does not necessarily follow from a  domain length increase. This behaviour is illustrated by Figure \ref{fig:divergence_eigenvalue} where it is seen that the fraction of $\lambda^3_{1j}$ reconstructed by $m=n=2$ decreases as the domain size is increased. Note that this does not mean that the number of Fourier modes needed for the reconstruction of a given eigenvalue is not converging in general. The current results could indicate (for this particular case) that there is a convergence towards the effective reconstruction of $\lambda^3_{1j}$ by the \textit{triplet} of conjugate Fourier pairs corresponding to $m=n\in[1:3]$. 
\begin{figure}[h]
\centering
\begin{subfigure}[h]{0.49\textwidth}
\includegraphics[width=\textwidth]{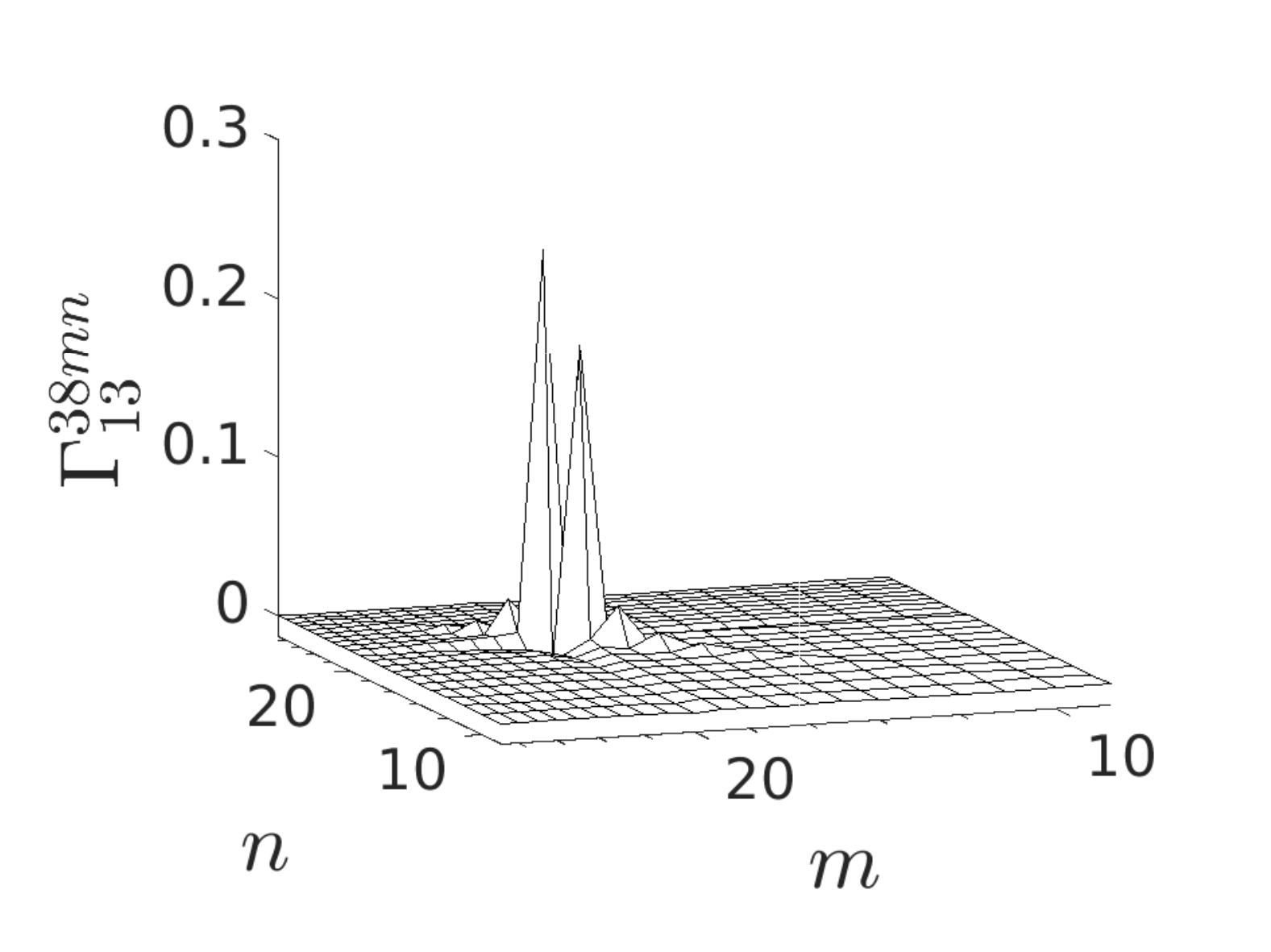}
\caption{\label{fig:evalue_reconstruction_domain_3_kernel_1_alpha_38}}
\end{subfigure}
\begin{subfigure}[h]{0.49\textwidth}
\includegraphics[width=\textwidth]{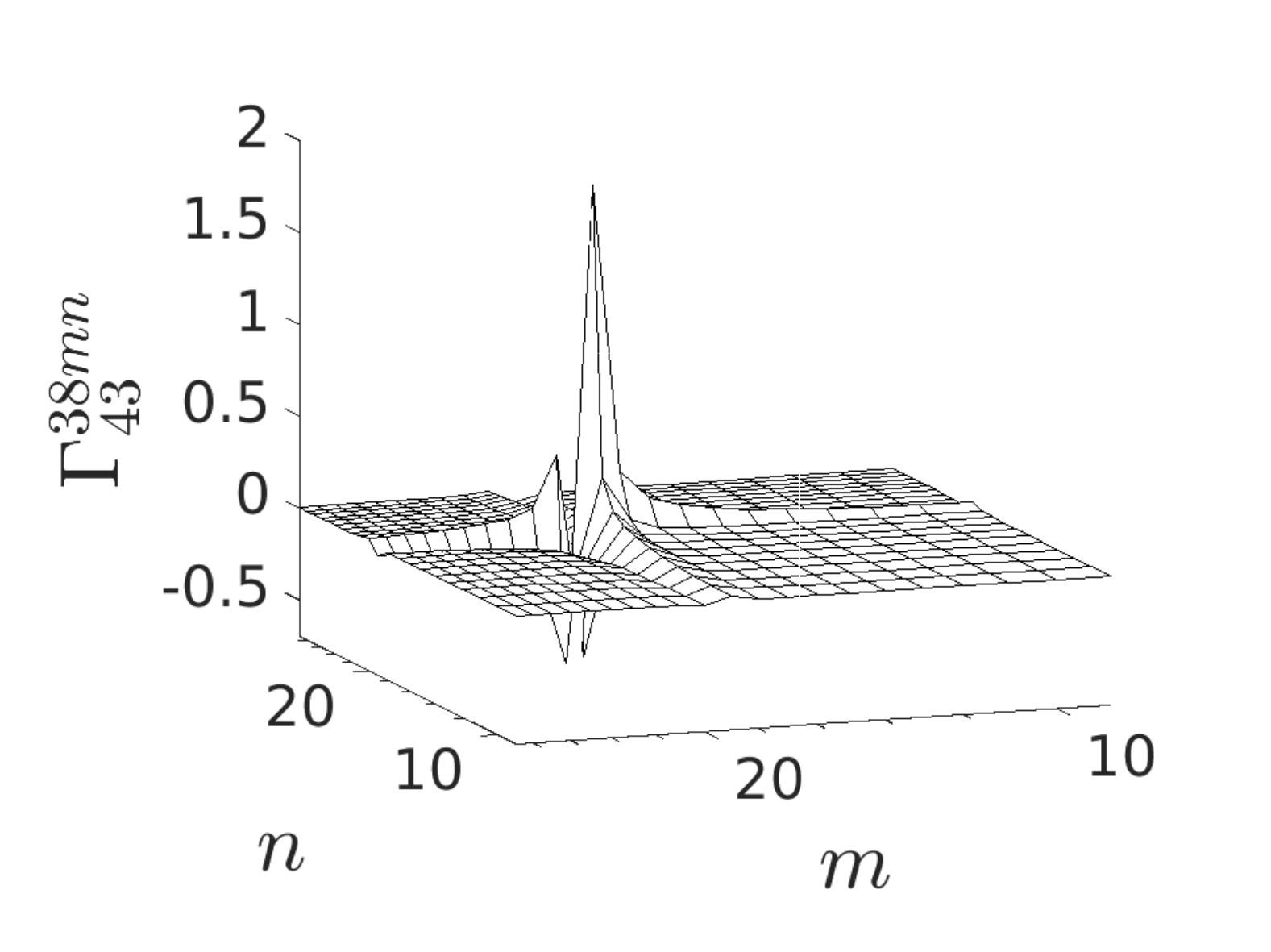}
\caption{\label{fig:evalue_reconstruction_domain_3_kernel_4_alpha_38}}
\end{subfigure}
\caption{Relative contributions, $\Gamma^{38 mn}_{i3}$, to the reconstruction of $\tilde{\lambda}^{38}_{i3}$ for the domain with length $L_{\Omega_3}=15\Lambda_f$. (a):  Results for $i=1$, (b): results for $i=4$. Multiple modes are seen to contribute to the reconstruction of the eigenvalue in question. \label{fig:reconstruction_of_eigenvalue_38}}
\end{figure}
\noindent
The results, therefore suggest that the convergence is simply not towards a \textit{single} Fourier mode (or Fourier conjugate pair), meaning that we cannot assume that POD eigenfunctions converge to Fourier modes for locally translationally invariant kernels as the domain size is increased - something that is commonly presumed in literature, \cite{George1988, Taira2017,Towne2018,Hodzic2018b,Schmidt2020b}. A more conservative hypothesis is that Fourier modes should be considered as approximations to POD eigenfunctions on aperiodic domains, especially given the discussions in Section \ref{sec:relations_between_Fourier_and_eigenspectra} and the fact that Fourier modes do not reside in $L^2(\mathbb{R},\mathbb{C})$ in relation to the homogeneous/stationary turbulence case. 

Generally, there appears to be a coupling between the deviation of the Fourier and eigenspectra at similar mode numbers and the number of Fourier modes needed to reconstruct a corresponding eigenvalue. Figure \ref{fig:reconstruction_of_eigenvalue_38} shows $\Gamma^{38 mn}_{13}$ and $\Gamma^{38 mn}_{43}$ for the domain length corresponding to $15\Lambda_f$. Given the larger deviation between the corresponding Fourier and eigenspectra seen in figure \ref{fig:e_values_fourier_spectrum_all_kernels} at $\alpha = 38$ for the domain length of $15\Lambda_f$, a larger deviation from a full eigenvalue reconstruction using a single Fourier pair may be expected. This tendency is reflected in Figure \ref{fig:reconstruction_of_eigenvalue_38} by the significantly larger deviations from peak values of $0.5$ in the case of $\Gamma^{38 mn}_{43}$ than for $\Gamma^{38 mn}_{13}$. A significant contribution to the reconstruction of these low-energy eigenvalues arises from the cross terms, $m\neq n$ in \eqref{eq:psi_expansion_of_lambda_fourier}, indicative of more complex relations between the Fourier and eigenspectra.
\subsection{Impact on reconstruction of the Taylor micro scale \label{sec:Taylor_micro_scale_reconstruction}}
Assuming that POD modes are Fourier modes results in enforcing a periodicity on the correlation function when the DFT is applied. In addition to the periodification of the correlation function, the correlation function is altered in more subtle ways, including its Taylor macro/micro scales. 
\begin{figure}[b]
\centering
\begin{subfigure}[h]{0.45\textwidth}
\includegraphics[width=\textwidth]{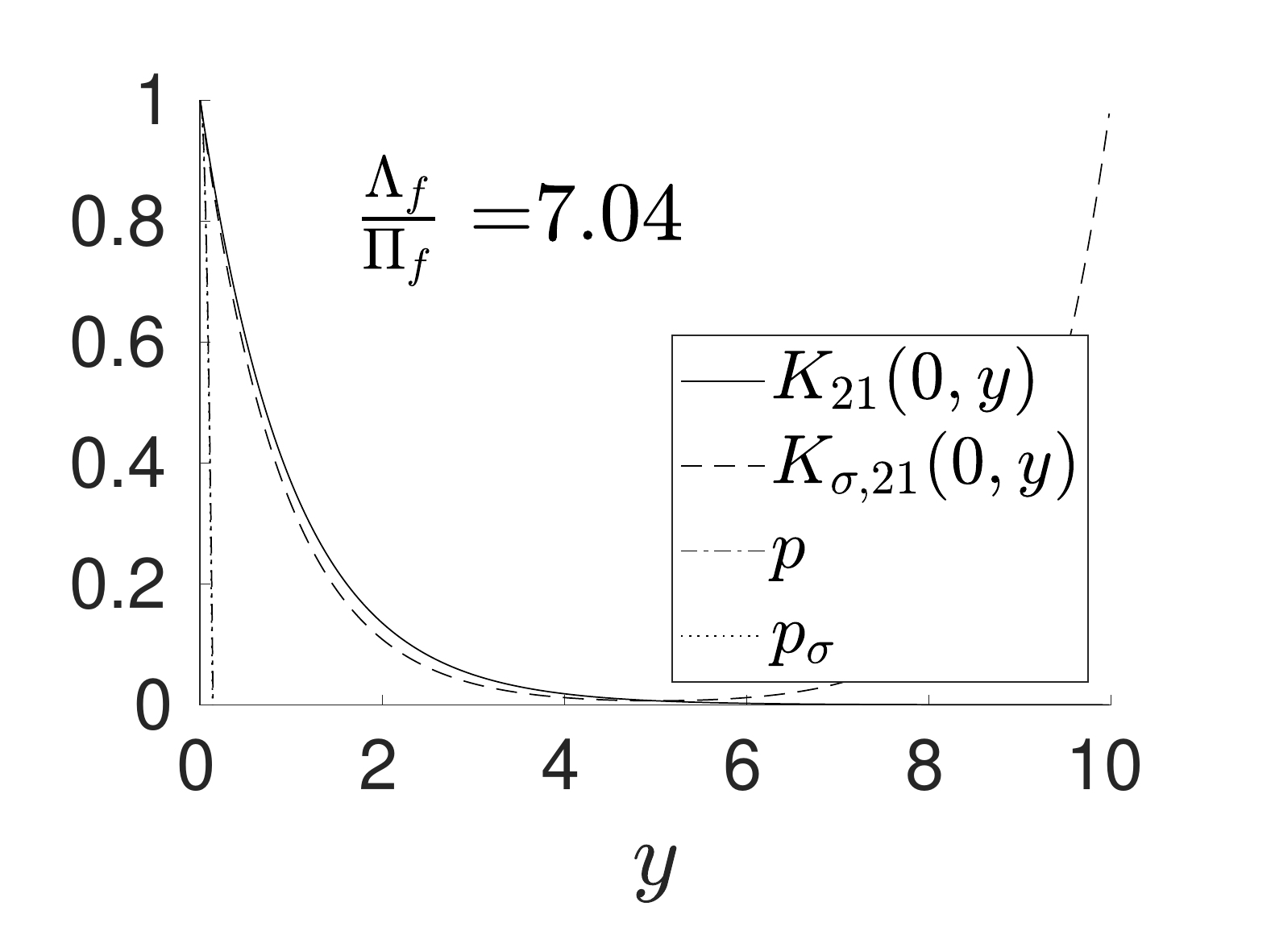}
\caption{\label{fig:correlation_functions_reconstructed_1}}
\end{subfigure}
\begin{subfigure}[h]{0.45\textwidth}
\includegraphics[width=\textwidth]{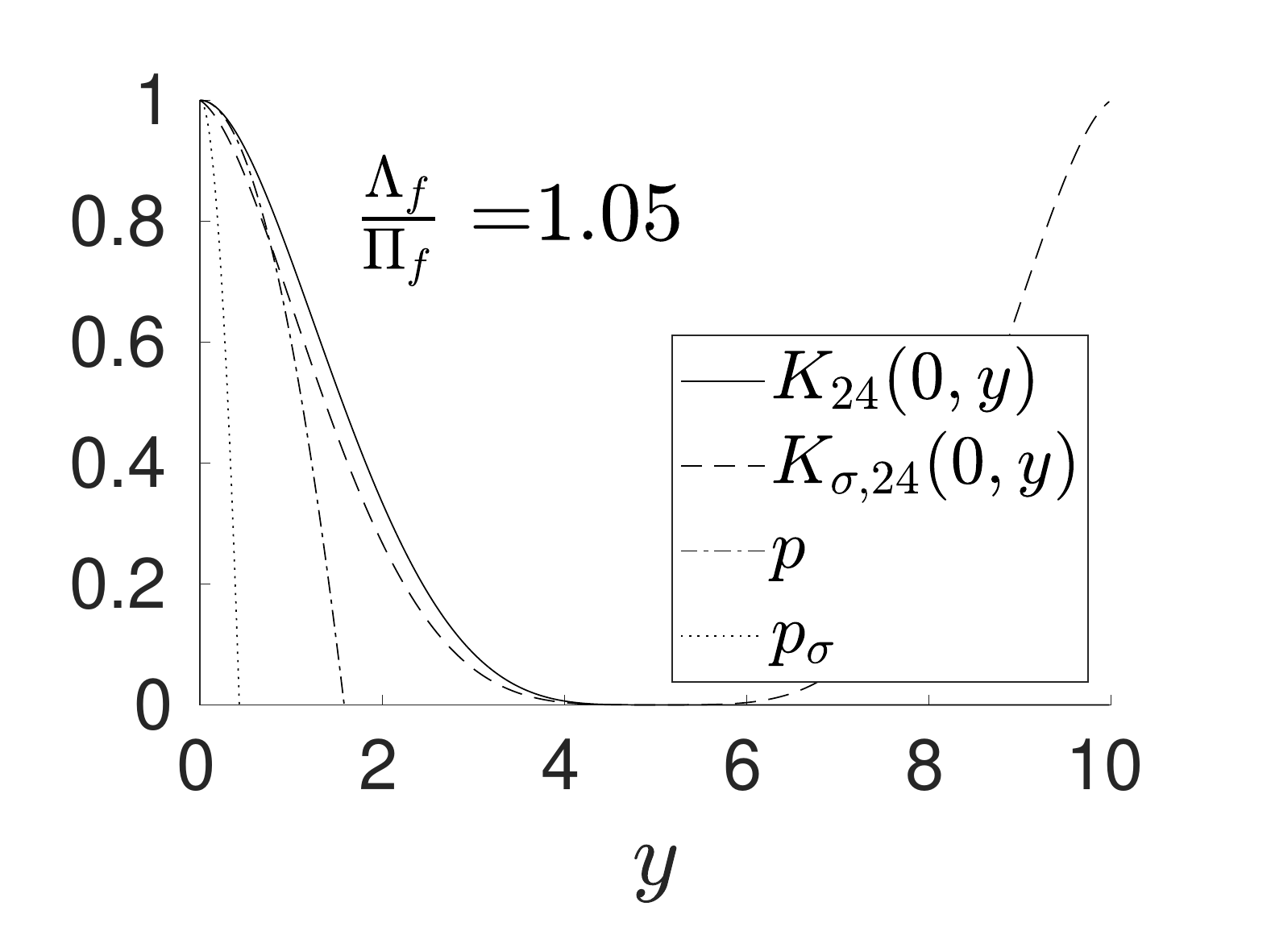}
\caption{\label{fig:correlation_functions_reconstructed_4}}
\end{subfigure}
\caption{Comparison between the Fourier reconstructed correlation functions $K_{21}$ and $K_{24}$, evaluated for $x=0$ on domain with lengths $L_{\Omega_2}=10\Lambda_f$.\label{fig:reconstructed_correlation_functions}}
\end{figure}
%

The objective is here to relate the spectral discrepancies between the Fourier and eigenspectrum for the low-energetic modes to the periodification of the correlation function. The metric used for these high mode number discrepancies is the Taylor micro scale, which is traditionally used as an estimate for the largest dissipative scale sizes in turbulent flows, \cite{Taylor1935_part_I}, \cite{Taylor1935_part_III}, \cite{Hinze1959}, or simply the estimate of velocity gradients. The Taylor micro scale, \eqref{eq:taylor_micro_scale}, is expanded using POD and Fourier bases by first expanding the second derivative of the correlation functions using \eqref{eq:R_recon_POD} and \eqref{eq:R_recon_fourier}
\begin{eqnarray}
\left.\frac{\partial^2 R(x,y)}{\partial x^2}\right\lvert_{y=x} &=& \sum_{\alpha=1}^N \lambda^\alpha\frac{d^2\varphi^\alpha(x)}{d x^2}\varphi^{\alpha*}\left(x\right)\,,\label{eq:dR_phi_exp}\\
&=& \sum_{\alpha,m,n,p,q}H^{\alpha mn}c^{\alpha,p}c^{\alpha,q*}\frac{d^2\psi^p(x)}{d x^2}\psi^{q*}(x)\,,\label{eq:dR_psi_exp}
\end{eqnarray}
where the second order derivatives were estimated numerically using a three-point parabolic fit around the diagonal elements of the partially reconstructed autocorrelation matrix. Then, assuming that $\varphi^\alpha=\psi^\alpha$, for all $\alpha\in[1:N]$ implies that all cross terms in \eqref{eq:dR_psi_exp} vanish (analogous to the step from \eqref{eq:R_recon_fourier} to \eqref{eq:R_recon_filtered}). This leads to the following approximation of the second derivative of the correlation function  
\begin{equation}
\left.\frac{\partial^2 R_\sigma(x,y)}{\partial x^2}\right\lvert_{y=x}  = \sum_{\alpha=1}^N\sigma^\alpha\frac{d^2\psi^\alpha(x)}{d x^2}\psi^{\alpha*}(x)\,,\label{eq:dR_sigma}
\end{equation}
which in form resembles \eqref{eq:dR_phi_exp}. However, using \eqref{eq:dR_sigma} in the case of aperiodic domains evaluates to a different estimate of the Taylor micro scale than using \eqref{eq:dR_phi_exp}. While \eqref{eq:dR_phi_exp} completely recovers the second derivative of the correlation functions used for the micro scale estimate, \eqref{eq:dR_sigma} yields a filtered estimate of the microscale, denoted by
\begin{equation}
\Pi_{f,\sigma} = \sqrt{\frac{-2}{\frac{d^2R_\sigma(x,y)}{dx^2}\vert_{y=x}}}\,.\label{eq:relative_error}
\end{equation}	
Figure \ref{fig:reconstructed_correlation_functions} illustrates the correlation functions $K_{2,1}(x,y)$ and $K_{2,4}(x,y)$ (corresponding to \eqref{eq:kernel_exp} and \eqref{eq:kernel_wendland}, respectively) along with the Fourier reconstructed correlation functions, $K_{\sigma,21}(x,y)$ and $K_{\sigma,24}(x,y)$, obtained from \eqref{eq:R_recon_filtered}. In addition to the periodification of the correlation function resulting from neglecting all cross terms in \eqref{eq:dR_psi_exp}, a deformation of the correlation function occurs corresponding to the functions being "compressed" towards the ends of the domain. The result is an overestimation of the second derivative at the correlation peak(s) and a thereby a underestimation of the Taylor micro scale. This is demonstrated by the differences in the parabolic fits, $p$ and $p_\sigma$, of the original correlation functions and the periodified ones, respectively (see also Appendix \ref{app:periodification}). 
\begin{figure}[t]
\centering
\begin{subfigure}[h]{0.49\textwidth}
\includegraphics[width=\textwidth]{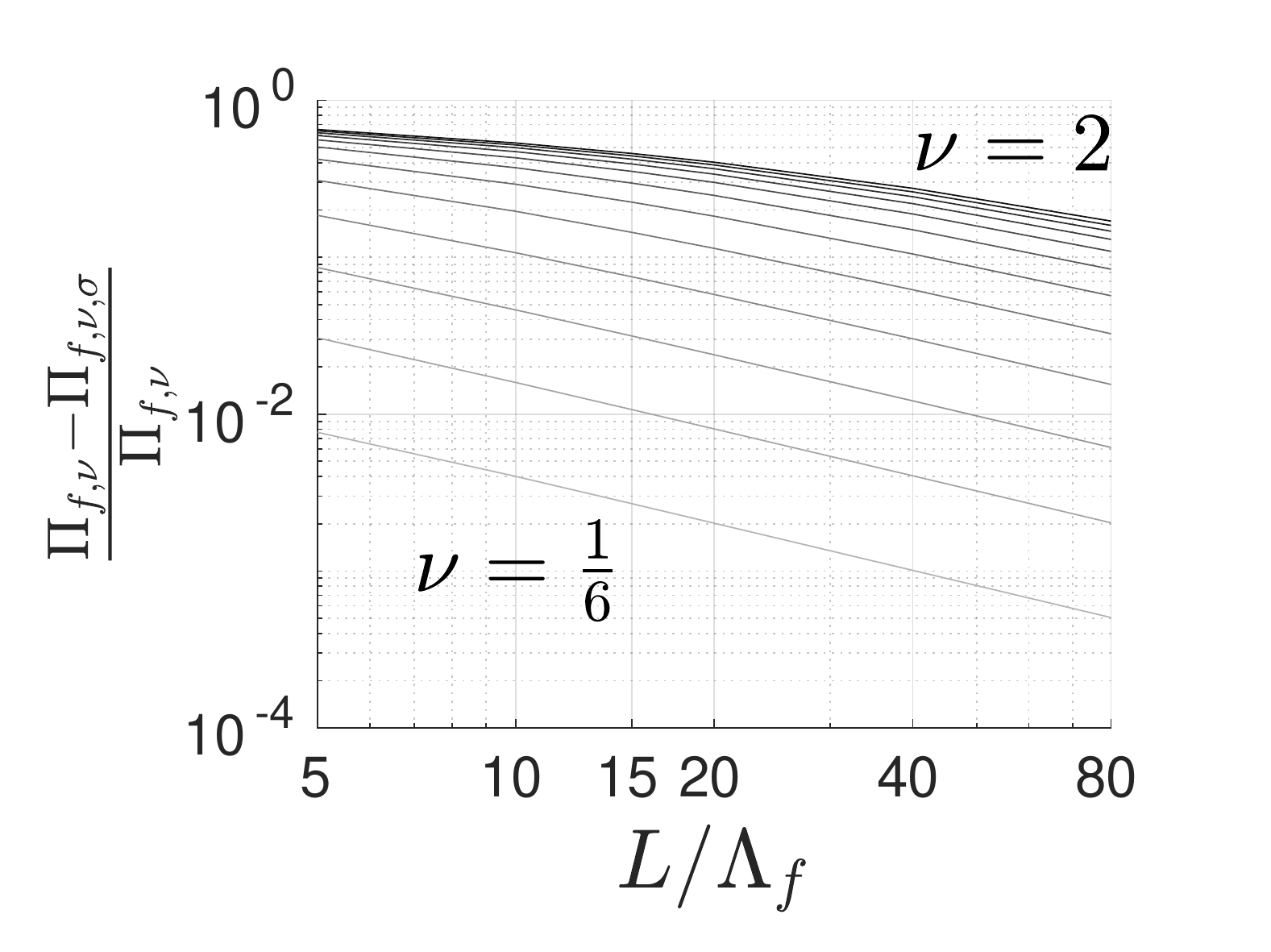}
\caption{\label{fig:micro_macro_ratio_recon_bessel}}
\end{subfigure}
\begin{subfigure}[h]{0.49\textwidth}
\includegraphics[width=\textwidth]{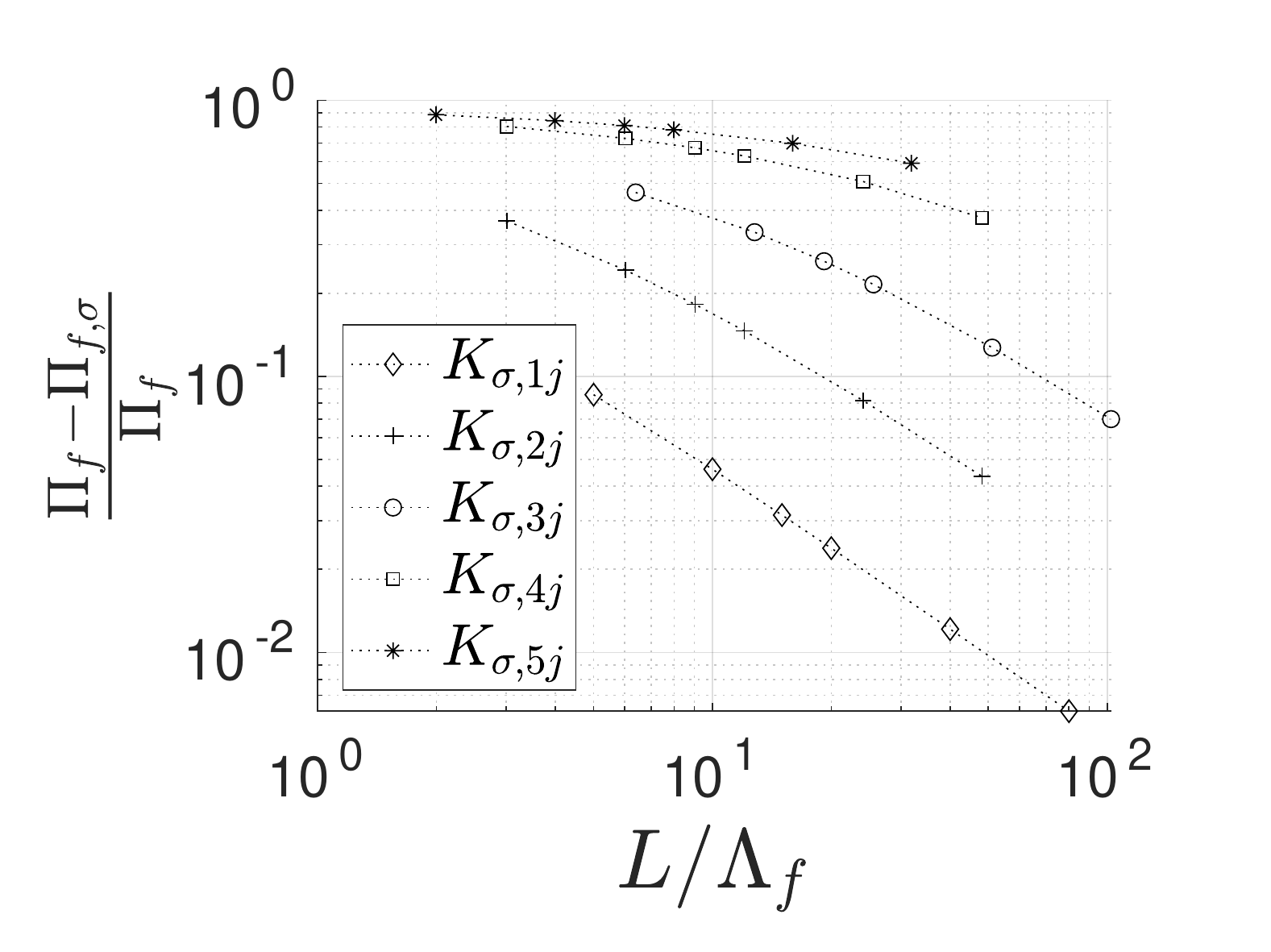}
\caption{\label{fig:micro_macro_ratio_recon_various}}
\end{subfigure}
\caption{Relative error of the Taylor micro scale estimate using the DFT for various correlation functions as a function of domain size. (a): Results for \eqref{eq:R_bessel}, (b): Results for \eqref{eq:kernel_exp}-\eqref{eq:kernel_gauss}. \label{fig:relative_error}}
\end{figure}
\noindent
Figure \ref{fig:relative_error} shows the relative error between \eqref{eq:relative_error} and \eqref{eq:taylor_micro_scale} for the two sets of correlation functions as a function of domain size. Note that these figures represent the deviations between the original Taylor micro scale, \eqref{eq:taylor_micro_scale}, and the micro scale obtained after the reconstruction of \eqref{eq:relative_error} using the complete Fourier basis. The results indicate a dependence on the micro scale size where the relative error is seen to increase for decreasing MMSRs. In all of the cases, the DFT modes underestimate the micro scale as a result of the implicit periodificiation and squeezing of the correlation function inherent to the use of DFT modes.

\begin{figure}[t]
\centering
\begin{subfigure}[h]{0.32\textwidth}
\includegraphics[width=\textwidth]{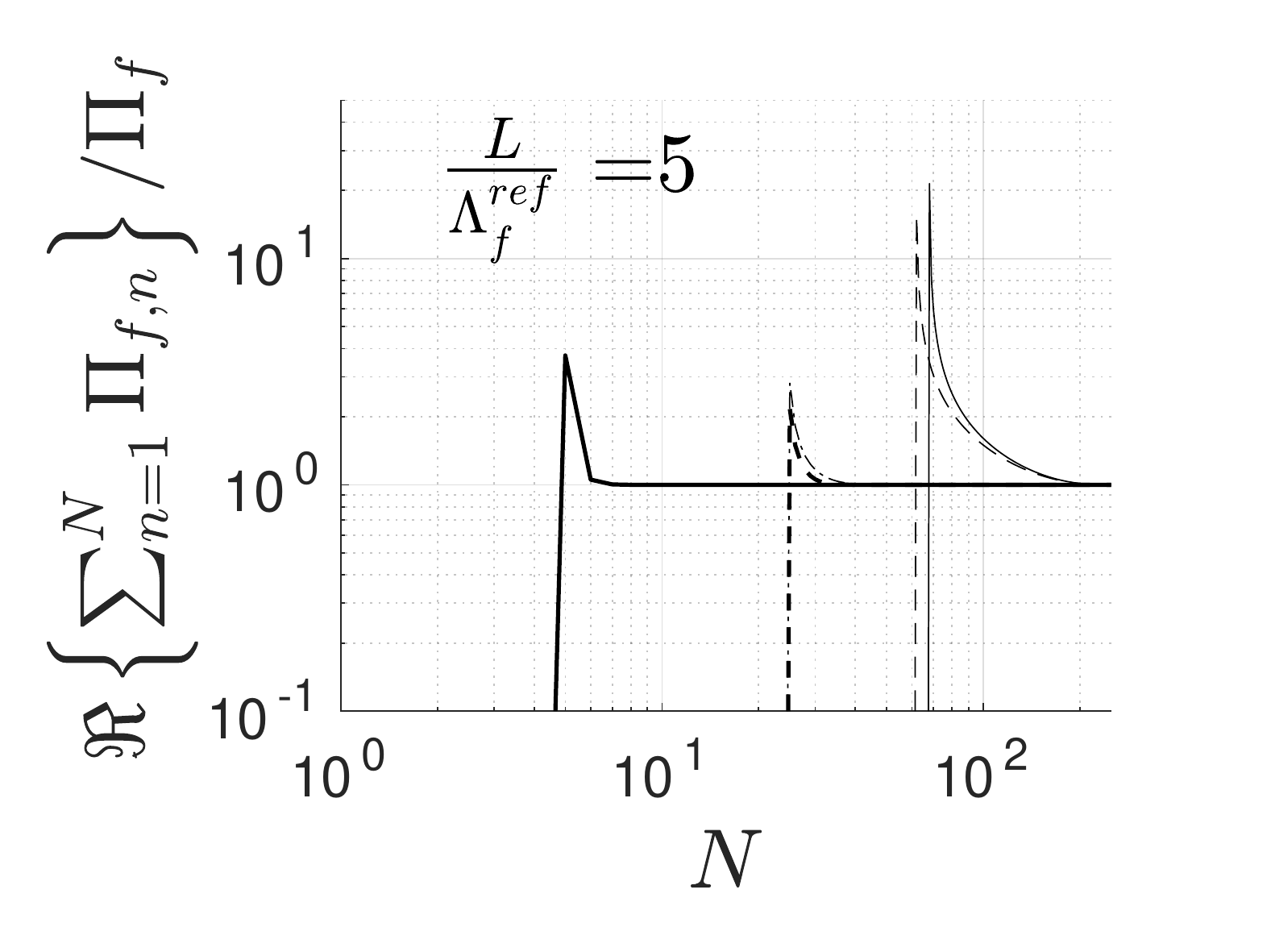}
\caption{\label{fig:micro_macro_ratio_evec_recon_various_domain_1}}
\end{subfigure}
\begin{subfigure}[h]{0.32\textwidth}
\includegraphics[width=\textwidth]{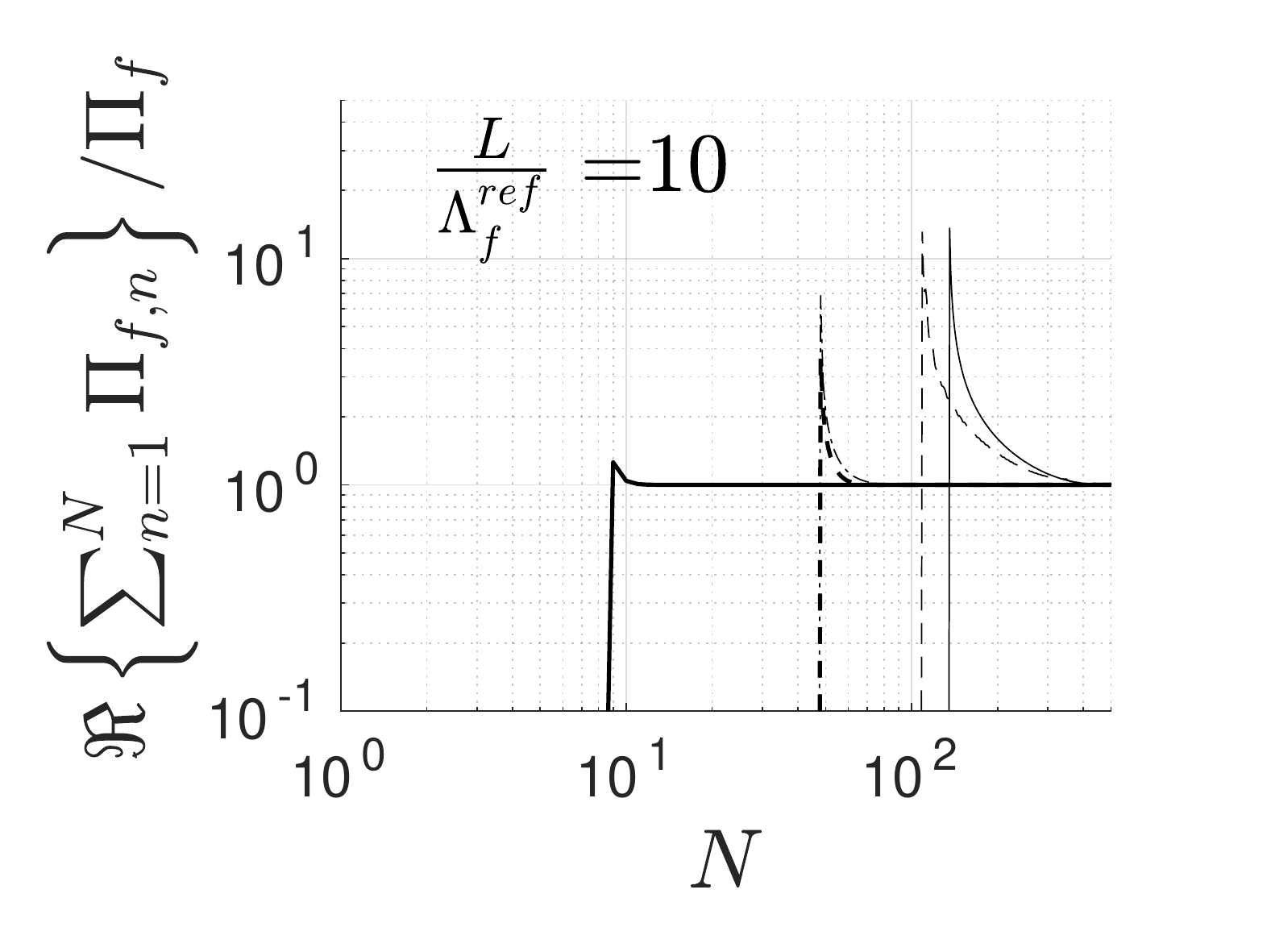}
\caption{\label{fig:micro_macro_ratio_evec_recon_various_domain_2}}
\end{subfigure}
\begin{subfigure}[h]{0.32\textwidth}
\includegraphics[width=\textwidth]{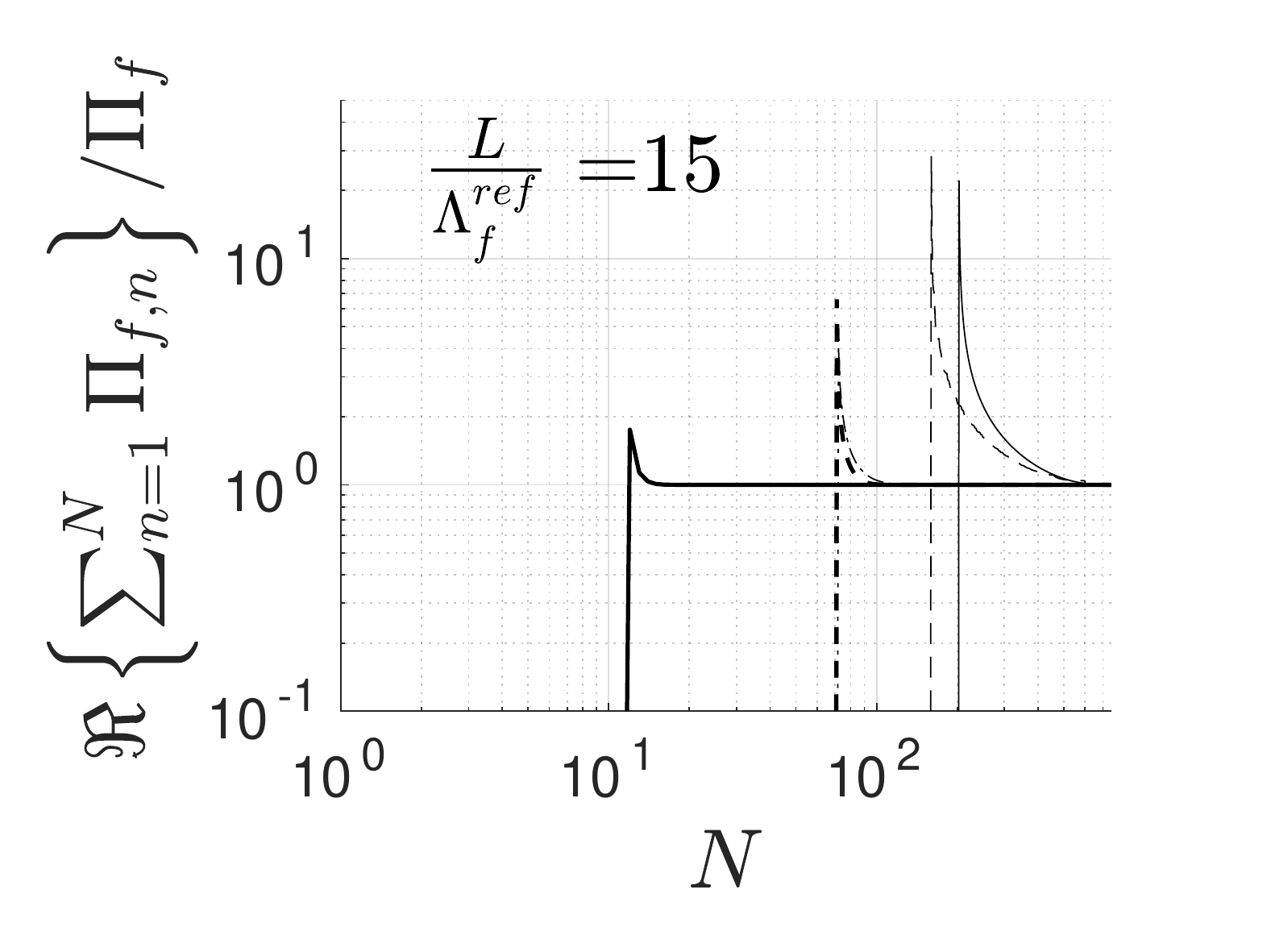}
\caption{\label{fig:micro_macro_ratio_evec_recon_various_domain_3}}
\end{subfigure}
\begin{subfigure}[h]{0.32\textwidth}
\includegraphics[width=\textwidth]{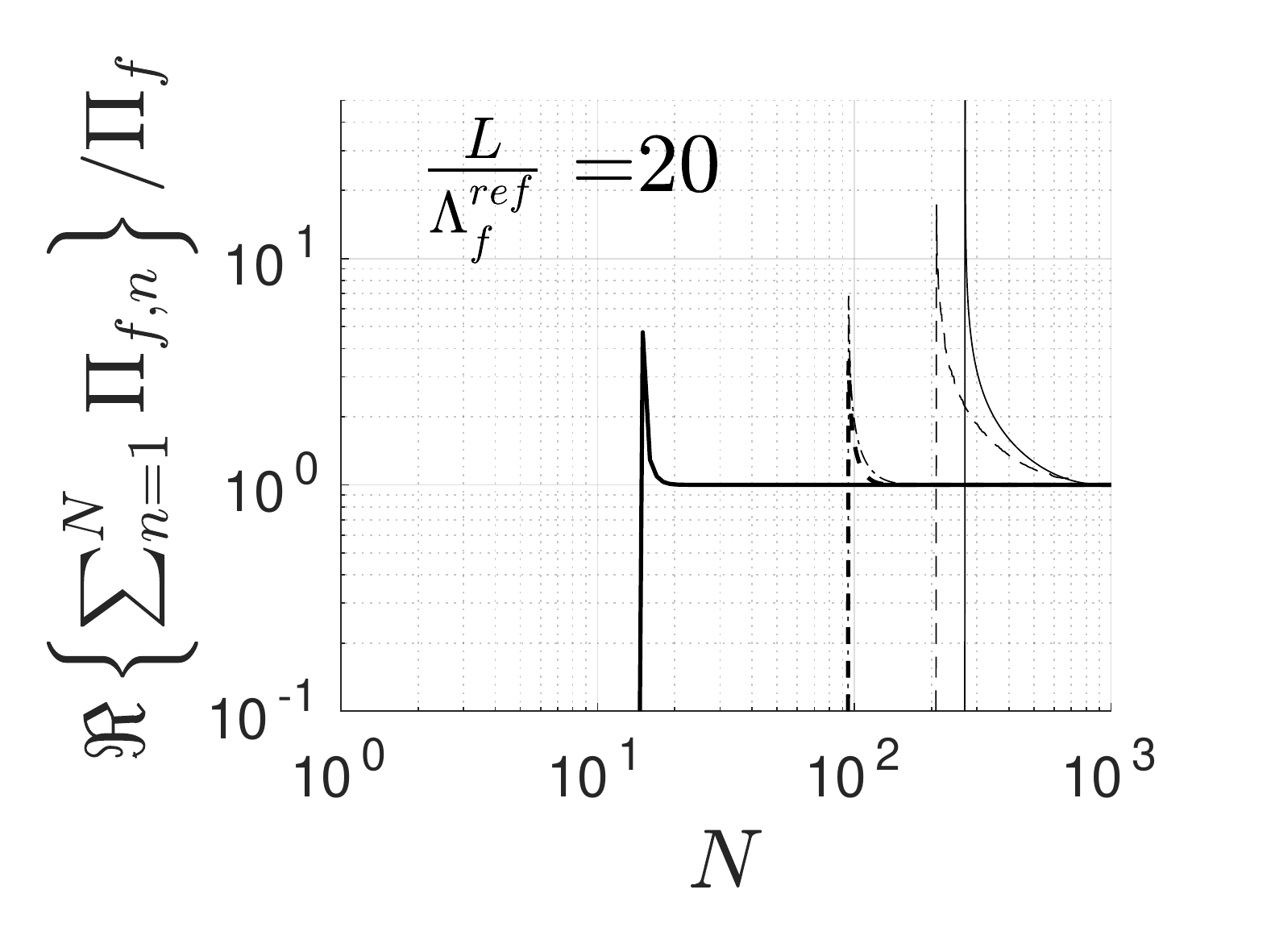}
\caption{\label{fig:micro_macro_ratio_evec_recon_various_domain_4}}
\end{subfigure}
\begin{subfigure}[h]{0.32\textwidth}
\includegraphics[width=\textwidth]{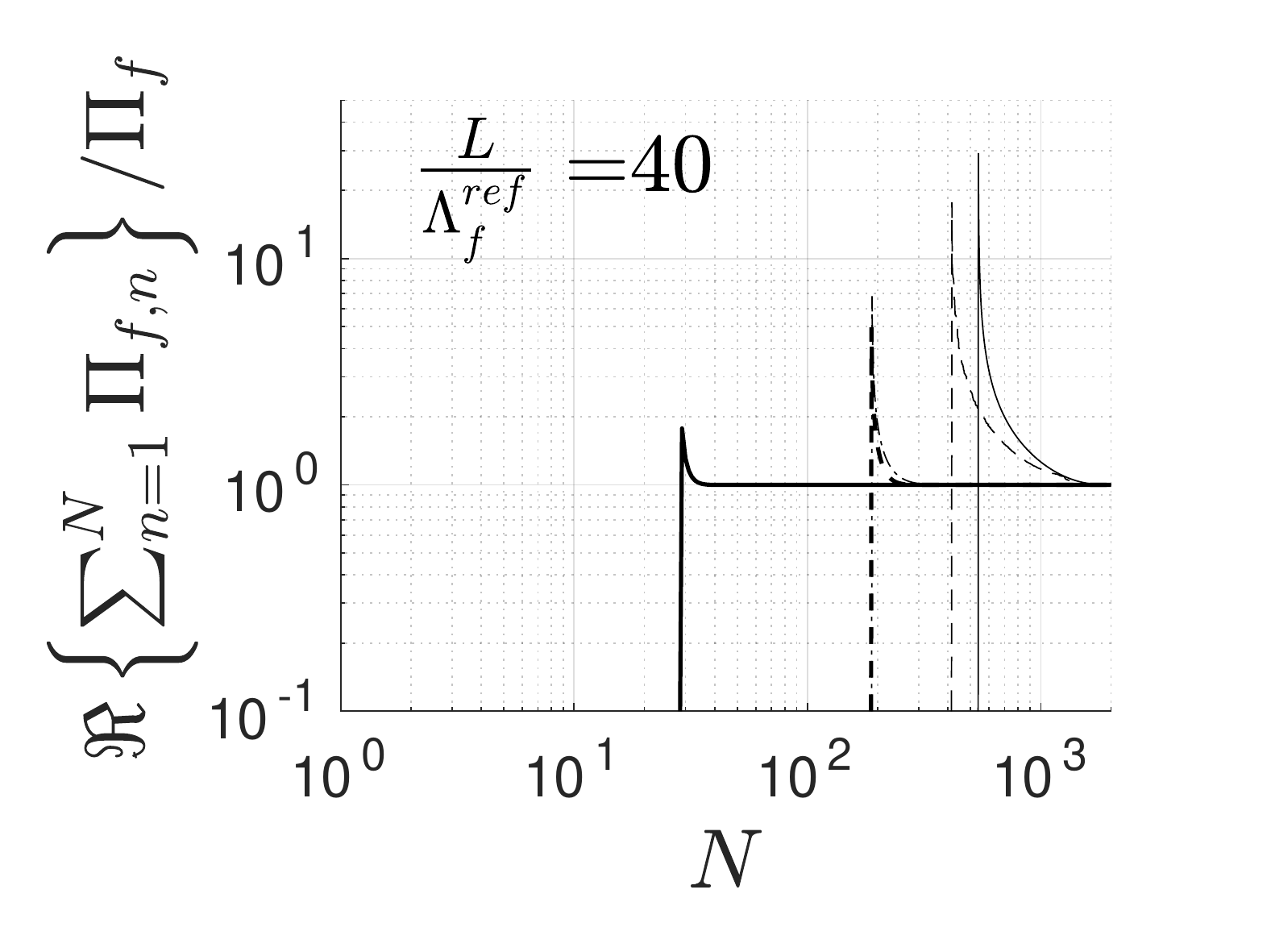}
\caption{\label{fig:micro_macro_ratio_evec_recon_various_domain_5}}
\end{subfigure}
\begin{subfigure}[h]{0.32\textwidth}
\includegraphics[width=\textwidth]{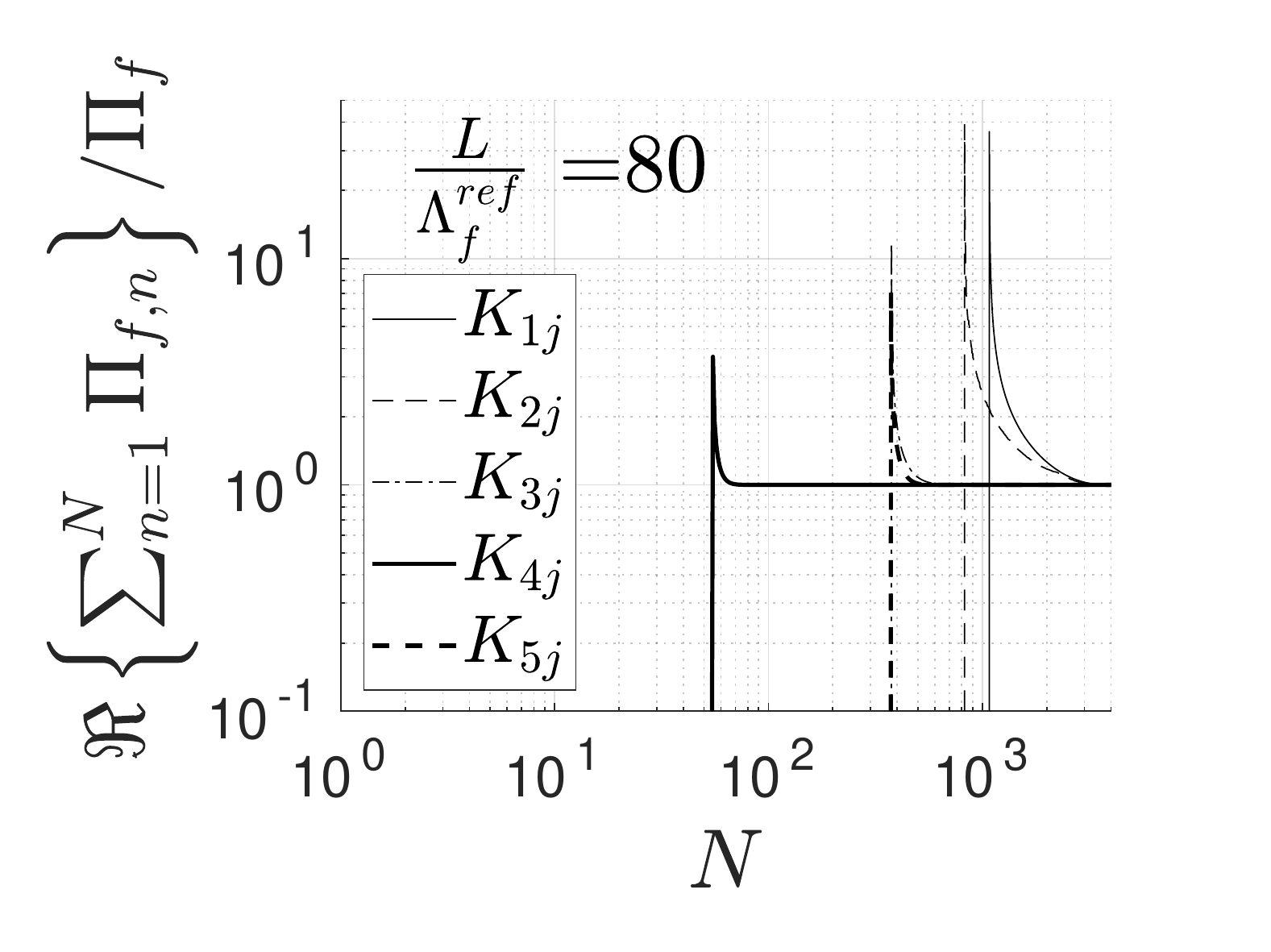}
\caption{\label{fig:micro_macro_ratio_evec_recon_various_domain_6}}
\end{subfigure}
\caption{Cumulative modal reconstruction of the Taylor micro scale using POD modes related to the correlation functions \eqref{eq:kernel_exp}-\eqref{eq:kernel_gauss} for various domain sizes \label{fig:relative_error_evec}.}
\end{figure}
\noindent

The full reconstruction of the micro scale is, however, possible using POD modes with much fewer terms facilitated by the exact reconstruction of the second derivative of the correlation function, \eqref{eq:dR_phi_exp}. Figure \ref{fig:relative_error_evec} shows the real part of the cumulative modal reconstruction of the micro scale related to \eqref{eq:kernel_exp}-\eqref{eq:kernel_gauss} as functions of domain size. The dominant contributions to the cumulative reconstructions of the micro scale leading up to maximum value at higher $N$ are in fact imaginary. Given that the Taylor micro scale is defined as the positive root of a second order polynomial fit, imaginary roots indicate that the partially reconstructed kernel has an off-diagonal peak - meaning that the partially reconstructed correlation function attains its maximum at $x\neq y$. The polynomial roots become real as an increasing number of modes is used in the reconstruction of the correlation function. A region of monotonic convergence is observed in all cases after the maxima in Figure \ref{fig:relative_error_evec}, due to sign change of the second derivative of the polynomial fit. 

The general monotonic relations exhibit similar behaviours across all correlation functions and domains in Figure \ref{fig:relative_error_evec}, where certain micro scale estimates are converging faster than others. For small MMSRs, where the Fourier modes perform worst in terms of the reconstruction of $\Pi_f$, the POD modes are seen to perform best - as seen in Figure \ref{fig:relative_error_evec}. The most extreme case is seen for the Gaussian kernel, $K_{5j}$, in Figure \ref{fig:relative_error_evec}, representing the self-similar Karman-Howarth solution, \cite{karman1938}, where a very small fraction of the POD modes reconstruct the micro scale down to an negligible error for all domain sizes. For the same kernel in Figure \ref{fig:micro_macro_ratio_recon_various} the relative error estimate is in the order of $100\%$ for all domain sizes after using all of the Fourier modes in the reconstruction. In addition, the error may be potentially even more substantial if the Fourier basis was used in a reduced-order model, where only a subset of these were used in the reconstruction. 
\\
\\
The analysis presented above demonstrates that several different kernels exhibit consistent trends in the relationship between, on one hand, the convergence of microscale reconstruction using Fourier modes, and on the other hand, MMSR and domain length. Further studies would be needed to address the extent to which these trends can be generalized, and whether the effect of the MMSR from sampled analytical kernels translates to a Reynolds number dependency in the case of kernels derived from empirical flows.

%
%
\FloatBarrier
\section{Conclusions}
The current work presented a theoretical and numerical analysis of the relation between Fourier and POD modes for locally translationally invariant kernels on aperiodic domains. The work asserted theoretically that trigonometric polynomials could not be considered solutions to the POD optimization problem on infinite spatial/temporal domains as they are not elements of $L^2(\mathbb{R},\mathbb{C})$, excluding them from spanning the same space. This fact leads to the numerical analysis of discrepancies between Fourier and eigenspectra on finite aperiodic domains as a function of kernel characteristics and domain size. A set of correlation functions was generated using modified Bessel functions of the second kind, enabling the variation of the Taylor macro/micro scale ratio. These correlation functions, characterized by their analytical Fourier transforms, were used to compare the spectral properties of the DFT and eigenspectra. The eigenspectra (unlike the DFT spectra) were shown to replicate the analytical spectral asymptotic behaviour to a very high degree, even for extremely small window sizes down to five integral length scales. The results indicated that the macro/micro scale ratio was of central importance for the observed spectral discrepancies between DFT and eigenspectra, and the spectral differences were increased for small macro/micro scale-ratios. These results were confirmed by the analysis of a second set of correlation functions - which included the Karman-Howarth solution of the fundamental equation for the propagation of the correlation function - confirming that for macro/micro scale-ratios approaching unity the deviation between the DFT and eigenspectra increased. The divergence between the first POD mode and the zeroth harmonic was analyzed, followed by the analysis of the Fourier expansion of the eigenspectrum, in order to couple the two spectral energy representations. Here a divergence between the POD and Fourier spectra was observed, measured in terms of the number of Fourier modes needed to reconstruct a given eigenvalue, as domain sizes were increased. This result challenges the implied notion of POD modes approaching Fourier modes for increasing domain sizes. Finally, it was demonstrated that the implicit periodification of the correlation function imposed by the use of DFT modes for aperiodic domains decreases the Taylor micro scale estimate obtained from the modified correlation function. The effect was largest for POD kernels characterized by small Taylor macro/micro scale-ratios and falls in line with the discrepancies observed in the corresponding spectral analyses performed earlier. While the Fourier basis was unable to fully reconstruct the micro scale for any combination of correlation function and domain size, the POD modes demonstrated the ability to reconstruct the aforementioned to a very high degree using only a subset of modes in the case of a small macro/micro scale-ratio. This ability demonstrates the advantageous properties of the POD modes compared to Fourier modes in the case of correlation functions with small macro/micro scale-ratios. 

\backmatter

\bmhead{Acknowledgments}

AH and CMV acknowledge financial support from the European Research council: This project has received funding from the European Research Council (ERC) under the European Unions Horizon 2020 research and innovation program (grant agreement No 803419). PJO acknowledges financial support from the Poul Due Jensen Foundation: Financial support from the Poul Due Jensen Foundation (Grundfos Foundation) for this research is gratefully acknowledged.

\section*{Declarations}
\subsection*{Conflict of interest}
The authors declare that they have no conflict of interest
\subsection*{Author's contributions}
AH performed the numerical analysis and wrote the main manuscript text. PJO and CMV supported the analysis of data. All authors reviewed the manuscript.
\subsection*{Data availability}
The datasets generated during and/or analyzed during the current study are available from the corresponding
author upon reasonable request.



\begin{appendices}
\section{Domain dependence on solutions\label{app:domain_dependence_on_solutions}}
A detailed analysis of the domain dependence of the POD operator on the relation between POD and Fourier modes is performed. The cases covered consist of translationally invariant kernels on infinite domains, locally translationally invariant kernels on finite aperiodic domains and translationally invariant kernels on weighted inner product spaces are discussed. Initially, however, the case of the one-dimensional POD integral eigenvalue problem with translationally invariant kernels on periodic domains is covered.
\subsection{Translationally invariant kernels on periodic domains\label{app:periodic_domains}}
Define the POD integral operator as the mapping~${R:L_w^2([-\frac{a}{2},\frac{a}{2}],\mathbb{C})\to L_w^2([-\frac{a}{2},\frac{a}{2}],\mathbb{C})}$, $0<a<\infty$ where $K:\mathbb{R}\times\mathbb{R}\to\mathbb{R}$ is a translationally invariant kernel
\begin{equation}
K(x,y) =\widetilde{K}(x-y)\,,\, x,y\in\mathbb{R}\,,
\end{equation}
which is periodic, with period $a$, i.e.
\begin{equation}
\widetilde{K}(x-y+a)=\widetilde{K}(x-y)\,,\,x,y\in\mathbb{R}\,.\label{eq:periodicity}
\end{equation}
The integral operator can then be evaluated for $\varphi(y)=e^{iky}$ with the substitution $z=x-y$
\begin{eqnarray}
\left(R\varphi\right)\left(x\right) &=& \int_{-a/2}^{a/2}\widetilde{K}(x-y)\varphi(y)dy\,,\\
&=& \underbrace{\int_{x-a/2}^{x+a/2}\widetilde{K}(z)e^{-ikz}dz}_{\lambda(k)}e^{ikx}\,,\\
&=& \lambda(k)\varphi(x)\,,
\end{eqnarray}
where $\lambda(k)$ is found to be invariant with respect to $x$ due to the condition \eqref{eq:periodicity}. Of the cases covered in the current work, condition \eqref{eq:periodicity} in combination with the finite domain, $\Omega=[-\frac{a}{2},\frac{a}{2}]$, is the only case where the Fourier basis can be deduced as the solution to the POD eigenvalue problem, given the restriction $\varphi\in L_w^2([-\frac{a}{2},\frac{a}{2}],\mathbb{C})$.
\subsection{Translationally invariant kernels on infinite domains}\label{app:infinite_domains}
It is often assumed, in cases of statistically stationary (aperiodic) turbulence, that the temporal eigenfunctions on a finite temporal domain are Fourier bases. For the case of the POD, this idea originated from \cite{lumley1967structure} who advocated the use of Fourier analysis in combination with the POD for homogenous fields of infinite extent.  In this capacity it is necessary to distinguish between Fourier transforms over the real line and basis expansions over the real line.

Having restricted the eigenfunctions of \eqref{eq:Lumley_Decomposition} to reside in $L^2_w(\Omega,\mathbb{C})$, we now focus on the strict limitations imposed on the functions a Fourier transform can be applied to. The Fourier transform can be defined as the mapping $\mathcal{F}:L^2\left(\mathbb{R},\mathbb{C}\right)\to L^2\left(\mathbb{R},\mathbb{C}\right)$. Even in the case when a translationally invariant kernel resides in $L^2\left(\mathbb{R},\mathbb{C}\right)$ for a homogeneous field of infinite extent, it is clear that the eigenfunction, $\varphi\neq e^{ikx}$, $x\in\mathbb{R}$, since $\varphi\notin L^2\left(\mathbb{R},\mathbb{C}\right)$. This leads to the conclusion that for homogeneous aperiodic fields the eigenfunction in \eqref{eq:Lumley_Decomposition} cannot be of the harmonic type, as the criterion of boundedness stated in \eqref{eq:L2_w} (in the case of $w(x)=1$, $x\in\mathbb{R}$) is not upheld. The homogeneous case is often referenced in literature when arguing that solutions to \eqref{eq:Lumley_Decomposition} are of the type $\varphi= e^{ikx}$ for aperiodic \textit{finite} fields, but we see that this argument is flawed, since the functions $\varphi(x)= e^{ikx}$, $x,k\in\mathbb{R}$, do not even reside in the vector space necessary for these to be eigenfunction candidates for $R$. More fundamentally, a basis spanning a \textit{normed vector space} necessarily consists of vectors that are within the normed vector space, cf.\ p.\ 41 in \cite{Christensen2010}.
\subsection{Locally translationally invariant kernels\label{app:locally_translational}}
We consider integral transforms $R: L^2(\Omega, \mathbb{C}) \rightarrow L^2(\Omega, \mathbb{C})$ given by
\begin{equation} 
	(R\varphi)(x) = \int_{\Omega} K(x,y) \varphi(y) dy\,, 
	\label{eq:integral_operator}
\end{equation}
where $K: \Omega\times\Omega \rightarrow \mathbb{C}$ is the kernel function. Let the kernel $K$ be translationally invariant within the subdomain $S \subset \mathbb{R}^2$ given by $S = \mathcal{L}_x \times \mathcal{L}_y$, where $\mathcal{L}_{x} = \left\{ x \vert -\frac{a}{2} \leq x \leq \frac{a}{2}\right\}$, $0 < a < \infty$; and zero outside $S$. This means that $K(x,y) = \widetilde{K}(x-y)$ for $(x,y)\in S$, where $\widetilde{K}: \left[-a, a \right] \rightarrow \mathbb R.$

Defining the window function $\chi(x)$ by
\begin{align}
	\chi(x) &= \left\{\begin{array}{c}
		1\,,\quad x \in \mathcal{L}_x\,, \\ 0\,, \quad x \notin \mathcal{L}_x\,,
	\end{array}\right.
\end{align}
we may write the kernel as $K(x,y) = \tilde{K}(x-y)\chi(x) \chi(y)$. We apply the operator $R$ defined by this kernel via \eqref{eq:integral_operator} to a candidate solution $\varphi(y) = e^{i\kappa y}$ with ${\kappa = 2\pi n/a, n \in \mathbb{Z}}$, yielding 
\begin{eqnarray}
	(R\varphi)(y) &=& \int_{-\frac{a}{2}}^{\frac{a}{2}} K(x, y) e^{i\kappa y}dy\,,\\ 
	&=& \int_{-\frac{a}{2}}^{\frac{a}{2}} \tilde{K}(x-y) \chi(x) \chi(y) e^{i\kappa y}dy\,,\\
	&=& \int_{x-\frac{a}{2}}^{x+\frac{a}{2}} \tilde{K}(z) \chi(x)\chi(x-z) e^{i\kappa (x-z)}dz\,,\\
	&=& \chi(x) \int_{x-\frac{a}{2}}^{x+\frac{a}{2}} \tilde{K}(z) \chi(x-z) e^{-i\kappa z}dz\, e^{i\kappa x}\,,\\
	&=& \zeta(x, \kappa) \varphi(x)\,,
\end{eqnarray}
where $z = x-y$ and
\begin{equation}
\zeta(x, \kappa) = \chi(x) \int_{x-\frac{a}{2}}^{x+\frac{a}{2}} \widetilde{K}(z) \chi(x-z) e^{-i\kappa z}\, dz\,.
\end{equation}
The candidate function $\varphi$ is an eigenfunction if $\zeta(x, \kappa)$ does not depend on $x$. Since $\varphi\in L^2(\mathcal{L}_x, \mathbb{C})$ we need only consider $x \in \mathcal{L}_x$, for which the first window function $\chi(x) = 1$. The limits on the integral correspond to $\mathcal{L}_{z-x}$, which is exactly the interval where the integrand's window function $\chi(x-z) = 1$. We need therefore not include any of the window functions in the expression for $\zeta$, leaving us with
\begin{align}
	\zeta(x, \kappa) &= \int_{x-\frac{a}{2}}^{x + \frac{a}{2}} \widetilde{K}(z) e^{-i\kappa z}\, dz\,. 
\end{align}
The extent to which this expression depends on $x$ is determined by the properties of $\widetilde{K}(z)$. If $\widetilde{K}(z)$ satisifies $\widetilde{K}(z\pm a) = \widetilde{K}(z)$ for $z\in \mathcal{L}_z$ we find that $\zeta(x, \kappa)$ is indeed independent of $x$; for example, for $-\frac{a}{2} \leq x \leq 0$ we have, with $z' = z+a$,
\begin{eqnarray}
	\zeta(x, \kappa) &=& \int_{x-\frac{a}{2}}^{x+\frac{a}{2}} \widetilde{K}(z) e^{i\kappa z} dz\,,\\ 
	&=& \int_{x-\frac{a}{2}}^{-\frac{a}{2}} \widetilde{K}(z) e^{i\kappa z} \, dz + \int_{-\frac{a}{2}}^{x + \frac{a}{2}} \widetilde{K}(z) e^{i\kappa z}dz\,, \\
	&=& \int_{x+\frac{a}{2}}^{\frac{a}{2}} \widetilde{K}(z+a) e^{i\kappa (z+a)}\, d(z+a) + \int_{-\frac{a}{2}}^{x + \frac{a}{2}} \widetilde{K}(z) e^{i\kappa z}dz\,,\\
	&=&   \int_{x+\frac{a}{2}}^{\frac{a}{2}} \widetilde{K}(z) e^{i\kappa z}\, dz + \int_{-\frac{a}{2}}^{x + \frac{a}{2}} \widetilde{K}(z) e^{i\kappa z}dz\,,\\
	&=&   \int_{-\frac{a}{2}}^{\frac{a}{2}} \widetilde{K}(z) e^{i\kappa z}dz = \lambda(\kappa)\,,
\end{eqnarray}
which does not depend on $x$. The same relation can be shown to hold for $0 \leq x \leq \frac{a}{2}$.

As expected, a periodic kernel produces harmonic eigenmodes (see Appendix \ref{app:periodic_domains}). A general kernel, however, does not.  Note that if the domain of integration would be set to the entire real line the $x$-dependency of the integral would vanish. In this case, however, the operator would be characterized by $R:L^2(\mathbb{R},\mathbb{C})\to L^2(\mathbb{R},\mathbb{C})$, but then $\varphi(x)=e^{ikx}$, $x\in\mathbb{R}$ would not qualify as a solution since $e^{ikx}\notin L^2(\mathbb{R},\mathbb{C})$, as discussed earlier
\subsection{Weighted translationally invariant kernels\label{app:weighted_invariant_kernels}}
An attempt to circumvent the above issues may be to apply a filter/window on the signal in order to ensure compactness of the kernel or in some way argue the attainment of periodicity. Common examples of windows used are Hamming, Hanning, and Bartlett windows to mention a few. In this approach we must introduce a new operator~${F:L_w^2(\mathbb{R},\mathbb{C})\to L_w^2(\mathbb{R},\mathbb{C})}$ with a kernel, $G:\mathbb{R}\times \mathbb{R} \to \mathbb{R}$. This filter/window can be represented by the inner product weight function, $w$. Note that $w:\mathbb{R}\to \mathbb{R}_{>0}$ can be chosen such that $e^{ikx}\in L^2_w(\mathbb{R},\mathbb{C})$. However, the introduction of $w$ means that $\varphi$ is now required to satisfy the POD integral eigenvalue problem with the filtered/weighted kernel. For a non-constant $w$, the effective kernel is given by
\begin{equation}
G(x,y)=\widetilde{G}(x-y)w(y)\,,
\end{equation}
from which it is clear that the resulting kernel, $G(x,y)$, is not translationally invariant despite the fact that $\widetilde{G}$ is. Because of this, $\varphi(x)= e^{ikx}\,,\,x\in\mathbb{R}$, is again disqualified from being a solution to the corresponding POD eigenvalue problem. 

The preceding theoretical considerations have led us to conclude that aperiodic domains do not admit to POD integral eigenfunctions of the form $e^{ikx}$, either due to the failure to attain true translational invariance in the kernel on finite domains, or in the case of infinite domains due to the fact that $e^{ikx}$ does not reside in $L^2(\mathbb{R},\mathbb{C})$. The introduction of a filter is also shown to modify the effective kernel such that it is not translationally invariant - disqualifying the use of filters as a strategy to conclude that $e^{ikx}$ are the eigenfunctions. These theoretical insights have therefore led us to the conclusion that we cannot expect that any numerical solutions to kernels on finite domains are Fourier bases.
\section{Reconstruction of Eigenspectra\label{app:reconstruction_of_eigenspectra_using_Fourier_modes}}
\begin{figure}[h!!!]
\centering
\begin{subfigure}[h]{\plotwidth\textwidth}
\includegraphics[width=\textwidth]{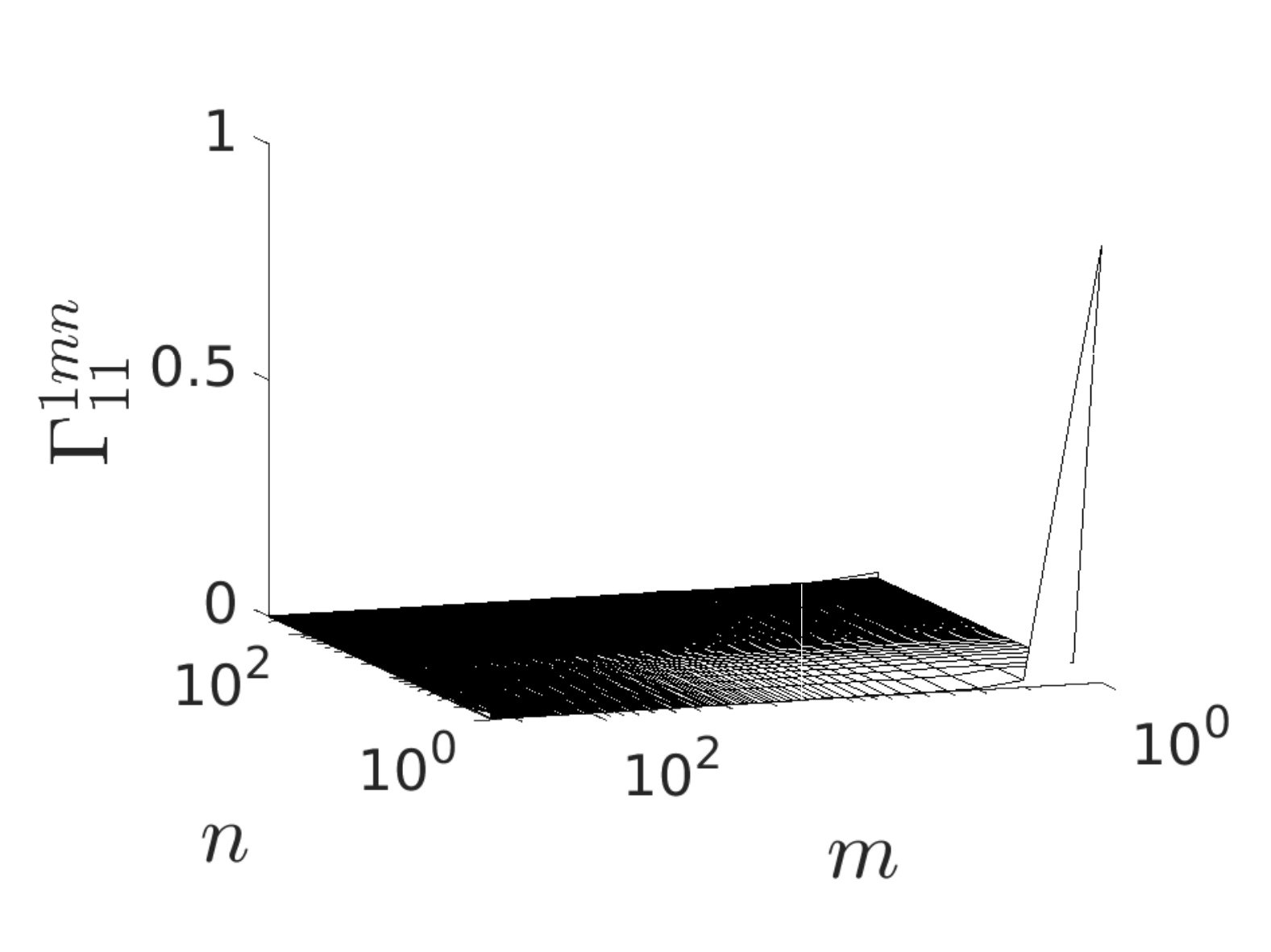}
\caption{\label{fig:evalue_reconstruction_domain_1_kernel_1_alpha_1}}
\end{subfigure}
\begin{subfigure}[h]{\plotwidth\textwidth}
\includegraphics[width=\textwidth]{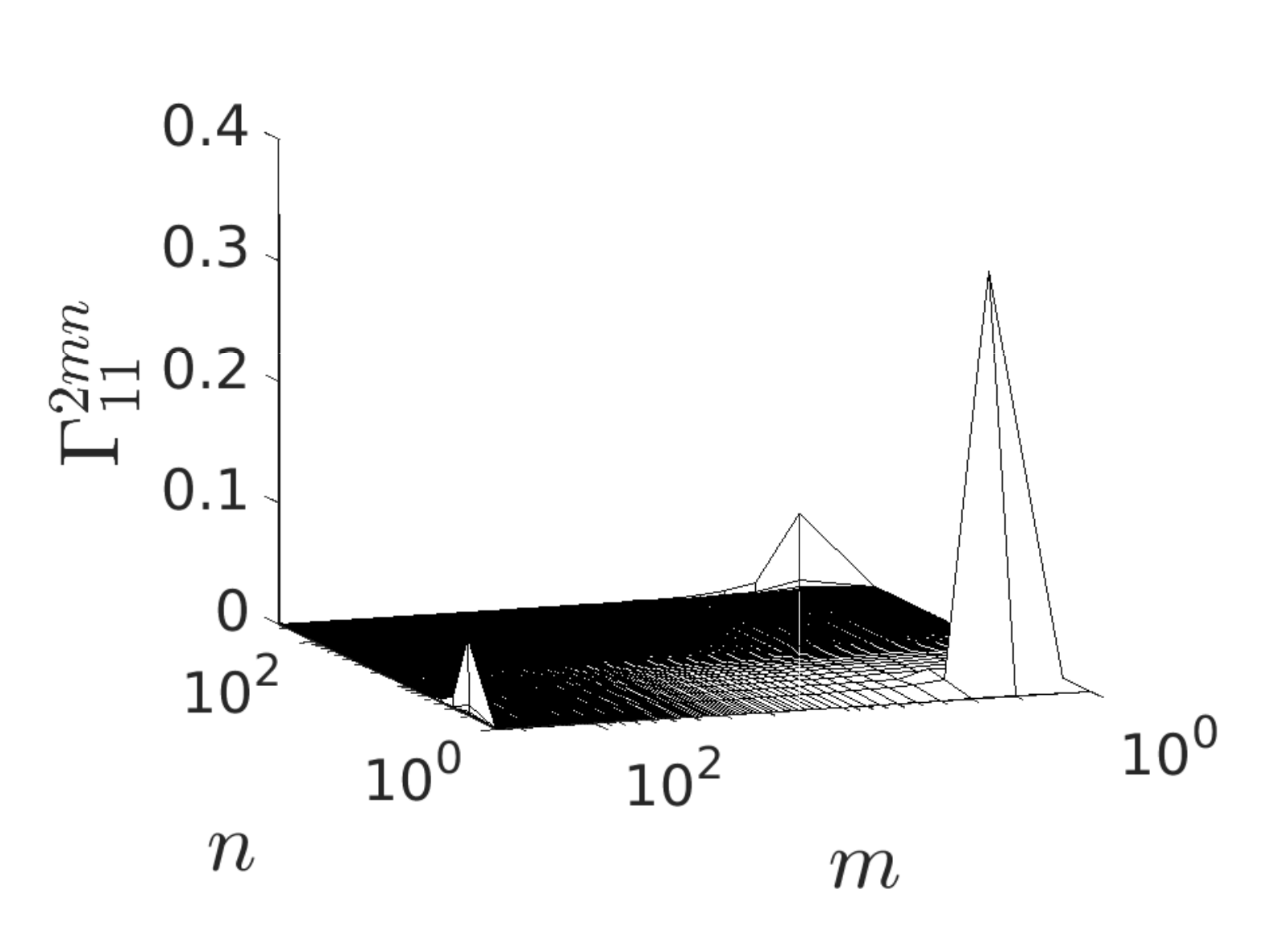}\caption{\label{fig:evalue_reconstruction_domain_1_kernel_1_alpha_2}}
\end{subfigure}
\begin{subfigure}[h]{\plotwidth\textwidth}
\includegraphics[width=\textwidth]{figs/pod/evalue_recon/evalue_reconstruction_domain_1_kernel_1_alpha_3}\caption{\label{fig:evalue_reconstruction_domain_1_kernel_1_alpha_3}}
\end{subfigure}
\begin{subfigure}[h]{\plotwidth\textwidth}
\includegraphics[width=\textwidth]{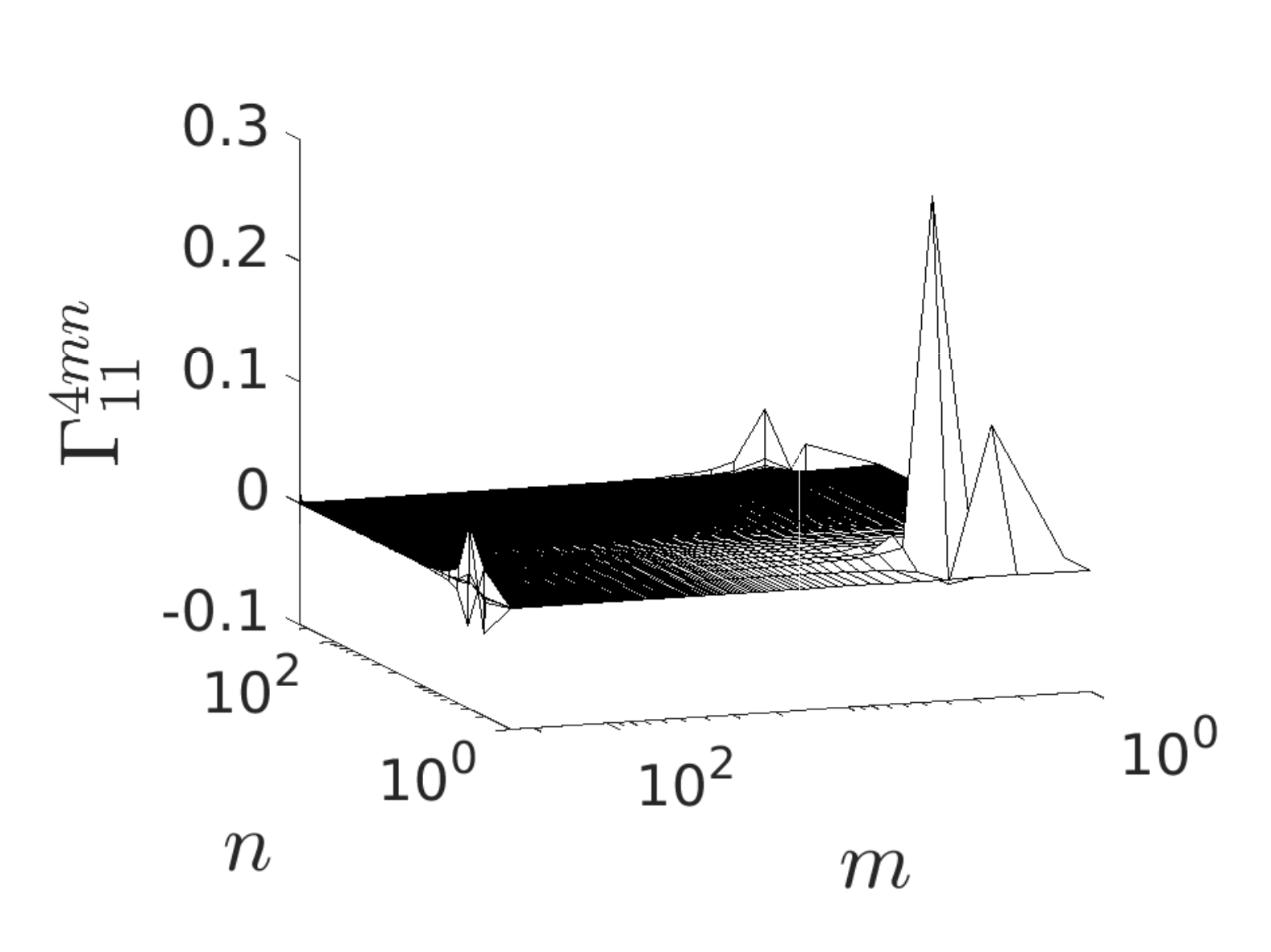}\caption{\label{fig:evalue_reconstruction_domain_1_kernel_1_alpha_4}}
\end{subfigure}
\begin{subfigure}[h]{\plotwidth\textwidth}
\includegraphics[width=\textwidth]{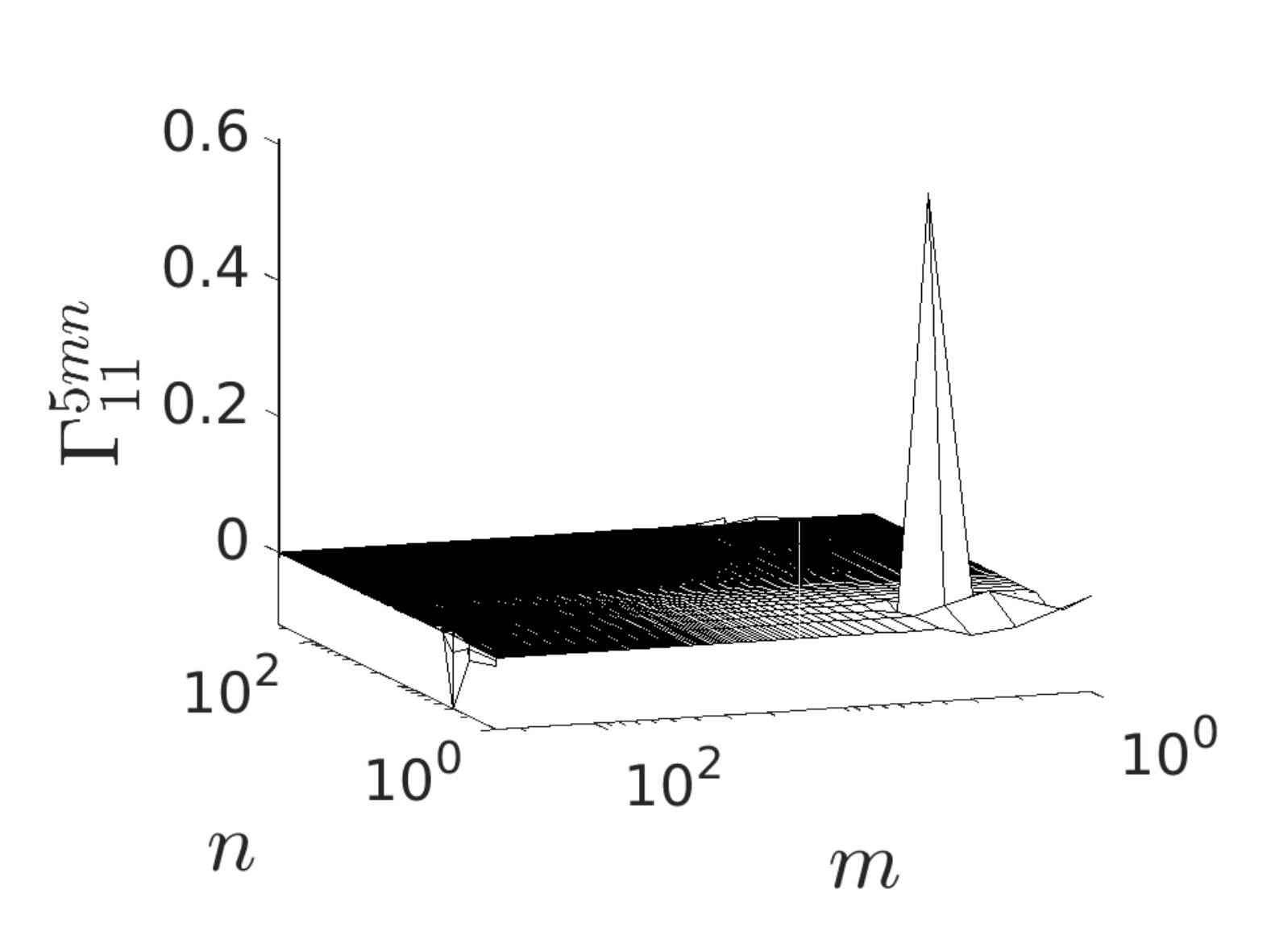}\caption{\label{fig:evalue_reconstruction_domain_1_kernel_1_alpha_5}}
\end{subfigure}
\begin{subfigure}[h]{\plotwidth\textwidth}
\includegraphics[width=\textwidth]{figs/pod/evalue_recon/evalue_reconstruction_domain_1_kernel_1_alpha_6}\caption{\label{fig:evalue_reconstruction_domain_1_kernel_1_alpha_6}}
\end{subfigure}
\caption{Contributions to the eigenvalue reconstruction of modes $\alpha=1:6$ using Fourier modes. (a)-(f): contributions for $K_{11}$.\label{fig:evalue_reconstruction_exp_domain_1}}
\end{figure}

Figures \ref{fig:evalue_reconstruction_exp_domain_1}-\ref{fig:evalue_reconstruction_kernel_5} show the reconstruction of the first six eigenvalues for the smallest and largest domain sizes.
%
%
\begin{figure}[h!!!]
\centering
\begin{subfigure}[h]{\plotwidth\textwidth}
\includegraphics[width=\textwidth]{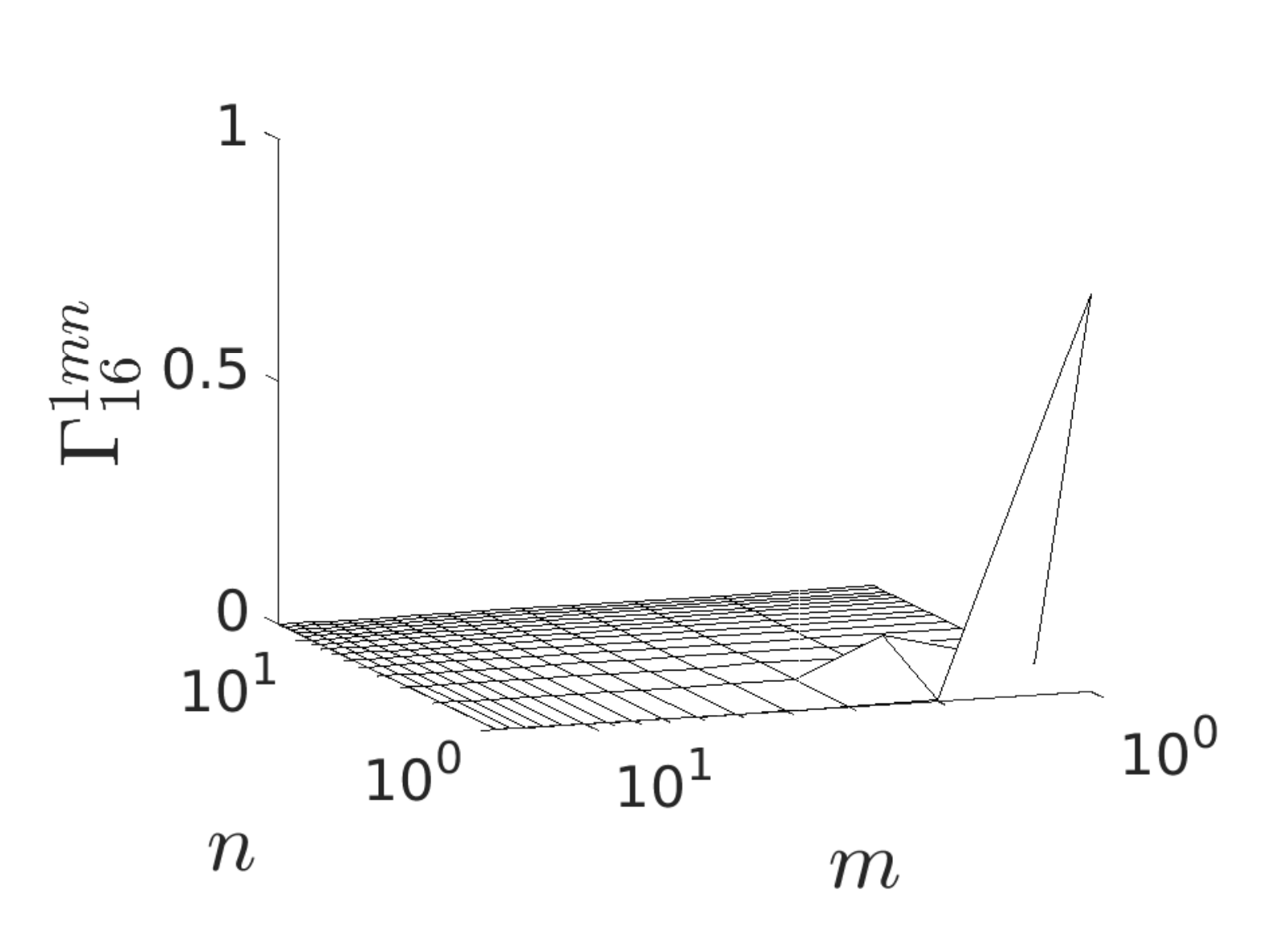}
\caption{\label{fig:evalue_reconstruction_domain_6_kernel_1_alpha_1}}
\end{subfigure}
\begin{subfigure}[h]{\plotwidth\textwidth}
\includegraphics[width=\textwidth]{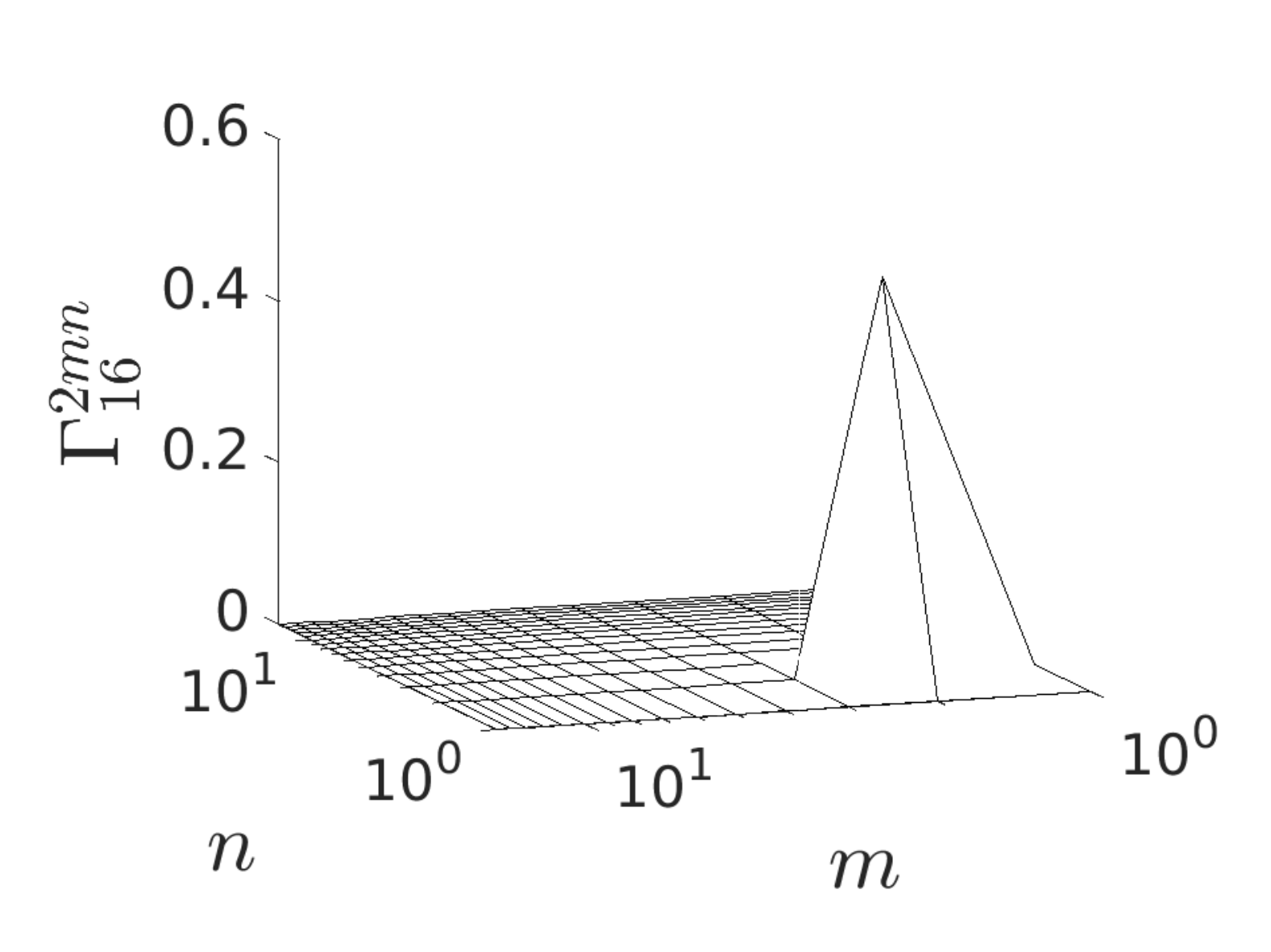}\caption{\label{fig:evalue_reconstruction_domain_6_kernel_1_alpha_2}}
\end{subfigure}
\begin{subfigure}[h]{\plotwidth\textwidth}
\includegraphics[width=\textwidth]{figs/pod/evalue_recon/evalue_reconstruction_domain_6_kernel_1_alpha_3}\caption{\label{fig:evalue_reconstruction_domain_6_kernel_1_alpha_3}}
\end{subfigure}
\begin{subfigure}[h]{\plotwidth\textwidth}
\includegraphics[width=\textwidth]{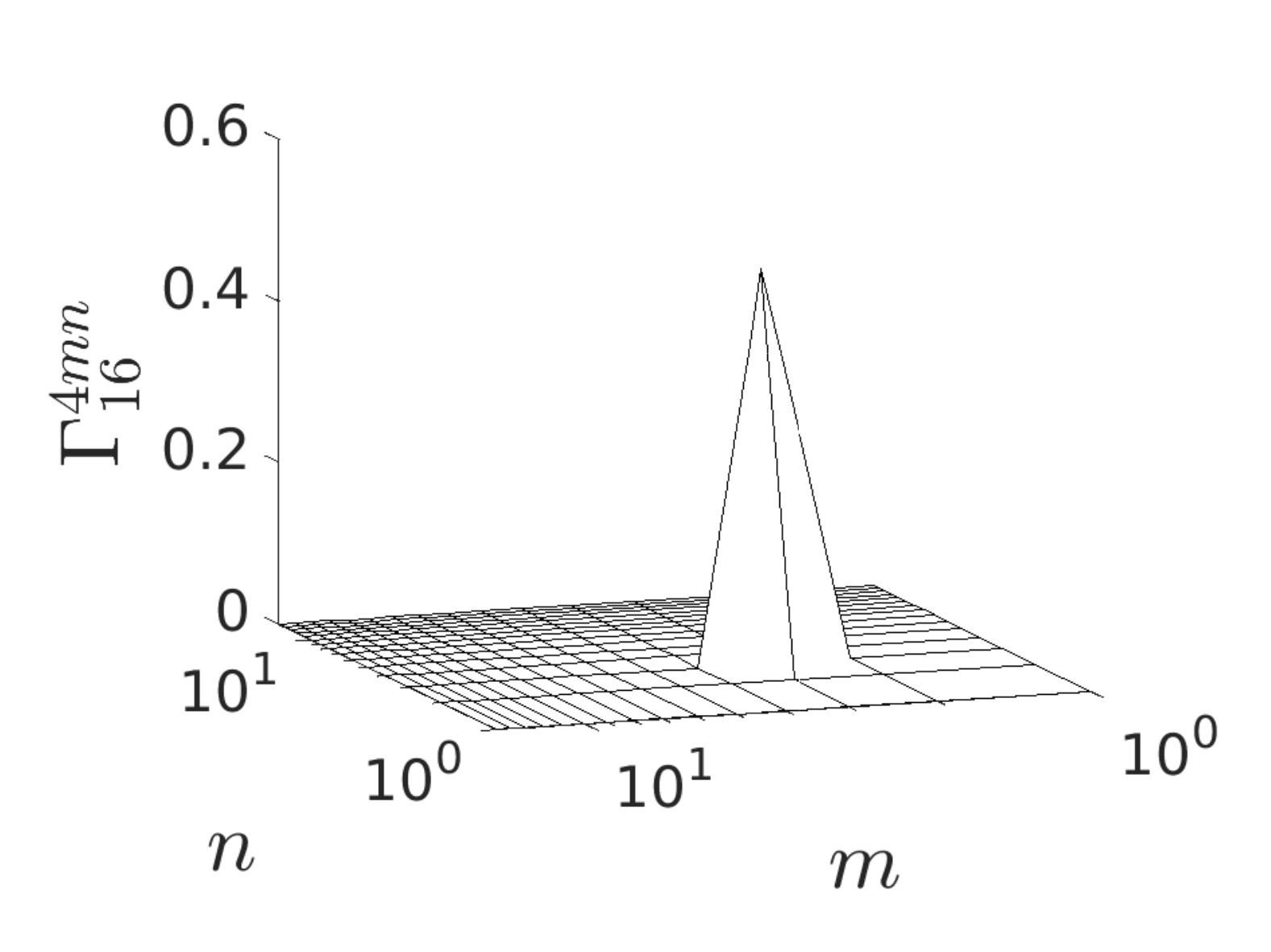}\caption{\label{fig:evalue_reconstruction_domain_6_kernel_1_alpha_4}}
\end{subfigure}
\begin{subfigure}[h]{\plotwidth\textwidth}
\includegraphics[width=\textwidth]{figs/pod/evalue_recon/evalue_reconstruction_domain_6_kernel_1_alpha_5}\caption{\label{fig:evalue_reconstruction_domain_6_kernel_1_alpha_5}}
\end{subfigure}
\begin{subfigure}[h]{\plotwidth\textwidth}
\includegraphics[width=\textwidth]{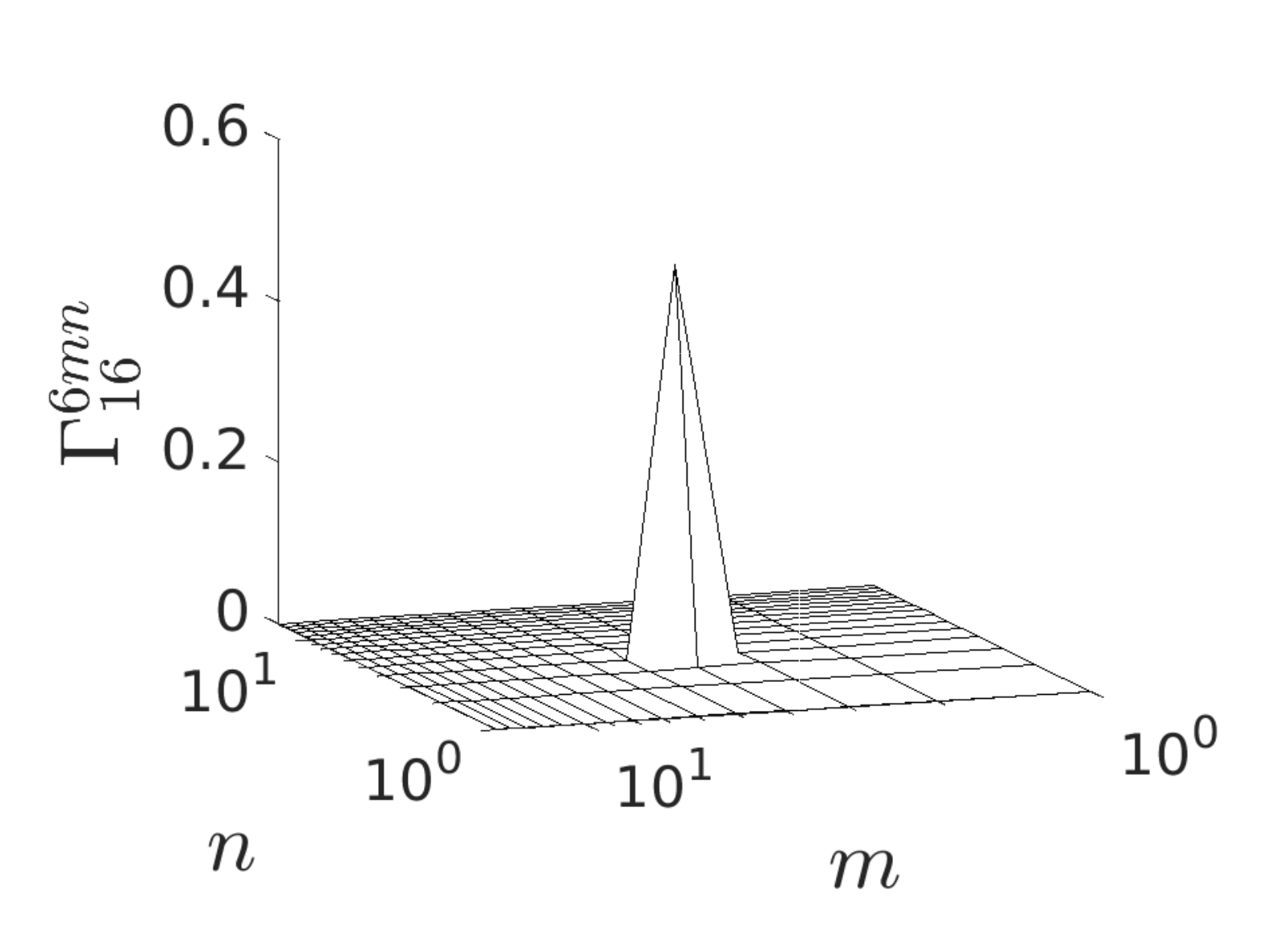}\caption{\label{fig:evalue_reconstruction_domain_6_kernel_1_alpha_6}}
\end{subfigure}
\caption{Contributions to the eigenvalue reconstruction of modes $\alpha=1:6$ using Fourier modes. (a)-(f):contributions for $K_{16}$.\label{fig:}}
\end{figure}
\FloatBarrier
\noindent
\begin{figure}[h!!!]
\centering
\begin{subfigure}[h]{\plotwidth\textwidth}
\includegraphics[width=\textwidth]{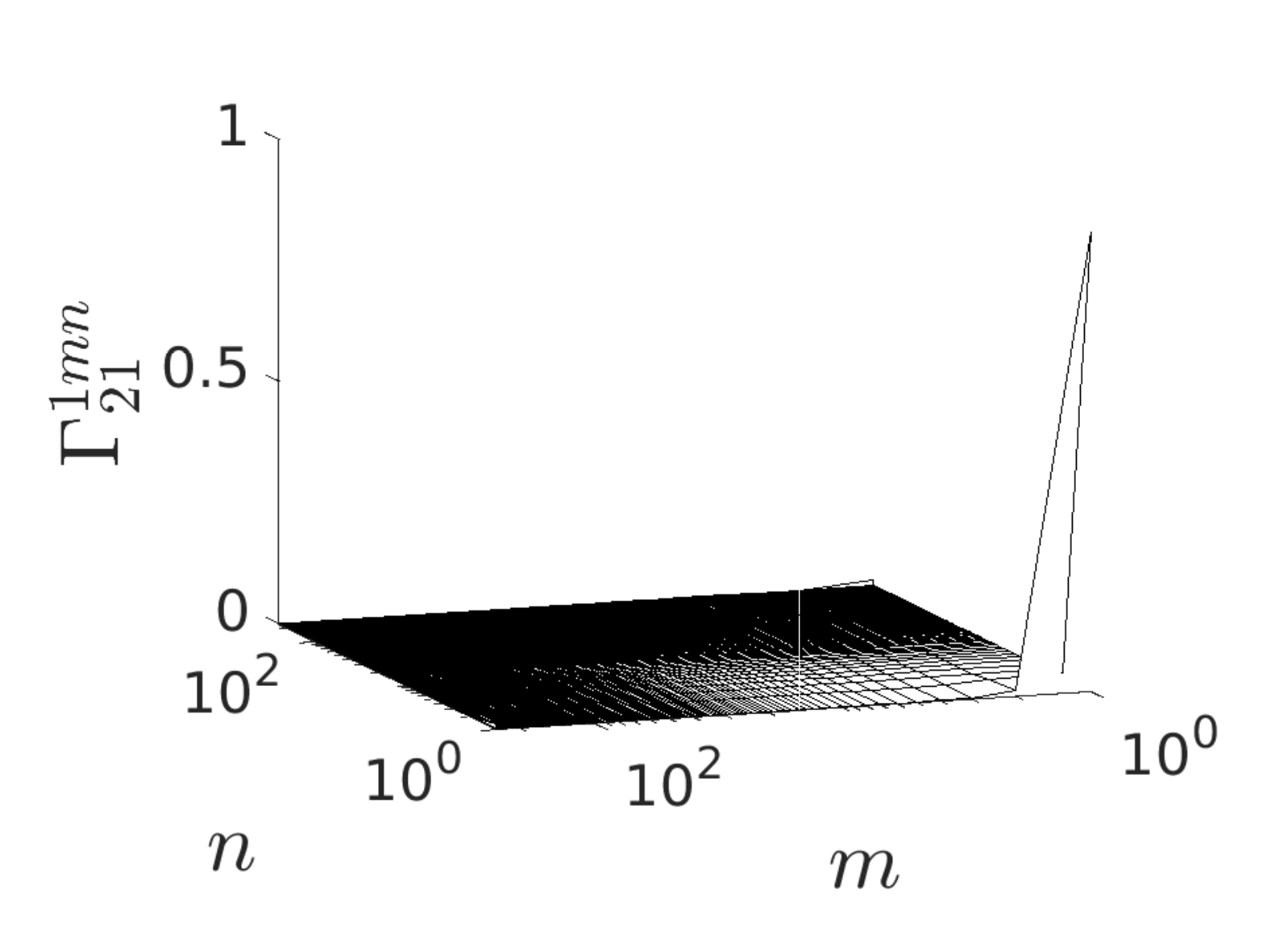}
\caption{\label{fig:evalue_reconstruction_domain_1_kernel_2_alpha_1}}
\end{subfigure}
\begin{subfigure}[h]{\plotwidth\textwidth}
\includegraphics[width=\textwidth]{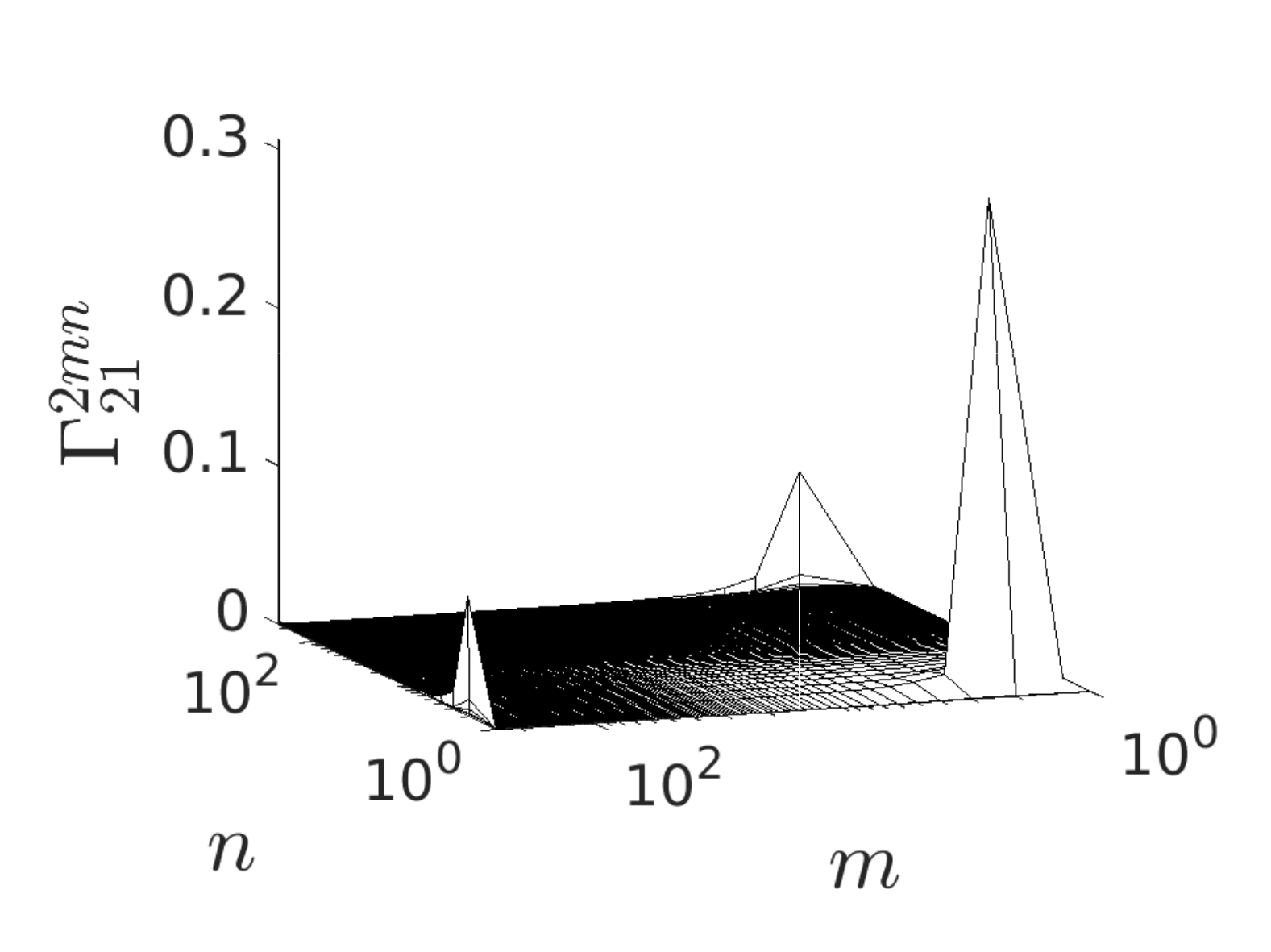}\caption{\label{fig:evalue_reconstruction_domain_1_kernel_2_alpha_2}}
\end{subfigure}
\begin{subfigure}[h]{\plotwidth\textwidth}
\includegraphics[width=\textwidth]{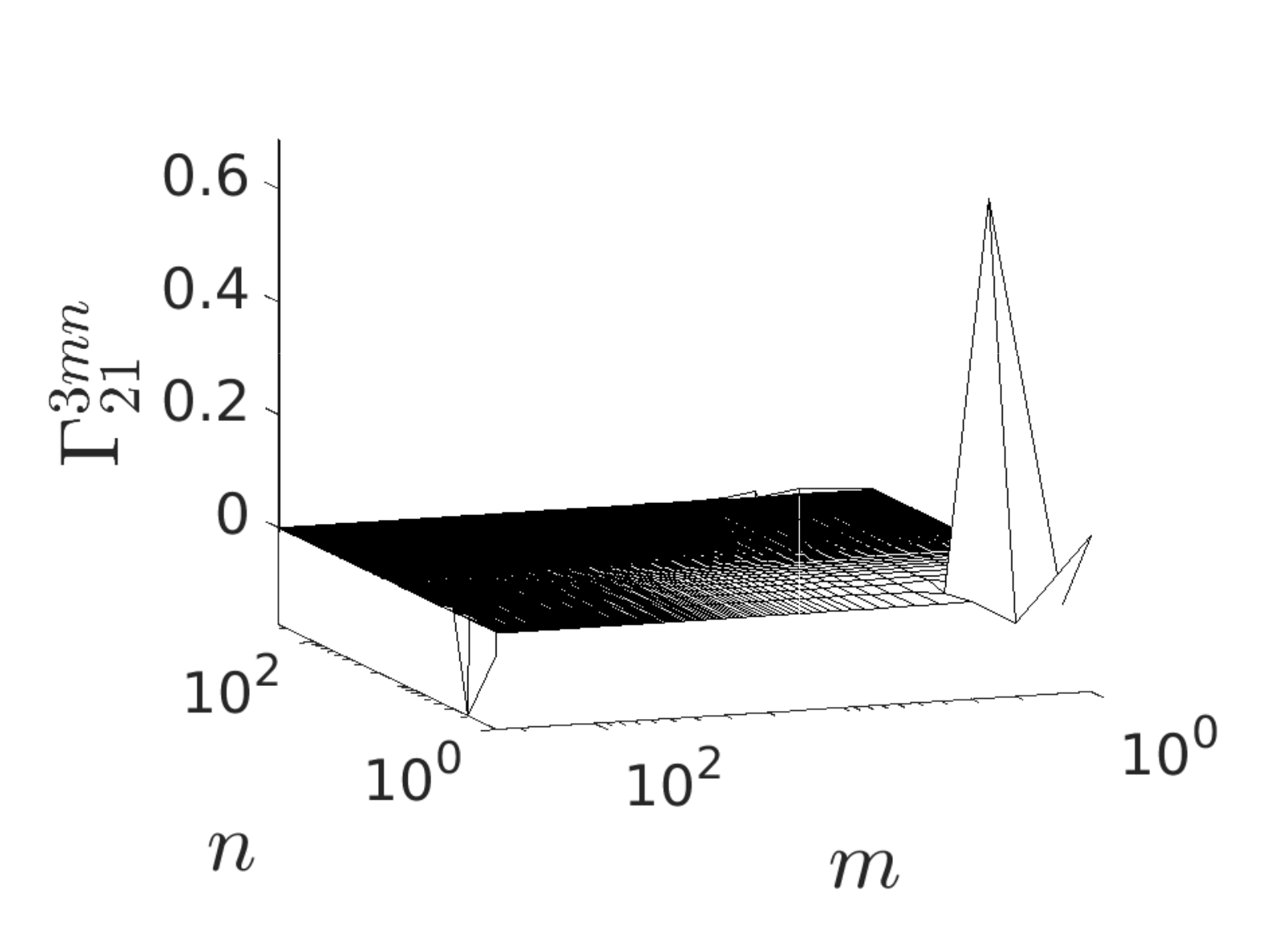}\caption{\label{fig:evalue_reconstruction_domain_1_kernel_2_alpha_3}}
\end{subfigure}
\begin{subfigure}[h]{\plotwidth\textwidth}
\includegraphics[width=\textwidth]{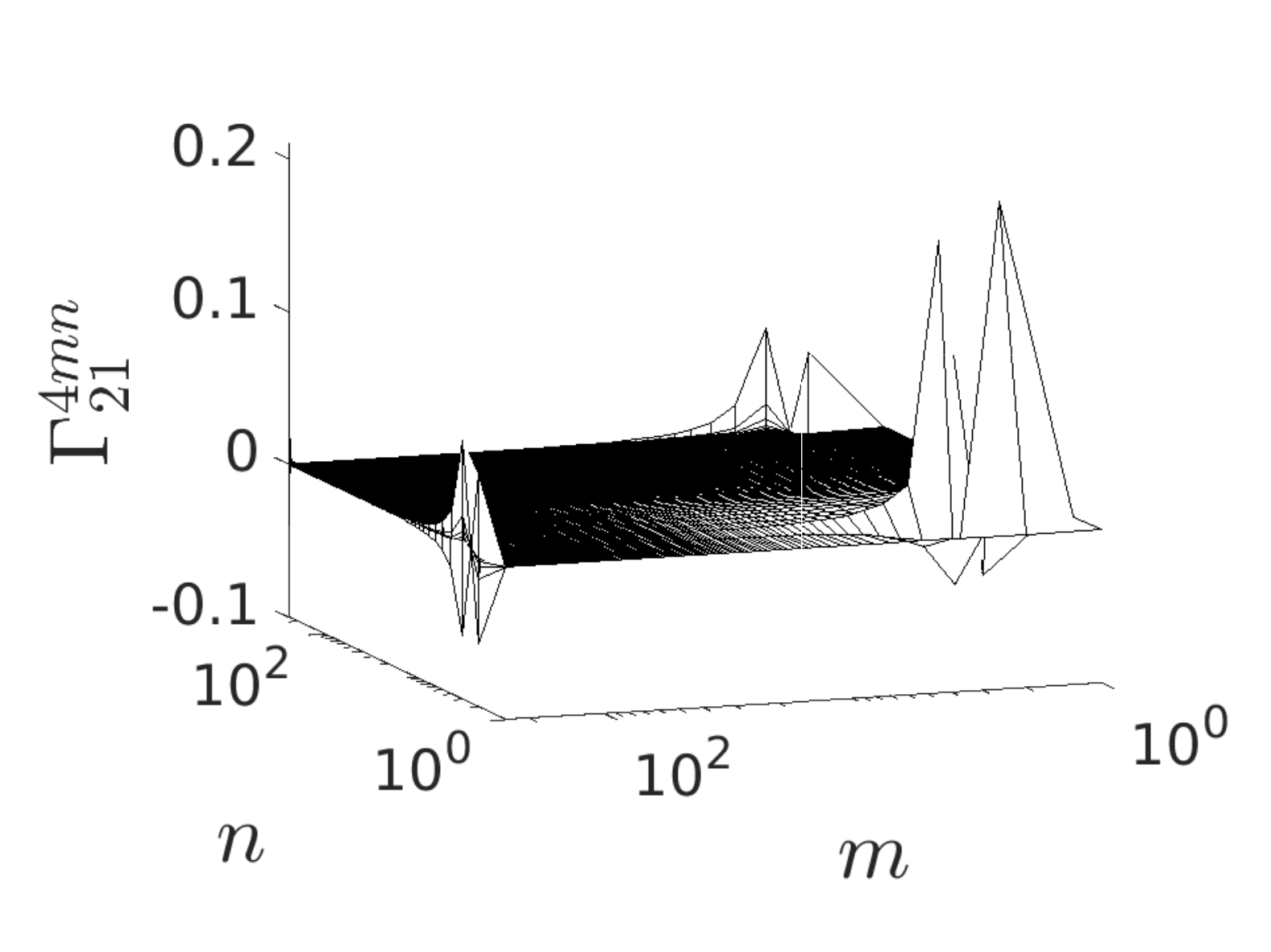}\caption{\label{fig:evalue_reconstruction_domain_1_kernel_2_alpha_4}}
\end{subfigure}
\begin{subfigure}[h]{\plotwidth\textwidth}
\includegraphics[width=\textwidth]{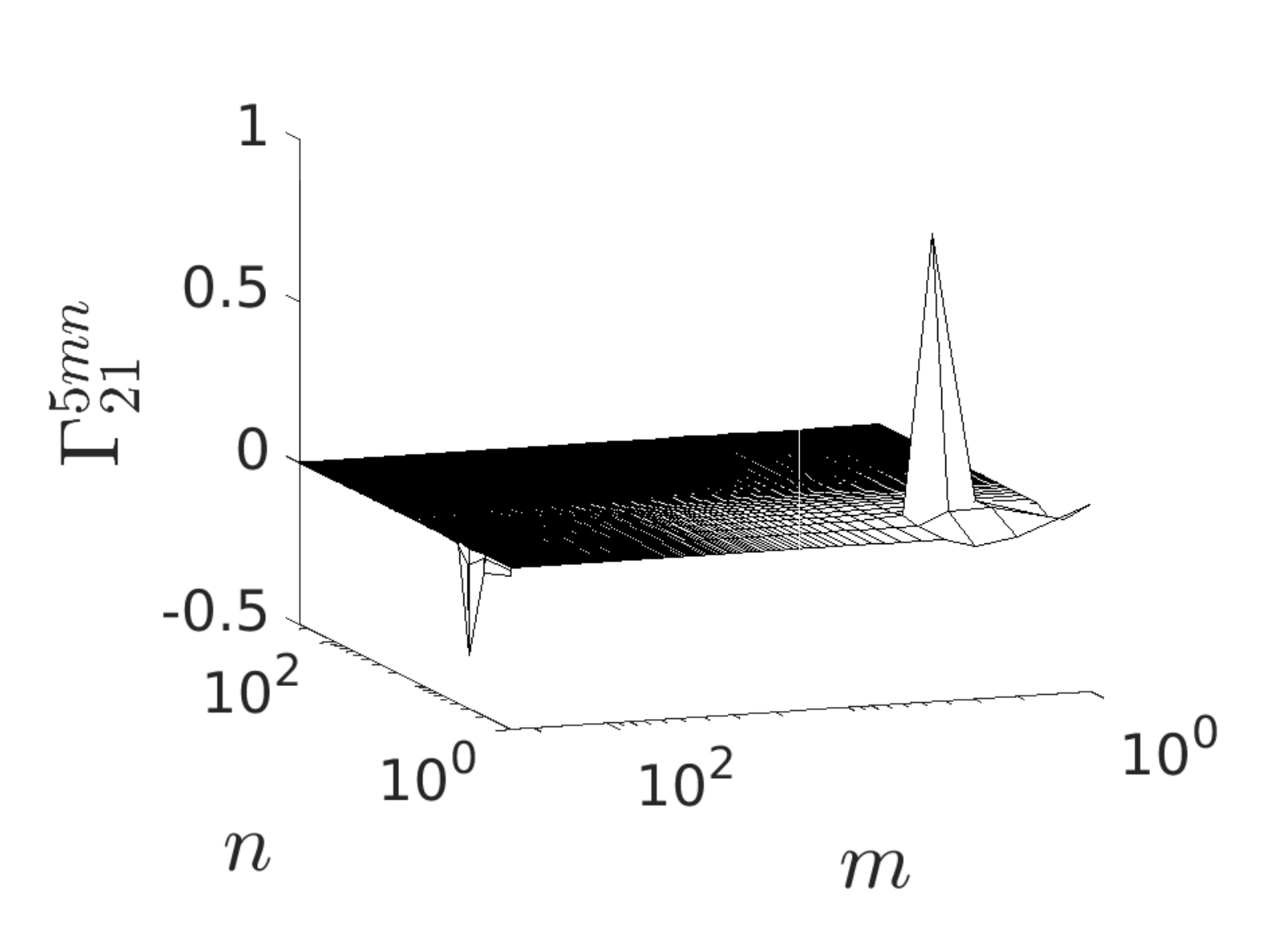}\caption{\label{fig:evalue_reconstruction_domain_1_kernel_2_alpha_5}}
\end{subfigure}
\begin{subfigure}[h]{\plotwidth\textwidth}
\includegraphics[width=\textwidth]{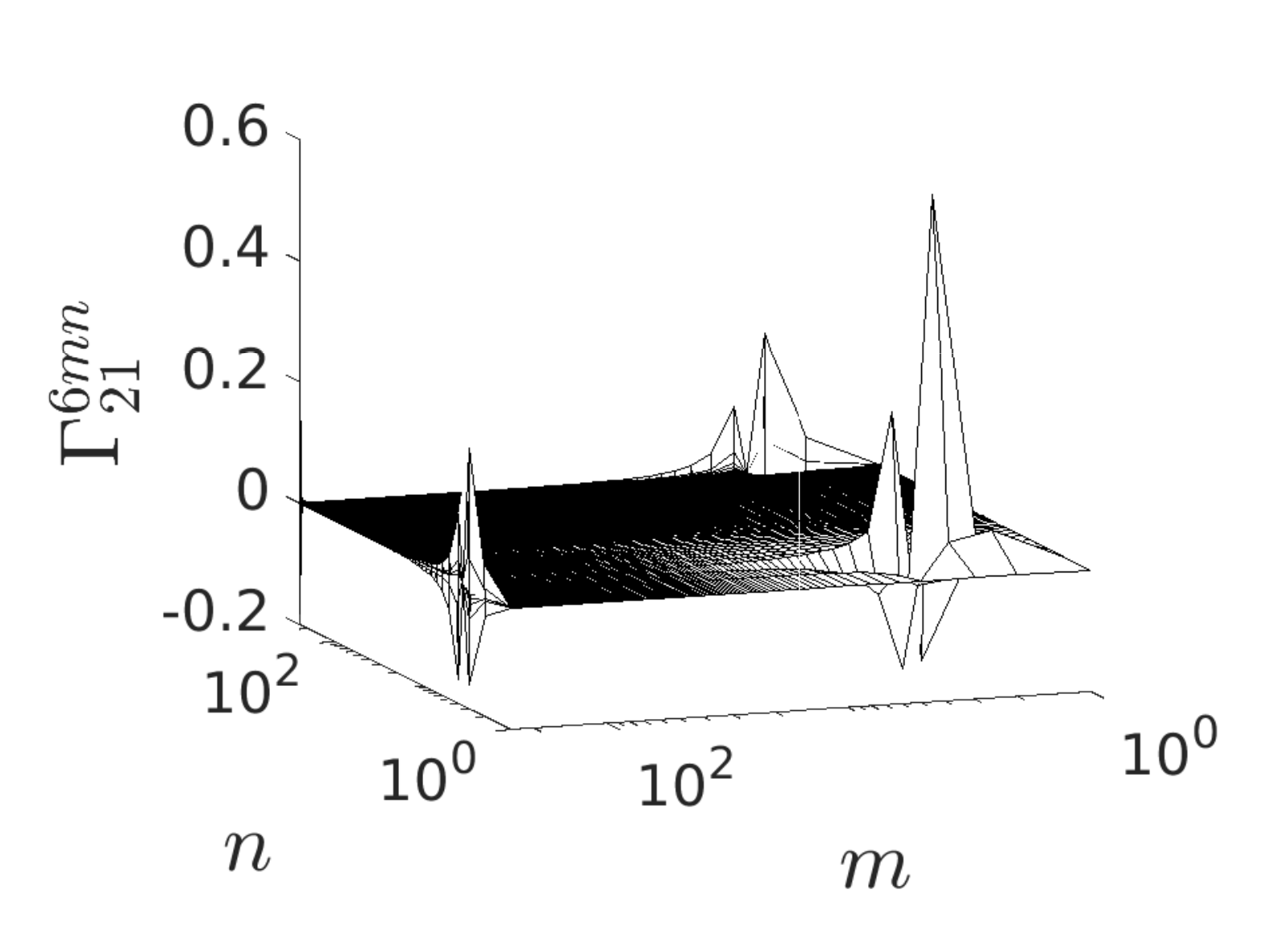}\caption{\label{fig:evalue_reconstruction_domain_1_kernel_2_alpha_6}}
\end{subfigure}
\caption{Contributions to the eigenvalue reconstruction of modes $\alpha=1:6$ using Fourier modes. (a)-(f): contributions for $K_{21}$.\label{fig:}}
\end{figure}
%
%
\begin{figure}[h!!!]
\centering
\begin{subfigure}[h]{\plotwidth\textwidth}
\includegraphics[width=\textwidth]{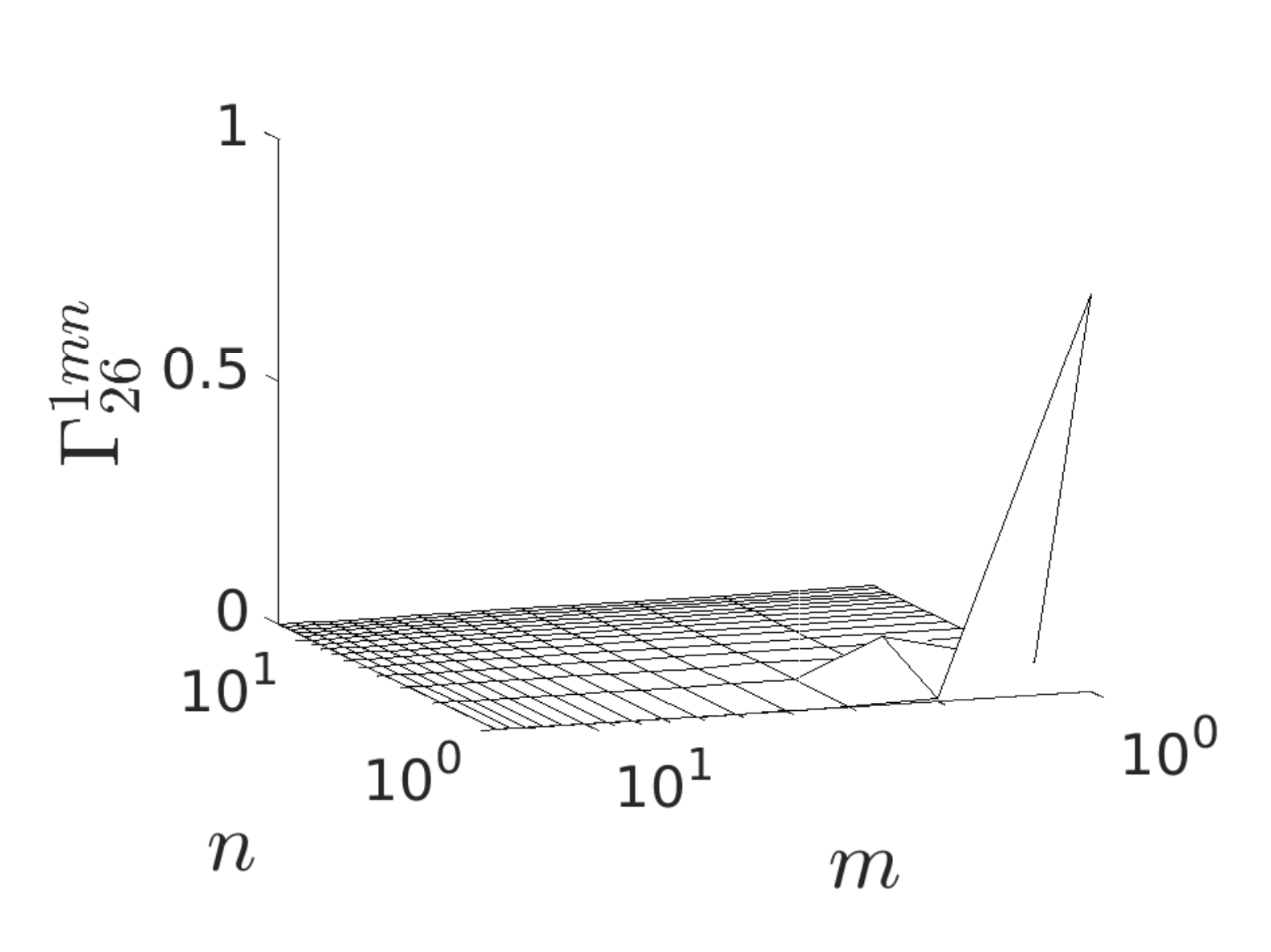}
\caption{\label{fig:evalue_reconstruction_domain_6_kernel_2_alpha_1}}
\end{subfigure}
\begin{subfigure}[h]{\plotwidth\textwidth}
\includegraphics[width=\textwidth]{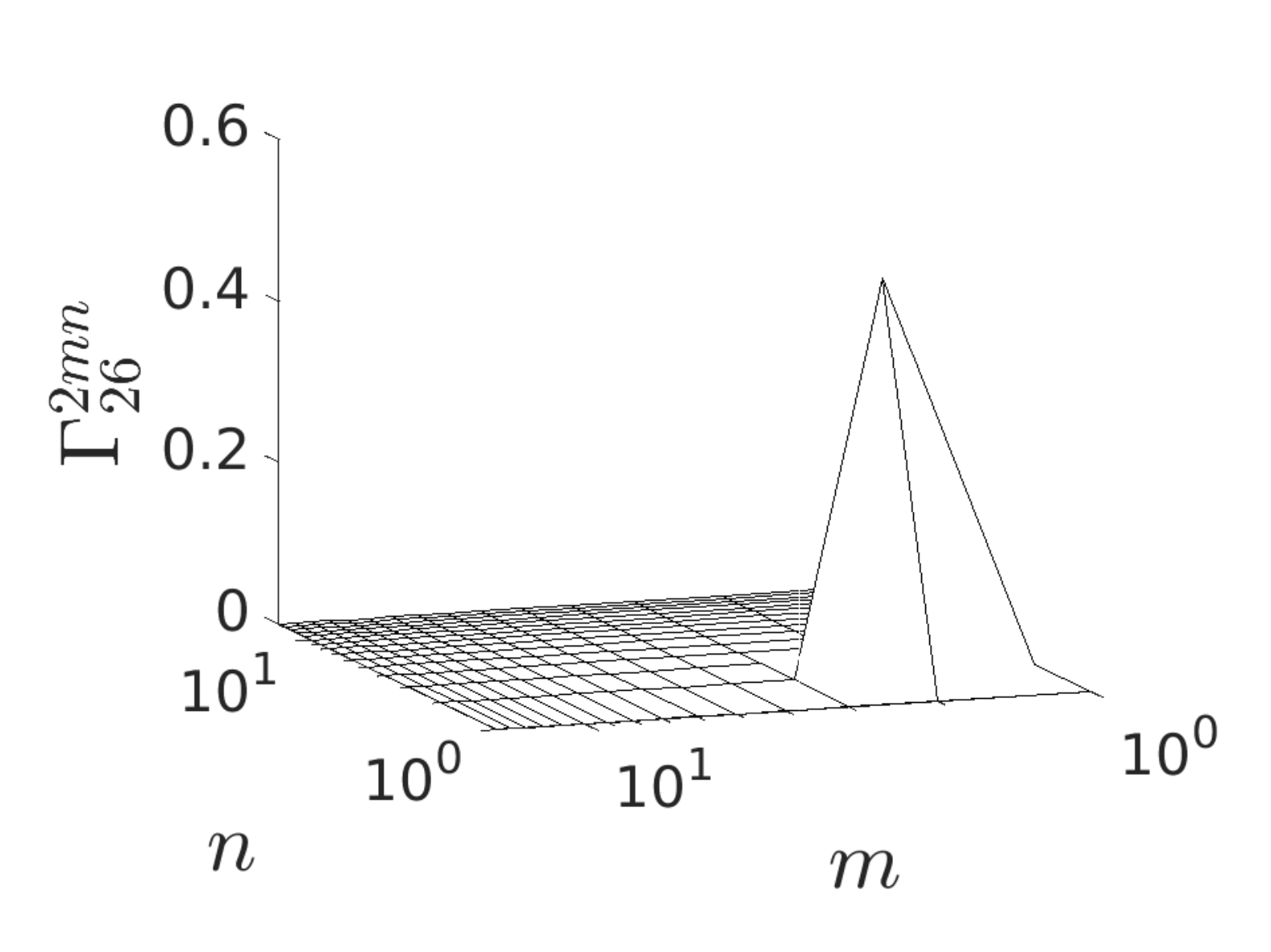}\caption{\label{fig:evalue_reconstruction_domain_6_kernel_2_alpha_2}}
\end{subfigure}
\begin{subfigure}[h]{\plotwidth\textwidth}
\includegraphics[width=\textwidth]{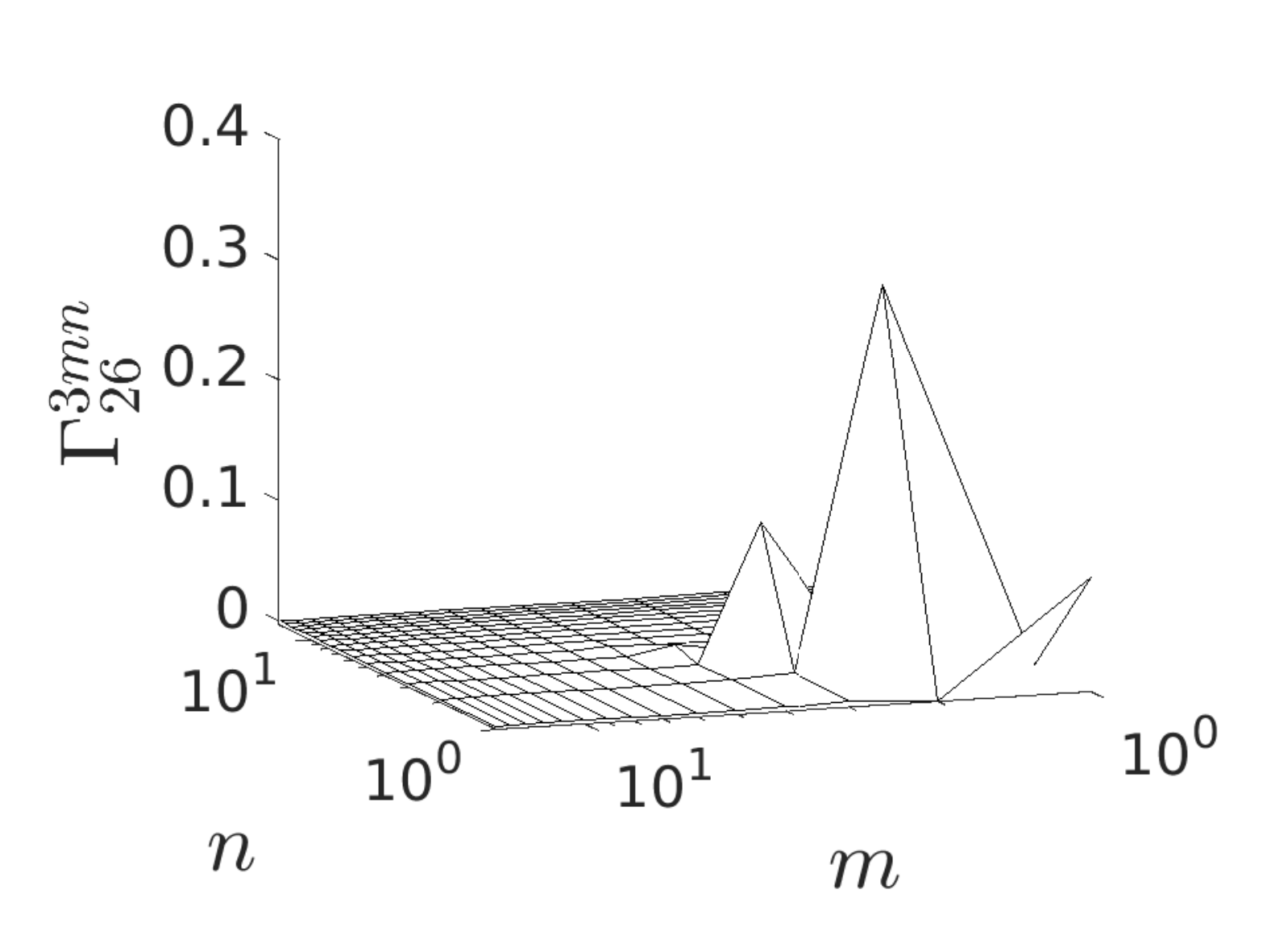}\caption{\label{fig:evalue_reconstruction_domain_6_kernel_2_alpha_3}}
\end{subfigure}
\begin{subfigure}[h]{\plotwidth\textwidth}
\includegraphics[width=\textwidth]{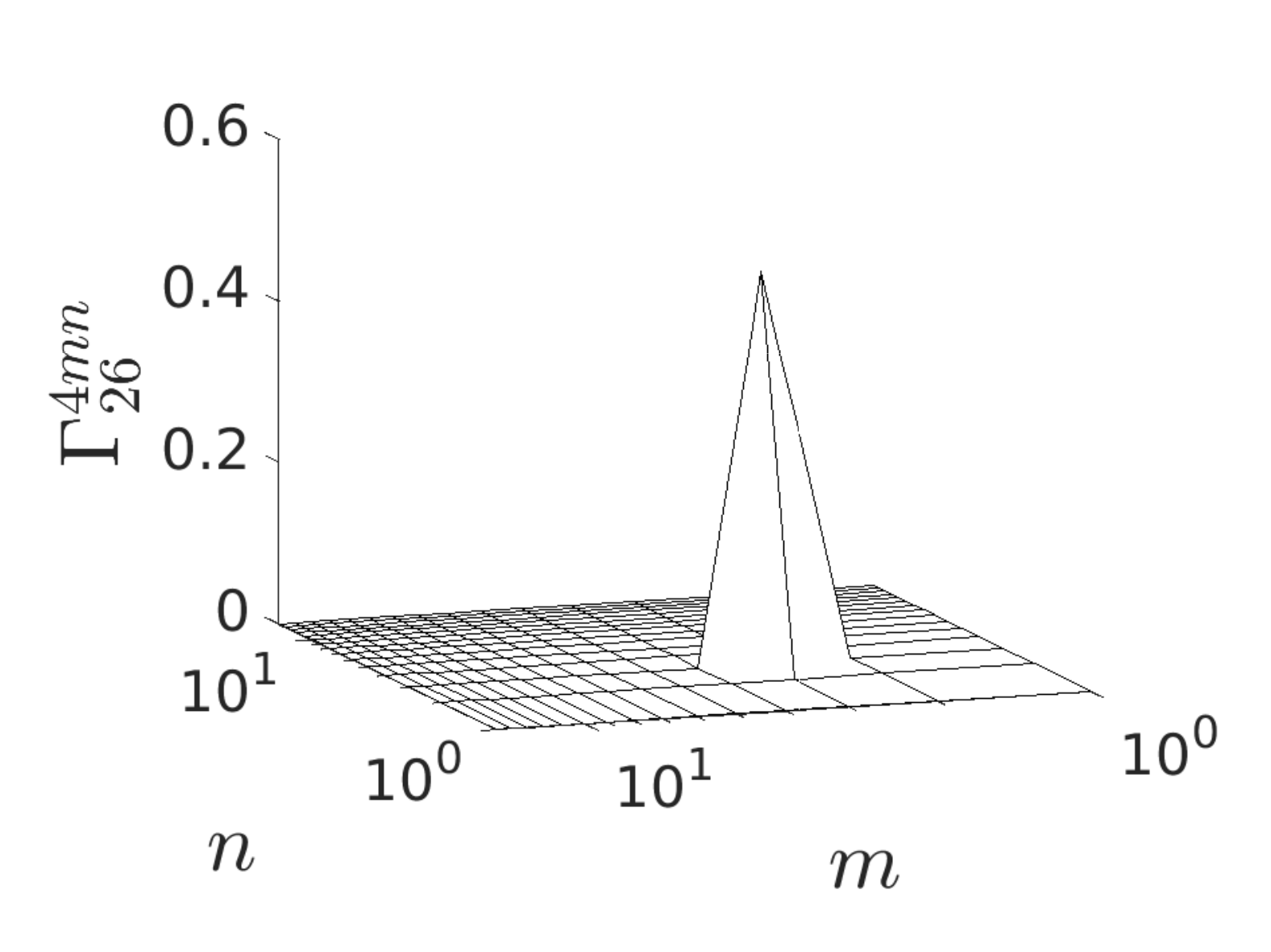}\caption{\label{fig:evalue_reconstruction_domain_6_kernel_2_alpha_4}}
\end{subfigure}
\begin{subfigure}[h]{\plotwidth\textwidth}
\includegraphics[width=\textwidth]{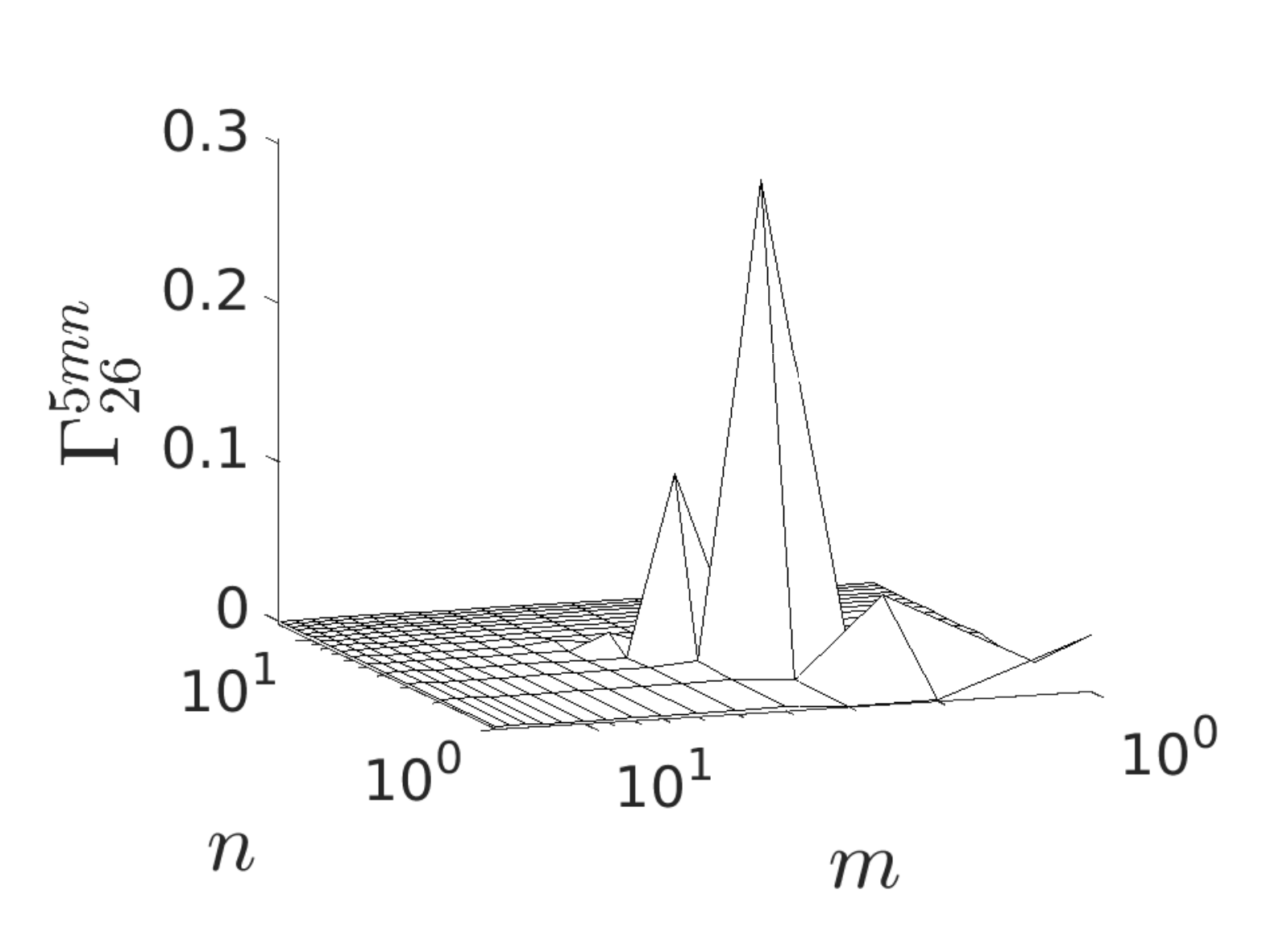}\caption{\label{fig:evalue_reconstruction_domain_6_kernel_2_alpha_5}}
\end{subfigure}
\begin{subfigure}[h]{\plotwidth\textwidth}
\includegraphics[width=\textwidth]{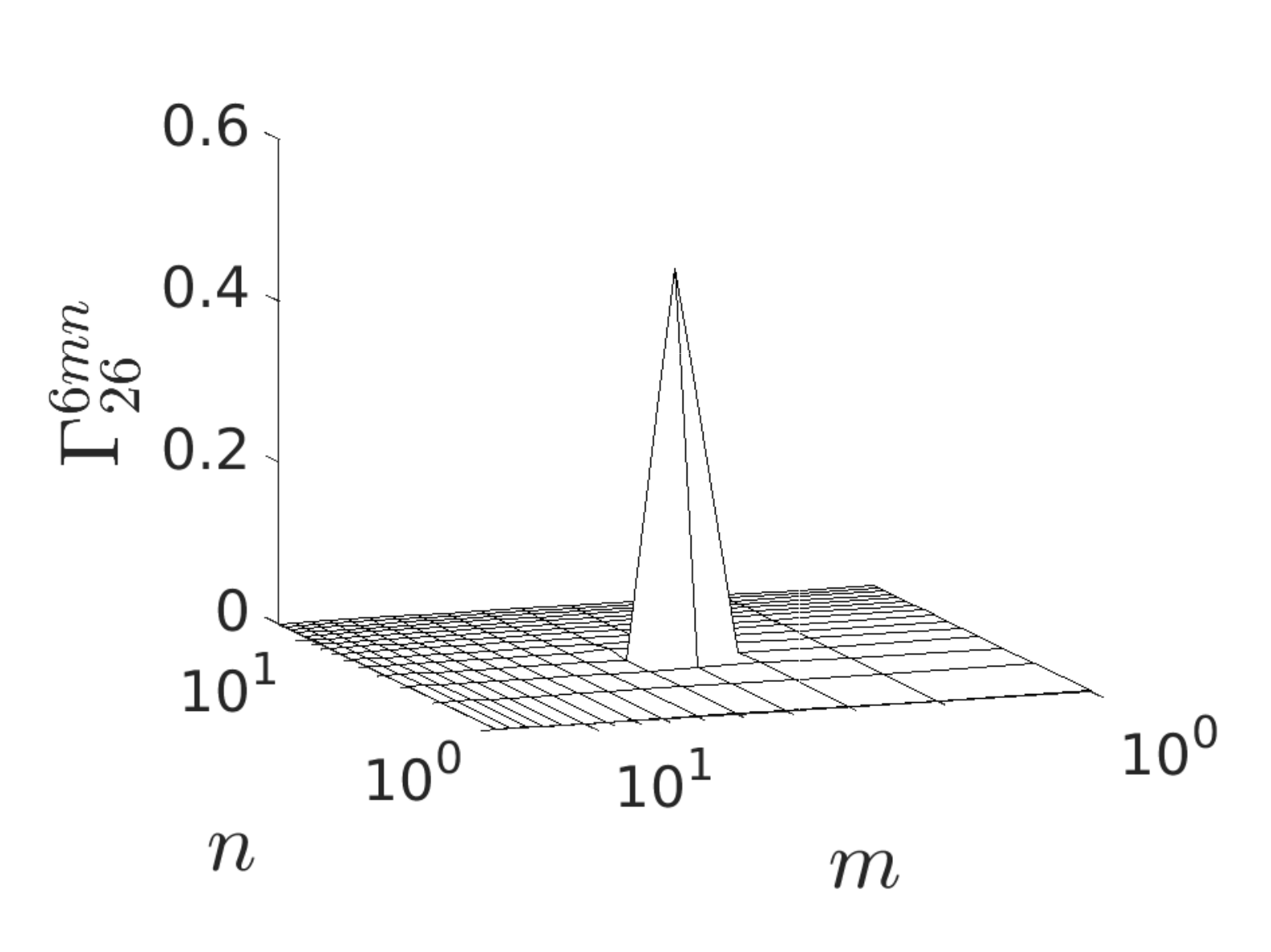}\caption{\label{fig:evalue_reconstruction_domain_6_kernel_2_alpha_6}}
\end{subfigure}
\caption{Contributions to the eigenvalue reconstruction of modes $\alpha=1:6$ using Fourier modes. (a)-(f): contributions for $K_{26}$.\label{fig:}}
\end{figure}
\FloatBarrier
\noindent
\begin{figure}[h!!!]
\centering
\begin{subfigure}[h]{\plotwidth\textwidth}
\includegraphics[width=\textwidth]{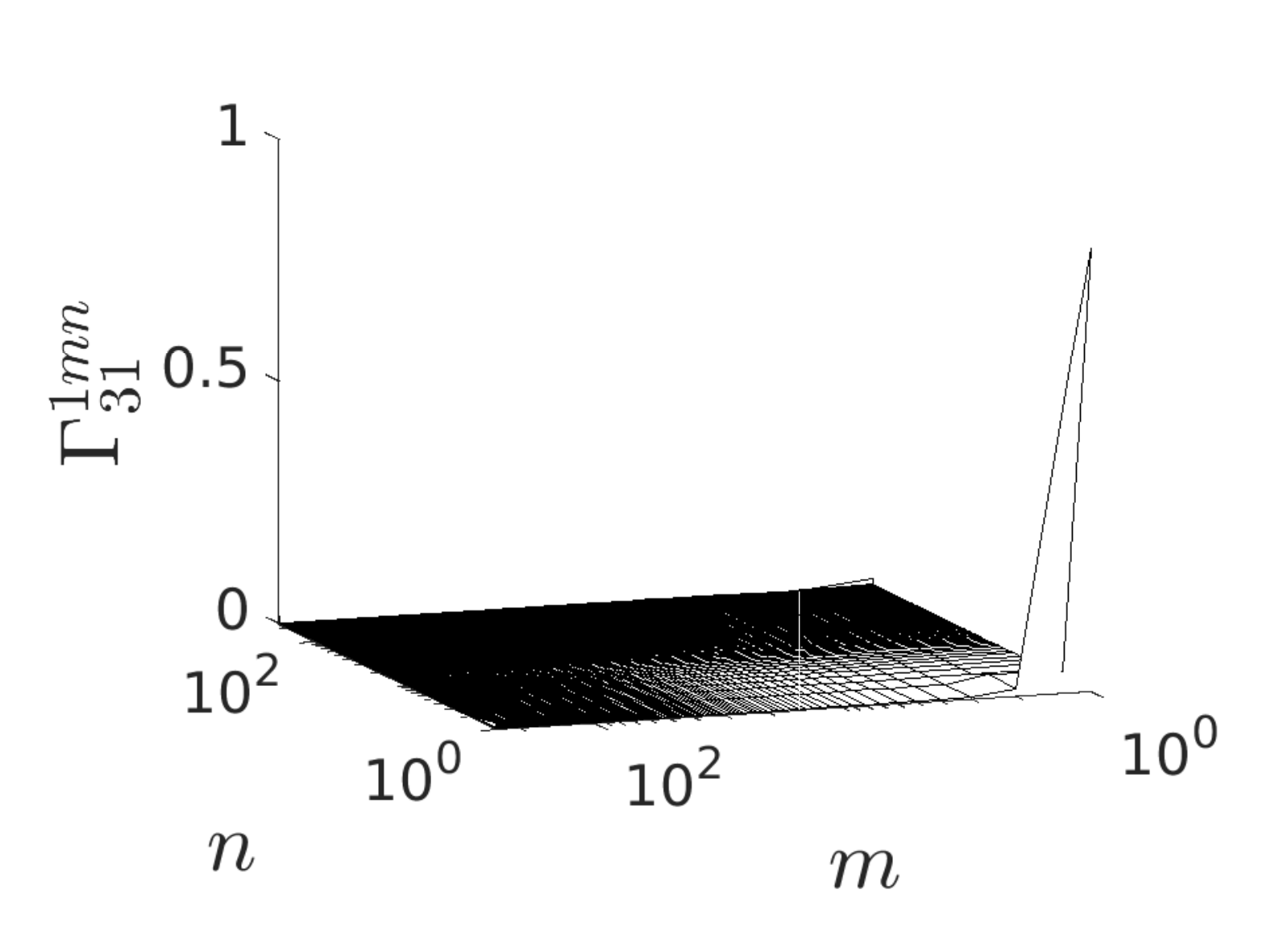}
\caption{\label{fig:evalue_reconstruction_domain_1_kernel_3_alpha_1}}
\end{subfigure}
\begin{subfigure}[h]{\plotwidth\textwidth}
\includegraphics[width=\textwidth]{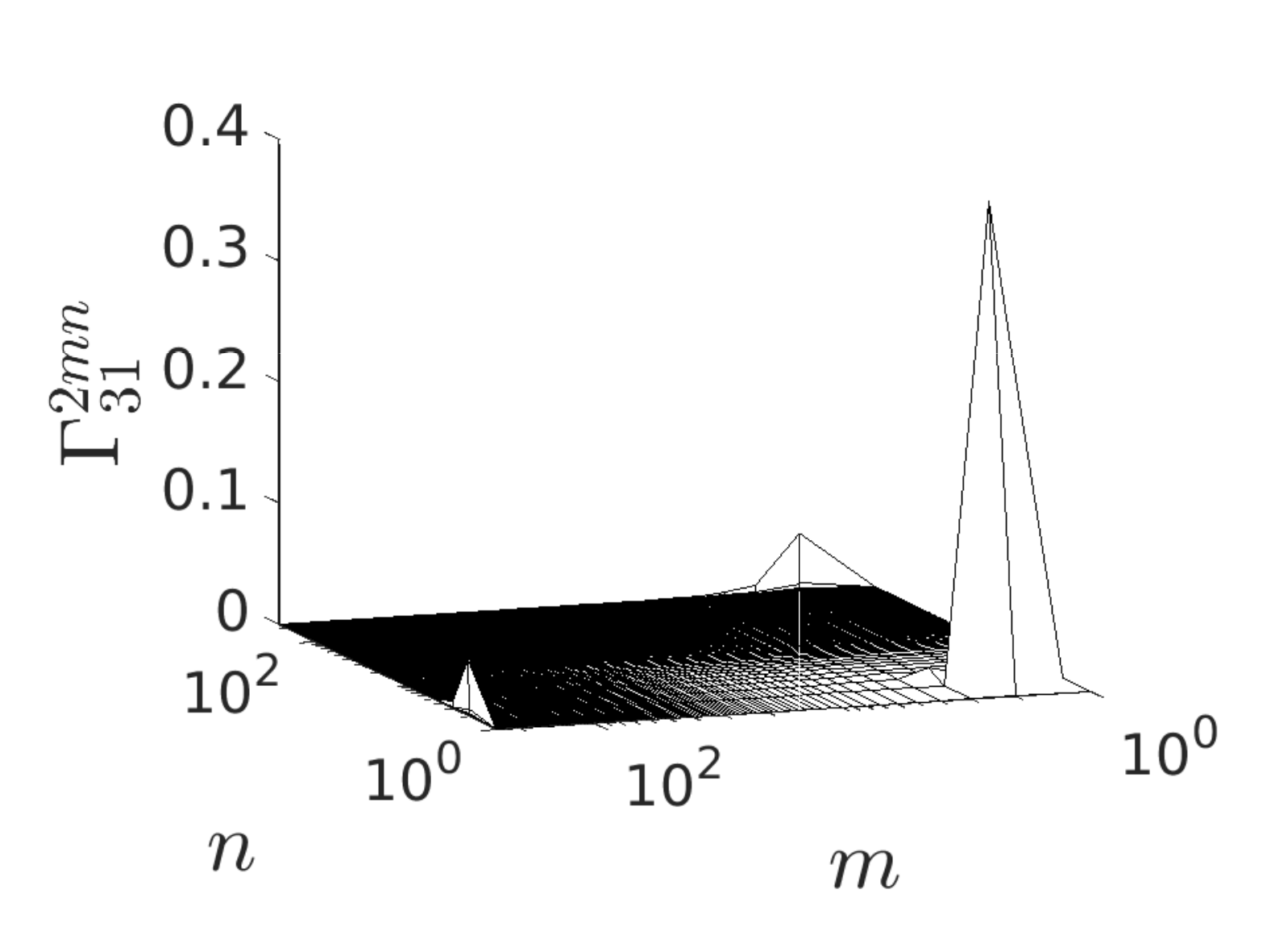}\caption{\label{fig:evalue_reconstruction_domain_1_kernel_3_alpha_2}}
\end{subfigure}
\begin{subfigure}[h]{\plotwidth\textwidth}
\includegraphics[width=\textwidth]{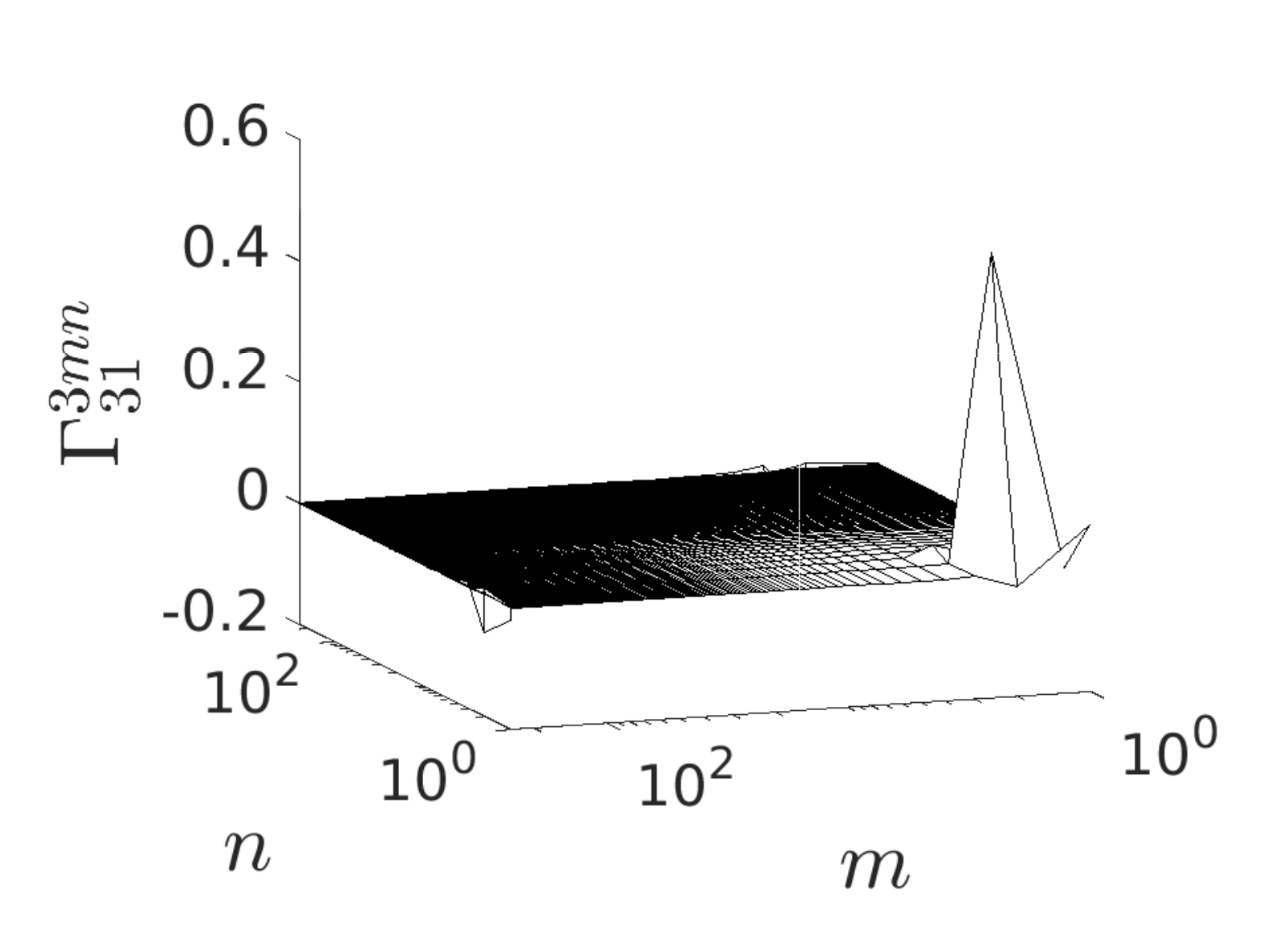}\caption{\label{fig:evalue_reconstruction_domain_1_kernel_3_alpha_3}}
\end{subfigure}
\begin{subfigure}[h]{\plotwidth\textwidth}
\includegraphics[width=\textwidth]{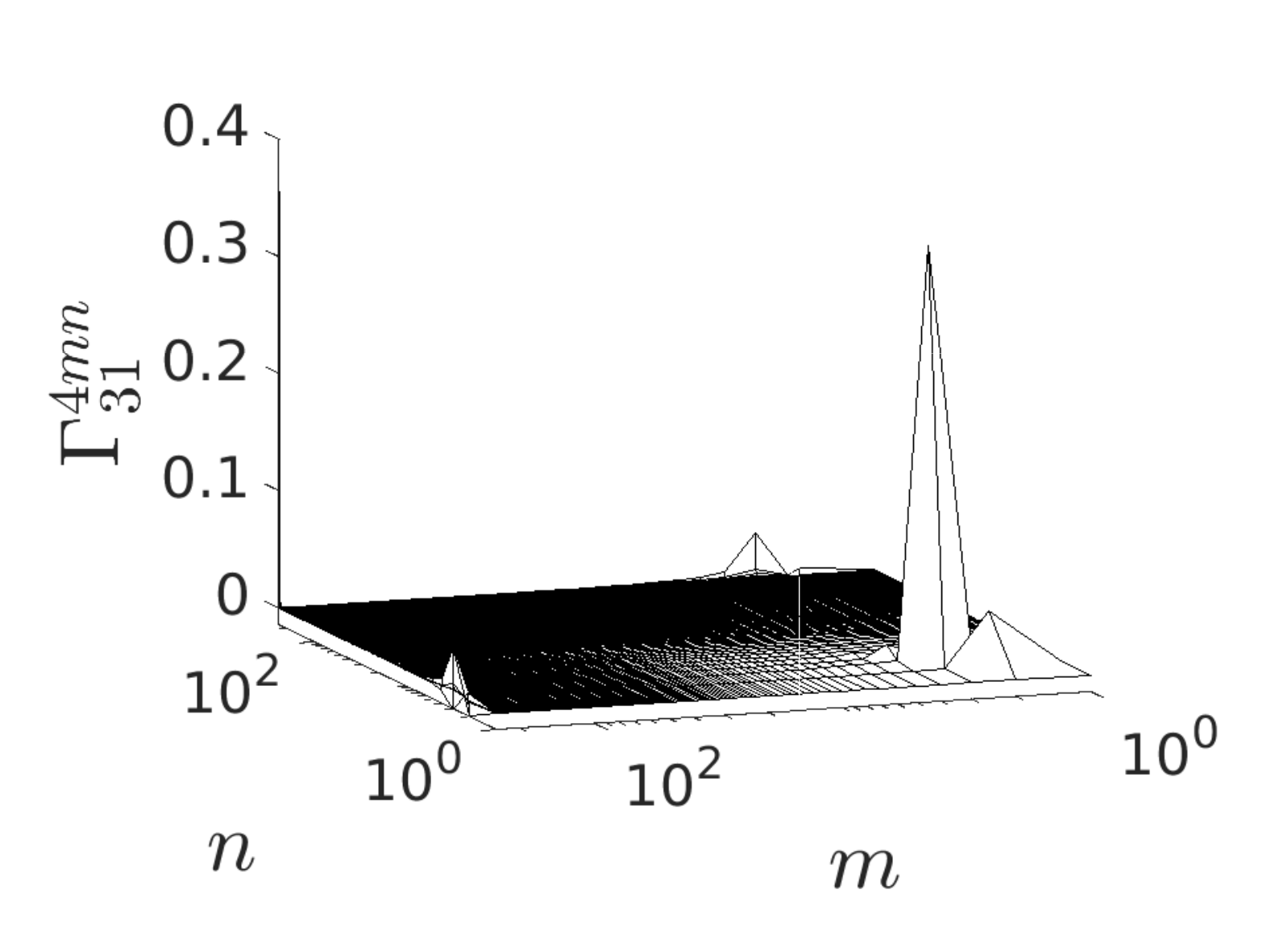}\caption{\label{fig:evalue_reconstruction_domain_1_kernel_3_alpha_4}}
\end{subfigure}
\begin{subfigure}[h]{\plotwidth\textwidth}
\includegraphics[width=\textwidth]{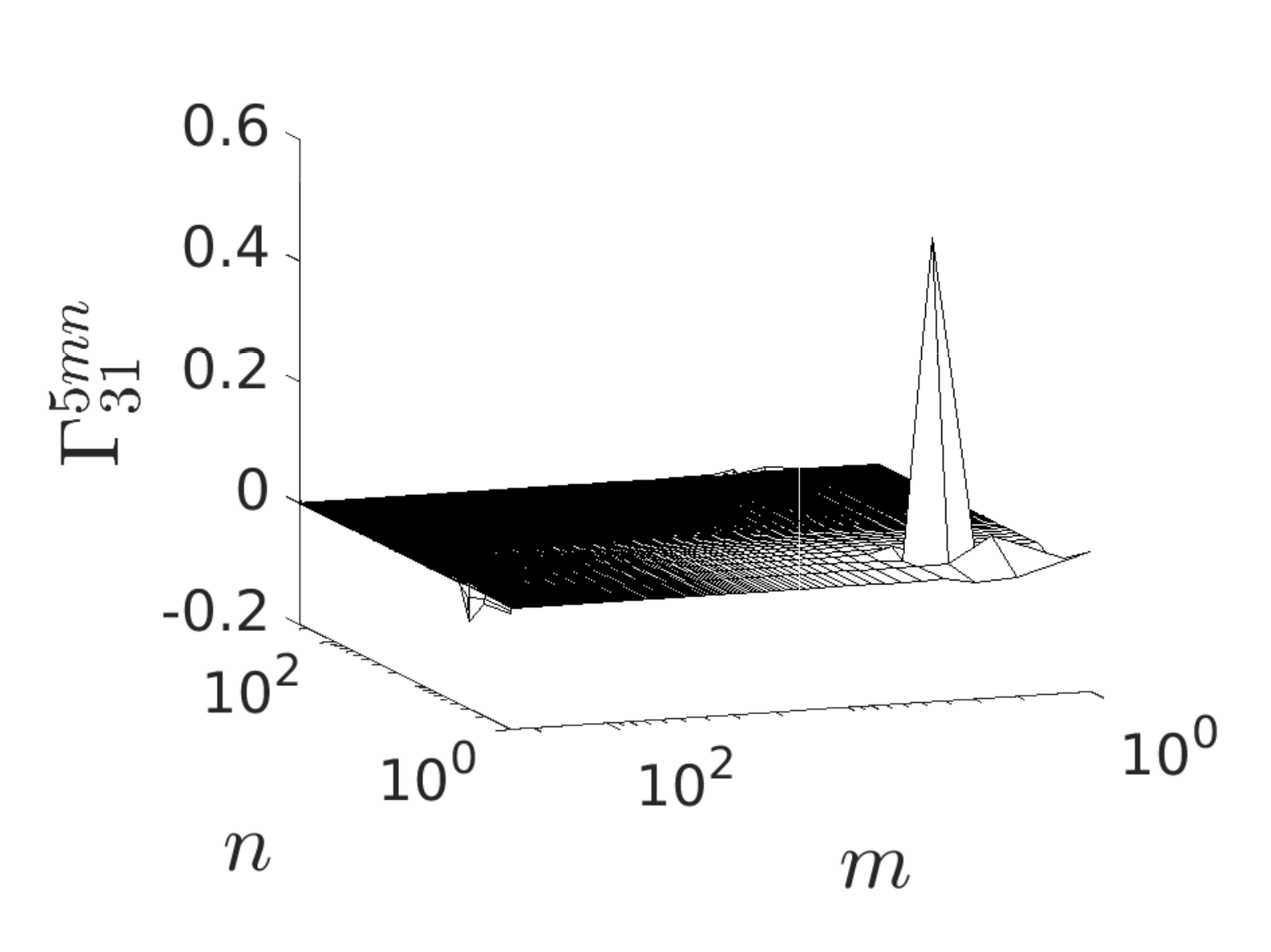}\caption{\label{fig:evalue_reconstruction_domain_1_kernel_3_alpha_5}}
\end{subfigure}
\begin{subfigure}[h]{\plotwidth\textwidth}
\includegraphics[width=\textwidth]{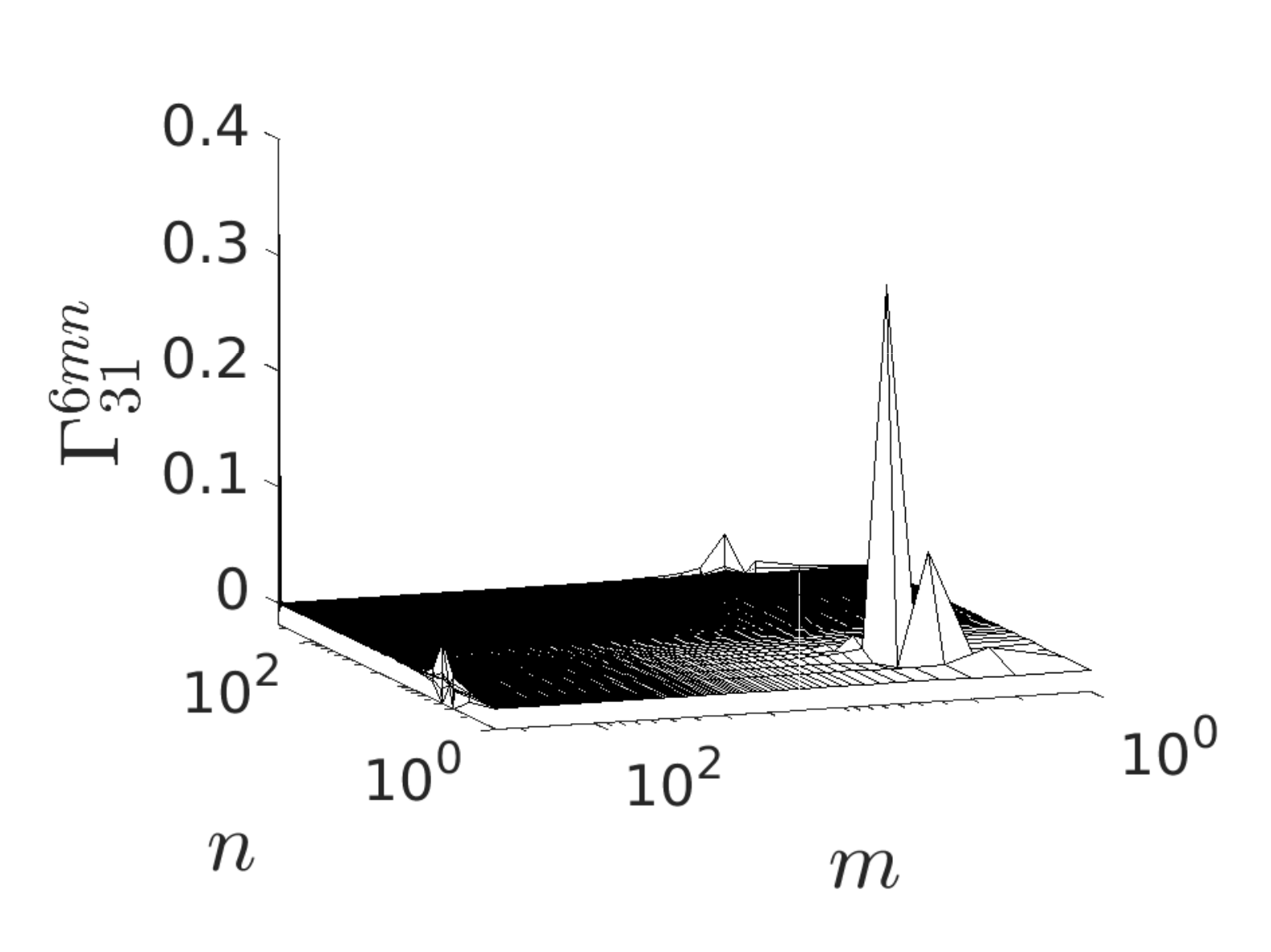}\caption{\label{fig:evalue_reconstruction_domain_1_kernel_3_alpha_6}}
\end{subfigure}
\caption{Contributions to the eigenvalue reconstruction of modes $\alpha=1:6$ using Fourier modes. (a)-(f): contributions for $K_{31}$.\label{fig:}}
\end{figure}
\FloatBarrier
\noindent
\begin{figure}[h!!!]
\centering
\begin{subfigure}[h]{\plotwidth\textwidth}
\includegraphics[width=\textwidth]{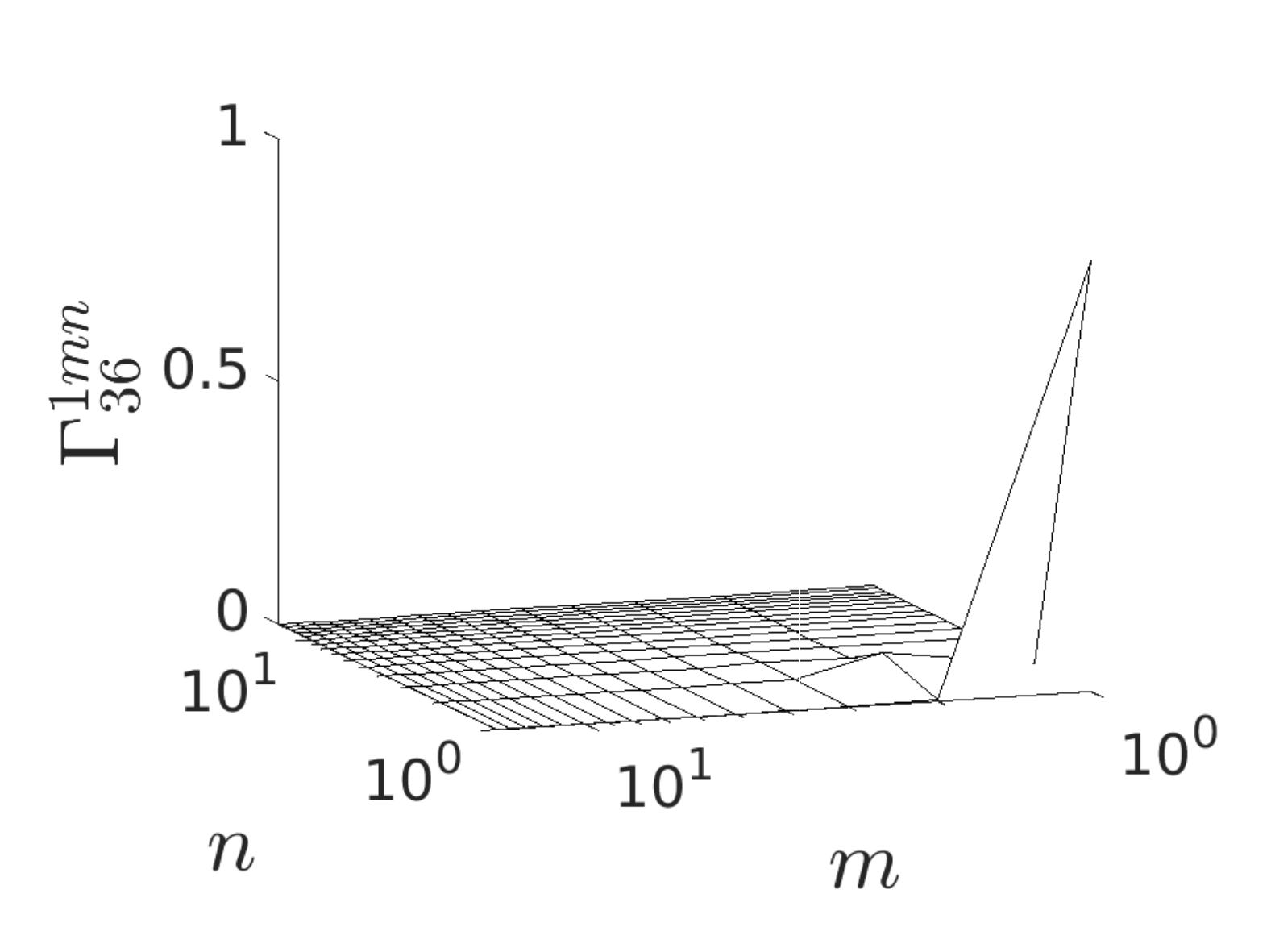}
\caption{\label{fig:evalue_reconstruction_domain_6_kernel_3_alpha_1}}
\end{subfigure}
\begin{subfigure}[h]{\plotwidth\textwidth}
\includegraphics[width=\textwidth]{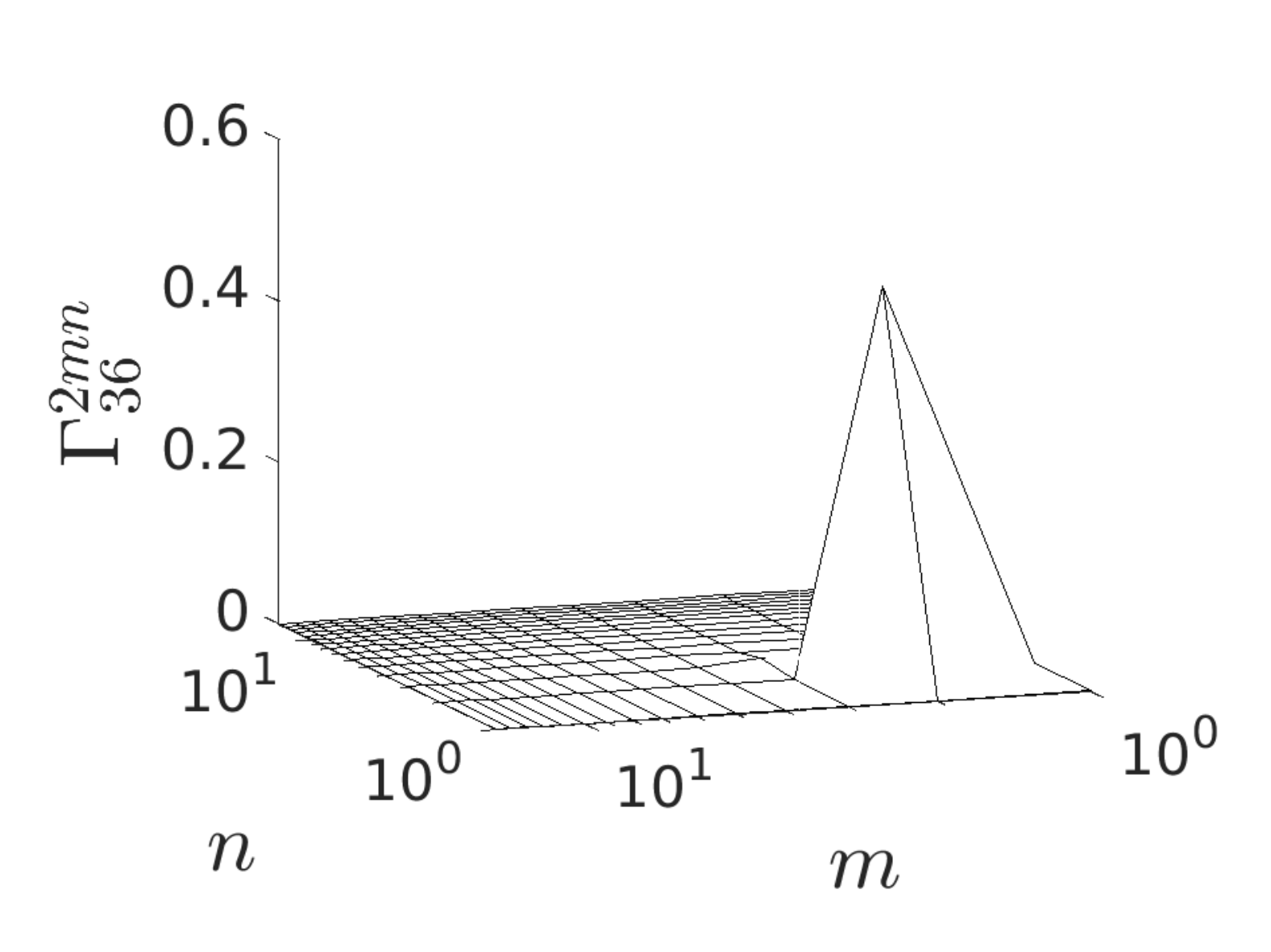}\caption{\label{fig:evalue_reconstruction_domain_6_kernel_3_alpha_2}}
\end{subfigure}
\begin{subfigure}[h]{\plotwidth\textwidth}
\includegraphics[width=\textwidth]{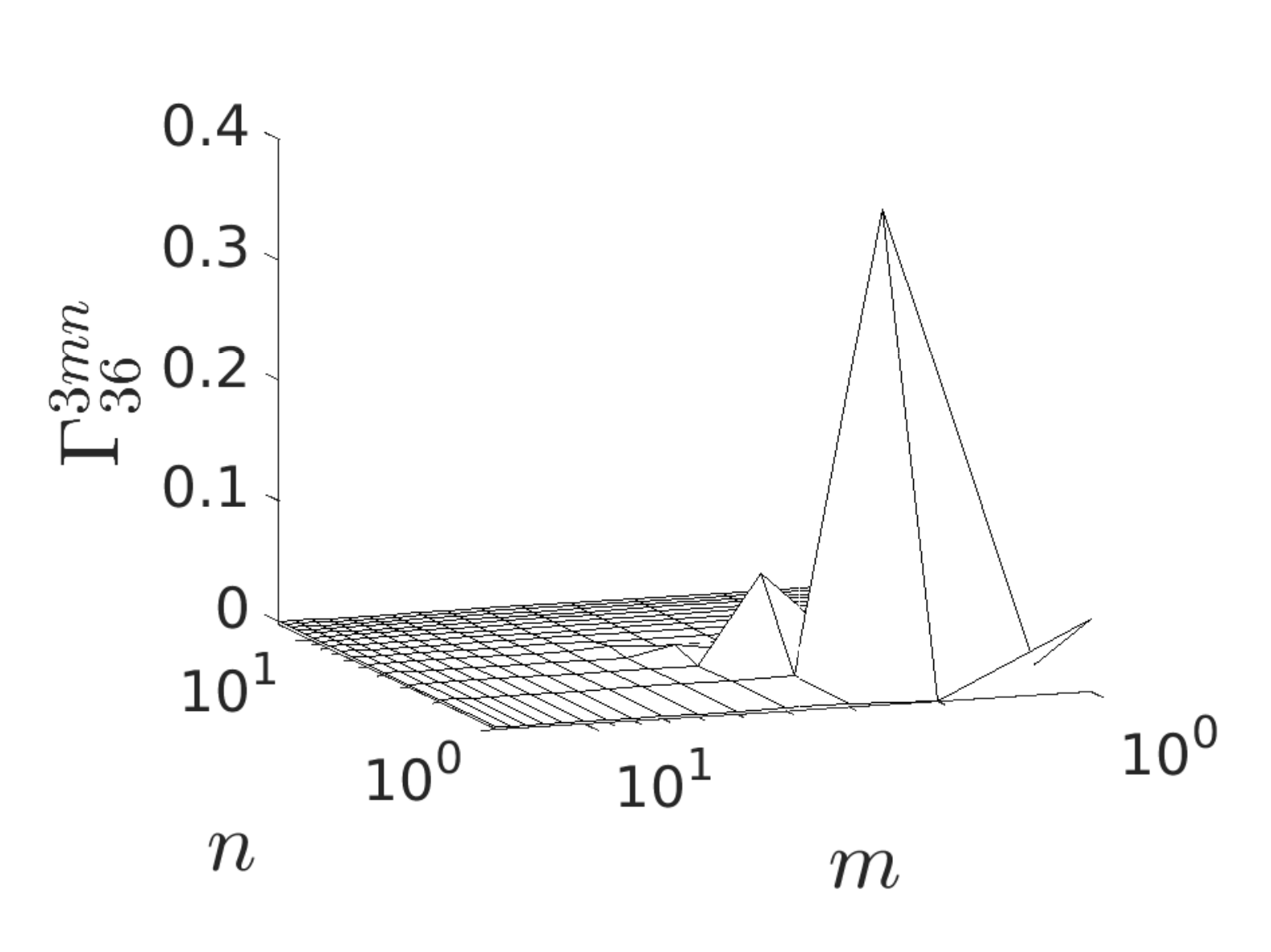}\caption{\label{fig:evalue_reconstruction_domain_6_kernel_3_alpha_3}}
\end{subfigure}
\begin{subfigure}[h]{\plotwidth\textwidth}
\includegraphics[width=\textwidth]{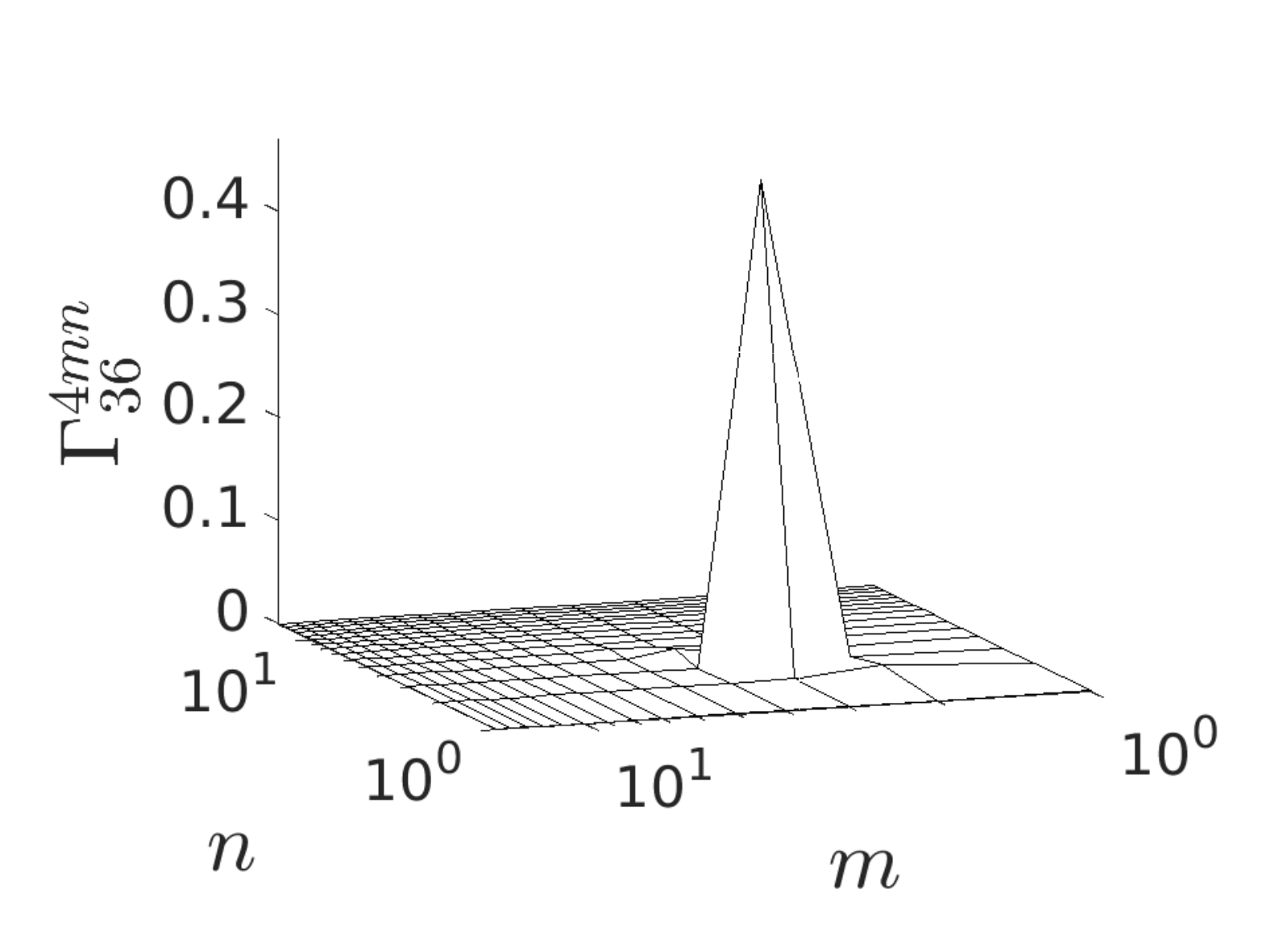}\caption{\label{fig:evalue_reconstruction_domain_6_kernel_3_alpha_4}}
\end{subfigure}
\begin{subfigure}[h]{\plotwidth\textwidth}
\includegraphics[width=\textwidth]{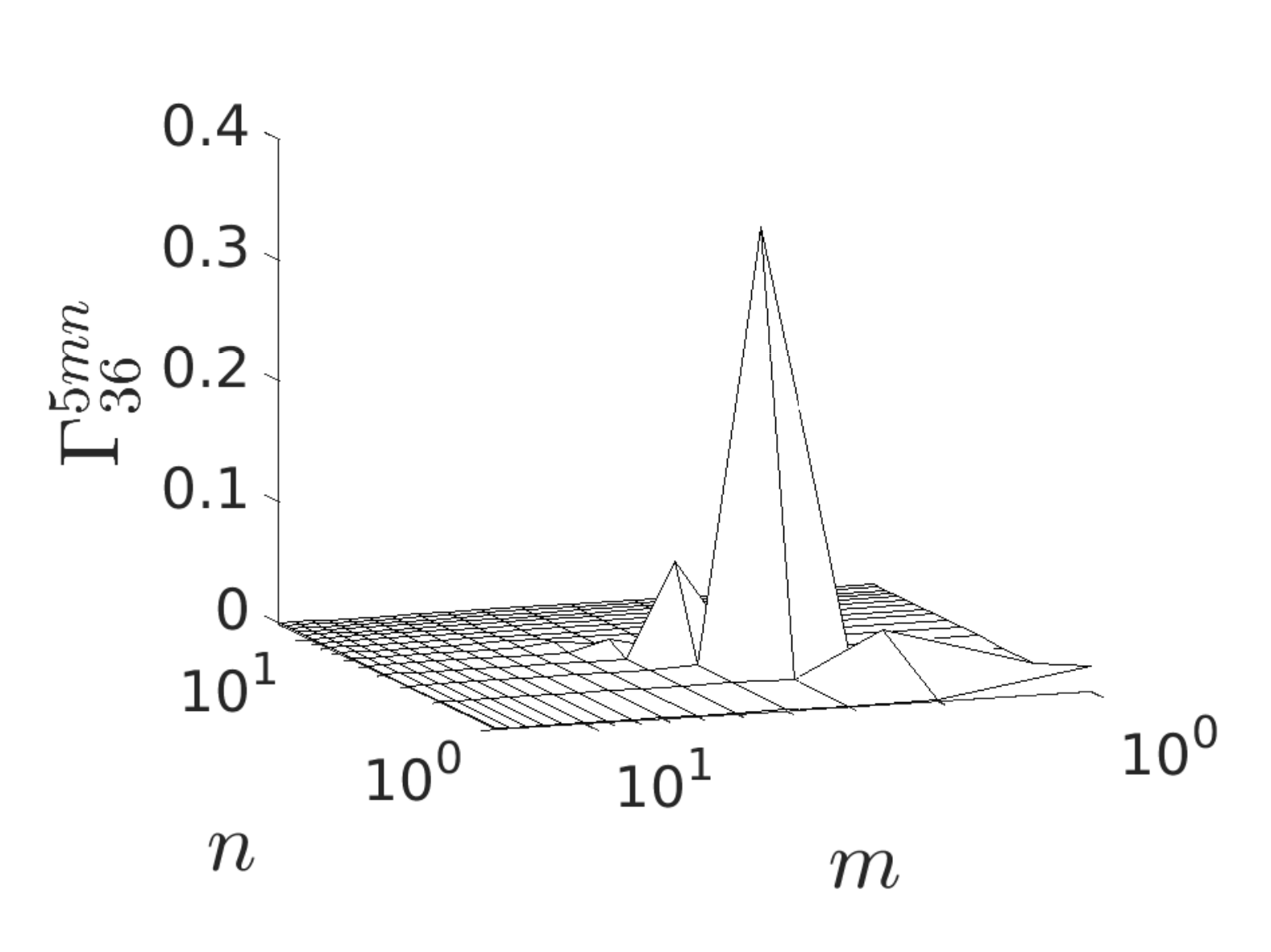}\caption{\label{fig:evalue_reconstruction_domain_6_kernel_3_alpha_5}}
\end{subfigure}
\begin{subfigure}[h]{\plotwidth\textwidth}
\includegraphics[width=\textwidth]{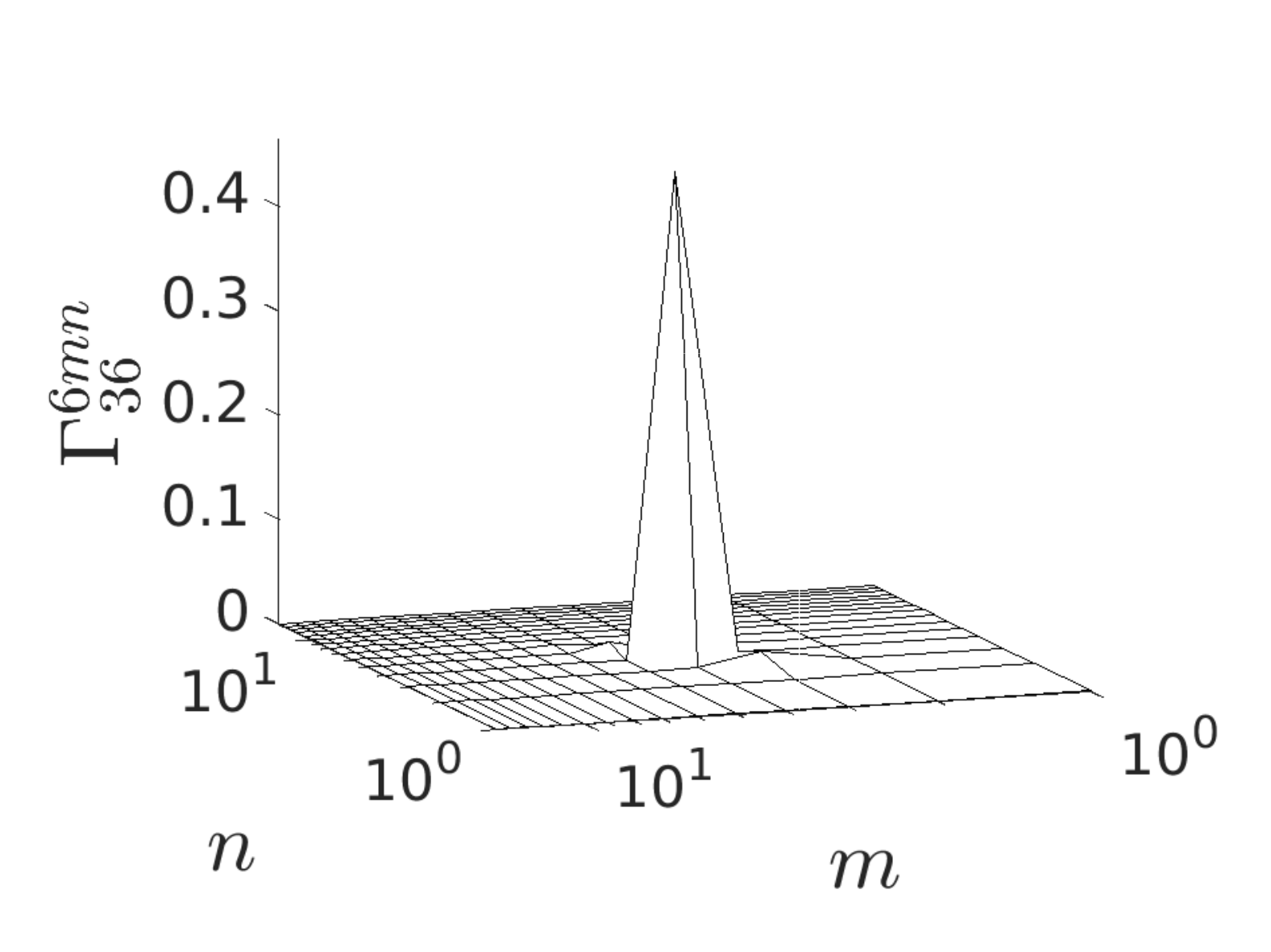}\caption{\label{fig:evalue_reconstruction_domain_6_kernel_3_alpha_6}}
\end{subfigure}
\caption{Contributions to the eigenvalue reconstruction of modes $\alpha=1:6$ using Fourier modes. (a)-(f): contributions for $K_{36}$.\label{fig:}}
\end{figure}
\FloatBarrier
\noindent
\begin{figure}[h!!!]
\centering
\begin{subfigure}[h]{\plotwidth\textwidth}
\includegraphics[width=\textwidth]{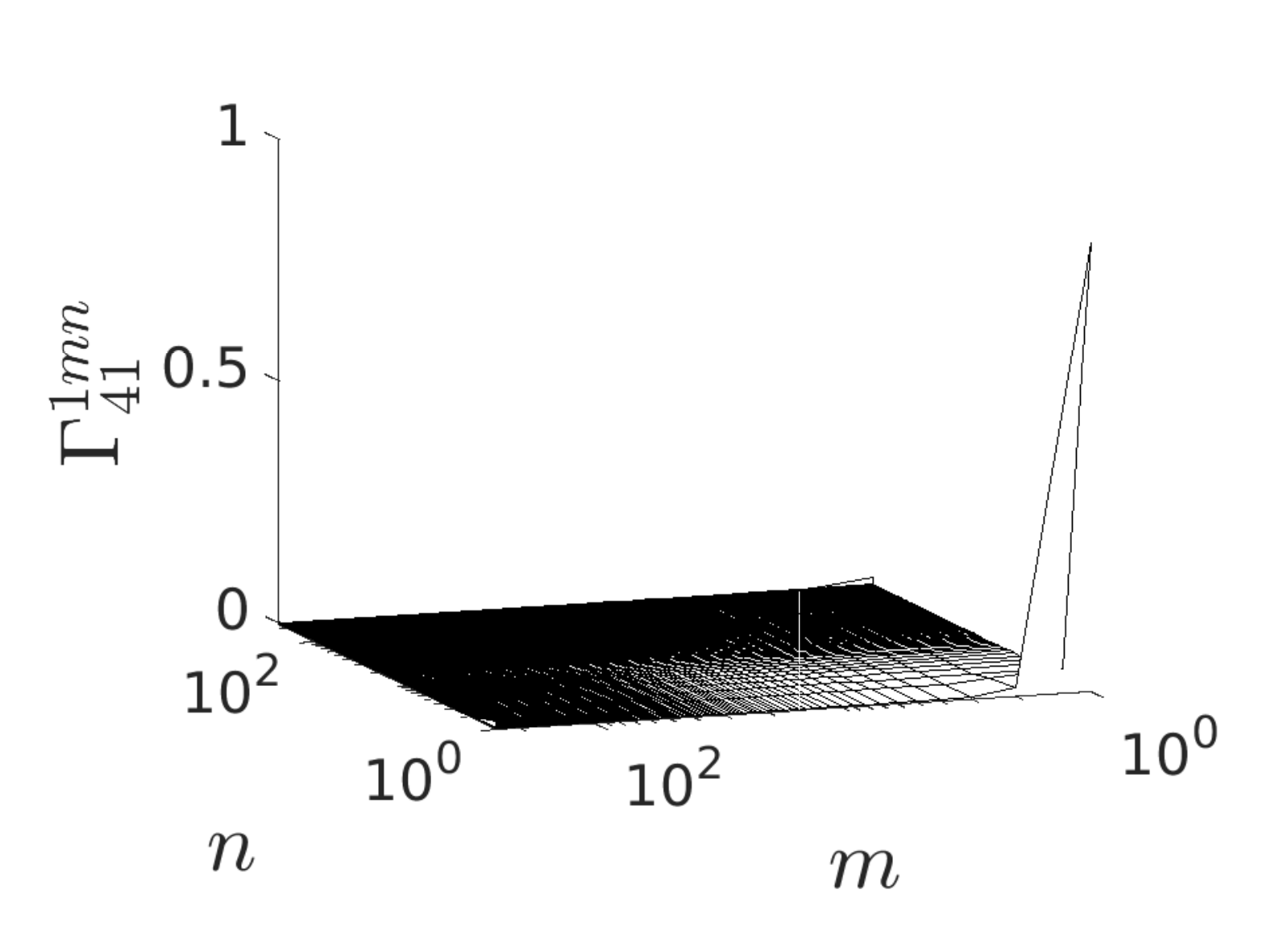}
\caption{\label{fig:evalue_reconstruction_domain_1_kernel_4_alpha_1}}
\end{subfigure}
\begin{subfigure}[h]{\plotwidth\textwidth}
\includegraphics[width=\textwidth]{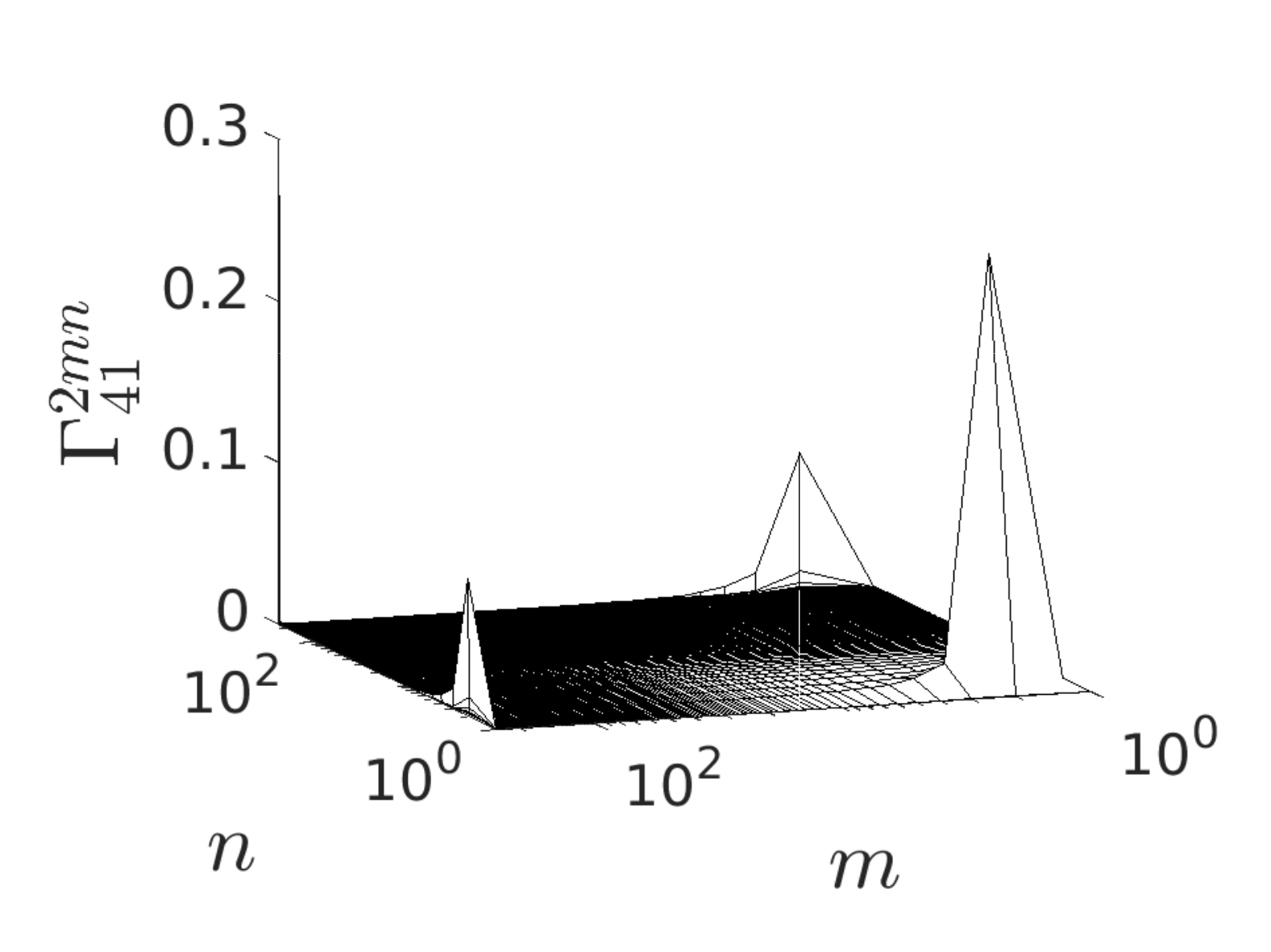}\caption{\label{fig:evalue_reconstruction_domain_1_kernel_4_alpha_2}}
\end{subfigure}
\begin{subfigure}[h]{\plotwidth\textwidth}
\includegraphics[width=\textwidth]{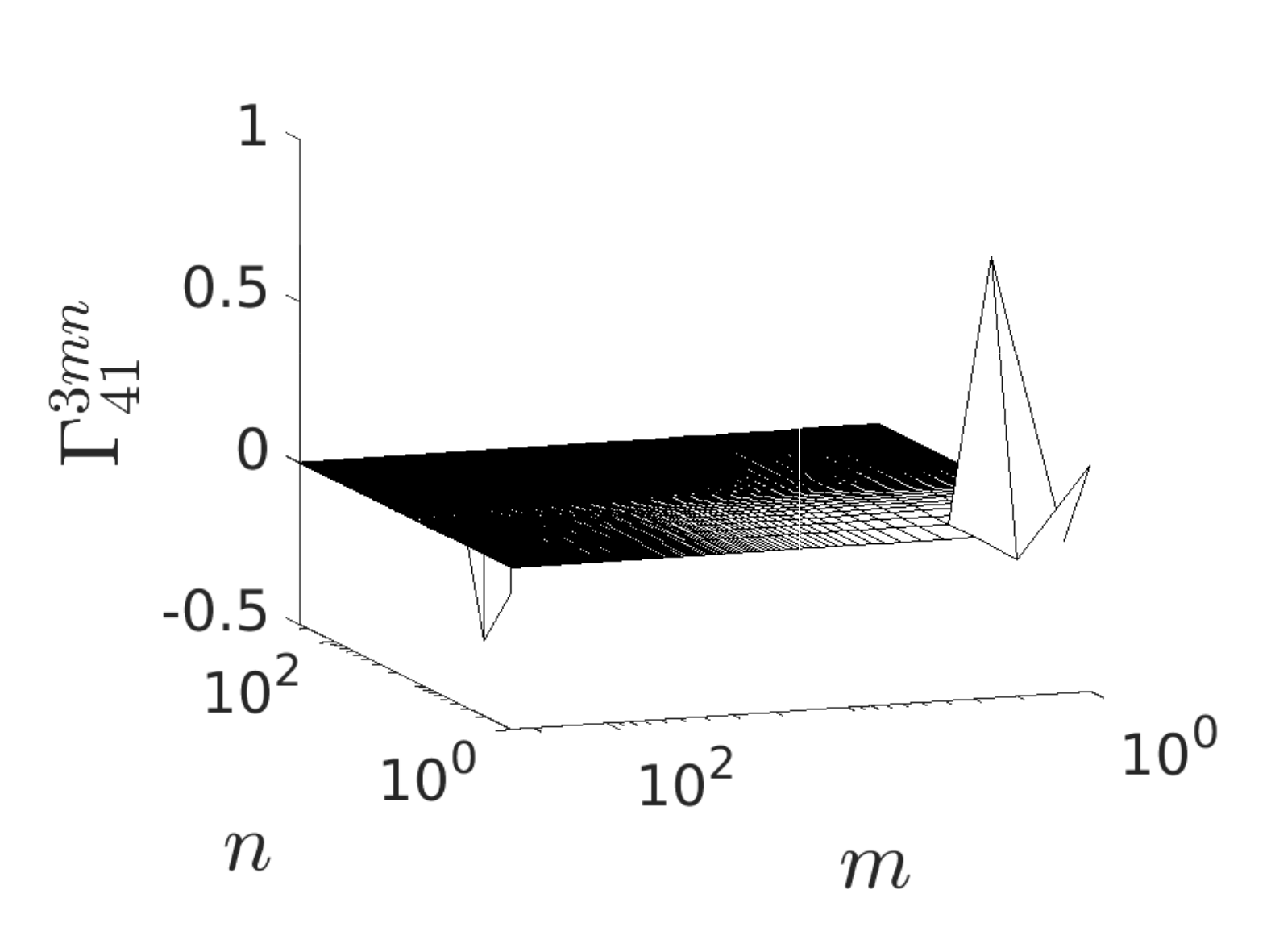}\caption{\label{fig:evalue_reconstruction_domain_1_kernel_4_alpha_3}}
\end{subfigure}
\begin{subfigure}[h]{\plotwidth\textwidth}
\includegraphics[width=\textwidth]{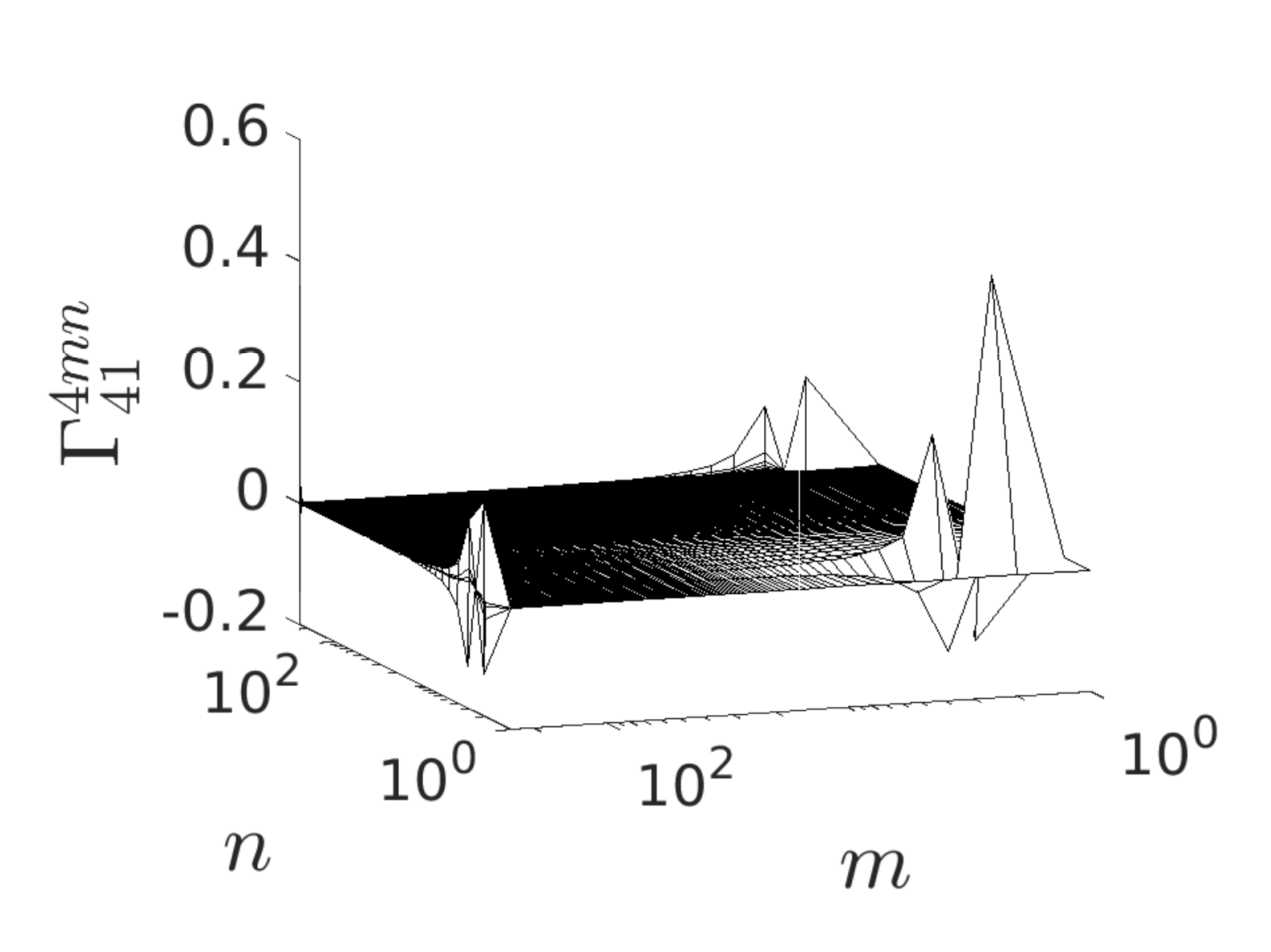}\caption{\label{fig:evalue_reconstruction_domain_1_kernel_4_alpha_4}}
\end{subfigure}
\begin{subfigure}[h]{\plotwidth\textwidth}
\includegraphics[width=\textwidth]{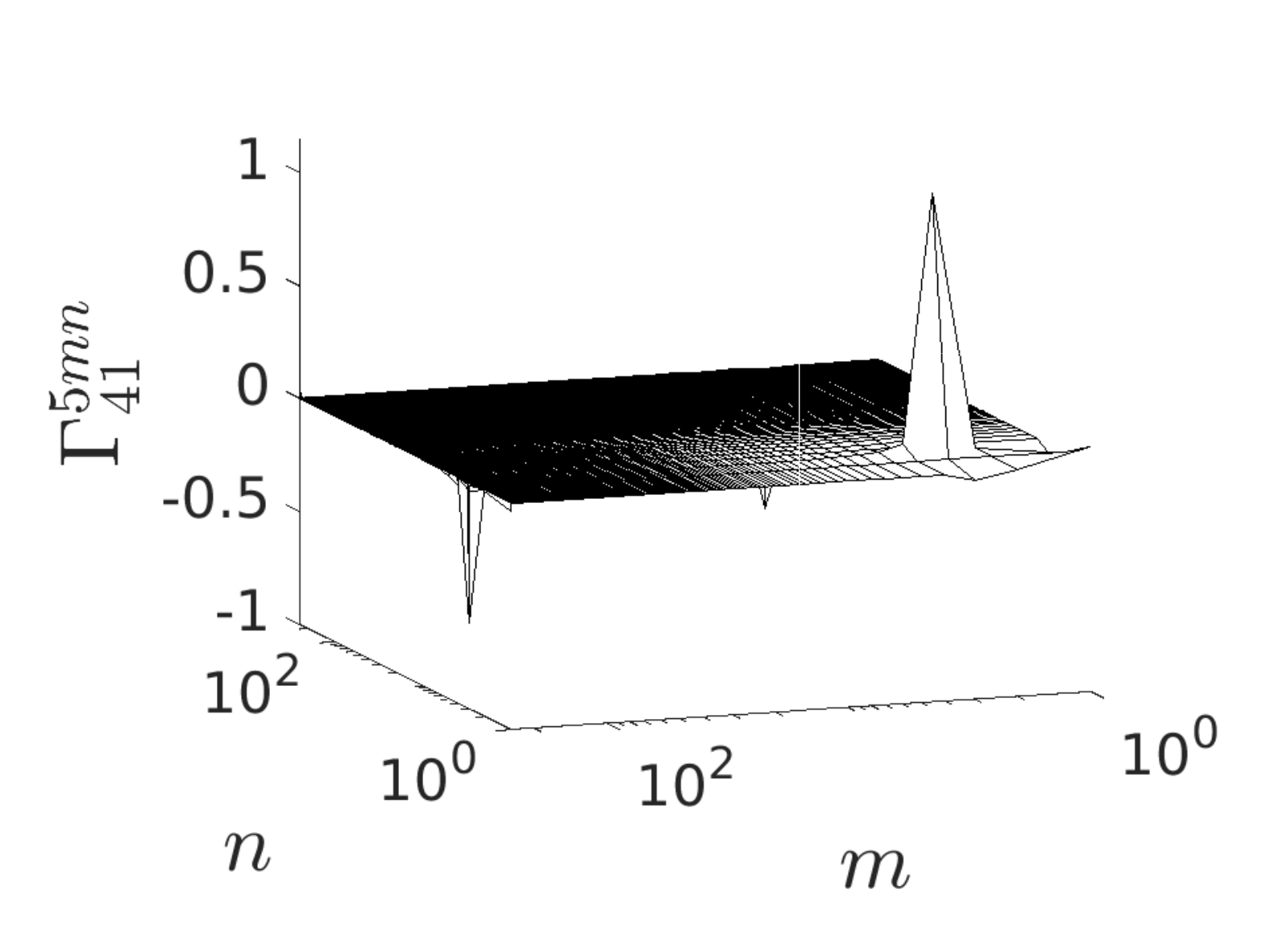}\caption{\label{fig:evalue_reconstruction_domain_1_kernel_4_alpha_5}}
\end{subfigure}
\begin{subfigure}[h]{\plotwidth\textwidth}
\includegraphics[width=\textwidth]{figs/pod/evalue_recon/evalue_reconstruction_domain_1_kernel_4_alpha_6}\caption{\label{fig:evalue_reconstruction_domain_1_kernel_4_alpha_6}}
\end{subfigure}
\caption{Contributions to the eigenvalue reconstruction of modes $\alpha=1:6$ using Fourier modes. (a)-(f): contributions for $K_{41}$.\label{fig:}}
\end{figure}
\FloatBarrier
\noindent
\begin{figure}[h!!!]
\centering
\begin{subfigure}[h]{\plotwidth\textwidth}
\includegraphics[width=\textwidth]{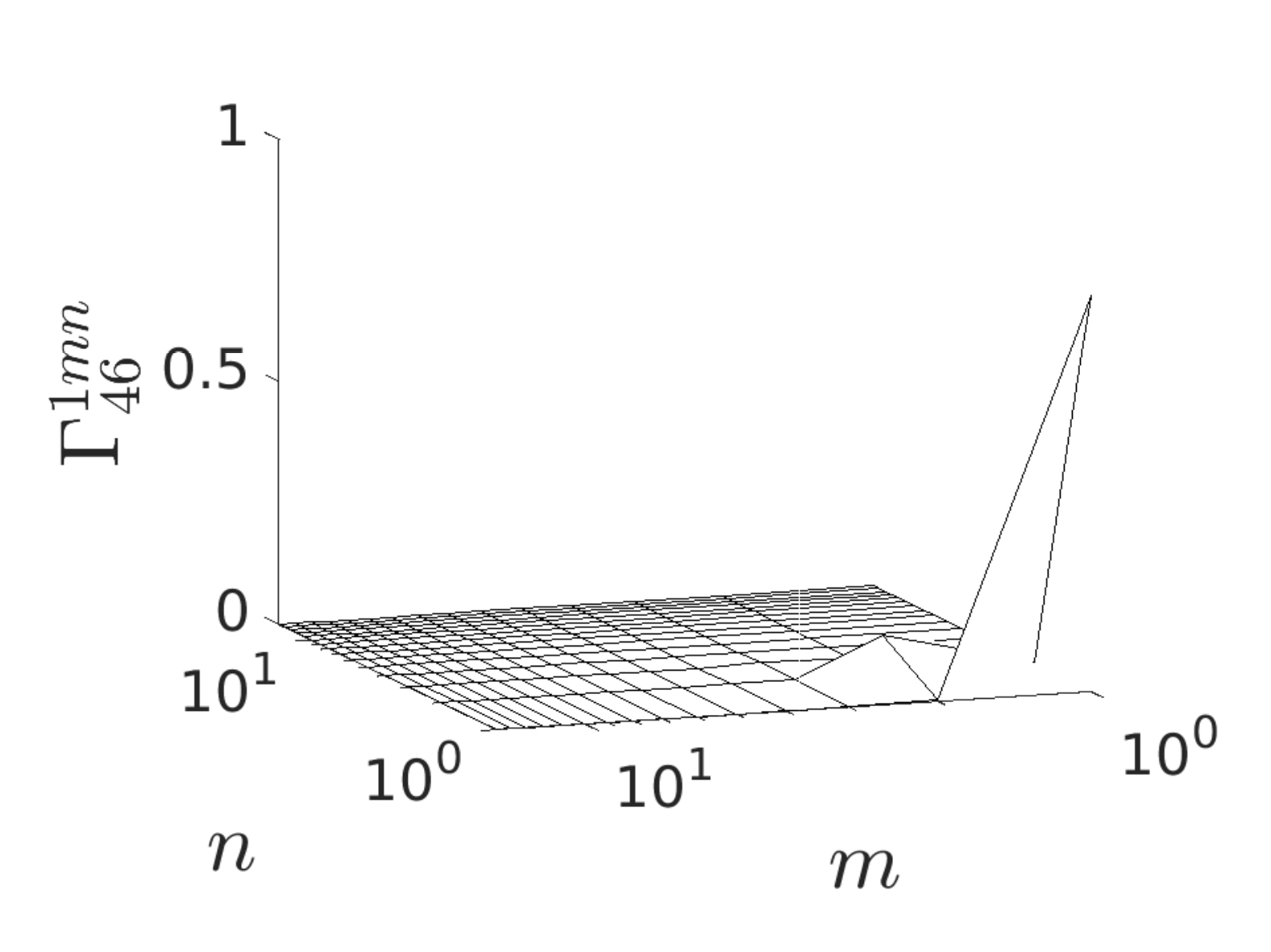}
\caption{\label{fig:evalue_reconstruction_domain_6_kernel_4_alpha_1}}
\end{subfigure}
\begin{subfigure}[h]{\plotwidth\textwidth}
\includegraphics[width=\textwidth]{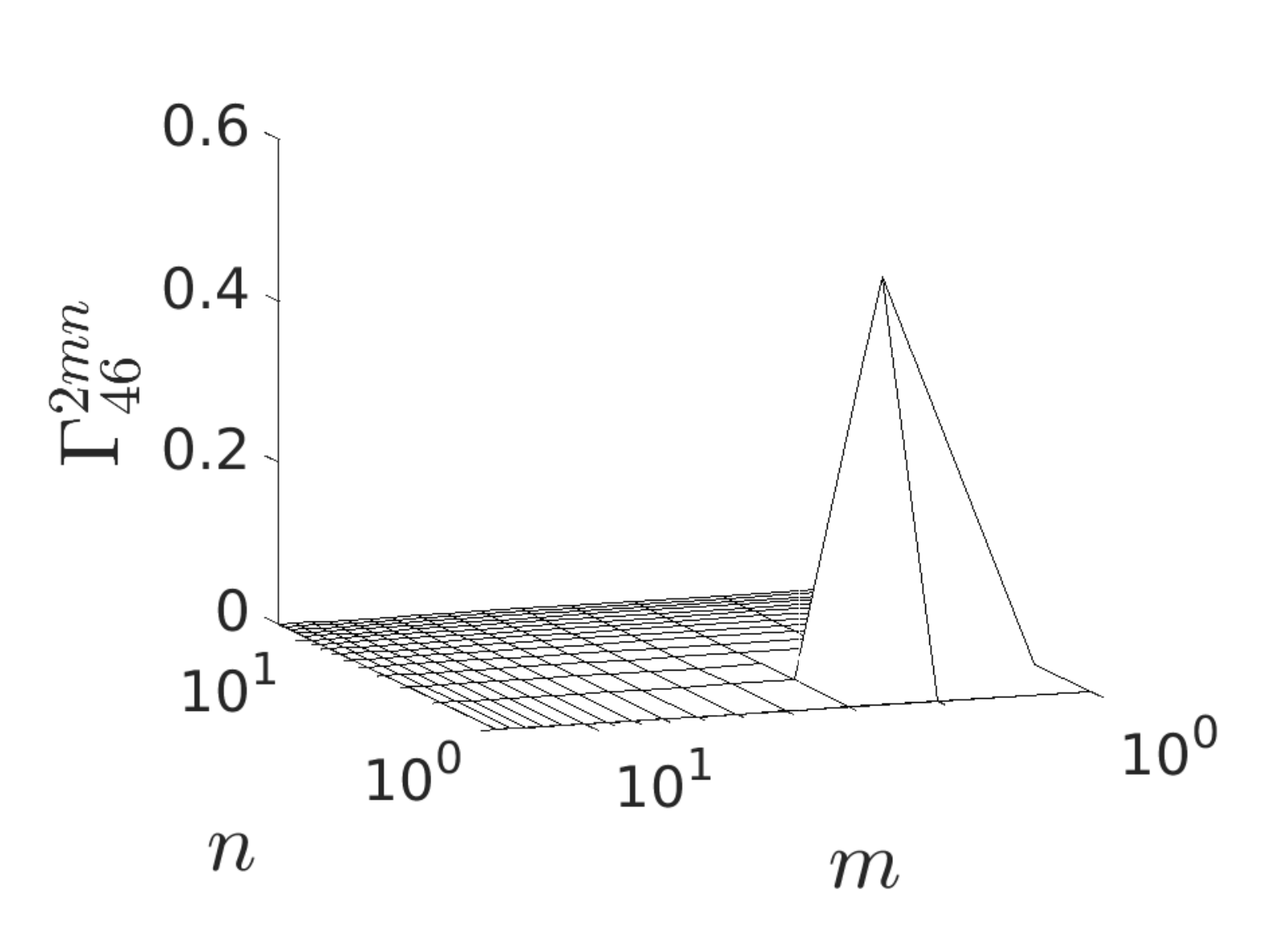}\caption{\label{fig:evalue_reconstruction_domain_6_kernel_4_alpha_2}}
\end{subfigure}
\begin{subfigure}[h]{\plotwidth\textwidth}
\includegraphics[width=\textwidth]{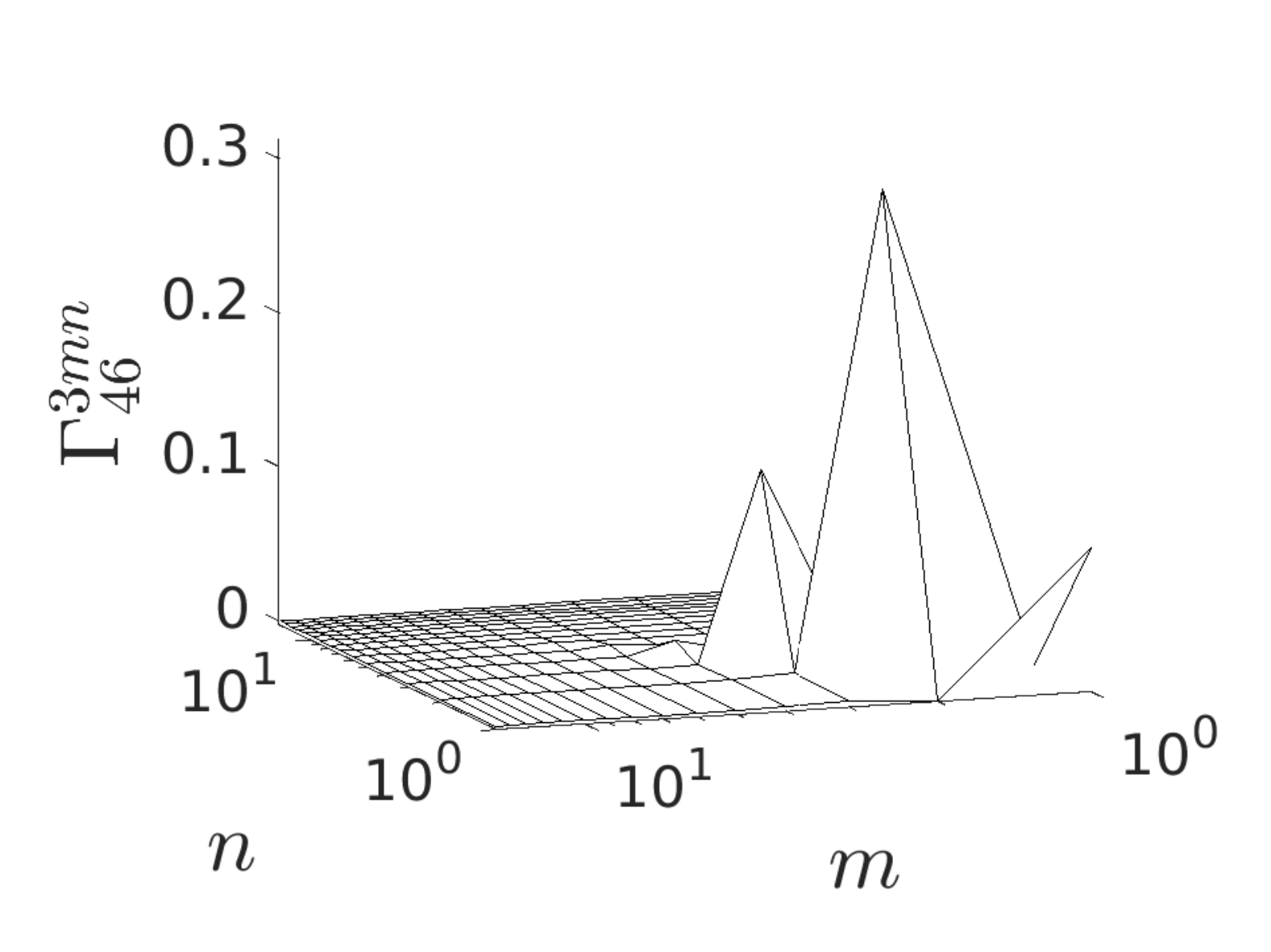}\caption{\label{fig:evalue_reconstruction_domain_6_kernel_4_alpha_3}}
\end{subfigure}
\begin{subfigure}[h]{\plotwidth\textwidth}
\includegraphics[width=\textwidth]{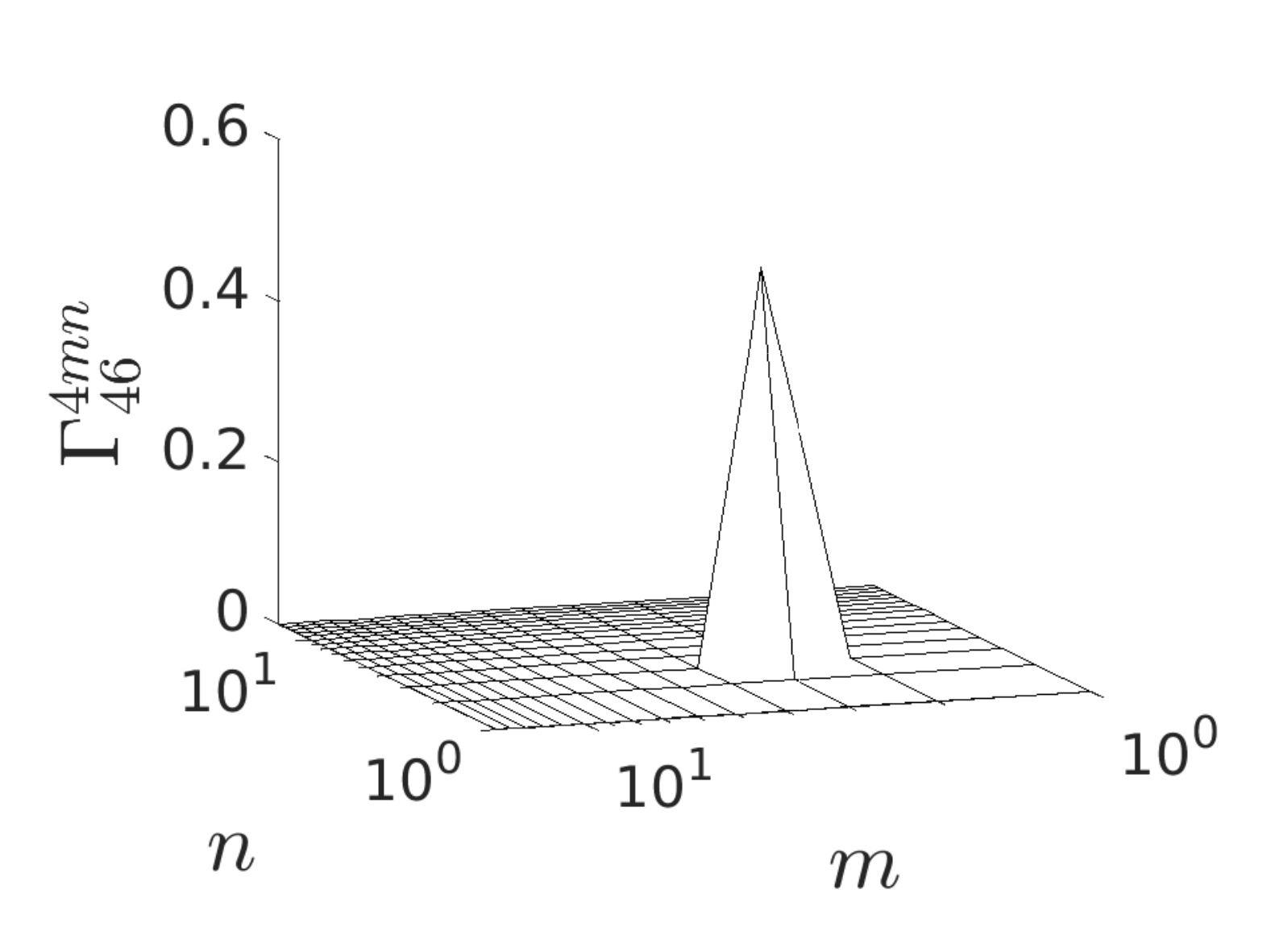}\caption{\label{fig:evalue_reconstruction_domain_6_kernel_4_alpha_4}}
\end{subfigure}
\begin{subfigure}[h]{\plotwidth\textwidth}
\includegraphics[width=\textwidth]{figs/pod/evalue_recon/evalue_reconstruction_domain_6_kernel_4_alpha_5}\caption{\label{fig:evalue_reconstruction_domain_6_kernel_4_alpha_5}}
\end{subfigure}
\begin{subfigure}[h]{\plotwidth\textwidth}
\includegraphics[width=\textwidth]{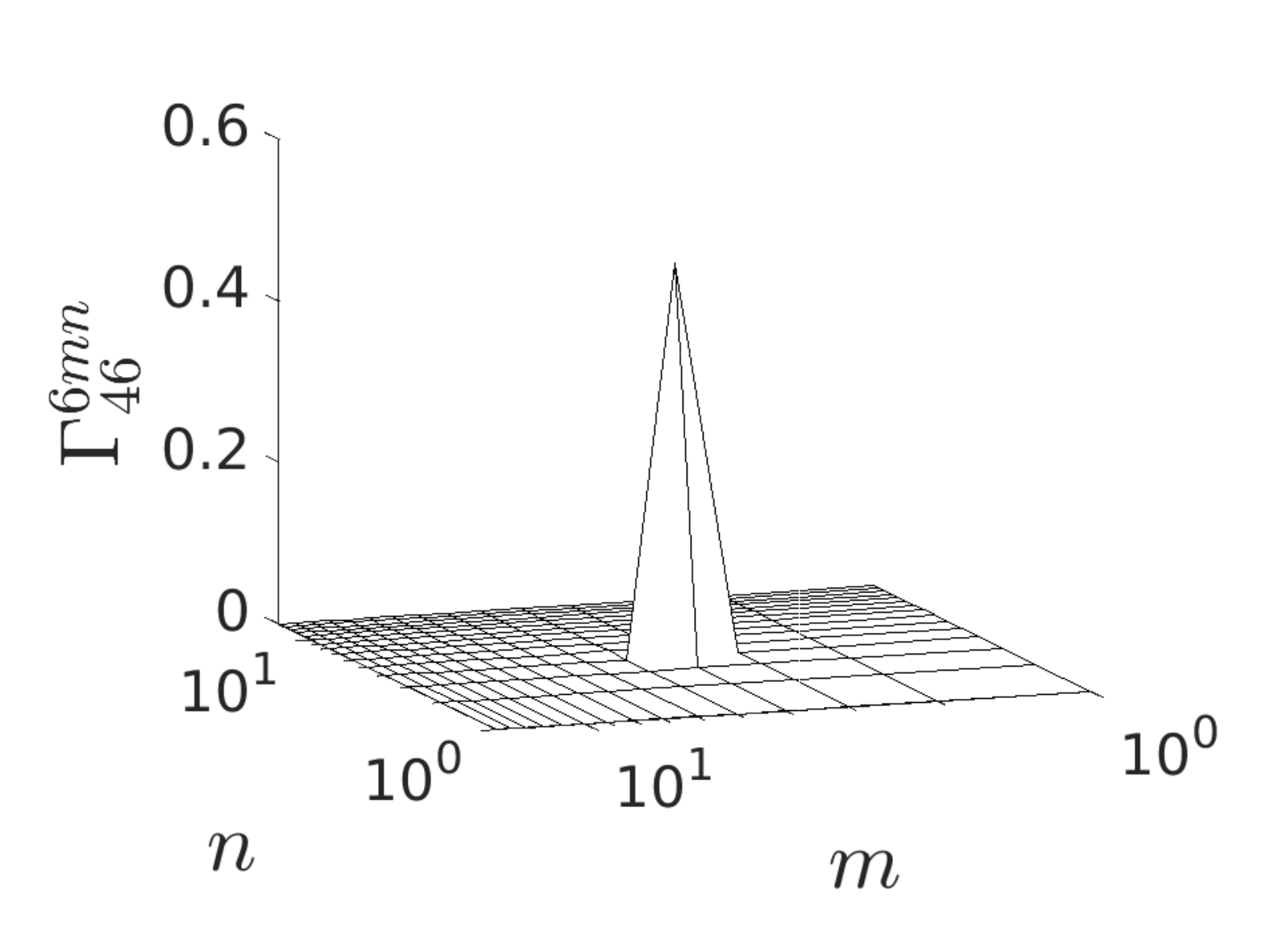}\caption{\label{fig:evalue_reconstruction_domain_6_kernel_4_alpha_6}}
\end{subfigure}
\caption{Contributions to the eigenvalue reconstruction of modes $\alpha=1:6$ using Fourier modes. (a)-(f): contributions for $K_{46}$.\label{fig:}}
\end{figure}
%
%
\begin{figure}[h!!!]
\centering
\begin{subfigure}[h]{\plotwidth\textwidth}
\includegraphics[width=\textwidth]{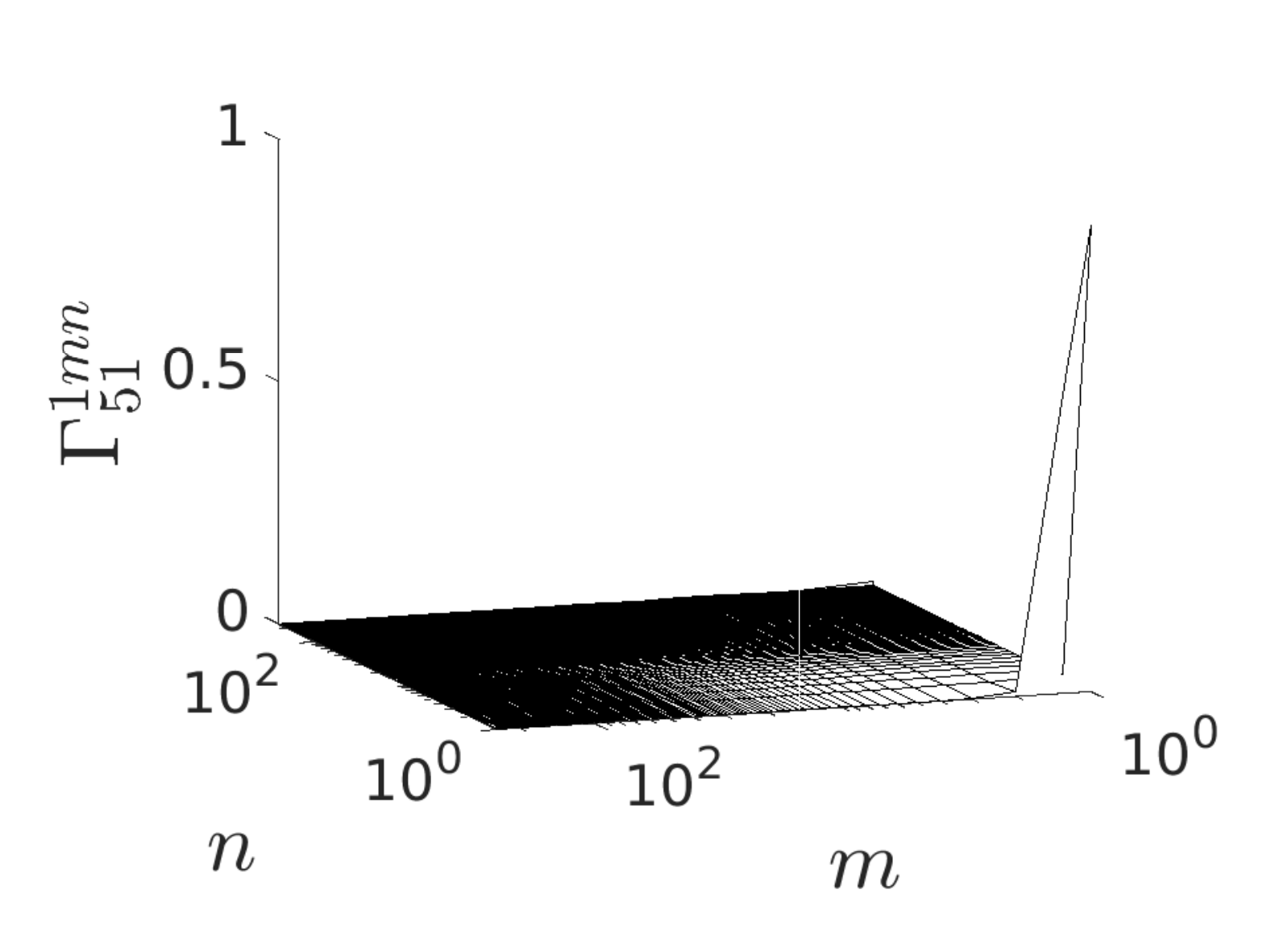}
\caption{\label{fig:evalue_reconstruction_domain_1_kernel_5_alpha_1}}
\end{subfigure}
\begin{subfigure}[h]{\plotwidth\textwidth}
\includegraphics[width=\textwidth]{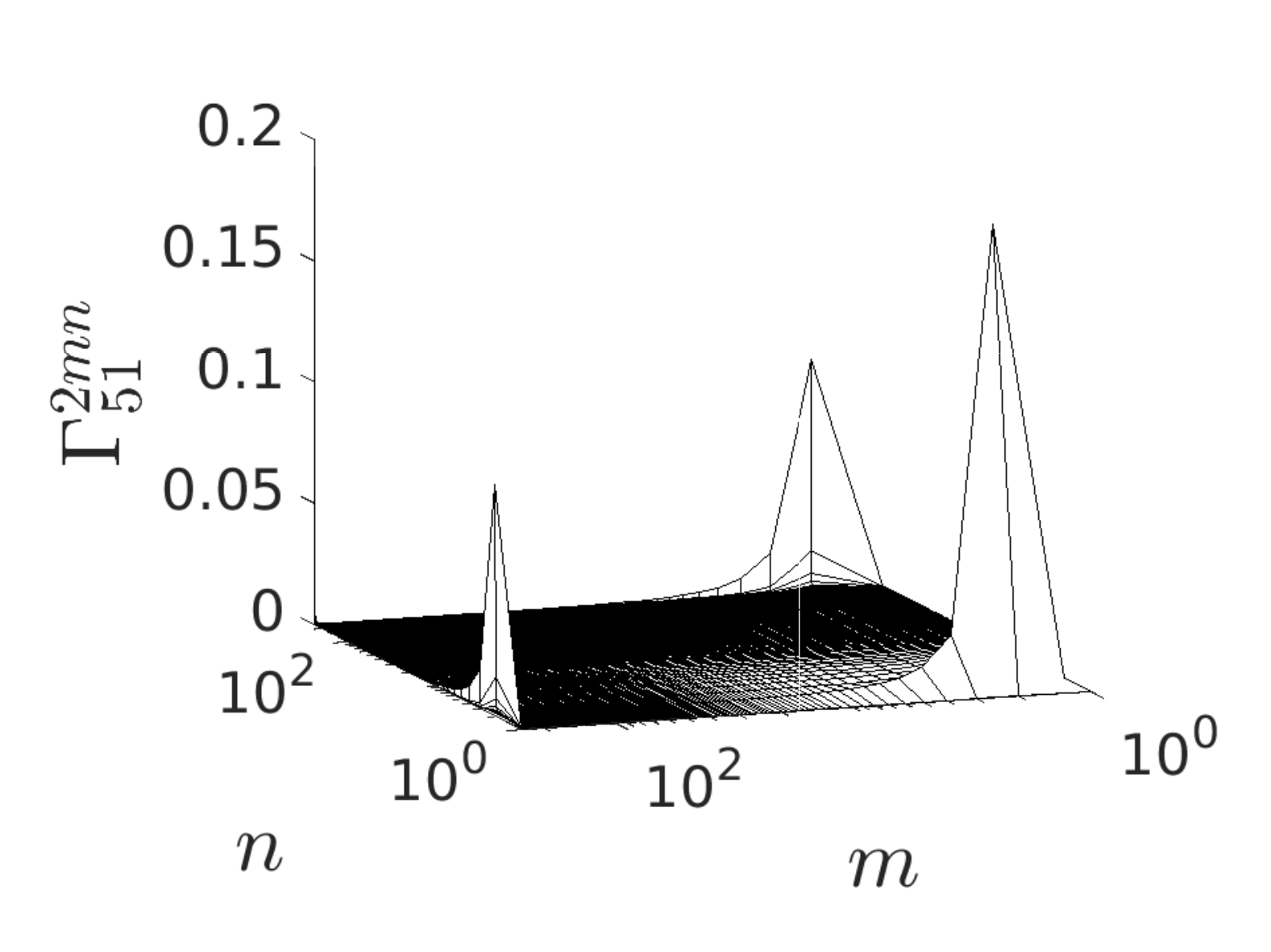}\caption{\label{fig:evalue_reconstruction_domain_1_kernel_5_alpha_2}}
\end{subfigure}
\begin{subfigure}[h]{\plotwidth\textwidth}
\includegraphics[width=\textwidth]{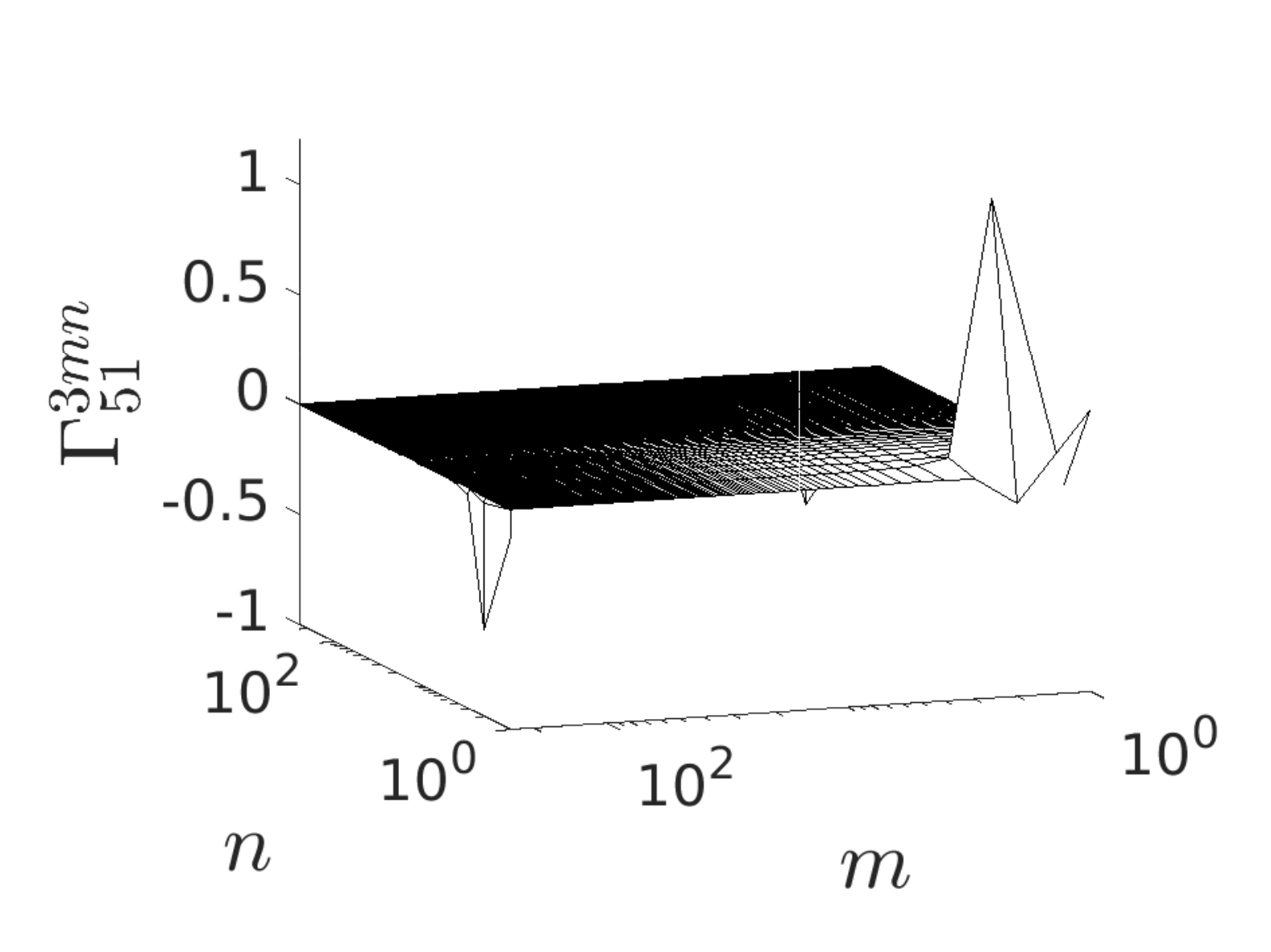}\caption{\label{fig:evalue_reconstruction_domain_1_kernel_5_alpha_3}}
\end{subfigure}
\begin{subfigure}[h]{\plotwidth\textwidth}
\includegraphics[width=\textwidth]{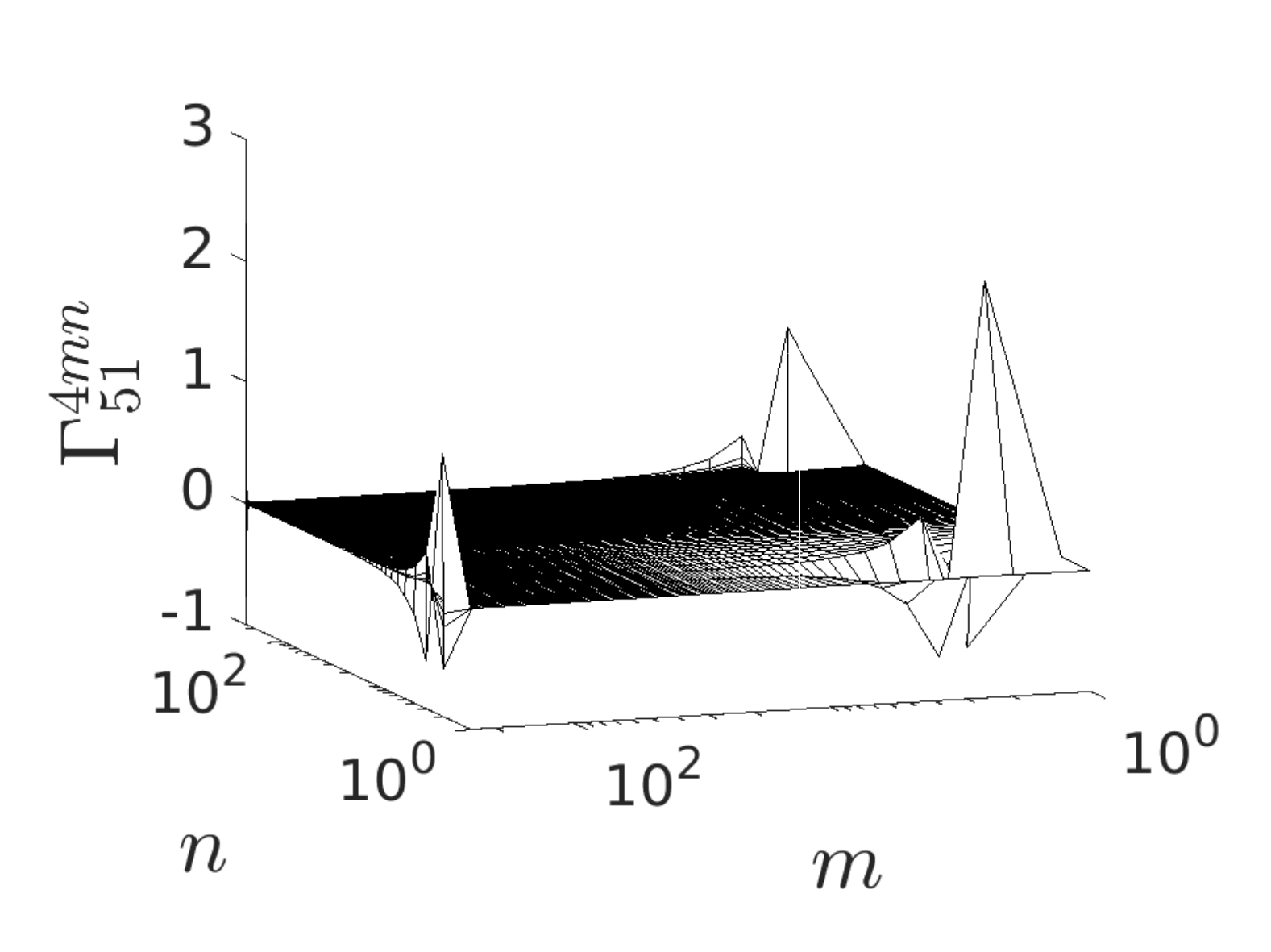}\caption{\label{fig:evalue_reconstruction_domain_1_kernel_5_alpha_4}}
\end{subfigure}
\begin{subfigure}[h]{\plotwidth\textwidth}
\includegraphics[width=\textwidth]{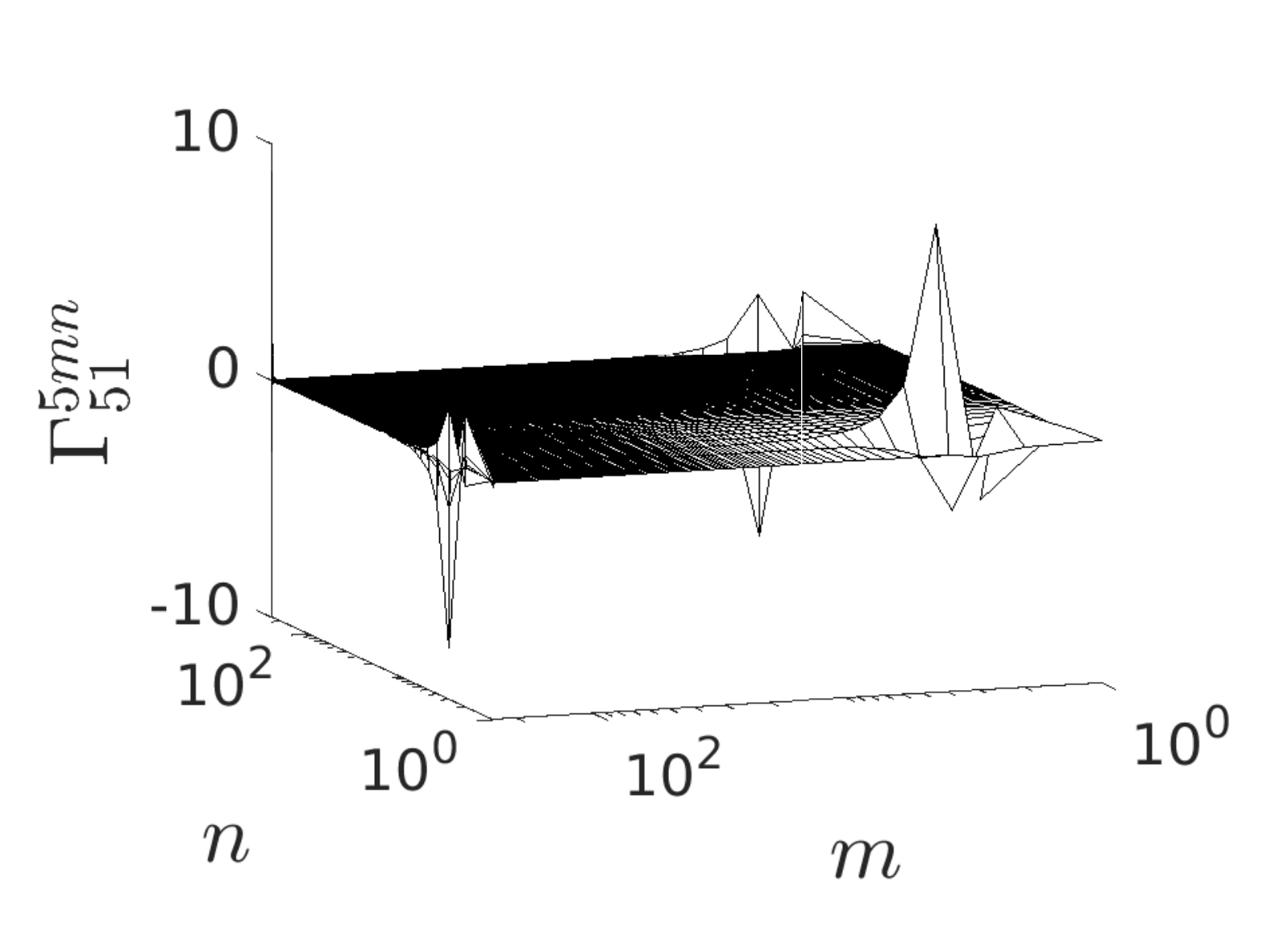}\caption{\label{fig:evalue_reconstruction_domain_1_kernel_5_alpha_5}}
\end{subfigure}
\begin{subfigure}[h]{\plotwidth\textwidth}
\includegraphics[width=\textwidth]{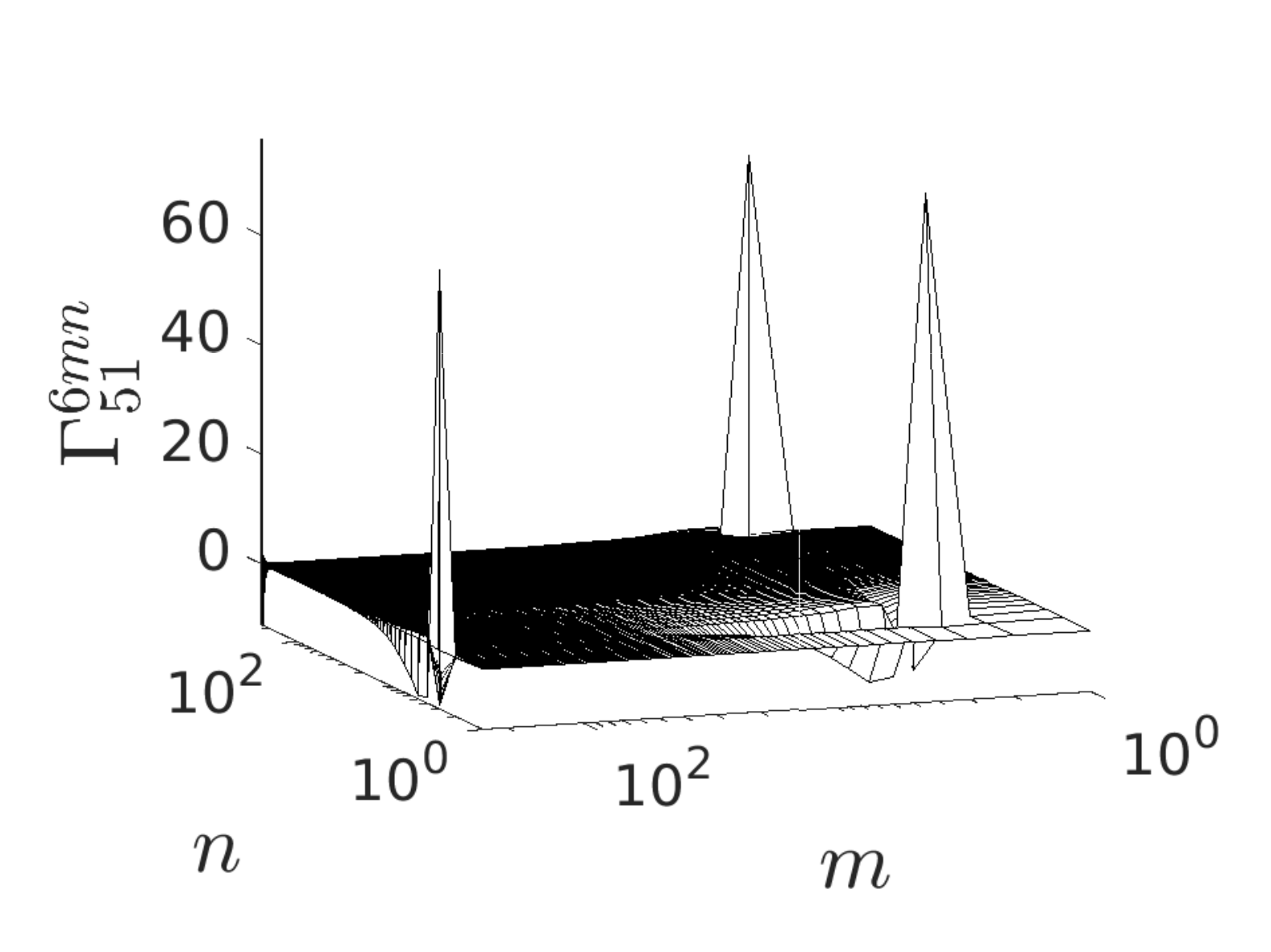}\caption{\label{fig:evalue_reconstruction_domain_1_kernel_5_alpha_6}}
\end{subfigure}
\caption{Contributions to the eigenvalue reconstruction of modes $\alpha=1:6$ using Fourier modes. (a)-(f): contributions for $K_{51}$.\label{fig:evalue_reconstruction_kernel_5}}
\end{figure}
%
%
\begin{figure}[t!!!]
\centering
\begin{subfigure}[h]{\plotwidth\textwidth}
\includegraphics[width=\textwidth]{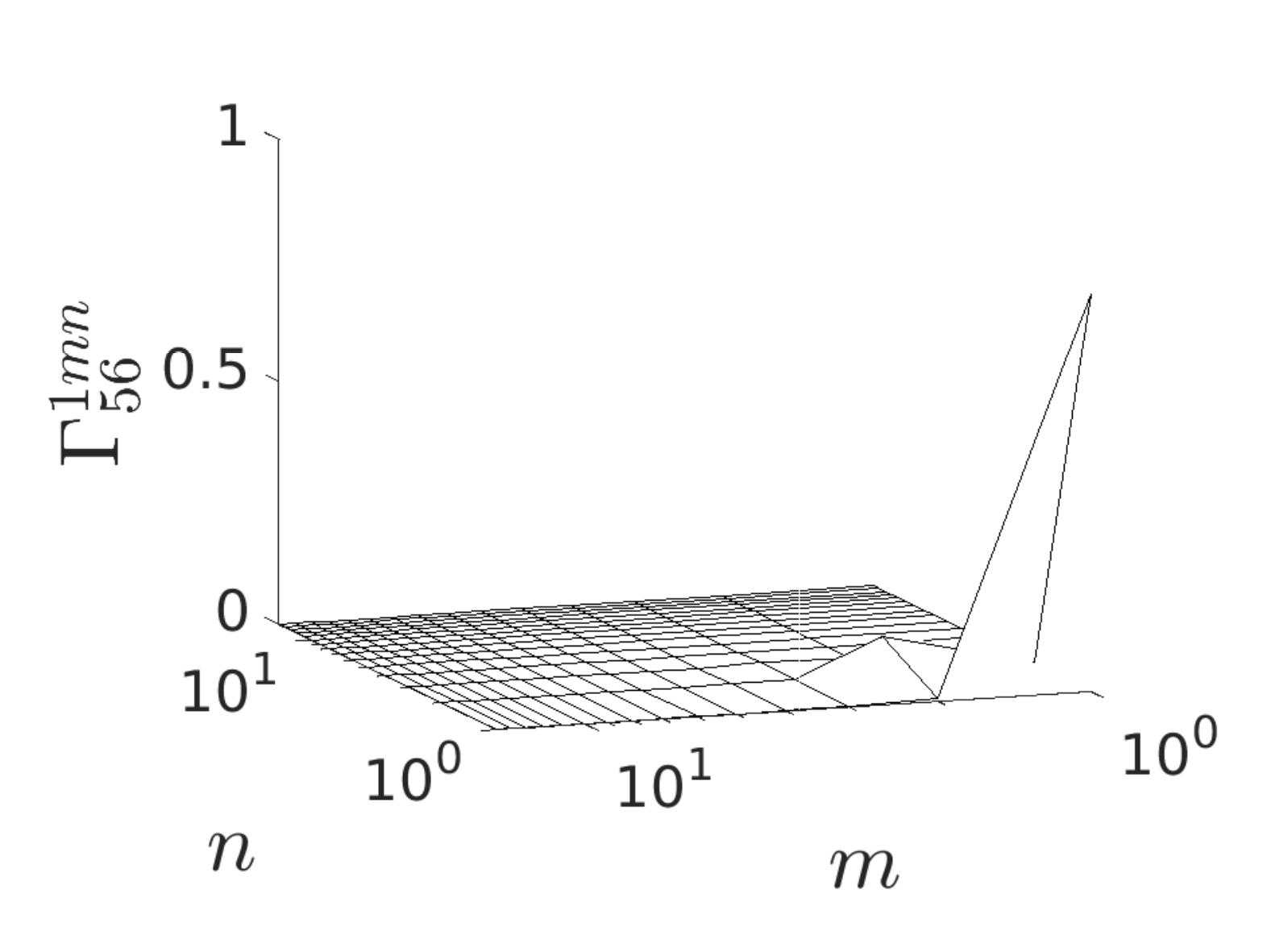}
\caption{\label{fig:evalue_reconstruction_domain_6_kernel_5_alpha_1}}
\end{subfigure}
\begin{subfigure}[h]{\plotwidth\textwidth}
\includegraphics[width=\textwidth]{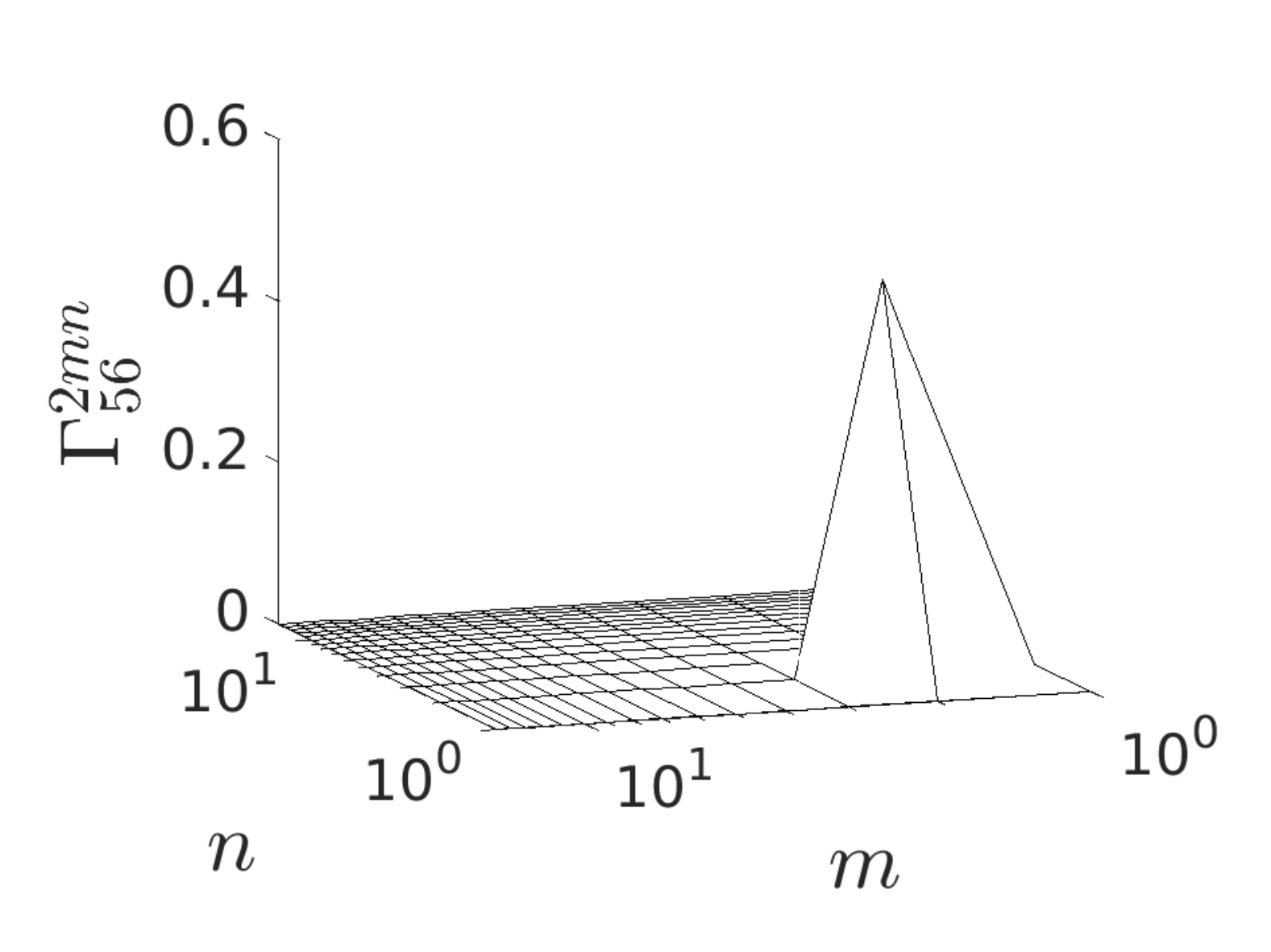}\caption{\label{fig:evalue_reconstruction_domain_6_kernel_5_alpha_2}}
\end{subfigure}
\begin{subfigure}[h]{\plotwidth\textwidth}
\includegraphics[width=\textwidth]{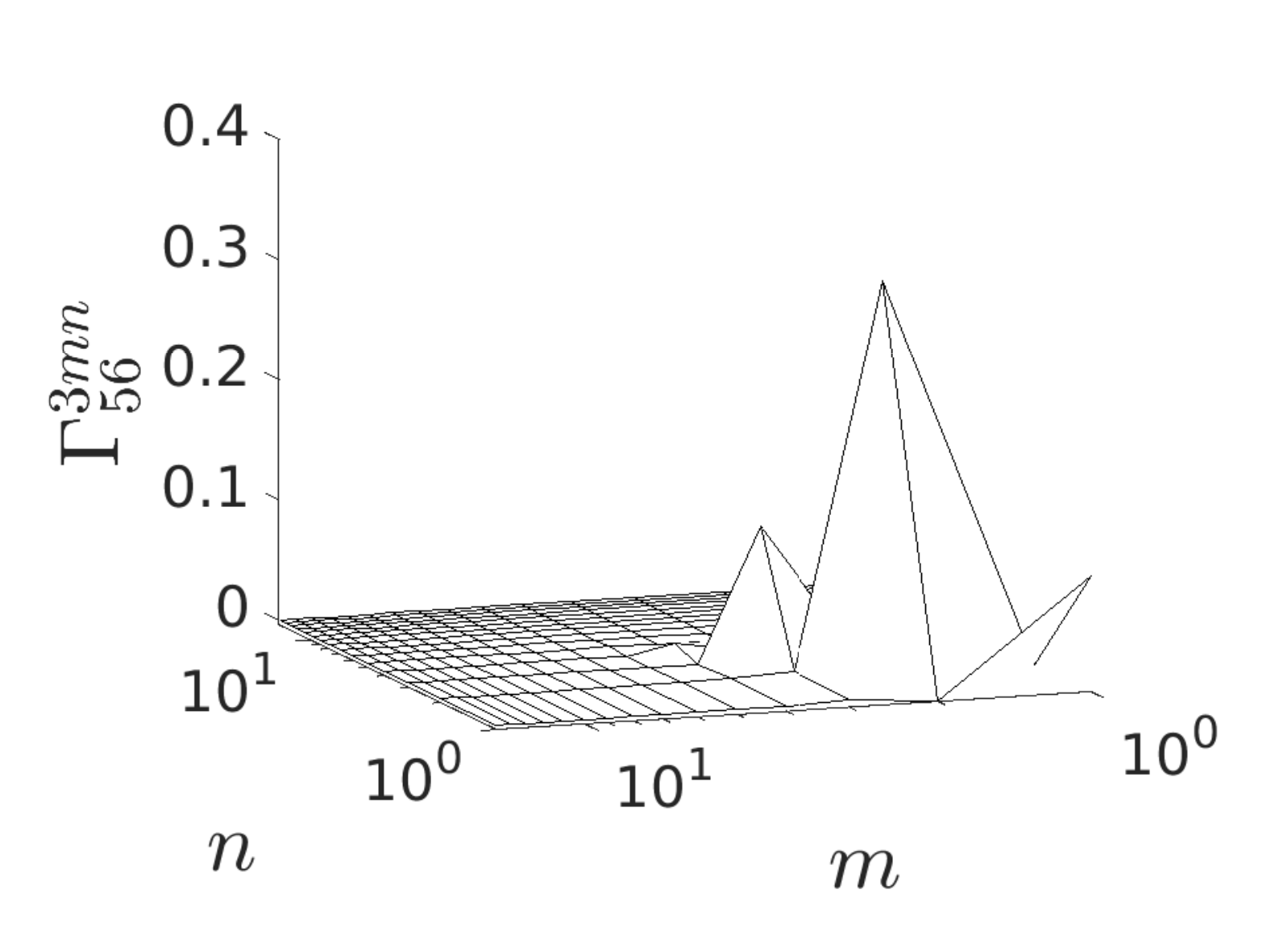}\caption{\label{fig:evalue_reconstruction_domain_6_kernel_5_alpha_3}}
\end{subfigure}
\begin{subfigure}[h]{\plotwidth\textwidth}
\includegraphics[width=\textwidth]{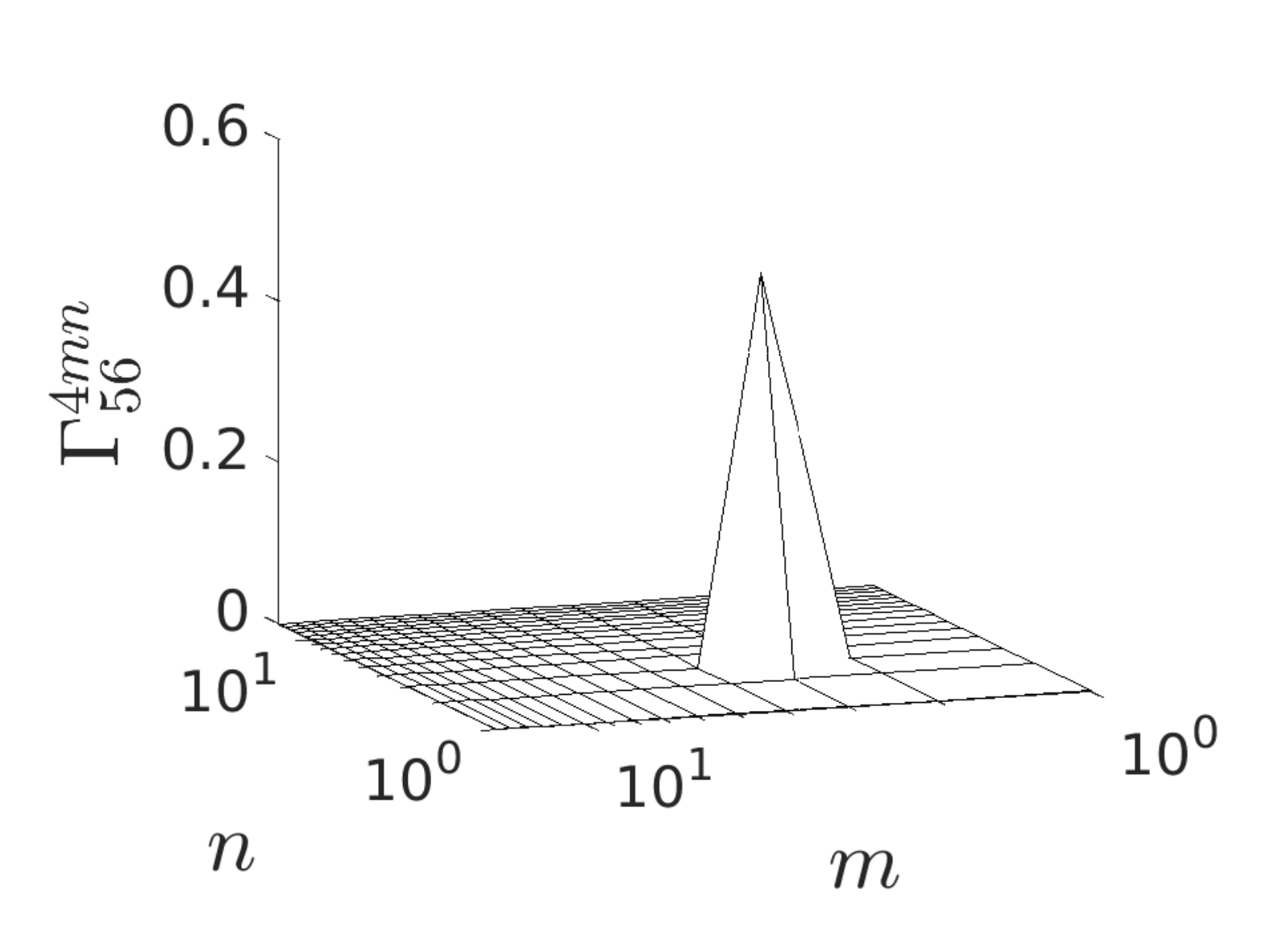}\caption{\label{fig:evalue_reconstruction_domain_6_kernel_5_alpha_4}}
\end{subfigure}
\begin{subfigure}[h]{\plotwidth\textwidth}
\includegraphics[width=\textwidth]{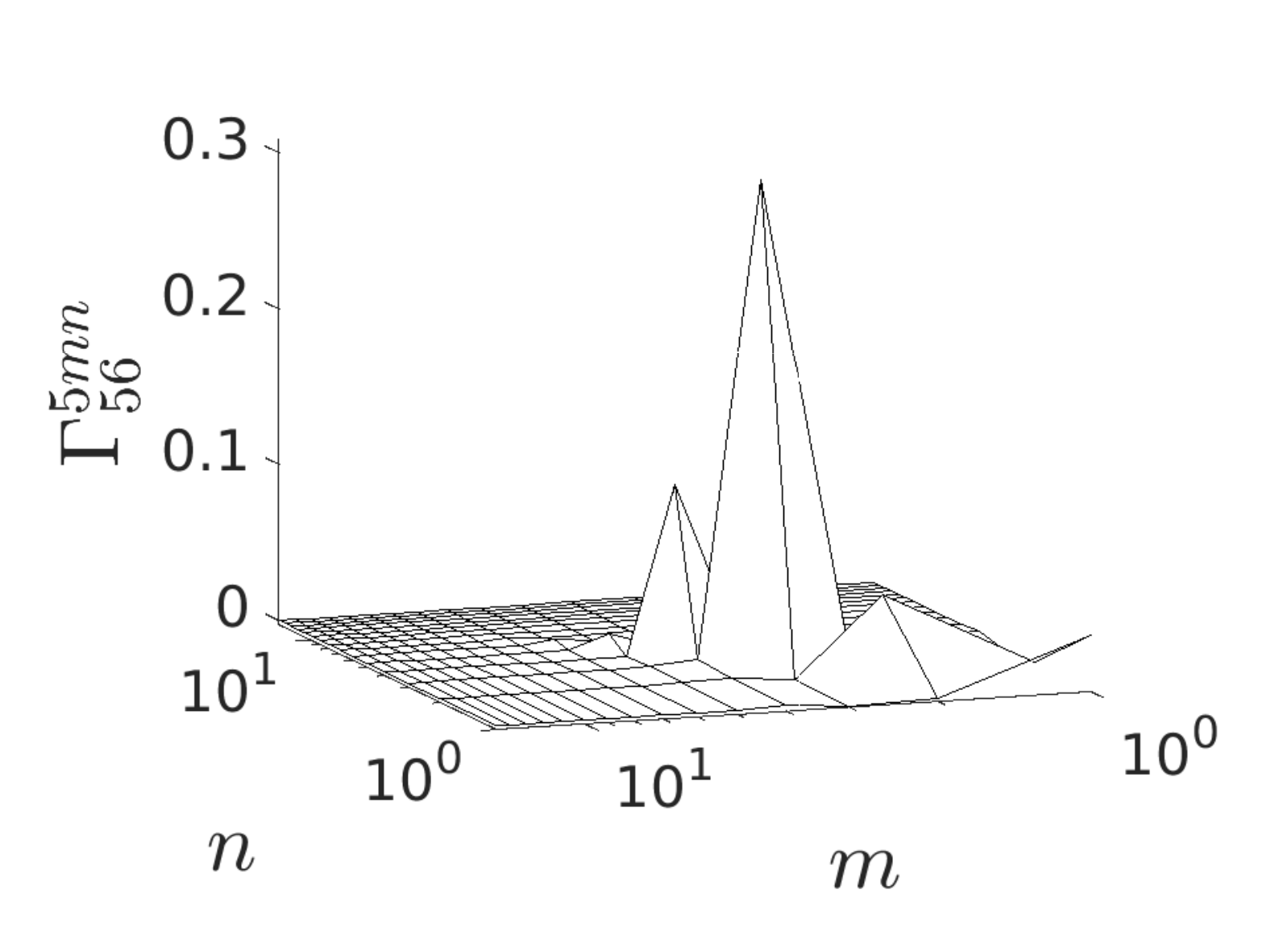}\caption{\label{fig:evalue_reconstruction_domain_6_kernel_5_alpha_5}}
\end{subfigure}
\begin{subfigure}[h]{\plotwidth\textwidth}
\includegraphics[width=\textwidth]{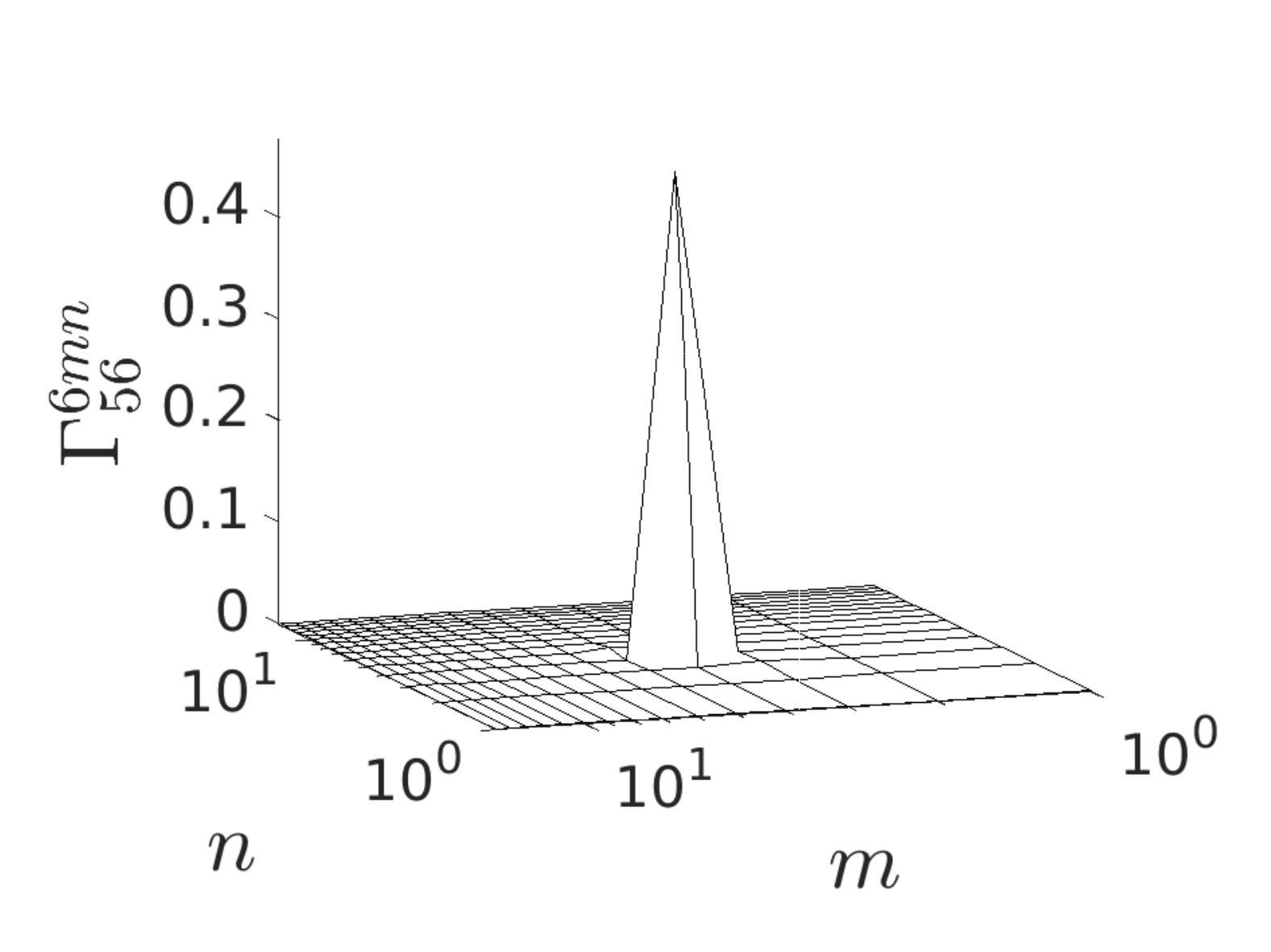}\caption{\label{fig:evalue_reconstruction_domain_6_kernel_5_alpha_6}}
\end{subfigure}
\caption{Contributions to the eigenvalue reconstruction of modes $\alpha=1:6$ using Fourier modes. (a)-(f): contributions for $K_{56}$.\label{fig:evalue_reconstruction_kernel_5}}
\end{figure}
\FloatBarrier
\noindent
\section{Periodification of the correlation function\label{app:periodification}}
\begin{figure}[h!!!]
\centering
\begin{subfigure}[h]{0.45\textwidth}
\includegraphics[width=\textwidth]{figs/standardplots/correlation_functions_reconstructed_1}
\caption{\label{fig:correlation_functions_reconstructed_1}}
\end{subfigure}
\begin{subfigure}[h]{0.45\textwidth}
\includegraphics[width=\textwidth]{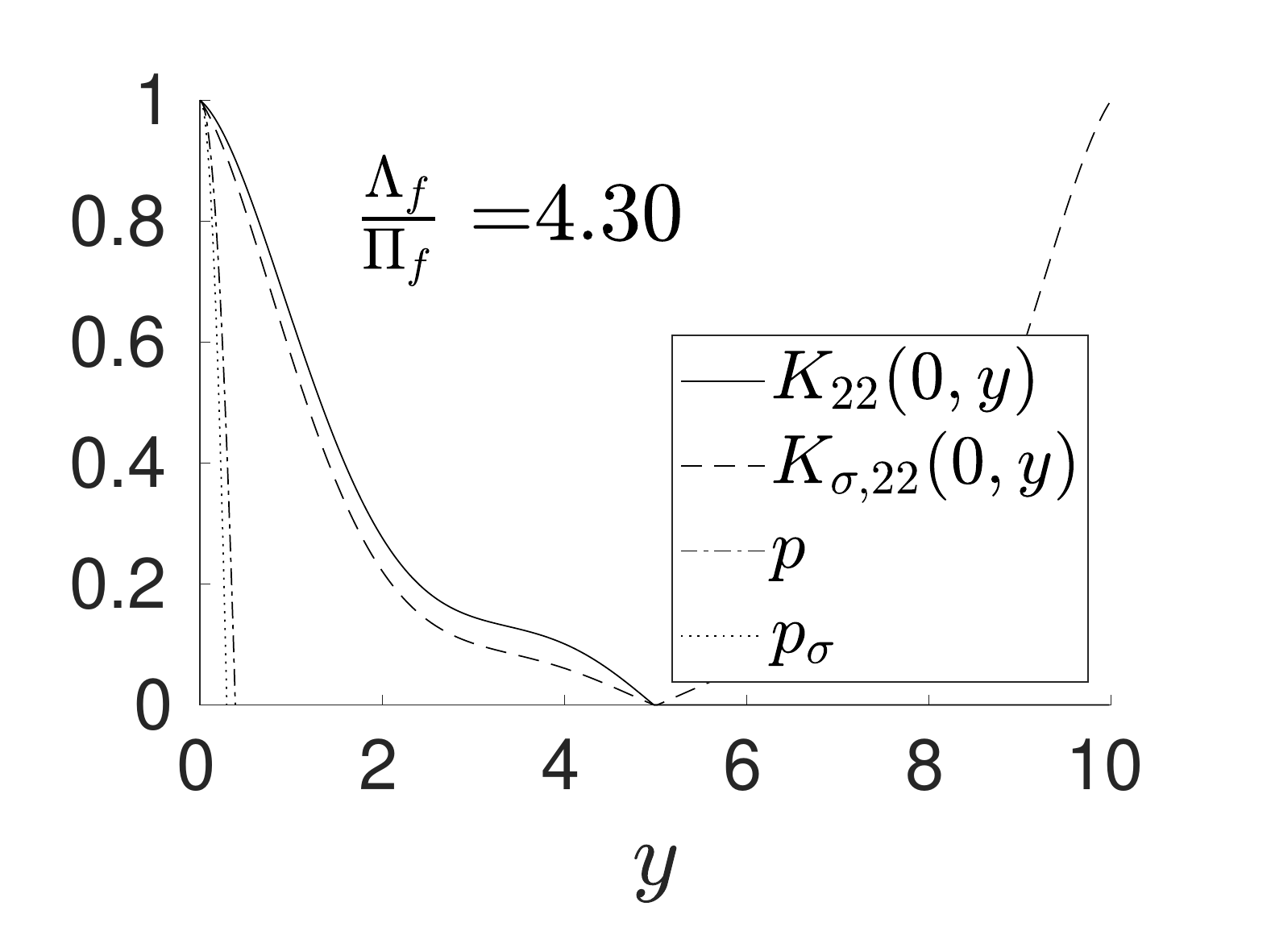}
\caption{\label{fig:correlation_functions_reconstructed_2}}
\end{subfigure}
\begin{subfigure}[h]{0.45\textwidth}
\includegraphics[width=\textwidth]{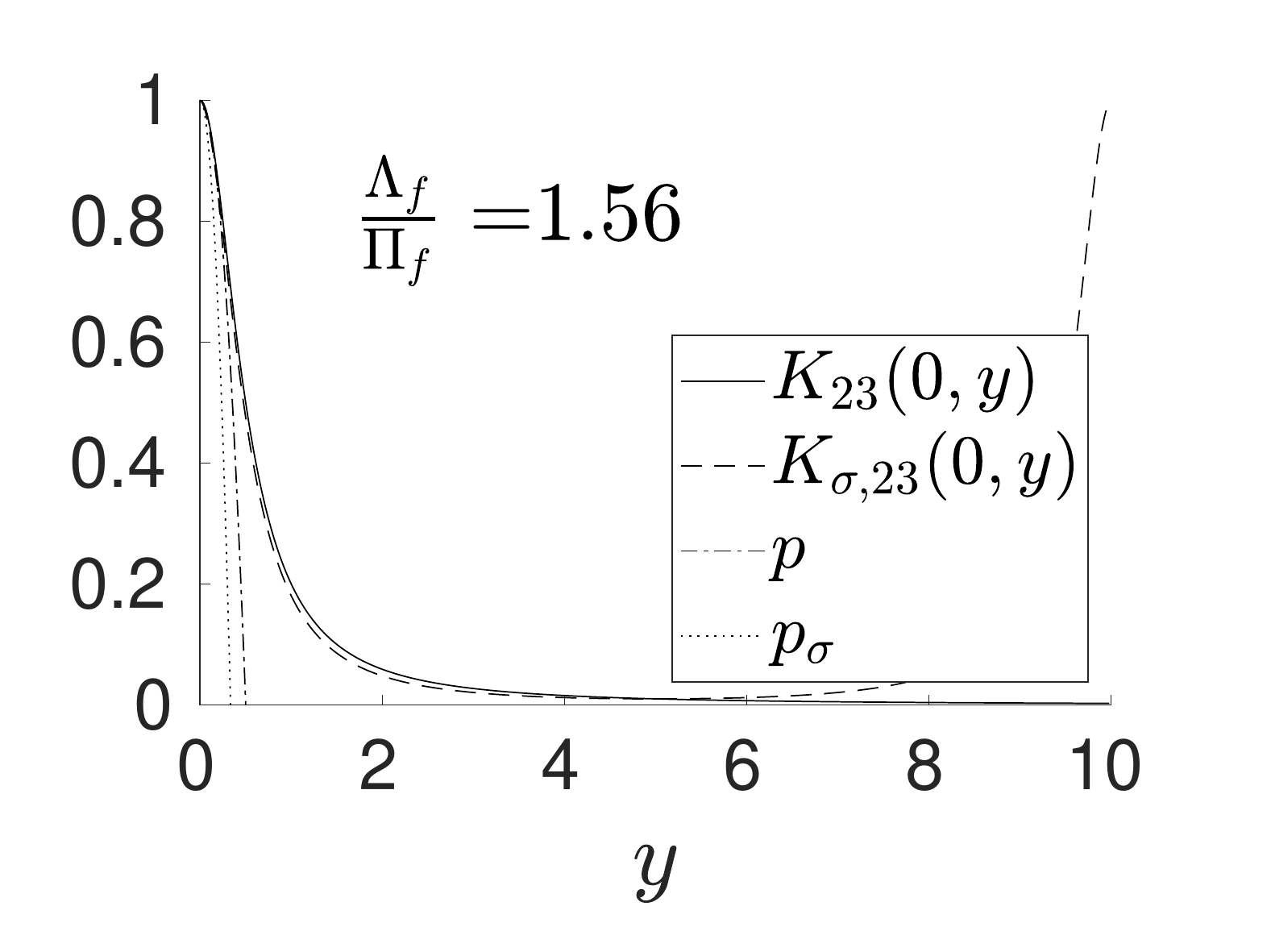}
\caption{\label{fig:correlation_functions_reconstructed_3}}
\end{subfigure}
\begin{subfigure}[h]{0.45\textwidth}
\includegraphics[width=\textwidth]{figs/standardplots/correlation_functions_reconstructed_4}
\caption{\label{fig:correlation_functions_reconstructed_4}}
\end{subfigure}
\begin{subfigure}[h]{0.45\textwidth}
\includegraphics[width=\textwidth]{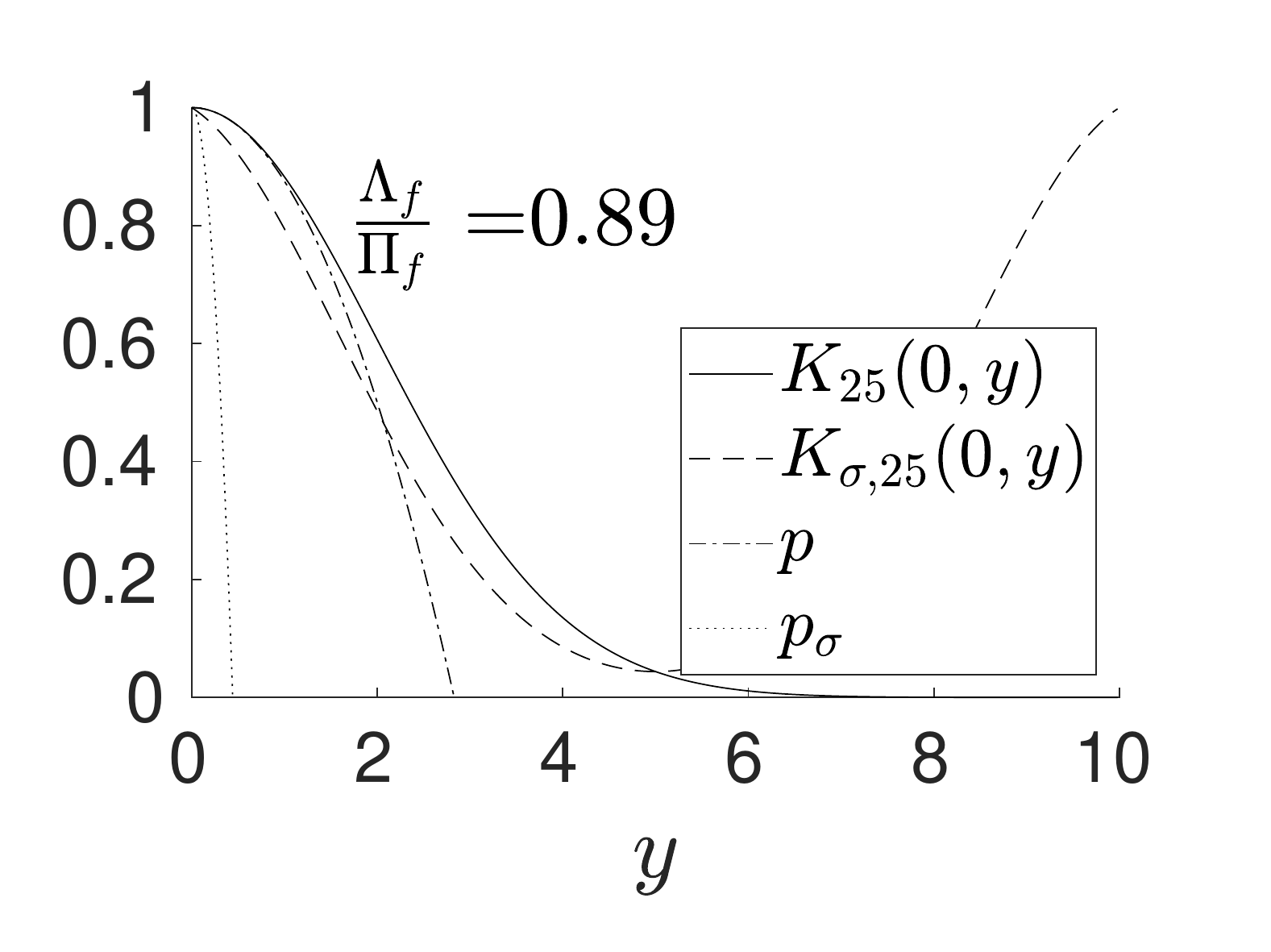}
\caption{\label{fig:correlation_functions_reconstructed_5}}
\end{subfigure}
\caption{(a)-(f): Illustration of the Fourier reconstructed correlation functions as a result of assuming that $\psi^\alpha=\varphi^\alpha$, for all $\alpha$ evaluated for $x=0$ on domains with lengths $L_{\Omega_2}=10\Lambda_f^{ref}$, respectively. $p$ and $p_\sigma$ denote the parabolic fits of $K_{2j}$, $j\in[1:5]$ and $K_{\sigma,2j}$, $j\in[1:5]$, respectively. \label{fig:correlation_functions_app}}
\end{figure}

Figure \ref{fig:correlation_functions_app} demonstrates the periodification of of the correlation functions $K_{\sigma,2j}$, $j\in[1:5]$ as a result of of assuming POD modes to be Fourier modes.
%
\noindent
%



\end{appendices}


\end{document}